\documentclass[a4paper,11pt]{article}

\usepackage{jcappub} 
\usepackage{epstopdf}

\title{\boldmath Spherical cosmological models: an alternative cosmology}

\author[a]{H. Dejonghe}

\affiliation[a]{Sterrenkundig Observatorium, Ghent University,\\Krijgslaan 281, S9, B9000 Gent, Belgium}

\emailAdd{Herwig.Dejonghe@UGent.be}

\abstract{The properties of universes are explored that are entirely in the interior of black holes in another universe, a `mother universe'. It is argued that these models offer a paradigm that may shed a new light on old cosmological problems. The geometry of such a universe is discussed including how it would appear to the observer. The Hubble parameter is direction dependent, but it is argued that the interpretation of any such dependence will be hard to separate from local inhomogeneities. The models do not originate from a big bang, but rather from an initial collapse and subsequent infall, that started probably a very long time ago, presumably much earlier than the accepted age of the universe. The relation to the concordance model is discussed and it is shown that a lot of the existing theory can be taken over into the proposed models. The universe has an edge, which is an ordinary spherical surface in 3 dimensions. That sphere acts as a gravitational mirror as seen from inside the universe, but it does not mirror redshift. The same object can thus be seen in direct sight and in reflection, although with different redshifts, different ages and different aspect angles. The models do not need dark energy, but they need dark matter, of course. Since the models are closed and neutrino's are nowadays believed to have mass, neutrino's can be reconsidered as candidates for the dark matter. As a bonus result from this paradigm, mass ejection from black holes is shown to be possible, which links that process to the controversial anomalous galaxy redshifts. Finally, we show that gravitational mass and inertial mass are proportional, and that the inertial acceleration scales as $c^2/M$, with $M$ a characteristic length scale of the universe.
}
\dedicated{I dedicate this paper to all non-expert people in the world who try to make sense of cosmology, and, with their own means and visions, long to probe its mysteries.}

\begin{document}

\def\cyc{{\rm cyc}}
\def\dacyc{{\rm dacyc}}

\maketitle
\flushbottom

\section{Introduction}

It is well-known that a closed universe with a Robertson-Walker metric oscillates from the big bang to a maximum expansion state and back to the big crunch. At the maximum expansion state the radius of the universe equals its Schwarzschild radius $2GM/c^2$, with $M$ the mass of the universe, $G$ the gravitational constant and $c$ the velocity of light.\footnote{See also subsection~\ref{sect_defgen} for the definition of $M$, and subsection~\ref{sect_synchronous}.} This suggests that one may explore universes that are the interiors of black holes. To this idea can be added that tidal forces generated by a spherical mass distribution scale as $M/r^3$, with $r$ the distance to the center of the distribution. This result holds for the Schwarzschild metric as well (\cite{MTW} 1973) in a local geodetic frame (freely infalling observer). Since the Schwarzschild radius scales with $M$, it follows that at that radius the tidal forces for an infalling observer scale as $M^{-2}$. Hence these tidal forces are negligible for a very massive black hole.\footnote{This fact is of no particular importance for this paper, but is mentioned here because that idea germinated the paper. One is used to think of black holes as fierce things that destroy everything that pass the event horizon. Very massive black holes, on the contrary, offer a gentle, but irreversible, welcome.} Of course, at the center of such a thing there waits disaster, but what if, as suggested above, inside the black hole there would be a distributed mass distribution and not a point mass? The density of such a distributed mass distribution with a constant mass density also scales as $M^{-2}$, hence could also be very small. The infalling observer would then simply enter a new part of spacetime, with the only consequence that he/she would never be able to leave it again. Such a thing could be called a universe, and it is the purpose of this paper to show that it actually is a universe much in the same sense as classical cosmology defines it.

In this paradigm, it comes natural to consider a 3D universe as embedded in another 3D universe. Since we will show that it can be the interior of a (spherical) black hole, we will adopt spherical symmetry. It is useful to make clear, from the outset, what the geometrical difference is between the universes in this paper and the standard Robertson-Walker (RW) universes. To that end, it helps to reduce the dimension by one, and consider 2D universes. A 2D RW universe is confined to the surface of a sphere, which is embedded in 3-space. In this paper, a 2D universe with positive curvature everywhere is the interior of a circle, though the surface inside the circle need not be flat. These universes therefore have a center (a point) and a boundary (a circle). Likewise, the 3D the universes we consider here have a center (a point) and a boundary (the surface of a sphere). 

As for the center, we can characterize it in an ideal, perfectly ordered universe, which we will call  a synchronous universe\footnote{According to this paper's definition of synchronous, the RW models are synchronous. The characterizations of the center such as given here, are generally valid though and do not depend on synchronicity.}.  It is the only place were a material particle can stay put and not partake in the expansion or contraction. In reality, collisions during the contraction phase will give some momentum to any structure formed, and hence all matter will have velocities with effects that add to the effects of the expansion or contraction.\footnote{The wording here is unusually convoluted, since we will argue that the term 'comoving velocity' is somewhat of a misnomer.} In the same ideal world we will show that the center is the only place where the Hubble parameter is the same in all directions, but we will argue that current observations are not good enough to disentangle local anisotropies from the effects of sphericity, let alone that we would, at present,  have a clue where we would be located in such a universe.  

We will devote considerable attention to the boundary. Suffice here to state that also at the boundary space-time is locally 4D Lorentzian in all directions, as it is everywhere, since we will show that at the boundary the singularity in the radial coefficient of the metric (this is one of the  characterizations of the boundary) can be transformed away.
  
In this paper we cannot exclude the now obsolete assumption that the universe could have gone through (many) cyclic phases of expansion and contraction, though we will try to define these contractions carefully and we will not need to assume any particularly neat periodic, nor global, behaviour. During contractions, the space time forced (or facilitated) structure formation, up to the merging of black holes and accretion of matter on them. The resulting dramatic increase in gravitational binding energy, together with increasingly intense background radiation, caused an immense radiation field, which therefore is a necessary companion of the next big bang, or rather, the next---relatively quiet---expansion phase. However, we will not, and need not, assume that a big crunch goes all the way to "the point"; a condensed state, possibly with multiple centres in non-synchronous evolution, suffices.

The spherical symmetry (or possible more complex geometries for which spherical symmetry is but the simplest of all models) needs to be reconciled with the isotropy of the Cosmic Microwave Background (CMB). While in standard cosmology inflation is one of the mechanisms to obtain isotropy, acting from the inside out, we will argue that isotropy can also come about by the mixing that the phases of contraction imply, more in particular during the dense phases. This mixing acts, so to speak, from the outside inwards. Hence the universes in this paper are, by design, homogeneously filled with photons (which happen to be microwave photons in the current state of the universe). Of course, some riddles can be present, but unlike in standard cosmology these riddles are probably not particularly informative on the history of the universe. 

Observations don't seem to indicate that pressure is an important player, and therefore we can suffice with considering pressureless (dusty) universes.

This paper is best read by first jumping to the summary.

\section{The model}
\label{chapt_LT}
\subsection{Definitions and general principles}
\label{sect_defgen}
Spherically symmetrical and pressureless solutions of the Einstein equations have been studied extensively in the past. For a comprehensive overview of the available material in the context of all kinds of cosmological settings, the book by Krasinski (\cite{kras}) is very instructive. According to Krazinsky, these models were (re)discovered at least 20 times. He calls them L-T models, after the first discoverers Lema\^itre (\cite{lemaitre}) and Tolman (\cite{tolman}). In this paper, we start from the paper by Bondi (\cite{bondi}), and adopt his notations for the metric:
\begin{equation}
\label{4metric}
ds^2 = dt^2 - X^2(r,t)\,dr^2 - R^2(r,t)\,d\Omega^2,
\end{equation}
with
\begin{equation}
\label{def_Omega}
d\Omega^2=d\vartheta^2+\sin^2\vartheta\,d\varphi^2
\end{equation}
the metric on the 2-sphere and
\begin{equation}
\label{defX}
X(r,t) = {\partial_r R(r,t)\over\sqrt{1-2e(r)}}.
\end{equation}
We use $\partial_r$ for the partial derivative with respect to $r$. We choose $R(r,t)\ge0$ and $\partial_rR(r,t)\ge0$. 

The 4-metric (\ref{4metric}) is the Gaussian extension (see, e.g. \cite{ABS}) to a fourth (time) dimension of the (spatial) 3-metric
\begin{equation}
\label{3metric}
ds^2 = X^2(r,t)\,dr^2 + R^2(r,t)\,d\Omega^2,
\end{equation}
with parameter $t$. This extension is constructed along geodesics in the fourth dimension for which $t$ is the arc length and that pass through $(r,\vartheta,\varphi)$.

The time coordinate
\begin{equation}
\label{deft}
t\equiv c\,{\rm t}
\end{equation}
has the dimension of a length and is the proper time for all particles with (geodetic) world lines $(t,r_0,\vartheta_0,\varphi_0)$, with $r_0$, $\vartheta_0$ and $\varphi_0$ constants. It is therefore a cosmic time that all such observers can agree on. These world lines also define the term 'comoving observers'. 

In order to make the distinction clear between a time (with the dimension of time) and the time coordinate with the dimension of a length, we write the former in plain text font, and italicize the latter, as already indicated in (\ref{deft}). We will express time t in units of 10 Ga. If we adopt as the unit of length 3.066 Gpc, the velocity of light equals unity, and therefore, $t$ and t have the same numerical value.

We will sometimes refer to the Cartesian $({\cal X}, {\cal Y}, {\cal Z})$ associated to the polar $(r,\vartheta,\varphi)$, and this we will do in the usual way: $\vartheta=0$ corresponds to the $\cal Z$-axis, and the positive $\cal X$-axis is given by $(\vartheta,\varphi)=(\frac\pi2,0)$. Contrary to the usual convention however, we will construct space by rotating meridional planes (i.e. planes through the $\cal Z$-axis with constant longitude $\varphi$) over the angle $\varphi$, with $\varphi$ in the interval $[0,\pi[$. In every meridional plane the polar angle $\vartheta$ is then defined in the interval $[0,2\pi[$. This choice has the advantage that any meridional plane is fine to stage the motion along geodesics (which is plane motion), while in the standard convention this is only practical in the equatorial plane $\vartheta=\pi/2$.\footnote{This choice implies a discontinuity in the orientation in the meridional plane $\varphi=0$ and $\varphi\to\pi$, but this is no inconvenience for this paper.}

The radial coordinate $r$, the dimension of which is actually undetermined and can (must) therefore be taken as dimensionless, has no direct quantitative relevance for radial distance. All points with the same $r$ we call a comoving shell, and accordingly we call $r$ the "shell label". The comoving shells are the reference spacetime surfaces that define the evolution of the universe. We can thus think of $r$ as the label that every comoving shell carries with it, and that can be read, by some unspecified means, by any traveler passing by. The shells can also be thought of as the rungs of a (growing or shrinking) ladder. The rungs have a unique label $r$. A comoving observer stands on a rung, a traveler ``climbs'' or ``descends'' the ladder. The actual distance between the rungs is determined at all times by the metric (\ref{4metric}), and equals $ds = X(r,t)\,dr$.

The function $R(r,t)$, on the other hand, has a geometrical interpretation, in the sense that (a) it has the dimension of a length and (b) an elementary distance perpendicular to the direction of the origin (which we will call henceforth a tangential distance along a tangential direction) is measured by the familiar $ds=R(r,t)d\vartheta$ (here in a meridional plane). Because of this geometrical property and the fact that the surface of a shell $r$ at time $t$ equals $4\pi R^2(r,t)$, we will call $R$ a "radius". A shell with radius $R(r_1,t)$ is interior to a shell with radius $R(r_2,t)$ if $R(r_1,t)<R(r_2,t)$. In that case shell $R(r_2,t)$ is exterior to shell $R(r_1,t)$ 

The condition $\partial_rR(r,t)\ge0$ means that $r$ does play its role as a qualitative radial marker, in the sense that, at any time, the order relations in both $r$ and $R$ are consistent with the relations "interior to" and "exterior to" for comoving shells, if that was the case at the start (in time) of the validity of the model (i.e. the initial conditions). This we will assume.

The dimensionless function $e(r)$ is arbitrary except for the obvious constraint $e(r)\le\frac12$ and some integrability conditions (see later); we call it the energy function.\footnote{The reader may notice that in this paper sometimes notations are used that do not conform with standard texts or practices. This is in order to avoid notational conflicts with other parts of the paper.} It is significant that it does not depend on $t$. Hence, we restrict our shell labels to any strictly monotonic non-singular time-independent function $r'(r)$ of a particular chosen radial coordinate. Using $d_r$ for the ordinary derivative with respect to $r$, this means $d_rr'(r)>0$. As a consequence, the class of valid shell labels leaves $e(r)$ independent of time. Note in particular that if $e(r)$ belongs to this class, we could adopt $e$ as the radial coordinate.

The local Lorentzian dust density is determined by (Bondi \cite{bondi}, Krasinski \cite{kras})
\begin{equation}
\label{defrho}
\kappa\rho(r,t)\equiv{4\pi G\over c^2}\rho(r,t)={d_rm(r)\over R^2(r,t)\displaystyle\partial_rR(r,t)}\ge0,
\end{equation}
with $m(r)$ another arbitrary function with the dimension of a length, whose physical meaning will be clarified later. The function $m(r)$ would be equal to $G/c^2$ times the total mass inside shell $r$ if space were Euclidean. Expressed in our units of length and time, we find
\begin{equation}
\label{rhounit}
\rho = 1.20{d_rm(r)\over R^2\displaystyle\partial_rR}\times10^{-26}\,{\rm kg}\,{\rm m}^{-3},
\end{equation}
and we adopt $10^{-26}\,{{\rm kg}\,{\rm m}^{-3}}$ as our unit of mass density, which is basically equal to the standard critical density of the universe 
\begin{equation}\label{def_rhoc}
\rho_{\rm crit} = {3{\rm H}_{\rm o}^2\over8\pi G}\sim 10^{-26}\,{{\rm kg}\,{\rm m}^{-3}}, 
\end{equation}
assuming for the current Hubble parameter
\begin{equation}
\label{def_Ho}
{\rm H}_{\rm o}=72\,\rm km/s/Mpc.
\end{equation}
Note that $\rho(r,t)$ will always appear in combination with $\kappa$ as the product $\kappa\rho(r,t)$ with the dimension of the inverse square of a length, since $\kappa$ has the dimension of a length divided by a mass. It follows thus from our choice of the units (time, length and mass density) and (\ref{rhounit}) that in numerical value
\begin{equation}\label{kapanum}
\kappa^{-1}=\left(4\pi G\over c^2\right)^{-1}=1.20\,\,.
\end{equation}

The volume element of the 3D metric equals
\begin{equation}
\label{defvolelem}
dV=XR^2|\sin\vartheta|\,dr\,d\varphi\,d\vartheta={R^2\partial_rR\,dr\over\sqrt{1-2e}}
|\sin\vartheta|\,d\varphi\,d\vartheta.
\end{equation}
Note that the notation $|\sin\vartheta|$ is actually shorthand for $\sqrt{\sin^2\vartheta}$, since $\vartheta$ is defined in the interval $[0,2\pi[$.

The cumulative mass function, which is the total mass interior to a comoving shell with label $r$ thus reads
\begin{equation}
\label{defcalM}
{\cal M}(r) = 4\pi\int_0^{r}{\rho\,R^2\partial_{r'}R\over\sqrt{1-2e(r')}}\,dr'={4\pi\over\kappa}\int_0^{r}{d_{r'}m(r')\over\sqrt{1-2e(r')}}\,dr',
\end{equation}
and it is, according to the second equation in (\ref{defcalM}), obtained via (\ref{defrho}), also independent of the time. That second equation can be readily inverted:
\begin{equation}
\label{relmM}
m(r) = {\kappa\over4\pi}\int_0^rd_{r'}{\cal M}(r')\sqrt{1-2e(r')}\,dr'.
\end{equation}
In our units, we find for the first equation of (\ref{defcalM})
\begin{equation}
\label{calMunit}
{\cal M}(r) = 5.35\times10^{22}M_\odot\int_0^{r}{\rho\,R^2\partial_{r'}R\over\sqrt{1-2e(r')}}\,dr'.
\end{equation}
The constancy in time of $m(r)$ and ${\cal M}(r)$ means that the constituent matter of the universe stays put on their shells, and hence, we can refer to that matter as comoving matter. 

It follows from (\ref{defrho}) that
\begin{equation}
\label{condmr}
d_rm(r)\ge0.
\end{equation}
We are now in the position to identify the shell exterior to which there is no comoving matter. That shell we call the outer boundary and its shell label we denote by $r_b$. Hence
\begin{equation}
\label{defrb}
d_rm(r)=d_r{\cal M}(r)=0 \qquad {\rm for} \qquad {r>r_b}.
\end{equation}

As is obvious from (\ref{defrho}) and (\ref{defvolelem}),  the occurrence of $\partial_rR(r,t)=0$ is special. We will assume it occurs at $R=0$ at all times, which therefore defines the inner boundary of the model, which we call the center, and we set $r=0$ there.\footnote{This is of course what is common practice in polar coordinates. In some strict mathematical sense, 'the origin', defined as the locus where $R=0$, is not part of the space that polar coordinates cover. The origin is part of the manifold, though, since an appropriate transformation (Cartesian coordinates) can eliminate the singularity.} We will see in section~\ref{sect_alternuniv} that the only shell with $r>0$ where $\partial_rR(r,t)$ can be $0$ is the boundary $r_b$.

If $\partial_rR(r,t)=0$ at some shell with label $r$ and at some time $t$ while $d_rm(r)>0$, the mass density reaches a singularity according to (\ref{defrho}).\footnote{Note that this condition is independent of the choice of the shell label.} This means that shells are colliding there, and a collision therefore marks the end of the validity of the metric. The geometrical model (i.e. the metric) is therefore only valid for these parts of spacetime that are either fully expanding or collapsing without any shell crossings.\footnote{Usually this is interpreted as consistent with the fact that the model is pressureless, since shells bumping into each other would create pressures. In a cosmological context this is less so the case, since we observe that colliding matter forms gravitationally bound structures such as (clusters of) galaxies and black holes, which do not necessarily increase the cosmic pressure significantly.} The term ``shell crossing'' was introduced by Hellaby \& Lake in \cite{hellaby}. Our use of the term ``shell collision'' refers to the point expressed in this paragraph: there is no geometrical model anymore at a shell collision, lest one introduces new physics and resolves the singularity.

It is also possible that for some shell $X(r,t)\to+\infty$. We will show in section~\ref{sect_alternuniv} that for the models we consider an appropriate transformation can eliminate this singularity, which must if the spacetime is to be Lorentzian. Hence that shell is also part of the manifold.
 
We will assume that, at all times for which the model is valid
\begin{equation}
\label{nocollision}
0<\partial_rR(r,t)(<+\infty) \quad {\rm for} \quad 0<r<r_b.
\end{equation}
Hence, we call the first inequality of (\ref{nocollision}) the no-collision condition, since it is extremely unlikely that, in view of the presence of collapse phases, $d_rm(r)=0$ over a range in $r$, except possibly around the center. The second inequality between brackets is an assumption. Should $\partial_rR(r,t)\to+\infty$ for some $r$ and $t$, we will assume that the singularity can be transformed away by another choice of the shell label.\footnote{This is only possible in the case that $R(r,t)$ is separable in $r$ and $t$. An example would be a singularity of the type
\[
R(r,t) = R_0(t)+(r-r_0)^{1/a},\qquad a>0.
\]
The singularity can be trivially removed by adopting a new shell label $r(r')= (r')^a+r_0$.}

The spacetime in which the metric (\ref{4metric}) is valid is the direct product of (a) all shells with comoving matter (\mbox{$0\le r\le r_{b}$}) for which the no-collision condition (\ref{nocollision}) is valid and (b) the time interval
\begin{equation}
\label{deftM}
t\le t_M\le+\infty.
\end{equation}
Equation (\ref{deftM}) defines the symbol $t_M$ and means that at $t=t_M$ some shells will be colliding. The lower bound of the interval is as yet unspecified.

In case $r_b<+\infty$ we will have occasion to use the normalized shell label
\begin{equation}
\label{def_rnorm}
0\le\tilde r={r\over r_b}\le 1.
\end{equation}
We denote
\begin{equation}
\label{defM}
M = m(r_{b})
\end{equation}
and assume that $M$ is finite for finite $r_b$.

If we integrate (\ref{defrho}) with an Euclidean volume element, we obtain
\begin{equation}
\label{relmMsimp}
4\pi\int_0^r\rho(r'\!,t) R^2(r'\!,t)\partial_{r'}R(r'\!,t)\,dr' = {c^2\over G} m(r),
\end{equation}
yielding an 'Euclidean total mass' for $r=r_b$ that is proportional to $M$ rather than to the total mass of the universe 
\begin{equation}
\label{def_calMtot}
{\cal M}_{\rm tot} = {\cal M}(r_{b}).
\end{equation}
The right hand side  equals $c^2M/G$ and is very reminiscent of the expression for the Schwarz\-schild radius. On these grounds $c^2M/G$ could actually be interpreted as the mass of the black hole as experienced by an observer 'outside' the universe (but see also section~\ref{sect_beyond} for a proof of this assertion). On the other hand, nothing more specific can be stated between $M$ and  ${\cal M}_{\rm tot}$ than what would follow from relations (\ref{defcalM}) or (\ref{relmM}), since $e(r)$ is (still) an arbitrary function.

\subsection{The evolution equation}
\label{sect_time_evolution}

The time evolution of $R(r,t)$ and $X(r,t)$ is governed by (Bondi \cite{bondi}, Krasinski \cite{kras})
\begin{equation}
\label{eqRt}
\left(\partial_tR\right)^2=-2e(r)+{2m(r)\over R} + {1\over3}\Lambda R^2,
\end{equation}
with $\Lambda$ the cosmological constant. It is actually already an integration of the equation
\begin{equation}
\label{eqRtdif}
\partial_t^2R=-{m(r)\over R^2} + {1\over3}\Lambda R.
\end{equation}
As is well known, the structure of these equations is essentially the same as the Newtonian analogues, apart from the presence of the cosmological constant. More particular differences include an overall factor of $c^{-2}$ and the quantity $m(r)$, since in the Newtonian case it would be $\kappa/(4\pi){\cal M}(r)=(G/c^2){\cal M}(r)$, according to (\ref{relmM}) and (\ref{defrho}). Following Bondi (\cite{bondi}) we call $m(r)$ the effective gravitating mass function and hence $M$ the total effective gravitating mass. 

We note in particular that the effective gravitating mass function does not explicitly enter the metric (\ref{4metric}), but has only impact on the time evolution through equation (\ref{eqRt}).
From the sign in front of $e(r)$ in (\ref{eqRt}) we conclude that the Newtonian analogue of $e(r)$ would be proportional to the negative of the total energy, and we will call $e(r)$ the energy function. The non-Euclidean character of the model resides solely in the time dependence of  $\partial_tR(r,t)$, since if $\partial_tR(r,t)=0$, a suitable radial shell marker transformation makes the metric (\ref{4metric}) Lorentzian. 

It is instructive to ponder the similarities and differences of the motion of the shells we consider here and the classical 2-body motion of celestial mechanics (barring the cosmological constant). In 2-body motion, the equations (\ref{eqRt}) or (\ref{eqRtdif}) describe the motion of 2 point masses on degenerate ellipses (i.e. straight lines). If we put the center of mass of the point masses in the origin and their linear orbits on the $\cal X$-axis, one point is moving on the positive $\cal X$-axis and the other on the negative $\cal X$-axis. They both reach their largest distance from the origin at the same time, and they collide (elastically) at the origin. The more common view is to place one of the point masses in the origin, and the other then oscillates on one side of the $\cal X$-axis, with periodic collisions. In this paper, the shells behave similarly in the sense that they oscillate from largest surface area to zero surface area, ideally back and forth. However, comoving observers will 'fly through' the center when reaching it, and if they were infinitesimally small, they wouldn't collide with anything. 

The velocity in the 2-body motion is entirely determined by the magnitude of the participating masses and, say, the distance of greatest elongation (or velocity at infinity). This is no different from what we have here: all motion is 'simply' due to gravitation. For shells with the same surface area, (a measure for) the expansion or contraction rate will be larger if the mass of the universe inside them is heavier. 
 
This is the place to point out that the dimensionless $\partial_tR$ is not (proportional to) the radial velocity of anything. Actually also here, as in the 2-body case, $|\partial_tR|\to+\infty$ for $R\to0$ if $r>0$, and thus, if $R$ is sufficiently small, $c\,\partial_tR$ can be larger dan $c$. While an infinite velocity is no problem in Newtonian dynamics, it clearly shows that here $\partial_tR$ is not a material velocity: it is a measure for the change in tangential distance of 2 comoving points, or, technically more precise, it is proportional to the geodetic deviation of 2 neighbouring comoving points at the same $R$.\footnote{The ratio $\partial_tR/R$ is the tangential Hubble parameter (see section~\ref{sect_H0}). In this context this ratio can also be characterised as follows. Be $\delta\vartheta_0=\upsilon\,\delta t/R_0$ the infinitesimal meridional angle subtended by the meridional tangential distance covered by a particle with meridional tangential velocity $\upsilon$ on shell $r_0$ at $t_0$, with $R_0=R(r_0,t_0)$, during an infinitesimal time interval $\delta t$. At time $t_1=t_0+\Delta t$ this angle would be $\delta\vartheta_1=\upsilon\,\delta t/R(r_0,t_1)=\upsilon\,\delta t/(R_0+\partial_tR_0\Delta t)$. The ratio of both angles equals 
\[
{\delta\vartheta_1\over\delta\vartheta_0}=1-{\partial_tR_0\over R_0}\Delta t
\]
for  $\Delta t\,(\partial_tR_0/R_0)\ll1$.  Hence in an expanding universe a particle or photon progresses less far from an angular perspective at the origin than it would in an Euclidean universe. We note in particular that there is no limit on the magnitude of $\partial_tR/R$.} We call $\partial_tR$ the tangential stretch (or shrinkage) rate, and denote it by the dimensionless
\begin{equation}
\label{def_ST}
S^T\!(r,t)=\partial_tR(r,t).
\end{equation}
We introduce this, at first sight somewhat pedantic, additional symbol for $\partial_tR$ in order to remind us that $\partial_tR$ is not a velocity. Hence the term 'comoving observer' is in a strict sense a misnomer, and the term co-stretching or co-shrinking would be more appropriate. This paper adheres to the common terminology, though. 

Clearly, we could also define a radial stretch rate $S^r=\partial_tX$, but this definition does not carry much added value for this paper.

As we will see in section~\ref{sect_beyond}, we will not adhere to the view that it is space that is expanding (created) or contracting (annihilated), but rather shells that are expanding or contracting in spacetime.

Returning to our ladder analogy, the distance between rungs is given by the Lorentzian
\begin{equation}
\label{Lorentz_dx}
d{\cal X} = {dR\over\displaystyle\left[1-{2m(R)\over R}+\bigl[S^T\!(R)\bigr]^2-\frac13\Lambda R^2\right]^{1/2}}
\end{equation}
at constant $t$ (hence $\partial_rR\,dr=dR$ and we can label the rungs with $R$ instead of $r$). We clearly see the Schwarzschild term \hbox{$1-2m(R)/R$} which makes $d{\cal X}$ larger with increasing $m(R)$, and therefore makes space more elliptic. The (longitudinal) stretching or shrinking rate of the rungs is proportional to $\bigl[S^T\!(R)\bigr]^2$.  The larger that rate (in absolute value), the closer the rungs are to one another, which is thus the opposite effect as $m(R)$, making space more hyperbolic. This is somewhat reminiscent of a Lorentz contraction. Finally there is the role of the cosmological constant which is obvious from (\ref{Lorentz_dx}) but not very intuitive. All 3 effects combine in such a way as to make the radicand in (\ref{Lorentz_dx}) independent of time. The radial Lorentz distance $d\cal X$ of course isn't, since it features the factor $\partial_rR(r,t)$.

We conclude that the non-Euclidean character is very closely connected with the mass content and the kinematics.
The only static model that is stable is the empty Euclidean space, with $e(r)=m(r)=\Lambda=S^T\!(r,t)=0$.\footnote{The only other static model has $R=r$, $m(r)=\Lambda r^3/3>0$ and $e(r)=\Lambda r^2/2$ as can be seen from  (\ref{eqRt}) and (\ref{eqRtdif}). It is the one for which Einstein introduced the cosmological constant. The repulsive force of a positive $\Lambda$ exactly balances the attraction caused by the effective gravitating mass, but this delicate balance is, of course, unstable.}

Finally, the no-collision condition (\ref{nocollision}) can always be realised for the initial conditions (at some cosmic time, to be discussed later), since shells can be ordered with respect to surface, and each shell can be given a label $r$ that preserves this order. An obvious choice would be 
\begin{equation}
\label{r_initial}
r=R,
\end{equation}
with $r$ only taking the numerical value of $R$ (and not its dimension).

\section{The current expansion phase}

\subsection{Definitions}
\label{sect_definitions}

Since we will be concerned in this paper only with the fully expanding universe without collisions, we will suffice with "simply" specifying $m(r)$ and $e(r)$ under certain conditions (see later), and subsequently solve the evolution equation (\ref{eqRt}) for $R(r,t)$.

The solutions of (\ref{eqRt}) are well known. They are found via an implicit form, as a function of $t$ that is, by the quadrature:
\begin{equation}
\label{soleqRtint}
\int_0^{R(r,t)\over p(r)}\!\!\!\!\!{\sqrt{\cyc'}\,\,d\cyc'\over\sqrt{1-\epsilon(r)\,\cyc'+a(r)\,{\cyc'}^3}} = \omega(r)[t+\phi(r)]\equiv\bar\psi(r,t).
\end{equation}
This expression introduces quite some functions. We start with defining
\begin{equation}
\label{def_psi}
\psi(r,t)=\omega(r)[t+\phi(r)]
\end{equation}
valid for all $r$ and $t$ (in a collision-free universe, that is), in contrast with the $\bar\psi(r,t)$ of equation (\ref{soleqRtint}) which is only valid for those $r$ and $t$ that realize the integral on the left hand side. We call the function $\psi(r,t)$ the state of the shell with label $r$ at time $t$. It is dimensionless. The quantity $\phi(r)$ we call the phase of the shell $r$, and is to be seen as an integration constant (as a function of $t$) in the solution of (\ref{eqRt}). The quantity $-\phi(r)$ is also called ``bang time'' for obvious reasons, but since we will assume that our universe actually does not realize the limit $\bar\psi(r,t)\to0$, this term is not very appropriate.

The radius is given by
\begin{equation}
\label{Rexplgen}
R(r,t) = p(r) \,\cyc[a(r),\epsilon(r),\psi(r,t)]
\end{equation}
with
\begin{equation}
\label{defp}
p(r) = {2m(r)\over2e(r)}\epsilon(r).
\end{equation}
We call $p(r)$ the shell parameter of the shell with label $r$. It has the dimension of a length. Next is
\begin{equation}
\label{defomega}
\omega(r) = \sqrt{2m(r)\over p^3(r)}={1\over 2m(r)}\left(2e(r)\over\epsilon(r)\right)^{3/2}
\end{equation}
which we call the shell 'frequency', with the dimension of a reciprocal length. The function
\begin{equation}
\label{defa}
a(r) = \frac13{\Lambda\over\omega^2(r)}
\end{equation}
incorporates $\Lambda$ into the expression (\ref{soleqRtint}) and is dimensionless.

Still to be defined is $\epsilon(r)$. We distinguish 2 cases, which we call representations.

\noindent\underline{(a) The function $\epsilon(r)$ is defined as a constant:}

\begin{equation}
\label{defepsilon1}
\epsilon = \left\{
\begin{array}{lcl}
1&\quad {\rm if} \quad& e(r)>0 \\
-1&\quad {\rm if} \quad& e(r)<0\\
0&\quad {\rm if} \quad &e(r)=0.
\end{array}
\right.
\end{equation}
The third case $\epsilon=0$ applies when $e(r)=0$ in a non-empty interval of $r$ (see appendix~\ref{sect_e0} for the details).

\noindent\underline{(b) The function $\epsilon(r)$ is the function:}

\begin{equation}
\label{defepsilon2}
\epsilon(r) = 2e(r)
\end{equation}
and hence
\begin{equation}
\label{defomega2}
p(r)=2m(r) \qquad {\rm and} \qquad \omega(r) = {1\over2m(r)}.
\end{equation}

Both representations (a) and (b) have their merits and drawbacks. For example, representation (a) cannot be used in the numerical calculations when $e(r)\to0$ at some $r$, as can be seen from (\ref{defp}). On the other hand, representation (b) introduces additional and strictly spoken unnecessary $r$-dependence in the function cyc. Obviously, in both representations, there remain only 2 independent functions: $m(r)$ and $e(r)$, while the quadrature (\ref{soleqRtint}) introduces the independent phase function $\phi(r)$.

The energy function $e(r)$ appears in the metric, and hence we must assume that it is sufficiently differentiable. Barring Dirac-$\delta$ behaviour in $\rho(r,t_0)$ but allowing discontinuity, it follows from (\ref{defcalM}) that ${\cal M}(r)$ is continuous (not necessarily differentiable) and non strictly monotonously increasing. With (\ref{relmM}), the same is true for the effective mass function $m(r)$. Since both functions $m(r)$ and $e(r)$ will need to be differentiated once in the orbit calculations, continuity for both functions suffices.\footnote{If only this minimum requirement is met at a certain shell $r$, one would need to account for this by stopping the integration at that shell with a certain set of (say, left) derivatives, and restarting it with a different set of (right) derivatives, causing a kind of refractive behaviour at that shell.} For flux calculations we will also need the second derivatives of $m(r)$ and $e(r)$, and thus continuity in their first derivatives. This means, with (\ref{defrho}), that in that case $\rho(r,t_0)$ must be continuous.

\begin{figure}
 \centering
   \includegraphics[width=150mm]{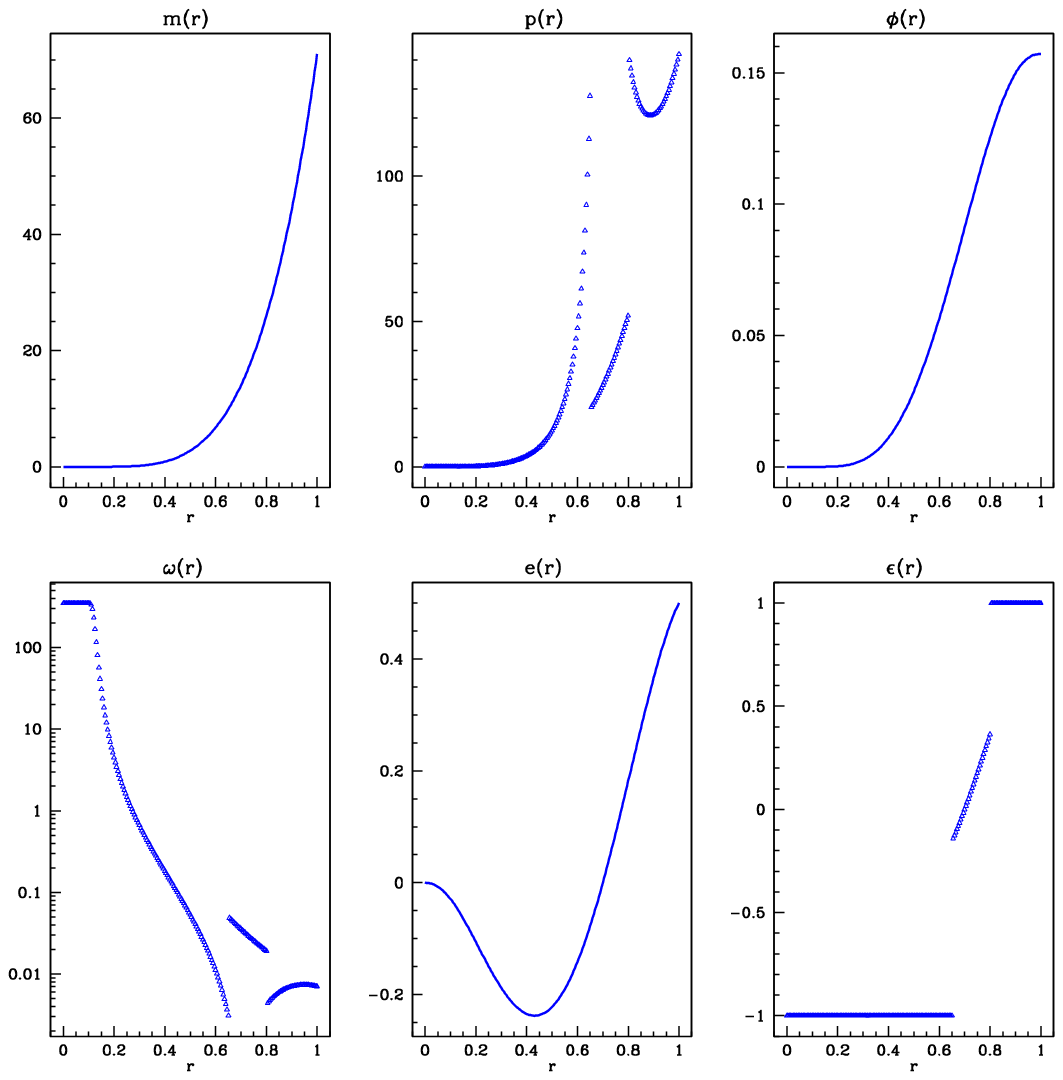}
      \caption{
An illustration of the functions introduced in this section. The functions $e(r)$, $m(r)$ and $\phi(r)$ are drawn in solid lines and must be sufficiently differentiable. In addition to the trivial zero $e(0)=0$ in $e(r)$ (see section~\ref{seccentral}), also $e(.7)=0$. The functions that are drawn with triangular points need not even be continuous. Representation (a) is used for $r<.7$ with $\epsilon=-1$, and for $r>.7$ with $\epsilon=+1$. Around $r=.7$ representation (b) is used. Clearly, if around $r=.7$ representation (a) would have been used, $p(r)$ would have been singular at $r=.7$. From $\omega(r)$ we see that this model happens to be synchronous (see sections~\ref{sect_synchronous} and \ref{sect_asynchronous} for more details) for $r\le.1$}
\label{fig_params}

   \end{figure}

The effective mass function features the parameter $M$, which we called the total effective mass. We can normalize $m(r)$, and define
\begin{equation}
\label{def_tildem}
m(r) = M\tilde m(r), \qquad \tilde m(r_b)=1.
\end{equation}
Likewise, we can explicitly introduce a scaling in the shell parameter
\begin{equation}
\label{def_tildep}
p(r) = 2M\tilde p(r),
\end{equation}
and the shell frequency
\begin{equation}
\label{def_tildeom}
\omega(r) = {1\over 2M}\tilde \omega(r).
\end{equation}

Equation (\ref{soleqRtint}) determines implicitly the function $\cyc$, and it is discussed at length in appendix~\ref{sect_cyc}. For $\Lambda=0$ and $\epsilon=1$ it is the familiar cycloidal function. The name "$\rm cyc$" makes explicit reference to the family of cyclic solutions of (\ref{eqRt}), but is here taken to be more general, also allowing for the monotonously expanding solutions.

When $\Lambda=0$, the total effective mass is a scaling parameter, both in space and $t+\phi(r)$. The scaling in space follows from the dependence of $R(r,t)$ on $p(r)$, while the (inverse) scaling in $t+\phi(r)$ follows from $\omega(r)$. When $\Lambda\ne0$ the proportionality is lost, since $\cyc(a,\epsilon,\psi)$ depends non trivially on $a$ which is also affected by the scaling.

In a globally expanding universes (i.e. all shells are expanding), we need only consider the branch of $\cyc$ for which $\partial_\psi\cyc(a,\epsilon,\psi)>0$ and thus $\psi=\bar\psi$. When $\cyc(a,\epsilon,\bar\psi)$ has a maximum as a function of $\bar\psi$, we denote by $\bar\psi_{\rm max}(a,\epsilon)$ the state at which that maximum is attained. The function $\bar\psi_{\rm max}$ is also discussed in appendix~\ref{sect_cyc}. If present,
\begin{equation}
\label{defpmax}
R_{\rm max}(r) = p(r)\,\cyc(a,\epsilon,\bar\psi_{\rm max}(a,\epsilon))\equiv p(r)\,w_0(a,\epsilon)
\end{equation}
is the expression for the maximum radius that a shell with label $r$ can attain. We omitted explicit reference to the dependence of $a$ and $\epsilon$ on $r$ for clarity. The expression (\ref{defpmax}) defines $w_0$, and $w_0$ is discussed in detail in appendix~\ref{sect_cyc}.\footnote{When $\Lambda=0$ and $\epsilon=1$ we have $\bar\psi_{\rm max}=\pi/2$ and $\cyc(\bar\psi_{\rm max})=w_0(0,1)=1$.}

As to the no-collision condition (\ref{nocollision}), it is hard to deduce any general statement as a condition that is relevant for $\partial_rR(r,t)>0$
because of the interdependence of the various functions appearing in (\ref{Rexplgen}). Hellaby \& Lake \cite{hellaby} produced such conditions for the case $\Lambda=0$, employing the particularly simple parametric equations for $R(r,t)$ in that case. In practice, when considering a particular model, we will rather first 'simply' verify (\ref{nocollision}) for all shells, and determine $t_M$ as indicated in connection with its definition (\ref{deftM}).

From table \ref{tab_cyc} in appendix~\ref{sect_cyc} we can deduce that if $\Lambda<0$ all shells will attain a maximum radius, irrespective of the sign of $e(r)$. On the contrary, if $\Lambda>(4/9)\omega^2(r)$ for shell $r$, that shell is unbound, again irrespective of the sign of $e(r)$. When $0\le\Lambda\le(4/9)\omega^2(r)$ boundedness depends on the sign of $e(r)$. Unbound means that the expanding shell $r$ will expand forever, that is, for as long as it does not bump into another shell. An upper bound on the maximum time span of validity of the metric is the minimum of all
\begin{equation}\label{deftb}
t_b(r)\equiv {\bar\psi_{\rm max}(r)\over\omega(r)}-\phi(r), \quad 0\le r\le r_b
\end{equation}
if bound shells are present.

\subsection{The phase function \texorpdfstring{$\phi(r)$}{phi(r)}}
\label{sect_phasefunction}

In order to interpret the phase function $\phi(r)$ in (\ref{soleqRtint}), we recall that, in our paradigm, at the end of the latest collapse, the universe was probably in a compressed, hot and chaotic state with shells bumping into or overtaking each other. The subsequent rebound also surely created instances and places where the no-collision condition (\ref{nocollision}) was violated. Since the current state of the universe seems to indicate that no major collisions occurred since a long time, there must have been a moment when all shells were again partaking in the expansion and condition (\ref{nocollision}) was satisfied throughout. The origin of time $t=0$ in a valid model is therefore that moment at the earliest, and the model therefore starts with shells on radii
\begin{equation}
R(r) = p(r) \,\cyc[a(r),\omega(r)\phi(r)],
\end{equation}
We will assume that the model is "maximal in history", which by definition means that the above equation at $t=0$ implies that the universe emerges at that time from a (temporarily last) collision of shells at some $r$. All states $\psi(r,t)$ are positive from that time on, and the global model is valid until, again, some shell bumps into another one. The phase function $\phi(r)$ is implicitly determined by the initial condition (\ref{r_initial}) and the inversion of
\begin{equation}
R= p(R) \,\cyc[a(R),\omega(R)\phi(R)].
\end{equation}

We must keep in mind, though, that our origin of time only marks the start of the validity of the metric and may very well be considerably smaller than the age of the oldest (compact) structures as we know them now. In other words, if the oldest structures are 13 Gyr old (in our units $t=1.3$), then that does not necessarily mean that the universe as a whole has been expanding for the last 13 Gyr according to the rules of our model, i.e. without colliding shells. If we call "the age of the universe" the duration of the validity of the model in the current expansion phase, as usually, we can very well consider values for this time span that are considerably shorter than 13 Gyr without a priori dealing with a model that is at variance with observations or generally accepted views. We should also keep in mind that the above defined non-collisionality is probably an (unnecessary) strong requirement from the physical point of view, but it is forced upon us because of the geometrical nature of a general relativistic cosmological model.

Lastly, but not in the least, a phase function that is increasing with increasing $r$ has the effect that the outer parts of the universe are more evolved than what the term $\omega(r)t$ could bring about, thereby effectively having an effect of acceleration, without having to invoke dark energy. This property has already been noticed and studied by Iguchi {\it et.al.} \cite{iguchi}. Since the observable properties of a universe are closely related to the paths of photons, we will have to defer the effects of the phase function to section~\ref{sect_light}, more in particular section~\ref{sect_comp_obs}.
 
\subsection{The central region}
\label{seccentral}

It seems reasonable to assume that the mass density in the center tends to a constant, as a function of $r$ that is.
In these conditions, integration of (\ref{defrho}) leads to
\begin{equation}
\label{mcenter}
m(r) \to {\kappa\over3}\rho(0,t) R^3(r,t) \qquad {\rm for} \qquad r\to0.
\end{equation}
The constancy of $\rho$ (as a function of $r$) is also consistent with the Newtonian limit (representation (a))
\begin{equation}
\label{pcenter}
p(r) \to p_c r \qquad {\rm for} \qquad r\to0
\end{equation}
since $m(r)$ must be proportional to $r^3$. We recall however that $r$ is a dimensionless shell label, and therefore the factor $p_c$ in (\ref{pcenter}) has the dimension of a length and merely realises the one-to-one relation of shell labels $r$ and physical radii in the central regions. Using (\ref{Rexplgen}), we find
\begin{equation}
\label{Rcenter}
R(r,t) \to p_cr \,\cyc[a(0),\epsilon,\psi(0,t)] \qquad {\rm for} \qquad r\to0
\end{equation}
and, in view of (\ref{mcenter}):
\begin{equation}
\label{rhocenter}
\rho(0,t) = {\rho_c\over \cyc^3[a(0),\epsilon,\psi(0,t)]},
\end{equation}
with $\rho_c$ a constant with the dimension of a mass density, and
\begin{equation}
\label{mcenter1}
m(r) \to {\kappa\over3}\rho_cp_c^3r^3=\frac12\omega_c^2p_c^3r^3\equiv m_cr^3\quad {\rm for} \quad r\to0.
\end{equation}
The second part is a consequence of the definition of $\omega(r)$ in (\ref{defomega}), which tends to the constant
\begin{equation}
\label{omisoch}
\omega_c=\sqrt{{2\kappa\rho_c\over3}}\cong0.745\sqrt{\rho_c},
\end{equation}
and the last equality of (\ref{mcenter1}) defines $m_c$ (dimension of length).

The energy function is defined in representation (a) as
\begin{equation}
e(r)=\epsilon {m(r)\over p(r)}
\end{equation}
and therefore tends quadratically to zero:
\begin{equation}
\label{ecenter}
e(r) \to e_c r^2 \qquad {\rm for} \qquad r\to0,
\end{equation}
with
\begin{equation}
\label{ecenterc}
e_c=\epsilon{\kappa\over3}\rho_cp_c^2=\epsilon\frac12(\omega_cp_c)^2.
\end{equation}
This is again consistent with the Newtonian limit. 


\subsection{Geometrical considerations}

\subsubsection{The Riemann scalar}

The Riemann curvature scalar of (\ref{4metric}) reads (Bondi \cite{bondi})
\begin{equation}
\label{4Riemgen}
{}^4R(r,t) = 2{\partial^2_tX\over X} + 4{\partial^2_tR\over R} +4{\partial_tX\over X}{\partial_tR\over R} + 2\left(\partial_tR\over R\right)^2 + {}^3R(r,t),
\end{equation}
with ${}^3R$ the Riemann curvature scalar of the 3-dimensional subspace (\ref{3metric}):
\begin{equation}
\label{3Riemgen}
{}^3R(r,t) = {2\over R^2} - {2\over X^2} \left[ 2{\partial^2_rR\over R} + \left(\partial_rR\over R\right)^2 - 2{\partial_rX\over X}{\partial_rR\over R} \right]
\end{equation}
We find, using (\ref{defX}),
\begin{equation}
\label{3Riem}
{}^3R(r,t) = {4\over R^2}{\partial_r(eR)\over\partial_rR} ={4\over R} \left({e\over R} + {d_re\over\partial_rR} \right).
\end{equation}
Clearly, for $e=0$, the universe has zero curvature. 

The curvature ${}^3R(r,t)$ tends towards the spatial constant $6\omega_c^2/\cyc^2[a(0),\epsilon,\psi(0,t)]$ in the center.

For the scalar ${}^4R(r,t)$ we find, using (\ref{eqRt}) and (\ref{eqRtdif}), the rather elegant expression
\begin{equation}
\label{4Riem}
{}^4R(r,t) = 4\Lambda+{2d_rm\over R^2\partial_rR}=4\Lambda+2\kappa\rho(r,t).
\end{equation}
In the center the 4-curvature scalar tends towards the value $4\Lambda+3\omega_c^2/\cyc^3[a(0),\epsilon,\psi(0,t)]$.

\subsubsection{The nature of the curved space}
\label{sect_nat_curv}
It is customary when discussing curved 3-spaces with spherical symmetry to lower the dimension by suppressing one angular coordinate and to consider the nature of such a surface. Here we will consider a (non-Euclidean) meridional surface with $\varphi=\varphi_0$ a constant, and thus we will suppress the longitudinal angle. The essential differences of such a surface compared to a meridional plane in Euclidean space is that, (a) while in the Euclidean plane 2 neighboring points on radial orbits $\Delta\vartheta$ apart have to travel a (Lorentzian) radial distance $dR$ in order to change their tangential distance by $dR\Delta\vartheta$ (independent of $R$!), on the meridional surface this (Lorentzian) distance would become $dR\Delta\vartheta/\sqrt{1-2e(r)}$, and that (b) this distance depends on the shell $r$.

As for the 3D universe, we can always (because of the symmetries) place the observer $O$ on the ${\cal Z}$-axis by means of a 3D rotation. For  $O$ not in the center $C$, the direction towards $C$ would be a direction on the celestial globe, say, a celestial pole. In the 2D cut $\varphi=\varphi_0$ on which we placed the observer ($\varphi_0$ is arbitrary since $O$ is on the ${\cal Z}$-axis) and to which we confined his/her view, this view translates into confinement to look along a great circle through $C$, and thus a meridian. In that 1D sky, there is mirror symmetry as seen from $O$ with respect to the direction $C$, because of the circular symmetry around $C$. In the full 2D sky as seen as the projection of the real 3D universe, this translates into the property that for any observable function $f(\alpha,\delta)$ on the sky that is the projection of a 3D function that satisfies the spherical symmetry condition around $C$, with $\alpha$ right ascension and $\delta$ declination, $f(\alpha,\delta)=f(\delta)$, since the declination of an object is the angular distance from that object to $C$. Put differently, the independence of $\alpha$ reflects the rotational symmetry of the meridians through the ${\cal Z}$-axis (the celestial poles), or the arbitrariness of the choice $\varphi=\varphi_0$.

\subsubsection{Embedding surfaces}
\label{sect_embed}

We choose the 2D cut $\varphi=0$, i.e. the $\cal XZ$ coordinate plane, and build on it a surface that bulges into the $\cal Y$ dimension. Because of the spherical symmetry in the plane $\cal XZ$ this will be a surface of revolution that can be described by a function ${\cal Y}(\cal Z)$. On that surface we consider the coordinates $(R,\vartheta)$.\footnote{Note that $R$ is now also a coordinate, which means that $t$ is a parameter. This is possible because at any given time $t$ we can choose $r=R$.} We construct a surface in Euclidean 3-space ($\epsilon=1$) or Minkowski 3-space ($\epsilon=-1$), both built on the chosen Euclidean meridional plane with metric 
$d{\cal X}^2+d{\cal Z}^2$ or $dR^2+R^2\,d\vartheta^2$:
\begin{equation}
ds^2 = dR^2 + R^2\,d\vartheta^2 + \epsilon\,d{\cal Y}_t^2,
\end{equation}
expressed in cylindrical coordinates $(R,\vartheta,{\cal Y})$. We require that on that surface the metric equals the metric of the 2D cut:
\begin{equation}
\label{defembedmet}
ds^2 = dR^2 + \epsilon\,d{\cal Y}_t^2 + R^2\,d\vartheta^2 = X^2(r,t)\,dr^2 + R^2(r,t)\,d\vartheta^2.
\end{equation}
This yields a surface of revolution ${\cal Y}={\cal Y}_t(R,\vartheta)={\cal Y}_t(R)$ around the ${\cal Y}$-axis in Euclidean 3-space or Minkowski 3-space, with the property that the $ds^2$ of the original metric is preserved, be it, of course, that for Minkowski space ($\epsilon=-1$) the distance in a plane of constant $\vartheta$ is given by $ds^2=dR^2-d{\cal Y}^2$ instead of the Euclidean $ds^2=dR^2+d{\cal Y}^2$ for $\epsilon=+1$. That surface thus 'bulges out' of the Euclidean meridional plane into an Euclidean 3-space or a Minkowski 3-space $(R,\vartheta,{\cal Y})$. This 3-space does not correspond to any reality however. It is simply the space we need to construct this isometric representation of the cut $\varphi=0$ through the original space. We find
\begin{equation}
\label{defembed}
{\cal Y}_t(R) = \int_{r(R,t)}^{r_{b}}\!\!\sqrt{2|e(r)|} \,X(r,t)\, dr \ge0.
\end{equation}

In a standard cosmology with uniform positive curvature the isometric representation is the surface of a sphere.\footnote{We come back to this case in section~\ref{sect_synchronous}.} At this point it needs to be remarked that the embedding surface we consider here only bulge out on one side of  the Euclidean plane (for an observer located at ${\cal Y}>0$), and are not continued 'below' it, in contrast with the common picture in cosmology. For now, this can be best appreciated by considering the limit $e(r)\to0$ which is likewise not a 'double' Euclidean plane. We return to this extensively later on.

The embedding surface not only depends on the particular choices for all 3 functions $m(r)$, $e(r)$ and $\phi(r)$, but also on the time evolution of $X(r,t)$ and $R(r,t)$. Equation (\ref{defembed}) clearly shows that the integrand is integrable when the integrability condition on the volume element (\ref{defvolelem}) is satisfied.



\subsection{Models that realise a given \texorpdfstring{$H_0$}{H0}}
\label{sect_H0}
\subsubsection{Definitions}
\label{sect_H0_def}
The Hubble parameter is defined for synchronous homogeneously filled universes (see section~\ref{sect_synchronous}). We denote
\begin{equation}
\label{defHtime}
{\rm H}(r,t) = {\partial_{\rm t} R\over R},
\end{equation}
where the plain text font for H and the partial derivative with respect to t (and not $t=c\,{\rm t}$) indicate that H has the dimension of a reciprocal time. The expression (\ref{defHtime}) reduces to the usual expression in the standard cosmological models. We call the quantity (\ref{defHtime}) the tangential Hubble parameter or simply Hubble parameter.\footnote{In section~\ref{sect_redshift} we will have occasion to define the radial Hubble parameter. See also the end of this section.} In our units, $c=1$, and we therefore redefine
\begin{equation}
\label{defH}
H(r,t) = {\partial_t R\over R},
\end{equation}
expressed in inverse length. We denote 
\begin{equation}
\label{defH0}
H_0=H(r_0,t_0)
\end{equation}
with $r_0$ and $t_0$ denoting a particular shell at a particular time. If $r_0$ and $t_0$ refer to our position and our epoch, we denote them by $r_{\rm o}$ and $t_{\rm o}$. With ${\rm H}_{\rm o}=72\,\rm km/s/Mpc$ as given by (\ref{def_Ho}), equation (\ref{defHtime}) implies ${\rm H}_{\rm o}^{-1}=13.6\,{\rm Gyr}$ or, with equation (\ref{defH}) $H_{\rm o}=0.736\,(3.066\,\rm Gpc)^{-1}$. Therefore, we can concentrate on models that realise $H_0\sim0.736$ somewhere, sometime. The rule-of-thumb conversion between $\rm km/s/Mpc$ and our units is easy: division by 100.

We can suffice in this section with representation (a), i.e. $\epsilon$ equal to $-1$, $0$ or $+1$. 

We find, with (\ref{eqRt}), or alternatively (\ref{dercyc})
\begin{equation}
\label{Hgen}
H = {\omega\over\cyc^{3/2}}\sqrt{1-\epsilon\,\cyc+a\,\cyc^3}
\end{equation}
valid for the expansion phase (a minus sign in front would appear in the contraction phase). For completeness, the deceleration parameter is given, from (\ref{der2cyc}), by
\begin{equation}
\label{defq}
q = -{R\partial_t^2R \over(\partial_tR)^2}={{1\over2}-a\,\cyc^3\over1-\epsilon\,\cyc+a\,\cyc^3}.
\end{equation}
In both equations, we omitted the dependence on $r$ and $t$ for clarity.

\subsubsection{Classification}
\label{sect_classH0}
In order to identify the models that realise a given $H_0$ somewhere, sometime, we need to solve equation (\ref{Hgen}) for $\cyc$. 

We first consider the bound shells, i.e. the shells that reach a maximum $R$. Since
\begin{equation}
\label{dtH}
\partial_tH(r,t) = \omega{\partial H\over\partial\,\cyc} ={\omega^2\over\cyc^{5/2}}{1\over\sqrt{1-\epsilon\,\cyc+a\,\cyc^3}}\left(2\epsilon\,\cyc-3\right)
\end{equation}
it is obvious that $\partial_tH(r,t)\le0$ if $\epsilon=-1$ or $\epsilon=0$. When $\epsilon=+1$,
one can prove, with the material provided in appendix~\ref{sect_cyc}, that \mbox{$\partial_tH(r,t)<0$} if $\cyc$ is bounded, since $0\le\cyc\le\frac32$. Since \mbox{$H(\cyc)\to+\infty$} for $\cyc\to0$ and $H(\cyc)\to0$ monotonically for $t$ approaching rebound time ($\cyc\to w_0$ in the notations of appendix~\ref{sect_cyc}), there is a unique solution for $\cyc$ and thus a unique state $\psi(H_0)$ that realizes a particular $H_0$ when $\cyc$ is bounded.

When $\cyc$ is unbound, $H(\cyc)\to\omega\sqrt{a}=\sqrt{\Lambda/3}$ for \mbox{$\cyc\to+\infty$} (recall (\ref{defa}). Hence there is always a unique solution $\psi(H_0)$ for finite $t$ when $\Lambda<3H_0^2$ and $\epsilon=-1$ or $\epsilon=0$ due to (\ref{dtH}). For $\epsilon=+1$, unboundedness means that $\Lambda\ge\frac49\omega^2$ (see table \ref{tab_cyc}). Equation (\ref{dtH}) learns that $H(\cyc)$ has a minimum equal to $\sqrt{\frac13\Lambda-\frac4{27}\omega^2}$ at $\cyc=\frac32$. Hence if $\sqrt{\frac13\Lambda-\frac4{27}\omega^2}\le H_0<\sqrt{\frac13\Lambda}$ and $\epsilon=+1$ there are 2 solutions, one for  $\cyc\le\frac32$ and one for $\cyc>\frac32$. This occurs therefore if $3H_0^2<\Lambda<3H_0^2+\frac49\omega^2$. If $\Lambda<3H_0^2$ only the one less than $\frac32$ remains. 
\begin{table}[h]
\caption{Summary of the classification of the number of shell solutions for $H_0$}
\label{tab_H0sol}
\renewcommand{\arraystretch}{1.5}
\begin{center}
\begin{tabular}{|l|l|l|}
  \hline
Shell & Condition & Solutions \\
\hline
\hline
Bound& None  & 1 \\
\hline
  Unbound &  $\Lambda\le3H_0^2$  & 1 \\ \cline{2-3}
   & $\epsilon=+1$ and $3H_0^2<\Lambda<3H_0^2+\frac49\omega^2$ & 2 \\ \cline{2-3}
   & $\epsilon=+1$ and $\Lambda=3H_0^2+\frac49\omega^2$ & 1 \\ \cline{2-3}
  &  $3H_0^2+\frac49\omega^2<\Lambda$  & 0 \\
\hline
\end{tabular}
\end{center}
\renewcommand{\arraystretch}{1.0}
\end{table}

\subsubsection{Solutions}
\label{sect_solH0}
In the cases where there is a unique solution, which we denote by ${\rm cc}_0$, the equation (\ref{Hgen}) which is to solve is similar to the one considered in appendix~\ref{sect_cyc}. Denoting
\begin{equation}
\label{defb}
b(r) = {\omega^2(r)\over H_0^2-\Lambda/3}={1\over \left[H_0/\omega(r)\right]^2-a(r)} \ge 0.
\end{equation} 
 we need to solve
\begin{equation}
\label{cycH0}
{\rm cc}_0^3+\epsilon b\,{\rm cc}_0-b=0
\end{equation}
This equation is already in the Cardano form, and we find readily
\begin{equation}
\label{solcycH0a}
{\rm cc}_0(b) = \sqrt[3]{b\over2}\left(\sqrt[3]{\sqrt{1+\epsilon\frac4{27}b}+1}-\sqrt[3]{\sqrt{1+\epsilon\frac4{27}b}-1}\right)
\end{equation}
if $1+\epsilon\frac4{27}b\ge0$ or
\begin{equation}
\label{solcycH0b}
{\rm cc}_0(b) = 2\sqrt{b\over3}\cos\left(\frac13\arctan\sqrt{\frac4{27}b-1}\right)
\end{equation}
otherwise. Equations (\ref{solcycH0a}), (\ref{solcycH0b}) and (\ref{defcyc}) define a function $\psi_0(b)$.

The integral (\ref{defcyc}) is always well-defined. This is obvious if $\cyc$ is unbound, since the radicand in the denominator is never zero, by definition. If $\cyc$ is bound, the inequality we have to verify is 
\begin{equation}
1-\epsilon\,{\rm cc}_0+a\,{\rm cc}_0^3>0.
\end{equation}
This is easily done, since the above inequality and (\ref{cycH0}) transform into
\begin{equation}
1+ab={\left[H_0/\omega(r)\right]^2\over\left[H_0/\omega(r)\right]^2-a(r)}>0.
\end{equation}

This does not necessarily mean that there is a time $t_0$ where $H_0$ is realised however, since the value for $\psi_0(b)$ may not lead to a positive $t_0$ because of a phase $\phi(r)$ that is too large. Hence we can maximise the number of models that realise some given $H_0$ sometime, somewhere, by setting $\phi(r)=0$.

The function $\psi_0(b)$ therefore defines a function $t_0(r)$, defined for $\omega\ge0$ through
\begin{equation}
\label{deftbrb}
t_0(r)={\psi_0[b(r)]\over\omega(r)}-\phi(r).
\end{equation}
When $b\to0$, and thus $\omega\to0$, expression (\ref{solcycH0a}) shows that \mbox{${\rm cc}_0\to0$}. In that case it follows from (\ref{Hgen}) that
\begin{equation}
\label{Hsmallt1}
H(r,t) \to {\omega(r)\over{\rm cc}_0^{3/2}} \quad{\rm for}\quad {\rm cc}_0\to0
\end{equation}
while, from equation (\ref{smallcyc}),
\begin{equation}
{\rm cc}_0 \to \left(\frac32\psi_0\right)^{2/3} \quad{\rm for}\quad \psi_0\to0.
\end{equation}
Hence, under these conditions,
\begin{equation}
\label{Hsmallt3}
H(r,t) \to \frac23 {1\over t+\phi(r)}
\end{equation}
which is the behaviour of the Hubble parameter at very early times, and thus, for $t_0=t_{\rm o}$
\begin{equation}
\label{Hsmallt2}
t_{\rm o}\to{2\over3H_{\rm o}}-\phi(r)\sim0.906-\phi(r).
\end{equation}
Since $\omega(r)$ scales inversely with the total effective mass, this is also the limit for large $M$. This can be understood because the larger $M$ the less evolved a model is for a given time, and thus the more the relation (\ref{Hsmallt3}), valid for small $t$, is approached. 
\begin{figure}[h]
   \centering
   \includegraphics[width=120mm]{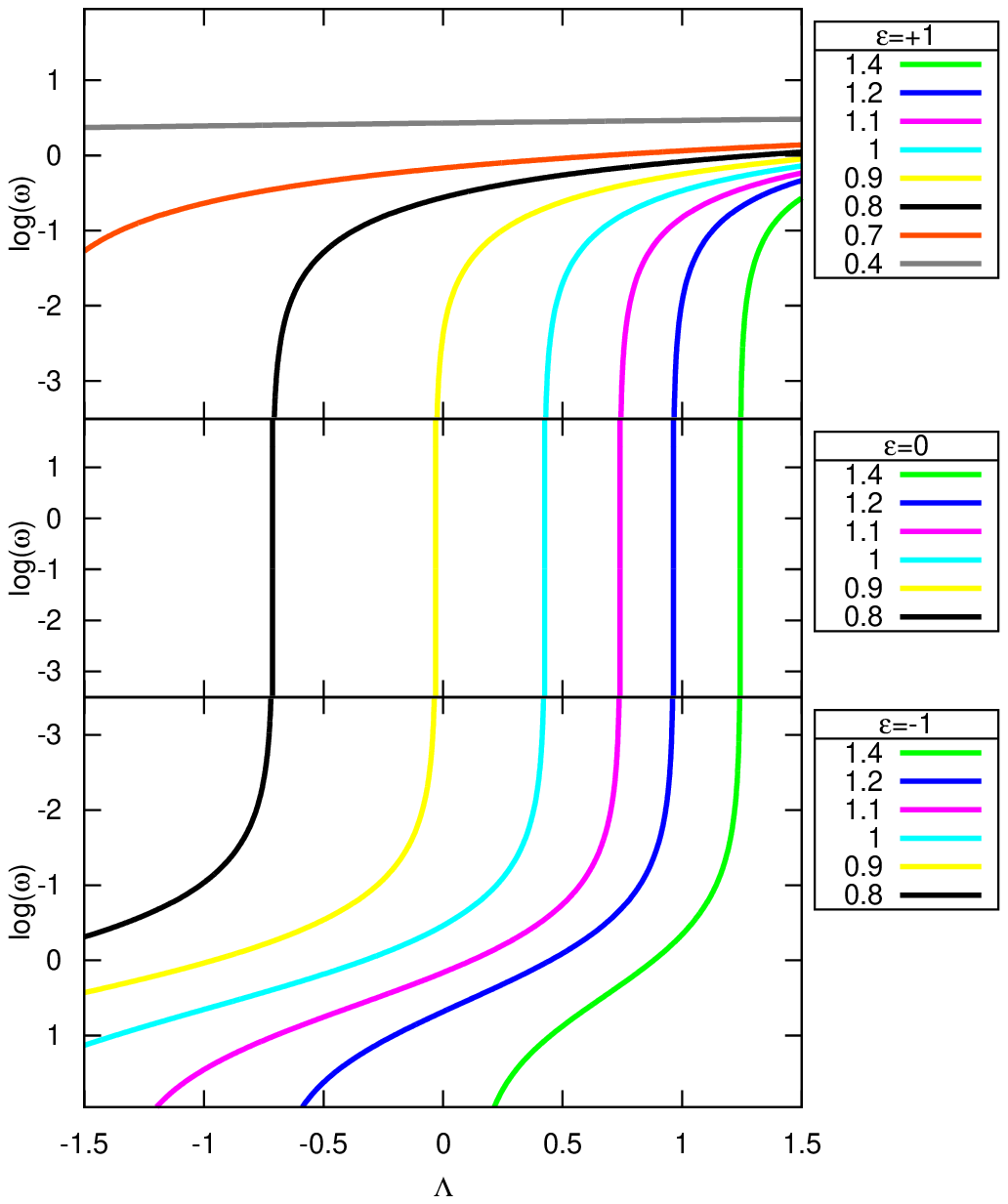}
      \caption{
      Contours of the function $t_{\rm o}(b)=t_{\rm o}(\omega,\Lambda)$ for reasonable ranges in $\Lambda<3H_{\rm o}^2\sim1.625$ and $\log(\omega)$. The figure should be interpreted as follows: a $\Lambda$-universe with $\omega(r_{\rm o})$ and $H_{\rm o}=H(r_{\rm o},t_{\rm o})$ at some shell $r_{\rm o}$, is $t_{\rm o}(\omega,\Lambda)\times10\,\rm Ga$ in the current global expansion phase. Note that for $\epsilon=0$, $t_{\rm o}(\omega,\Lambda)$ is independent of $\omega$, as is shown explicitly in appendix~\ref{sect_e0}. Clearly ages in the interval $[8,14]$ Ga are indicated. Again, as explained at the end of section~\ref{sect_phasefunction}, these 'ages' have not necessarily much to do with the the age of the universe, since there could very well be structures that were formed earlier.}

   \end{figure}

In the case when there are 2 solutions, which we denote by ${\rm cc}_{0,1}<{\rm cc}_{0,2}$,
we obtain
\begin{equation}
\label{solcc01}
{\rm cc}_{0,1}(b) = 2\sqrt{|b|\over3}\cos\left(\frac13\arctan\sqrt{{4|b|/27-1}} +\frac\pi3\right)
\end{equation}
and
\begin{equation}
\label{solcc02}
{\rm cc}_{0,2}(b) = 2\sqrt{|b|\over3}\cos\left(\frac13\arctan\sqrt{{4|b|/27-1}} +{5\pi\over3}\right).
\end{equation}
Which of the 2 is realized depends on the details of the particular universe considered.

\subsubsection{On the radial Hubble parameter}

The tangential Hubble parameter is the Hubble parameter that is measured in a tangential direction, i.e. in a plane perpendicular to the direction of the center of the universe. In section~\ref{sect_redshift} we define the radial Hubble parameter $I(r,t)$ as the Hubble parameter that is measured in the direction of the center (equation (\ref{radHubble})). Clearly, because of the derivatives with respect to $r$ that are involved, we cannot make easily general statements such as the ones we obtained for the tangential Hubble parameter. E.g., for large $M$ and thus small $\psi$, we can use the approximation 
\begin{equation}
R(r,t)=p(r)\cyc(a,\epsilon,\psi(r,t))\to p(r)\left(3\psi/2\right)^{2/3}
\end{equation}
to obtain
\begin{eqnarray}
I(r,t)&=& {\partial_tX(r,t)\over X(r,t)}={\partial^2_{rt}R(r,t)\over\partial_rR(r,t)}\nonumber\\
&\to&\frac23{1\over t+\phi(r)}{\displaystyle d_rp+\frac p3\left(2{d_r\omega\over\omega}-{d_r\phi\over t+\phi}\right)\over
\displaystyle d_rp+p\left({d_r\omega\over\omega}+{d_r\phi\over t+\phi}\right)}.
\end{eqnarray}
This little enlightening expression for a special case shows that general statements will be hard to formulate. In the limit $r\to0$ however, it follows from (\ref{Rcenter}) that $I\to H$, which means that if we choose shells sufficiently close to the center, the difference between $I$ and $H$ will never be an issue. Also, we will see in section~\ref{sect_synchronous} that in standard cosmologies $\omega(r)$ and $\phi(r)$ are constant functions, while $p(r)\sim r$. Hence in these cases $I=H$.

\subsection{Connection with standard notation}
\label{sect_conn_standard}

Equation (\ref{eqRt}) can be rewritten as
\begin{equation}
\label{defOmegas}
-{2e\over R^2H^2} + {2m\over R^3H^2} + {\Lambda\over3H^2} = \Omega_k + \Omega_M + \Omega_\Lambda = 1,
\end{equation}
defining the familiar $\Omega_k$, $\Omega_M$ en $\Omega_\Lambda$ in terms of our notations. We omitted the dependence on $r$ and $t$ for clarity.
In terms of the function cyc, we find, preserving the order of the terms in (\ref{defOmegas}),
\begin{equation}
\label{defOmegas1}
-{\epsilon\,\cyc\over1-\epsilon\,\cyc+a\,\cyc^3} + {1\over1-\epsilon\,\cyc+a\,\cyc^3} + {a\,\cyc^3\over1-\epsilon\,\cyc+a\,\cyc^3} = 1.
\end{equation}
In the universes we consider the above functions depend on $r$ and $t$, while in standard cosmologies they only depend on $t$ (see also later in section~\ref{sect_synchronous}). 

We know continue with a standard analysis in order to solve (\ref{eqRt}). We first multiply (\ref{defOmegas}) with $H^2/H_{\rm o}^2$:
\begin{equation}
\label{HoverH0}
{H^2\over H_{\rm o}^2} = {R_{\rm o}^2\over R^2}\Omega_{k,{\rm o}} + {R_{\rm o}^3\over R^3}\Omega_{M,{\rm o}} + \Omega_{\Lambda,{\rm o}},
\end{equation}
and transform this expression with the well-known relations\footnote{We will see in section~\ref{sect_synchronous} that the $R$ in this paper is not what in many texts is called $a$, the radius of the universe. Equations (\ref{cosmozt}) remains valid though, because our definition of $H$ is the same as the standard one.}
\begin{equation}
\label{cosmozt}
{R(t)\over R_{\rm o}} = {1\over1+z(t)} \quad {\rm and} \quad H=-{1\over1+z}\partial_tz
\end{equation}
into
\begin{equation}
-{1\over H_{\rm o}}{1\over1+z}\partial_tz=\sqrt{(1+z)^2\Omega_{k,{\rm o}} + (1+z)^3\Omega_{M,{\rm o}} + \Omega_{\Lambda,{\rm o}}}.
\end{equation}
Integration yields, with $x=z^{-1}$,
\begin{equation}
(t_{\rm o}-t)\,H_{\rm o} = \int_{[z(t)+1]^{-1}}^1{\sqrt{x}\,dx\over \sqrt{\Omega_{M,{\rm o}} + \Omega_{k,{\rm o}}x + \Omega_{\Lambda,{\rm o}}x^3}}.
\end{equation}
The above integral is equivalent with (\ref{soleqRtint}) or (\ref{defcyc}) upon transforming
\begin{equation}
u = {|\Omega_{k,{\rm o}}|\over\Omega_{M,{\rm o}}}x={|e|\over m}{R_{\rm o}x}={R_{\rm o}\over p}x=\cyc(t_{\rm o})x,
\end{equation}
taking account of the definition of $p$ in (\ref{defp}) 
and a shift of the origin, since in (\ref{soleqRtint}) the origin of time, which is the start of the validity of the metric, is left undefined, while $t_{\rm o}$ here is the current epoch.

\subsection{On the flatness problem}
From the second term in (\ref{defOmegas1}) we see that $\Omega_M\to1$ at early times (the flatness problem). In our paradigm, this feature rather appears as a peculiarity of the decomposition in the $\Omega$'s instead of something particularly fundamental that needs to be explained. This comes about because here we do not expect the expansion phase of the universe to start from a point, and any phase term $\omega(r)\phi(r)$ would make this particular feature disappear. We will return to a discussion of the early times in section~\ref{sect_early}.

\section{Beyond the mass distribution}
\label{sect_beyond}

In this section, we consider a finite matter distribution that finds itself in an era of universal expansion or contraction. Denoting
\begin{equation}
\label{defP}
0<P = p(r_{b})<+\infty,
\end{equation}
the radius
\begin{equation}
\label{defrmaxt}
R_{\rm max}(t)=P\,\cyc[a(r_{b}),\epsilon(r_b),\psi(r_{b},t)]
\end{equation}
encloses all effective gravitating mass $M$. If the matter distribution is embedded in empty space, beyond this physical radius we should find ourselves in the Schwarzschild-$\Lambda$ metric
\begin{equation}
\label{Schwarz}
ds^2 = \left(1\!-\!{2M\over R}\!-\!{\Lambda R^2\over3}\right)d{\bar t}{\,}^2 - {dR^2\over\displaystyle1\!-\!{2M\over R}\!-\!{\Lambda R^2\over3}} - R^2d\Omega^2.
\end{equation}
If the coefficient
\begin{equation}\label{deffR}
1-{2M\over R}-{\Lambda R^2\over3}\equiv V_0(R)
\end{equation}
is positive (it need not be), then we can interpret $d\bar t$ as proportional to the familiar Schwarzschild-$\Lambda$ coordinate time $\bar{\rm t}$, 
\begin{equation}
\label{def_bar_t}
\bar t \equiv c\,\bar{\rm t}
\end{equation}
using the same convention as (\ref{deft}) and $M$ a measure for the total gravitating mass. That these assertions can be made follows from the classical theorem that a time dependent spherical mass distribution of constant total mass creates a gravitational field beyond its boundary that is the same as the field of a point mass with a mass equal to the total gravitating mass of the mass distribution. We will see shortly that it is correct to use the same symbol $M$ as it was defined in (\ref{defM}).

The coefficient $V_0(R)$ has 1 or 2 positive roots $(M\ne0)$. In the case of 1 root, the root is the Schwarzschild radius $R_S$. In the case of 2 roots we denote the roots $R_S<R_2$. The notation $V_0(R)$ follows the notation of appendix~\ref{sect_Novrad} and refers to the effective potential of a radial orbit.
We now have to distinguish 2 different cases.

\subsection{Particles can reach the space beyond the mass distribution}
\label{sect_dustball}

If for all $0\le r\le r_{b}$ the energy function $e(r)<\frac12$, then there is no singularity in (\ref{4metric}) and there is no $R_S$ (see also later). We are in the case of a dust ball. When calculating orbits, the transition from the inner metric (\ref{4metric}) to the outer metric (\ref{Schwarz}) and vice versa is done at the moment the particle passes the outermost shell $r_{b}$ at radius $R_{\rm max}(t)$ given by (\ref{defrmaxt}) with coordinates $(r_b,\Omega)$ and coordinate velocity $(d_t r,d_t\Omega)$. The integration is stopped at that moment, and resumed after transforming coordinates and velocities to the Schwarzschild-$\Lambda$ metric. This can be done by noting that the universe metric at that point can be seen as an Novikov metric, and therefore the procedure outlined in appendix~\ref{sect_NoviSchwarz} is applicable. The stationary Novikov observer is the comoving observer at $(r_b,\Omega)$. The velocity that this observer should be assigned to from the perspective of a stationary Schwarzschild-$\Lambda$ observer is therefore given by (\ref{novschwarzupsilon1}): 
\begin{equation}
-1<\upsilon_N=\pm\sqrt{1-{V_0(R)\over{\tilde E}^2}}<1.
\end{equation}
The sign of $\upsilon_N$ depends on whether we are in an expansion phase $(+)$ or a contraction phase $(-)$. The effective potential $V_0(R)>0$ is known since $R$ is known, and the normalized relativistic energy $\tilde E>\sqrt{V_0(R)}$ is known, by definition, as a characteristic of the dust ball in the Schwarzschild-$\Lambda$ world. 
We can subsequently perform the Lorentz transformation from the velocity of particle
\begin{equation}
\label{expr_VXN_VYN}
(V_{{\cal X},N},V_{{\cal Y},N})=(Xd_tr,Rd_t\Omega)
\end{equation}
in the local Lorentzian frame of the shell $r_b$ to the velocity in the local Lorentzian frame of the stationary Schwarzschild-$\Lambda$ observer $(V_{x,S},V_{y,S})$ with the inverse Lorentz transformation of (\ref{Lorentz_transfo}). Finally we can pass on to
\begin{equation}
\label{expr_VXS_VYS0}
(d_{\bar t}R,d_{\bar t}\Omega)=\left(V_0(R)V_{{\cal X},S},{\sqrt{V_0(R)}\over R}V_{{\cal Y},S}\right)
\end{equation}
and coordinates $(R(r_b,t),\Omega)$ at some (to be defined) time $\bar t$, and resume the orbit calculation in the Schwarzschild-$\Lambda$ frame.

For light, we also need to calculate the refraction angle, so that the wavelength change can be determined (subsection~\ref{sect_NoviSchwarz}).

All this occurs irrespective of the time evolution of the dust ball (oscillation, permanent expansion, or mixed forms), and without special consideration of expansion due to $\Lambda$, since $\Lambda$ is common to both representations.

The procedure for a particle that falls into the dust ball from outside is similar (mutatis mutandis).

\subsection{Particles remain confined inside the outermost massive shell at \texorpdfstring{$r_{b}$}{rb}}
\subsubsection{The definition of a universe}
\label{sect_defuniverse}

In the case $e(r)\to\frac12$ for some shell, we can take that shell to be $r_b$, since the metric (\ref{4metric}) is only defined for shells with comoving matter. If we assume that $w_0(a(r_b),1)$ is finite (see table \ref{tab_cyc}),
\begin{equation}
\label{defrmax}
R_{\rm max}=Pw_0(a(r_b),1)
\end{equation}
is the largest possible value of $R$ in the valid spacetime of the metric (\ref{4metric}). Inserting this value for $R_{\rm max}$ in
\begin{equation}
\label{schwarzg00}
{1\over3}\Lambda R^3 - R + 2M = -RV_0(R)
\end{equation}
we recover the definition of $w_0$ as the root of $1-w_0+aw_0^3=0$ introduced in appendix~\ref{sect_cyc}, with $a$ given by (\ref{novRt}), if and only if $2m(r_b)=2M=p(r_b)=P$. This proves that the case of a universe from which nothing can escape, and therefore with an outer solution (\ref{Schwarz}) which is a Schwarzschild-$\Lambda$ black hole, has $e(r_{b})=\frac12$ and fills completely its allowed volume at maximum expansion. Hence, the Schwarzschild radius $R_S$, solution of (\ref{schwarzg00}), is also given by
\begin{equation}
\label{defschwarzRS}
R_S = p(r_{b})w_0[a(r_{b}),1)]=2Mw_0(4M^2\Lambda/3,1),
\end{equation}
and at all other times the maximum radius $R(r_b,t)\le R_S$. Spacetime with $R(r_b,t)<R\le R_S$ has the signature of an inner Schwarzschild-$\Lambda$ metric.

We are now in the position to state a necessary condition for the type of universe we will explore in the rest of this paper: given the metric (\ref{4metric}), and irrespective of the behaviour of $e(r)$, $r<r_{b}$, at $r=r_{b}$ we have
\begin{equation}
\label{defebound}
e(r_{b})={M\over P}=\frac12
\end{equation}
or, with (\ref{def_tildep}),
\begin{equation}
\tilde p(r_{b})=1.
\end{equation}

\subsubsection{Infalling particles}
\label{sect_infall}

We consider infalling particles from outside the black hole. Since we need to cross the horizon, we adopt the Novikov frame (appendix~\ref{sect_Novikov}, and more in particular section~\ref{sect_NoviSchwarz}).

Parallel to the events in the mother universe, the universe inside the black hole is also evolving. Any particle (massive or a photon) that ventures from outside $R_S$ inside $R_S$ must follow a course with decreasing $R$ (see also section~\ref{sect_Novi_orbit}), and will eventually reach the edge of the universe inside $R_S$, at $R_{\rm max}(t_c)$ as given by (\ref{defrmaxt}), $t_c$ a cosmic time that is indeterminate, and at $(P,\Omega)$ in the Novikov frame. From that time on we can, and must, consider the orbit in the universe.

Novikov coordinates and cosmic coordinates are strictly disjunct, because everything in the Novikov frame is traveling with decreasing $R$ while the boundary of the universe on the contrary is expanding and therefore cannot be described in the ingoing Novikov metric: the boundary behaves as a comoving shell in an outgoing Novikov metric.  Therefore we need a metric that encompasses both metrics, and the obvious choice is the inner Schwarzschild-$\Lambda$ metric ($V_0(R)<0$)
\begin{equation}
\label{InnerSchwarz}
ds^2 = {dR^2\over [-V_0(R)]} - [-V_0(R)]\,d{\bar t}^{2} - R^2d\Omega^2.
\end{equation}
As is well known, the roles of $\bar t$ and $R$ have switched. An infinitesimal time interval is given by
$dR/\sqrt{-V_0(R)}$ while a radial length is 
$d{\bar t}\sqrt{-V_0(R)}$.
The Lorentzian radial velocity in the inner Schwarzschild-$\Lambda$ metric therefore is 
\begin{equation}
\label{vradinschwarz}
\upsilon_{iS} = {d{\bar t}\over dR}[-V_0(R)]={d{\bar t}\over dt}{dt\over dR}[-V_0(R)].
\end{equation}
The second equation is relevant for Novikov shells in their coordinates $(t,P,\Omega)$. Using (\ref{novgeod}), we obtain\footnote{In section~\ref{sect_Novrad} it is quite irrelevant whether we deal with the inner or the outer Schwarzschild-$\Lambda$ metric: the analysis is valid outside $R_S$ and inside $R_S$} 
\begin{equation}
\label{vradinschwarz1}
\upsilon_{iS} = -\tilde E_\infty{dt\over dR}.
\end{equation}
The boundary shell of the universe follows the outgoing Novikov orbit (\ref{defrmaxt}), which we now can write as
\begin{equation}
R_{\rm max}(t_c)=2M\,\cyc[4M^2\Lambda/3,1,\psi(r_{b},t_c)].
\end{equation}
This boundary will eventually touch $R_S$, and since it is an outgoing Novikov shell turning into an infalling one, it has $\tilde E_\infty=0$ in the Novikov swarm, with the appropriate $M$. We thus find that the boundary of the universe has zero Lorentzian radial velocity in the inner Schwarzschild-$\Lambda$ metric. This is in concordance with the results of section~\ref{sect_mass_ejection}.

The other infalling Novikov shells on the contrary have \mbox{$\tilde E_\infty>0$}. The expression for the radial velocity (\ref{novdRdt}) in the outer Schwarzschild-$\Lambda$ metric now reads in the inner Schwarzschild-$\Lambda$ metric as
\begin{equation}
d_tR = \sqrt{{\tilde E_\infty}^2- V_0(R)},
\end{equation}
with the positive sign, since in the inner Schwarzschild-$\Lambda$ metric $R$ has the character of a time, and $d_tR$ is therefore a relation between 2 times, which is always positive (and irrespective of the circumstance that the universe is expanding or contracting). Note also that, since $V_0(R)<0$, there is no a priori limit on the magnitude of $d_tR$. This is no problem because $d_tR$ is not a velocity. 

Insertion of (\ref{novdRdt}) in (\ref{vradinschwarz1}) yields
\begin{equation}
\label{radvelrb}
(\upsilon_N)_{iS} = -{\tilde E_\infty\over\sqrt{\tilde{E_\infty}^2- V_0}}=-\left(1- {V_0\over\tilde{E_\infty}^2}\right)^{-1/2}.
\end{equation}
The effective potential $V_0<0$, and thus $|\upsilon_N|\le1$, as should.\footnote{Note that standard treatises denote our $V_0$ as $V_0^2$, which is clearly not appropriate inside the Schwarzschild radius.}
 
The transition from Novikov coordinates to cosmic coordinates is similar to the transformation explained in section~\ref{sect_NoviSchwarz}. The infalling observer at $(P,\Omega)$ with coordinate velocity $(\dot P,\dot\Omega)$ encounters at Novikov time $t$ the boundary shell $r_b$, which happens to be at cosmic time $t_c$. The relation between $t$ and $t_c$ is undetermined. The particle coordinate velocity transforms into the local Lorentzian Novikov frame velocity according to
\begin{equation}
\label{entervelocity}
(V_{{\cal X},N},V_{{\cal Y},N}) = \left({|\partial_PR_P(t)|\over\sqrt{1-\epsilon_N(2M/P)}}d_tP,Rd_t\Omega)\right).
\end{equation}
At the boundary of the universe, the Novikov frame has radial velocity $\upsilon_N$ given by (\ref{radvelrb}).
The Lorentz transformation of subsection~\ref{sect_NoviSchwarz} then yields $(w,R\dot\Omega)$, with $w$ the local Lorentzian radial velocity component that will be needed in section~\ref{sect_orbits} (it is defined in (\ref{defwvel}))  in order to integrate the orbit at $(r_b,\Omega)$ and at $t_c$ further in the universe.

For light, the procedure is the same, but we also need to calculate the refraction angle in order to determine the wavelength change (subsection~\ref{sect_NoviSchwarz}).

We also note that inspection of (\ref{entervelocity}) learns that, since $d_t P$ can be positive, an infalling particle or photon can enter the universe with an outward velocity! 

\section{The universe before the present expansion phase}
\label{chapt_early}
\subsection{The latest collapse and structure formation}
\label{sect_collapse}
In our paradigm it would be grotesque to assume that the universe somehow originated from a point, with at the time $t=0$ of 'creation' $\partial_tR(0,0)=0$ and $\partial_tR(r,0)=+\infty$, $r\ne0$. We must assume that the universe has gone through at least one collapse, and therefore it is instructive to outline how such a collapse could be modeled. 

Since we are concerned here with the latest collapse, we will assume that the universe is cold enough, so that the matter is localized on a number of thin shells $R_i(t)$ with surface density $\sigma_i(t)$, $i\ge1$. Allowing for the spherical symmetry, this is certainly not contrary to the observations, since we can think of the shells as a kind of galaxy clusters, with one dimension that is small on a cosmological scale. We order them according to increasing $R_i(t)$ and the $i$'s are thus the shell labels. The clusters have mass  $\Delta{\cal M}_i$ and tangential stretch rate $S^T_i(t)$.

In order to establish the initial state of the universe, given the initial conditions $R_i(t_0)$, $\Delta{\cal M}_i$ and $S^T_i(t_0)$, we first recast the relation  (\ref{defrho}) into a difference equation
\begin{equation}
\label{mpacket}
\Delta m_i = m_i -m_{i-1}= \kappa\sigma_{i-1}(t)[R_{i-1}(t)]^2.
\end{equation}
This choice of the indices ensures that the gravitating mass on shell $i$ has no impact on the dynamics of shell $i$. Likewise, the cumulative mass (\ref{defcalM}) gives rise to
\begin{equation}
\label{calMpacket}
\Delta m_i={\kappa\over4\pi}\Delta{\cal M}_{i-1}\sqrt{1-2e_{i-1}}.
\end{equation}
The energy $e_i$ at shell $i$ reads:
\begin{equation}
\label{epacket}
e_i = {m_{i-1}+\Delta m_i\over R_i(t)} - \frac12 [S^T_i(t)]^2 + \frac16\Lambda [R_i(t)]^2.
\end{equation}

From (\ref{calMpacket}) we calculate $\Delta m_i$ in a recursive way since the $\Delta{\cal M}_i$ are given, and from (\ref{epacket}) and the known $R_i(t_0)$ and $S^T_i(t_0)$ we find $e_i$. The recursive scheme starts with $m_0=e_0=S^T_0(t_0)=R_0(t_0)=0$, and $\Delta{\cal M}_0={\cal M}_0$ is an (optional) mass in the center. Note that the solution does not depend on the sign of $S^T_i$, reminding us of the same feature in the Lorentz contraction. This also shows that the scheme we have outlined is also valid for an expanding universe, or even for a mixed universe with parts in collapse and parts in expansion.

As for the metric, we may set $r=i$. We define
\begin{equation}
\label{mr_step}
m(r)=m_i \quad {\rm for} \quad i-1<r\le i .
\end{equation}
In contrast to (\ref{mr_step}), the energy function $e(r)$ is undefined as long as we do not specify the tangential stretch rate $S^T(r)$ for $r\ne i$. It is useful to think of the space between shells as being filled with a dynamically unimportant `fog' of test particles that can be used to embody $S^T(r)$ for $r\ne i$. Clearly $e(i)=e_i$ and $e(r)$ must be differentiable for $r\ne i$, and we will assume, but need not necessarily to, that $e(r)$ is monotonous for $r\ne i$.

The initial metric being established, we can run the model in the cosmic time, with (\ref{eqRt}) recast in the form
\begin{equation}
\bigl(d_tR_i(t)\bigr)^2 =  -2e_i + {2m_i\over R_i(t)}  + \frac13\Lambda [R_i(t)]^2.
\end{equation}
Note that (\ref{eqRt}) does not require any differentiability with respect to the shell label $r$ (equation (\ref{defrho}) is the one that requires differentiability). The equation is thus compatible with discrete shell labels $i$ and associated constants $e_i$ and $m_i$. The solutions are discussed extensively in section~\ref{sect_definitions} and appendix~\ref{sect_cyc}:
\begin{equation}
\label{shell_evol}
R_i(t) = p_i \,\cyc[a_i,\epsilon_i,\psi_i(t)]=p_i \,\cyc[a_i,\epsilon_i,\omega_i(t+\phi_i)],
\end{equation}
with $\psi_i(t)$, $\omega_i$, $a_i$ and $p_i$ the obvious discrete analogs from the continuous ones defined in (\ref{soleqRtint}), (\ref{defomega}), (\ref{defa}) and (\ref{defp}). It is most convenient to use representation (b). The sign of $S^T_i$ determines whether cyc is to be taken on the upward or downward branch, and the value of $|S^T_i|$ determines $\phi_i$.
 
When a collision occurs between comoving clusters (no radial velocity!), i.e. two functions attaining $R_{i_c}(t)=R_{i_c+1}(t)$ at the same time, one can assume that this results in (a) new structure(s). This collision/merger/reorganization is local and therefore a Newtonian process in Euclidean space. It could be taken to happen instantaneously. At that time and radius
\begin{equation}
S^T_{i_c}=\partial_tR_{i_c}\ne\partial_tR_{i_c+1}=S^T_{i_c+1}
\end{equation}
and one needs a prescription for the new $\Delta {\cal M}_i$, $R_i$ and $S^T_i$, where the number of new shells depends on the details of the collision. After the process, the new shell(s) are moving more 'in sync', from a cosmological perpective.

Be $i_r$ the new shell label of the outermost resulting shell after collision, one needs to relabel the shells $R_i>R_{i_r}$. Hence the equations (\ref{calMpacket}), (\ref{mpacket}) and (\ref{epacket}) need to be solved again for $i>i_r$. After that, the model can be restarted until the next collision. 

In setting up the new initial conditions after a collision, it is possible that for $i=i_b\ge i_r$ the energy $e_{i_b}\ge\frac12$. This is no problem for the time evolution of shell $i_b$ according to (\ref{shell_evol}), but we cannot use expression (\ref{calMpacket}) for $\Delta m_{i_b+1}$. Be $i$ the outermost shell inside $i_b$ for which $e_i<\frac12$. Then we can locate in the `fog' between shell $i$ and shell $i_b$ a Schwarzschild radius where $e(r)=\frac12$ by virtue of continuity of $e(r)$, and a new black hole is born. In that case, the effective gravitating mass $m_{i_b+1}=m_i$.

\begin{figure}[h]
   \centering
   \includegraphics[width=150mm]{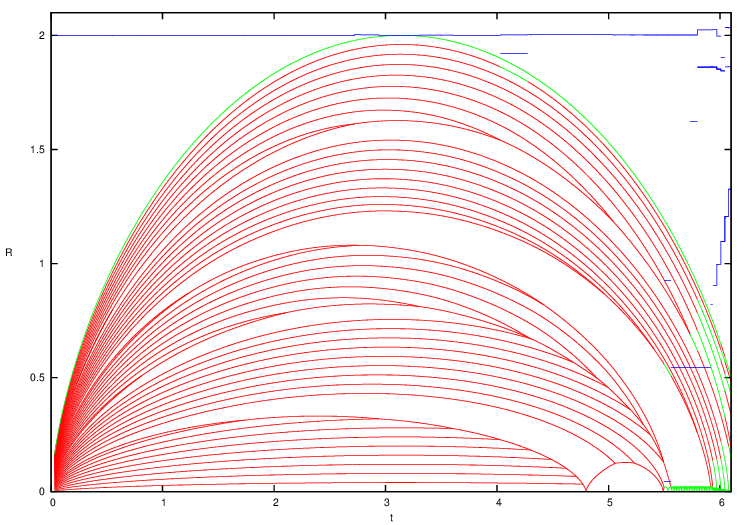}
      \caption{An example of the scheme outlined in this subsection. A universe with $\Lambda=0$ and $M=1$ is represented by 50 shells (in red). One oscillation period equals $2\pi$. The initial conditions are such that the universe is synchronous, except for shells 10 and 30 which are given somewhat less tangential stretch, and shells 20 and 40 that have somewhat more tangential stretch than the synchronous values. We see that the number of shells diminishes as time goes on, especially in the central region, where 1-9 crash into 10, which survives a (collisionless) rebound. The green shells are those for which $e_i\ge\frac12$. Clearly 50 is in this case. At late times, also other shells turn green. As 10 collides with 13, a small central black hole is created. The blue lines 'above' the green shells are the corresponding $2m_i$, and thus the corresponding Schwarzschild radii. The shells in between are in the Novikov zone. Note that all shells in the Novikov zones are infalling.}
         \label{fig_coll}
   \end{figure}

In order to make this scheme more concrete, we will now briefly elaborate on a toy model. We assume that collisions are fully inelastic with conservation of tangential stretch. Be the colliding shells $i-1$ and $i$, then
\begin{equation}
S^T =  {\Delta{\cal M}_{i-1}S^T_{i-1} + \Delta{\cal M}_iS^T_i \over \Delta{\cal M}} \quad {\rm and} \quad \Delta{\cal M} =\Delta{\cal M}_{i-1}+\Delta{\cal M}_i,
\end{equation}
with $\Delta{\cal M}$ and $S^T$ the resulting mass and tangential stretch. Of course, energy is not conserved, and the energy loss equals 
\begin{equation}
\Delta E = \frac12{{\cal M}_{i-1}{\cal M}_i\over{\cal M}}(S^T_i-S^T_{i-1})^2.
\end{equation}
The gravitating mass of shell $i-1$ will not contribute any more to the motion of shell $i$ after collision, and we have
\begin{equation}
\Delta m = m-m_{i-2}={\kappa\over4\pi}\Delta{\cal M}\sqrt{1-2e_{i-2}}
\end{equation}
and
\begin{equation}
e = {m_{i-2}+\Delta m\over R_i} - \frac12(S^T)^2 + \frac16\Lambda R_i^2.
\end{equation}
The resulting $S^T_i$, ${\cal M}_i$, $m_i$ and $e_i$ are equal to the just calculated unsubscripted ones. In figure \ref{fig_coll}, an impression is given of how these simple prescriptions affect the evolution of a discrete shell universe.

Three more remarks are in order. Firstly, the scheme we outlined is not at variance with the second law of thermodynamics, since entropy clearly increases at collision time. Even in the theoretical and "ideal" situation that no collisions occur anymore and the universe is cyclic\footnote{As indicated earlier, we do not advocate this scenario, be it only because in a contraction phase the universe will rebound well before "the point" is reached and at rebound time entropy will always increase.}, the dynamics of the universe, which is basically radial 2-body motion save for the effect of the cosmological constant, is not at variance with the second law, for the same reason as it is not an issue in cyclic 2-body motion.

Secondly, it is also clear that this scheme has important limits: there is no reason to assume that spherical symmetry is even indicated in the collapse phase. Be it for that reason only, the model in figure \ref{fig_coll} is largely a toy model, illustrating some general principles, and should not be taken too seriously, certainly at later times, when the shells come close together and some chaos sets in. Surely other physics should be put in beyond the collisionless rather naive dynamical model that is presented here. Colliding shells would almost certainly break the spherical symmetry. 

This touches upon a more general third remark. Cosmological models assume isotropy. There is a well-known rationale for this, and geometrical elegance and simplicity are important properties of the current cosmological models. If one breaks this perfect symmetry, then there is no particular reason to hold on to any other symmetry, such as the spherical one. Spherical symmetry is the first generalisation away from isotropy, and as such it has its merits, but one should not take the spherical non-equilibrium modeling too far.



\subsection{The early history}
\label{sect_early}

A scenario for the early universe is of course completely speculative. In our paradigm, the universe may have originated as a black hole, in a "mother universe" of which we will never be able to say anything, except for the fact that, perhaps, still now, material from that universe could enter our universe. In that early phase, at times very much longer ago than the currently accepted age of the universe, the universe must have been extremely hot, and filled with matter decomposed down to elementary particles and radiation. It would be a state similar to what is now our current view of the early universe, at a time when radiation decoupled. The big difference is that this state of complete chaos and mixing came about naturally. Matter and radiation of course did interact continuously, and the universe was initially too hot for any bound structure to develop. Hence entropy was maximized and the universe was in a homogeneous state in which particles whiz around, only, but irrevocably, confined by the horizon. The radiation field was, and remained, completely isotropic. This radiation was to become now the cosmic microwave background radiation (CMB). Note that in this scenario the CMB is in zeroth order isotropic by design, as must be any other field of weakly interacting particles. 

Over (very long) time, the universe kept accreting matter, including probably vast amounts of elementary particles, and possibly also black holes. Once the latter were part of our universe, they may have become effective sinks of kinetic energy and matter. They did therefore their part in sweeping the chaos clean, and thus could have been a source of rarefaction. Pressure thus gradually decreased.

The larger the mass of the universe, the larger the event horizon. Matter, even bound structures, would eventually be able to enter the universe without being torn apart to elementary particles by the tidal forces at the event horizon, even without 'feeling' the transition altogether . This could conceivably also have been a source of injection of bound structures.

Once our universe was sufficiently rarefied, filled with black holes of various sizes surrounded by matter, it entered the state in which we can start talking about structures, and hence comoving matter. At that stage our universe could be modeled with the scheme outlined in section~\ref{sect_collapse}. 

\subsection{The initial conditions of the present expansion phase}

Of course, we may not bother about the earlier phases of our universe, and be content with the specification of continuous initial conditions. 
This option is also indicated if one takes the view, as seems to be strongly born out by the observations, that the constituent (dark) matter of the universe is unseen, and that a visible structure is but the top of an iceberg. In that case, the dark matter could very well be smoothly distributed, or at least much more smoothly than the visible matter. The CMB now can, of course, contain signatures of a previous phase (see e.g. Gurzadyan \& Penrose \cite{gurzadyan}, though we do not invoke or need Conformal Cyclic Cosmology).

Equation (\ref{eqRt}) can now be seen as a relation determining $e(r)$ if ${\cal{M}}(r)$ or $m(r)$ and the initial conditions (\ref{r_initial}) and $S^T\!(R)$ at some $t_0$ are given. 

In terms of the effective gravitating mass, we get the trivial recast of (\ref{eqRt}):
\begin{equation}
\label{eoutofinitconda}
e(R)={m(R)\over R} -\frac12[S^T\!(R)]^2 + {1\over6}\Lambda R^2,
\end{equation}
showing dat $e(R)$ increases with increasing gravitating mass and decreasing $S^T\!(R)$. If  $e(R)=\frac12$ at a finite $R=r_b$, the initial conditions cause a universe. Note that in that case $R=r_b$ will be smaller than the Schwarzschild radius unless \mbox{$S^T\!(R)=0$} there, since $r_b$ here is the solution of 
\begin{equation}
{1\over3}\Lambda R^3 - \bigl[1+[S^T\!(R)]^2\bigr]R + 2M = 0
\end{equation}
to be compared with $R_S$ which is the solution of (\ref{schwarzg00}). In general, we thus obtain a universe embedded in an inner Schwarzschild metric. 

Alternatively, the expression of $e(R)$ in terms of the mass function leads to a first order differential equation:
\begin{equation}
\label{eoutofinitcond}
d_R(Re) - {\kappa\over4\pi}d_R{\cal M}\sqrt{1-2e} +\frac12d_R\bigl(R[S^T\!(R)]^2\bigr)=\frac12\Lambda R^2,
\end{equation}
which can be solved for the energy $e(R)$, given e.g. the boundary condition $S^T\!(0)=e(0)=0$. The integration can be carried out for $0\le R$ under the constraint $e(R)\le\frac12$. 

\subsection{On the energy function \texorpdfstring{$e(r)$}{e(r)} and the effective gravitating mass \texorpdfstring{$M$}{M}}

Here is the place to speculate on the behaviour of $e(r)$. We have seen in section~\ref{seccentral} that $e(0)=0$, while from section~\ref{sect_defuniverse} we have $e(r_b)=\frac12$. This last condition is hard to change once established, unless one would invoke massive collisions of dynamically important (not necessarily comoving) shells, on scales of the size of the universe. As before, we forego for simplicity the likely breakdown of the spherical symmetry that this would entail. This reshuffling of shells and their kinetic energies would, according to equation (\ref{eoutofinitcond}), cause another $e(r)$, and probably also another $r_b$ and $M$ (even assuming that the total mass ${\cal M}(r_b)$ remains the same), much like the (simplified) scheme we outlined in section~\ref{sect_collapse}. It seems unlikely that the new $e(r)$ would differ importantly from the old one, though, because this would lead to massive instabilities that probably would have been dealt with in the chaotic phases of the history of the universe. 

The above considerations point to the possibility that the mass of a black hole can change with time, not only because mass is falling in, but also because what the outside world is measuring is effective gravitating mass, which depends on the kinetic state of the shells inside. We will return to this process from another point of view in section~\ref{sect_anom_red}, where we will argue that there is evidence that it actually occurs.

In this paper we will adhere to the working hypothesis that the energy $e(r)$ is rather close to 0, as seems to be indicated by the observations. Our definition of a universe then imposes that $e(r)$ rises rather sharply to $\frac12$ at the edge $r_b$. Note that, according to the analysis in section~\ref{sect_e0}, this does not exclude universes that are largely homogeneous.

\section{A classification of the universes}

\subsection{Preamble: the synchronous homogeneously filled universes}
\label{sect_synchronous}
\subsubsection{General formulas}
In the case that $\omega(r)=\omega_c=\omega$ and $\phi(r)=\phi_c=\phi$ are constant functions throughout (note that we can drop the subscript $c$ in this section), we can lift all time limitations on the model since the initial condition $\partial_rR(r,t)>0$ will remain valid. Note that also $a$ and $\epsilon$ (representation (a)) are constant in that case, which means that we can also omit the explicit dependence of $\cyc$ on $r$ as indicated in (\ref{Rexplgen}), and we obtain
\begin{equation}
\label{Rsync}
R(r,t) = p_cr\,\cyc[a,\epsilon,\omega(t+\phi)],
\end{equation}
which is separable in $r$ and $t$. All relations that are specific for the central regions, as presented in section~\ref{seccentral}, hold now for all shells at all times.

The metric can be written as
\begin{equation}
\label{isochmetric}
ds^2 = dt^2 - \left[{\cal R}(t)\right]^2\bigl[{d{\tilde r}^2\over1-\epsilon{\tilde r}^2} + {\tilde r}^2(d\vartheta^2+\sin^2\vartheta\,d\varphi^2)\bigr],
\end{equation}
with
\begin{equation}
\label{radius_sync}
{\cal R}(t) = {\cyc[a,\epsilon,\omega (t+\phi)]\over\omega},
\end{equation}
which is obtained from (\ref{4metric}) by changing the radial coordinate from $r$ to 
\begin{equation}
\tilde r=\omega p_cr={r\over r_b},
\end{equation}
using (\ref{defomega}), (\ref{pcenter}), (\ref{omisoch}) and (\ref{ecenter}). We recognise the familiar Robertson-Walker metric, of course.
We can identify
\begin{equation}
\label{e_iso}
e_{\rm sync}(\tilde r) = {\epsilon\over2}{\tilde r}^2.
\end{equation}
In the case $\epsilon=+1$ we recall the well-known
\begin{equation}
\label{isochmetrichi}
ds^2 = dt^2 - \left[{\cal R}(t)\right]^2\bigl[d\chi^2+\sin^2\chi(d\vartheta^2+\sin^2\vartheta\,d\varphi^2)\bigr],
\end{equation}
where the radial coordinate $\chi$ was introduced which is inspired by the spherical embedding model.
This implies 
\begin{equation}
\label{def_chi}
\tilde r = |\sin\chi|,
\end{equation}
while the metric coefficients in the original metric (\ref{isochmetric}) read
\begin{equation}
\label{sync_R_chi}
R(\tilde r,t) = {\cal R}(t)|\sin\chi|,\qquad X(\tilde r,t) = {{\cal R}(t)\over|\cos\chi|}.
\end{equation}
From the first of these equations also follows
\begin{equation}
\label{def_radius_universe}
{\cal R}(t) = R(r_b,t) = R_{\rm max}(t)
\end{equation}
in line with definition (\ref{defrmaxt}) of the radius of a universe.

The case $\epsilon=-1$ follows immediately from the same formulae by substituting the trigonometric functions with hyperbolic functions, and the case $\epsilon=0$ by the limit $\chi\to0$ in first order in $\chi$. Hence, inside these universes, these synchronous spherical models are identical, locally, to the models used in standard cosmologies.

The effective mass function reads, with equation (\ref{mcenter1}):
\begin{equation}
\label{eff_mass_sync}
m(\tilde r) = {\kappa\over3}\rho_{c,{\rm sync}}(2M)^3{\tilde r}^3=M\,{\tilde r}^3\equiv m_{c,{\rm sync}}r^3.
\end{equation}
As for the $\Omega$'s defined in (\ref{defOmegas}), we recover, with the formulae in \ref{seccentral}, the well known
\begin{equation}
\label{defOmegaM}
\Omega_M(t)={2m(r)\over R^3H^2}={2\kappa\rho\over3H^2}={8\pi G\rho\over3c^2H^2}={8\pi G\rho\over3{\rm H}^2}\equiv{\rho(t)\over\rho_{\rm crit}}.
\end{equation}
The cumulative mass function can be calculated explicitly with (\ref{defcalM}). We obtain
\begin{equation}
\label{cum_mass_sync}
{\cal M}(\tilde r) = {3c^2M\over G}\left\{
\begin{array}{ll}
\frac12\left[\arcsin(\tilde r)-\tilde r\sqrt{1-{\tilde r}^2}\right]&\qquad(\epsilon=+1) \\[2mm]
{\tilde r}^3/3&\qquad (\epsilon=0)\\[1mm]
\frac12\left[\tilde r\sqrt{1+{\tilde r}^2}-{\rm asinh}(\tilde r)\right]&\qquad(\epsilon=-1).
\end{array}
\right.
\end{equation}
These expressions, taken in $\tilde r=1$, yield the total mass of the closed model, and approximations for the total mass of models with $e(\tilde r)\le0$ that are synchronous over most of the $\tilde r$ range and that have close to $\tilde r$ a steeply rising $e(\tilde r)$ so as to have $e(1)=\frac12$. We obtain
\begin{eqnarray}
\label{tot_mass_sync}
{\cal M}_{\rm tot}\!\!&=&\!\! 4\pi\rho_{c,{\rm sync}}(2M)^3\!\left\{
\begin{array}{ll}
\displaystyle\pi/4=0.785& (\epsilon=+1) \\[2mm]
1/3=0.333& (\epsilon=0)\\[1mm]
\frac12\left[\!\sqrt2-\ln(1\!+\!\!\sqrt2)\right]=0.266\,&(\epsilon=-1)
\end{array}
\right.\nonumber\\
\end{eqnarray}
in units of $5.35\times10^{22}M_\odot$ according to (\ref{calMunit}).

\subsubsection{The concordance model}
In the current standard model, one finds that $\Omega_{k,{\rm o}}\sim0$, $\Omega_{M,{\rm o}}\sim0.3$ and $\Omega_{\Lambda,{\rm o}}\sim0.7\,$. With an age $t_{\rm o}\sim13\,{\rm Gyr}$ and a Hubble parameter of 0.74 (in our units), this implies, with equations (\ref{defOmegas}), (\ref{defOmegas1}) and the formulas in appendix~\ref{sect_e0}, that $\Lambda=1.14$, $\phi(r)=\phi_c=0.010$, $q_{\rm o}=-0.508$ and ${\cal R}=2.20=6.75$ Gpc. The start phase can be interpreted as the end of an inflationary epoch. The shell label $\tilde r$ is, of course, undetermined.

\subsubsection{The closed universes}
\label{sect_sync_closed}
We now turn to the synchronous universes that conform to our formal definition of a universe as embedded in an inaccessible outer space (section~\ref{sect_beyond}), which in this case implies $\epsilon=1$, together with the relation (\ref{defebound}) which states that $P=2M$. This expression provides an additional relation between the parameters.

The isometric representation is a hemisphere (${\cal Y}\ge0$) with radius ${\cal R}(t)$.
Note that this representation is the same as for the standard closed cosmological model, except at the boundary $r_{b}$. As already remarked upon in section~\ref{sect_embed}, in this paper  the hemisphere is not continued for ${\cal Y}<0$ to a full sphere as in the standard model: this universe has a center, and for $\chi>\pi/2$ we are revisiting the shell $\pi-\chi$. It bulges out over the 'black disk' with Euclidean surface area $\pi R^2$, yielding a hemisphere with the double surface area $2\pi R^2$. Yet another way to appreciate the correctness of the picture without a 'lower hemisphere' is to keep in mind that at $r_b$ the junction of the universe with the (Euclidean) outer space is continuous but not differentiable.  However, one must envisage this universe as the limit of the dust ball discussed in section~\ref{sect_dustball}, which has a junction with the outer world that is continuous and differentiable. A small (local) disturbance could then 'pinch off' the dust ball from the rest of space, thereby creating a universe. In that case, it is quite obvious that one doesn't get a lower hemisphere 'for free'. In section~\ref{sect_orbitbound} we will come back to the special status of the shell $r_{b}$.

It follows from equations (\ref{pcenter}), (\ref{mcenter1}) and (\ref{defebound}) that
\begin{equation}
\label{p_c_iso}
p_{c,{\rm sync}}=\sqrt{3\over2\kappa\rho_c}{r_{b}}^{-1}=\sqrt{1.8\over\rho_c}{r_{b}}^{-1}=(2M){r_{b}}^{-1},
\end{equation}
\begin{equation}
\label{rho_0_iso}
\rho_{c,{\rm sync}}={3\over2\kappa}{1\over(2M)^2}={0.45\over M^2}.
\end{equation}
These relations imply that out of the three parameters $p_c$, $\rho_c$ and $M$ (the parameter $r_b$ is an unimportant scaling factor of the dimensionless shell label) only one can be freely chosen. The most obvious choice is $M$, as it is related to the Schwarzschild radius of the universe, but also $\rho_{c,{\rm sync}}$ would be a good choice, as it has the unit of mass density (which is one of our 3 primary units) and it is instrumental in fixing the amount of matter via (\ref{rhocenter}):
\begin{equation}
\label{rhosync}
\rho(t) = {\rho_{c,{\rm sync}}\over \cyc^3[a,\epsilon,\omega(t+\phi)]}.
\end{equation}
Finally, with (\ref{omisoch}),
\begin{equation}
\label{omega_0_M}
\omega_{\rm sync}= {1\over2M}.
\end{equation}
The smaller $\omega$, and hence, the larger the mass, the larger the expansion time. Note also that the above relations imply that $0\le\tilde r\le1$ as should.

An interesting property of these models is that, contrary to the classical cosmological models, the magnitude of the mass density has no bearing on $\epsilon$, since $\epsilon=+1$ anyway. It follows from (\ref{rho_0_iso}) that the density is merely inversely proportional to the square of the size of the model. 

The total effective gravitating mass $M$ is poorly constrained by the Hubble parameter. The larger $M$, and thus the smaller $\omega_{\rm sync}$ with (\ref{omega_0_M}), the closer the limit (\ref{Hsmallt2}) is reached. For example, if we take $\Lambda=0$ and $M\sim100$, we find $\Omega_k\sim-0.04$, $\Omega_M\sim1.04$, $q\sim0.52$, $t_b\sim314=\pi M$, $t_{\rm o}\sim0.9$ and $\rho$ close to the critical density, since the rebound is very far in the future. This is to be compared with the case $M\sim1$, where we find $\Omega_k\sim-1.3$, $\Omega_M\sim2.3$, $q\sim1.2$, $t_b\sim3.14=\pi M$, $t_{\rm o}\sim0.75$ and $\rho\sim2.3$. Note that these two very different values for $M$ yield similar $t_{\rm o}$'s. Hence, there is, with $H_{\rm o}$ as the only constraint at least, no useful upper limit on $M$.

If we combine the expression for the radius of the synchronous universe (\ref{radius_sync}) and
the expression for the Hubble parameter (\ref{Hsmallt1}), valid for large $M$, we obtain
\begin{equation}
{\cal R}_{\rm o} = {\cal R}(t_{\rm o}) \to H_{\rm o}^{-2/3}\omega^{-1/3}=H_{\rm o} ^{-2/3}(2M)^{1/3}.
\end{equation}
Hence, the larger $M$, the smaller the fraction of the universe that will be visible for a given look back time, since the latter implies a fixed light travel distance. The horizon, measured relative to the size of the universe, therefore scales as $(t_{\rm o}/M)^{1/3}$. This result remains valid qualitatively for the more general universes we will discuss next.

\subsection{The functions \texorpdfstring{$e(r)$}{e(r)}, \texorpdfstring{$m(r)$}{m(r)} and \texorpdfstring{$\phi(r)$}{phi(r)}}
\label{sect_asynchronous}

In the general case that $\omega(r)$ or $\phi(r)$ are not constant functions of $r$, the expression for $R(r,t)$ is not separable anymore into a factor that depends only on $t$ and a factor that depends only on $r$. This complicates the mathematics considerably. Also the relations (\ref{p_c_iso}), (\ref{rho_0_iso}) and (\ref{omega_0_M}) do not hold anymore: the 3 parameters $p_c$, $\rho_c$ and $M$ now form a 3-dimensional parameter space. In the sequel, we will adopt as our basic parameters $e_c$, $m_c$ and $M$, which is an equivalent choice due to (\ref{ecenterc}). But there is much more additional freedom: in order to make the metric fully explicit, we need also to specify the free functions that appear in the radial metric coefficient $X(r,t)$.

Taking account of the boundary conditions at $r_b$ (subsection~\ref{sect_defuniverse}) and in the center (subsection~\ref{seccentral}), we now have to choose functional forms for $m(r)$, $e(r)$ and $\phi(r)$.

We adopt
\begin{equation}
\label{defer}
e(r) = \left\{
\begin{array}{ll}
\displaystyle e_c r^2&\quad {\rm for}\,\,\,r\le r_{ea} \\[2mm]
\displaystyle \frac12 - \sum_{i\ge0} {\alpha_i\over2}(r_{b}-r)^{2\lambda+i}&\quad {\rm for}\,\,\, r_{ea}\le r\le r_{b},\\
\end{array}
\right.
\end{equation}
with $\lambda>0$. We will find it useful to also include the possibility to require that 
\begin{equation}
\label{def_re}
e(r)=0 \quad {\rm for\,\, some\,\, finite\,\, interval} \quad 0\le r\le r_e>0,
\end{equation}
and additionally to include the possibility that 
\begin{equation}
\label{def_re0}
e(r_{e_0})=0 \quad {\rm at\,\, some\,\, isolated\,\, shell}\,\,r_{e_0}.
\end{equation}
Note that the synchronous $e(r)$ is included if these additional conditions are absent and $r_{ea}=r_b$. The coefficients $\alpha_i$ are determined in number and value by the optional requirements and the condition that $e(r)$ is sufficiently smooth at $r_{ea}$. There are therefore 4 to 6 parameters: $e_c$, $r_{ea}$, $\lambda$, the order of the smoothness of $e(r)$ at $r_{ea}$, and possibly $r_e$ and $r_{e_0}$.

As for $m(r)$, we also have to take account of the condition (\ref{condmr}). We adopt
\begin{equation}
\label{defmr}
\tilde m(r) = {m(r)\over M} = \left\{
\begin{array}{ll}
\displaystyle {m_c\over M}\,r^3& {\rm for}\,\,\, r\le r_{ma} \\[4mm]
\displaystyle 1 - \sum_{i\ge0} \beta_i(r_{b}-r)^{\lambda+\mu+i}& {\rm for}\,\,\, r_{ma}\le r\le r_{b},\\
\end{array}
\right.
\end{equation}
with $\lambda+\mu>0$. Again, the synchronous $m(r)$ is included if $r_{ma}=r_b$. There are therefore 5 parameters: $m_c$, $r_{ma}$, $M$, $\mu$ and the order of the smoothness at $r_{ma}$. 

Finally
\begin{equation}
\label{defphir}
\omega(r)\phi(r) = \left\{
\begin{array}{ll}
\displaystyle \omega_c\phi_c& {\rm for}\,\,\, r\le r_{\phi a} \\[2mm]
\displaystyle {\Phi\over2M} - \sum_{i\ge0}\gamma_i(r_{\phi b}-r)^{\lambda+\nu+i}\,& {\rm for}\,\,\, r_{\phi a}\le r\le r_{\phi b}\\[2mm]
\displaystyle {\Phi\over2M} & {\rm for}\,\,\, r_{\phi b}\le r\le r_{b},\\
\end{array}
\right.
\end{equation}
with $\lambda+\nu>0$. The factor $(2M)^{-1}$ appears because $\omega(r_b)=(2M)^{-1}$. The synchronous $\phi(r)=\phi_c$ has $r_{\phi a}=r_{\phi b}=r_b$ or $\Phi=2M\omega_c\phi_c$. There are 6 parameters: $\phi_c$, $r_{\phi a}$, $r_{\phi b}$, $\Phi$, $\nu$ and the order of the smoothness at $r_{\phi a}$.

Inside
\begin{equation}
\label{defrsync}
r_{\rm sync} = \min(r_{ea},r_{ma},r_{\phi a})
\end{equation}
the model is synchronous. When $r_{\rm sync}=r_{b}$, the model is fully synchronous.

Though the above parameterisations seem rather necessary in order to include non-trivial functions $e(r)$, $m(r)$ and $\phi(r)$, they imply already a rather daunting number of models that could be explored. In the sequel, we will assume $r_b=1$, which is no restriction. For the asynchronous models, we will take $r_{ea}=r_{ma}=r_{\phi a}=.1$ and $\phi_c=0$, since it seems reasonable to assume synchronicity in (the vicinity of) the center. Neither will we vary the order of smoothness of the junctions at shell $r=.1$. We also assume $r_{\phi b}=r_b$. This leaves us, apart from $e_c$, $\rho_c$ and $M$, with $\lambda$, $\mu$, $\nu$ ($\nu$ only relevant if $\Phi\ne0$) and $r_e$ or $r_{e_0}$.

\begin{figure}[h]
   \centering
   \includegraphics[width=150mm]{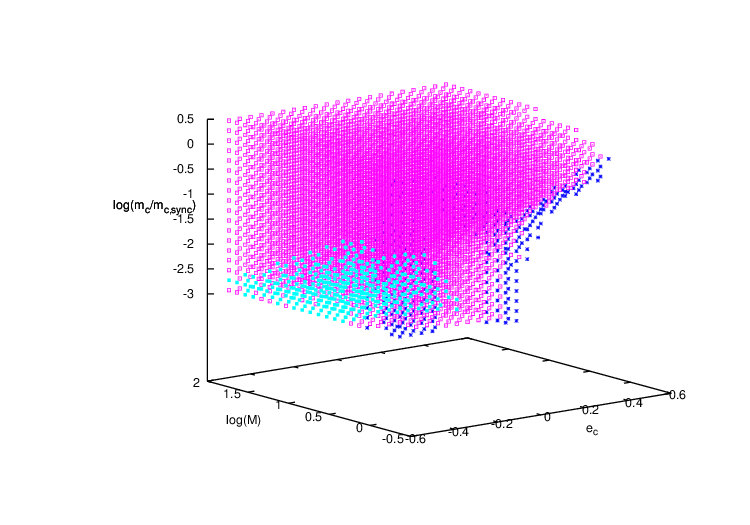}
      \caption{A rendition of parameter space for the parameters $m_c$, $e_c$ and $M$.
The various dots represent models that realize $H(r,t)$ and $I(r,t)$ within 10\% of $H_{\rm o}$ somewhere, sometime. For all the models shown there are at least some shells that realize this at cosmic time that is larger than 8 Ga. The blue dots are the models that realize this for less that 50\% of the shells, the magenta dots for more than 50\% of the shells. The cyan dots are models that allow this for at least 75\% of the shells, at a cosmic time that, of course, varies with the shell, but for which some shells allow an 'age' of more than 13 Ga. These models are somehow the 'best'. }
         \label{fig_paramscan}
   \end{figure}

As to the parameters $\rho_c$, $e_c$ and $M$, there is considerable freedom. Figure \ref{fig_paramscan} shows a typical parameter space. There is essentially little constraint on $M$ on the high side, and the parameters $m_c$ and $e_c$ have there 'home base' around the isotropic values.

\subsection{The behaviour of the metric coefficient \texorpdfstring{$X(r,t)$}{X(r,t)} at the boundary \texorpdfstring{$r_{b}$}{rb}}
\label{sect_behaviourX}

We define
\begin{equation}
\label{defx}
u = r_b - r
\end{equation}
and write, for $r\to r_{b}$, and thus $u\to0$
\begin{equation}
\label{deferlim}
2e(r) = 1 - \alpha u^{2\lambda},
\end{equation}
with $\alpha=\alpha_0>0$,
\begin{equation}
\label{defmrlim}
m(r) = M - \beta u^{\,\mu+\lambda}
\end{equation}
with $\beta=\beta_0>0$, and
\begin{equation}
\label{defphirlim}
\omega(r)\phi(r) = {\Phi\over2M} - \gamma u^{\nu+\lambda},
\end{equation}
with
\begin{equation}
\label{defgamma}
\gamma = \left\{
\begin{array}{lcl}
0&\quad {\rm if} \quad& r_{\phi b}<r_{b} \\
\gamma_0&\quad {\rm if} \quad& r_{\phi b}=r_{b}.\\
\end{array}
\right.
\end{equation}
The above expressions make $\partial_rR$ at the boundary explicit, and we obtain, after some calculations,
\begin{equation}
\label{Rasymp}
\partial_rR(u,t) \to R_\lambda(t) u^{2\lambda-1} + R_\mu(t) u^{\lambda+\mu-1} + R_\nu(t) u^{\lambda+\nu-1}
\end{equation}
with
\begin{eqnarray}
\label{drRasymp}
R_\lambda(t) &=& \alpha\lambda\left( -4M\cyc+3t\,\partial_\psi\cyc-12Ma\partial_a\cyc\right) \nonumber\\
R_\mu(t) &=& \beta(\lambda+\mu)\left( 2\cyc-{t\over M}\partial_\psi\cyc-4a\partial_a\cyc\right) \nonumber\\
R_\nu(t) &=& 2\gamma(\lambda+\nu)\partial_\psi\cyc.
\end{eqnarray}
Note that $\epsilon=1$ in the vicinity of $r_b$ by our definition of a universe, that all 3 coefficients  $R_{\{\lambda,\mu,\nu\}}(t)$ behave well at the boundary and that the three exponents $\lambda$, $\mu$ and $\nu$ are independent of time.

As a consequence
\begin{equation}
\label{Xasymp}
X(u,t) \to X_{\lambda}(t)u^{\lambda-1} + X_{\mu}(t)u^{\,\mu-1} + X_{\nu}(t)u^{\nu-1}
\end{equation}
with
\begin{equation}
\label{Xasympcoef}
X_{\{\lambda,\mu,\nu\}}(t) = \alpha^{-1/2}R_{\{\lambda,\mu,\nu\}}(t).
\end{equation}
It follows that the volume element (\ref{defvolelem}) of the metric, which equals $4\pi XR^2du$ upon integration over the angles,  is integrable if 
\begin{equation}
\label{condlamunu}
\lambda>0, \quad \mu>0, \quad \nu>0,
\end{equation}
with the proviso that every statement about or inclusion of $\nu$ here and in the sequel is only relevant if $\gamma\ne0$. This shows that also volume is a relative notion, since the stretch in the radial coordinate can become arbitrarily large. In the limit, the space geometry at the boundary therefore approaches a cylindrical symmetry (see also subsection~\ref{sect_limit}).

The inequalities in the no-collision condition (\ref{nocollision}) and the conditions (\ref{condlamunu}) imply that either
\begin{equation}
\label{lambda_frac12}
\lambda=\frac12,  \quad   \mu\ge\frac12, \quad \nu\ge\frac12   
\end{equation}
if the term in $R_\lambda$ in (\ref{Rasymp}) ensures that $\partial_rR\ne0$, 
or
\begin{equation}
\label{mu_frac12}
\frac12<\lambda<1,  \quad   \min(\mu,\nu)=1-\lambda
\end{equation}
if either the term in $R_\mu$ or $R_\nu$ ensure that $\partial_rR\ne0$. 

We note, for instance, that for the standard synchronous case $e(r)$ and $m(r)$ behave linearly at $r=r_b$, and thus $\lambda=\mu=\frac12$. 

In fact, we can transform the radial coordinate into any other increasing function of it (see section~\ref{sect_defgen}). Since the function $e(r)\to\frac12$ for $r\to r_b$ while $e(r)<\frac12$ for $r<r_b$, we could, locally, adopt $e(r)$ as the new shell label. In that case $\lambda=\frac12$. Hence parameter space could be reduced to $(\mu,\nu)$.

Similarly, $m(r)$ is a strictly increasing function in the vicinity of $r_b$ (recall that $r_b$ is the largest shell label on which comoving matter resides). Therefore we could, alternatively, adopt $m(r)$ as the new shell label locally, and thus $\lambda+\mu=1$. Hence parameter space could be reduced to $(\lambda,\nu)$ or $(\mu,\nu)$.
  
We denote
\begin{equation}
\label{defdeltahat}
0<\delta\equiv\min(\lambda,\mu,\nu).
\end{equation}
From the conditions (\ref{lambda_frac12}) and (\ref{mu_frac12}) we deduce
\begin{equation}
\delta=1-\lambda\le\frac12.
\end{equation}
We retain the symbol $\delta$, since in section we will assume different conditions from (\ref{lambda_frac12}) and (\ref{mu_frac12}).

The asymptotic behaviour of $X(r,t)$ equals 
\begin{equation}
\label{Xasymp1}
X(u,t) = X_{\rm lim}(t)u^{\delta-1}\to+\infty \quad{\rm for}\quad r\to r_{b},
\end{equation}
with
\begin{equation}
\label{defXlim}
X_{\rm lim}(t) = \sum_{\rho=\lambda,\mu,\nu=\delta} X_\rho(t).
\end{equation}
The above expressions are independent of the conditions (\ref{lambda_frac12}) or (\ref{mu_frac12}). The parameter $\delta$ is thus a measure of the steepness of the singularity of $X$ at the boundary. The smaller $\delta$, the more 'extra space' there is in the radial direction as compared to the Euclidean case. We also note that the upper limit for $\delta$, being $\frac12$, is only attained in case $\lambda=\frac12$.

As a consequence of (\ref{Rasymp}), (\ref{defdeltahat}) and (\ref{Xasymp1}) we have
\begin{equation}
\label{drRasymp1}
\partial_rR(u,t) \to \sqrt{\alpha}X_{\rm lim}(t)u^{\lambda+\delta-1}
\quad{\rm for}\quad r\to r_{b}.
\end{equation}
The conditions (\ref{lambda_frac12}) or (\ref{mu_frac12}) ensure that 
$\partial_rR(u,t) \to \sqrt{\alpha}X_{\rm lim}(t)\ne0$.

For the mass density, we find, using (\ref{defrho}), for $r\to r_b$:
\begin{equation}
\label{rhoasympshort}
\rho(u,t) \to {1.2\over R^2(u,t)}{\beta(\lambda+\mu)u^{\mu-\delta}\over \sqrt{\alpha}X_{\rm lim}(t)}.
\end{equation}
We briefly examine a few consequences in the case that $\rho(r_{b},t)=0$ for all $t$, thus $\mu>\delta$, under the conditions (\ref{lambda_frac12}) or (\ref{mu_frac12}). If $\lambda=\frac12$, then $\mu>\frac12$. If $\lambda>\frac12$, then $\nu<\mu>1-\lambda$. If, however, the phase function $\omega(r)\phi(r)$ is a constant in the vicinity of $r_b$, then the first inequality disappears, $\mu=\delta$, hence $\rho(r_{b},t)\ne0$. 

\subsection{Alternative characterisation of a universe}
\label{sect_alternuniv}

The fact that $X(r,t)\to+\infty$ for $r\to r_b$, for all $t$, points towards yet another choice of shell marker. In the radial direction
\begin{equation}
d{\cal X} = Xdr = -Xdu =-X_{\rm lim} u^{\delta-1}du = -\delta^{-1} X_{\rm lim}\,d\bigl(u^{\delta}\bigr)
\end{equation}
showing that a change of radial shell marker proportional to \mbox{$(r_b-r)^{\delta}$} will transform the metric locally at the boundary to a Lorentz metric. We will need such a transformation in section~\ref{sect_orbitbound}.

We could thus adopt a new variable with this behaviour at the boundary as an alternative shell marker. A possible choice would be 
\begin{equation}
\label{def_chiregular}
\chi(\tilde r) = \int_0^{\tilde r}\!\!{d\tilde r'\over\bigl[1-(\tilde r')^2\bigr]^{1-\delta}}=
\tilde r\,\,{}_2F_1\left(1-{\delta},\frac12;\frac32;{\tilde r}^2\right)
\end{equation}
with $0\le\tilde r\le1$ given by (\ref{def_rnorm}) and ${}_2F_1$ the hypergeometric function. The transformation preserves order, and $\chi(0)=0$. Moreover we find
\begin{equation}
0\le\chi\le{\sqrt\pi\over2}{\Gamma(\delta)\over\Gamma(\delta+\frac12)}.
\end{equation}
It reduces to (\ref{def_chi}) in the synchronous case ($\delta=\frac12$) for $\epsilon=+1$. 

With the transformation (\ref{def_chiregular}) we find that at the boundary $\tilde r=1$
\begin{eqnarray}
\partial_\chi R(\chi(1,t)) &=&\lim_{\tilde r\to 1} \left[\partial_{\tilde r}R(\tilde r,t){d\tilde r\over d\chi}\right]\nonumber\\
&=&\lim_{u\to0}\left[X(u,t)(1-{\tilde r}^2)^{1-\delta}\right]\lim_{r\to r_b}\left[\sqrt{1-2e(r)}\right]\nonumber\\
&=&2^{1-\delta}X_{\rm lim}\lim_{r\to r_b}\left[\sqrt{1-2e(r)}\right]=0.
\end{eqnarray}
This points to an alternative definition of a universe: its metric is such that the radial coefficient of the metric is regular up to the shell $r_b$ beyond which there is no mass, and for that shell $\partial_rR(r_b,t)=0$, for all $t$. Note that, because of the existence of the above transformation, the boundary is part of the manifold. 

Though little would change in the paper (since $r$ is an arbitrary shell marker anyway), we do not make this choice here, and we leave the radial marker quite generally defined. However, it sheds another light on the condition $\partial_rR(r,t)=0$ if it occurs at some $t$, such as happened in the early phases of the universe (see also section~\ref{sect_orbitbound}). Such a shell divides the universe into an inner and an outer universe, for as long as $\partial_rR(r,t)=0$ occurs.

\begin{figure}[ht]
   \centering
   \includegraphics[width=110mm]{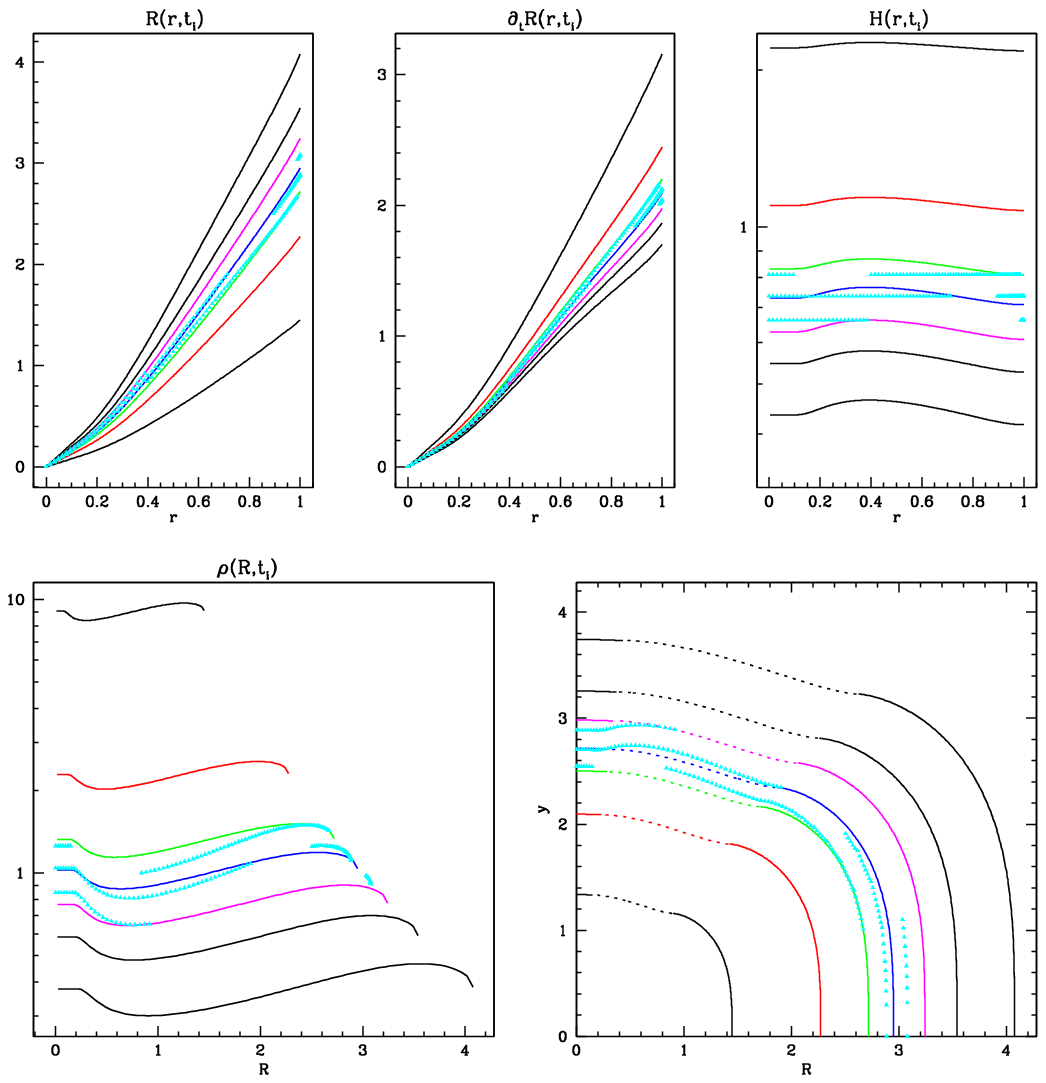}
      \caption{A few characteristics of a model with $\Lambda=0$, $\log(M)=1$, $e_c=-.15$, $\log(m_c/m_{c,\rm{sync}})=-2.8$ and $r_{e_0}=.7$. The boundary parameters are $\lambda=\mu=\frac12$ and $\gamma=0$, hence $\delta=\frac12$. In the coding of fig. \ref{fig_paramscan}, this would be a cyan model. The curves are drawn for times $t_i= $ \{0.3 (black), 0.6 (red), 0.8 (green), 0.9 (blue), 1.2 (magenta), 1.35 (black), 1.5 (black)\}. The cyan dots are plotted where and when $H(r,t)=0.736=H_{\rm o}$ and $H(r,t)=H_{\rm o}\pm 10\%$, yielding at most 3 dots for a single shell, but in addition $I(r,t)$ should not deviate more than 10\% from $H_{\rm o}$. \textbf{Top row: left panel:} the radius $R(r,t_i)$. These curves depict the transformation from shell label $r$ to radius $R$, which is time dependent. Note that the central shells hardly expand. \textbf{Middle panel:} the tangential stretch rate $\partial_tR(r,t_i)$, showing that this rate monotonically decreases with time for all shells (which is also obvious from (\ref{der2cyc})). Note also that $\partial_tR(r,t)$ can be larger than 1, which therefore shows that the tangential stress rate cannot be interpreted as a material velocity. \textbf{Right panel:} the Hubble parameter $H(R,t_i)$. Clearly also $H$ decreases monotonically with time, for all shells, as proven in section~\ref{sect_H0}. Here it can be seen best that for $r\le.1$ the model is synchronous (by design), and there $H=I$. The sharp drop at $r=.1$ is an artifact of the model, and has no particular significance. For shells larger than .2 the $H=H_{\rm o}-10\%$ are absent, which means that there $I$ deviates more than 10\% from $H_{\rm o}$. The observed $H_{\rm o}$ is (within 10\%) realised in the center at $t=1.35$ and at the boundary for $t=0.8$. At all other epochs outside this range, $H(r,t)$ is everywhere different from the observed value. This means that the expansion time (or 'age', see discussion in subsection~\ref{sect_time_evolution}) in this particular model must be between 8 and 13.5 Gyr. \textbf{Bottom row: left panel:} $\rho(R,t_i)$. The mass density is everywhere non zero, and decreases monotonically with time for all shells. \textbf{Right panel:} embedding surfaces, as defined in (\ref{defembed}), for the times $t_i$, here shown in their meridional sections. The embedding surfaces are dotted where $e(r)<0$ in which case they are to be interpreted as embedded in a Minkowski space. Note that close to the boundary these surfaces tend to spheres, because of the choice of $\lambda$, $\mu$ and $\nu$. Models with larger $M$ are very similar, and differ essentially from this one in their size.
                    }
\label{fig_model1}
   \end{figure}

 \begin{figure}[h]
   \centering
   \includegraphics[width=110mm]{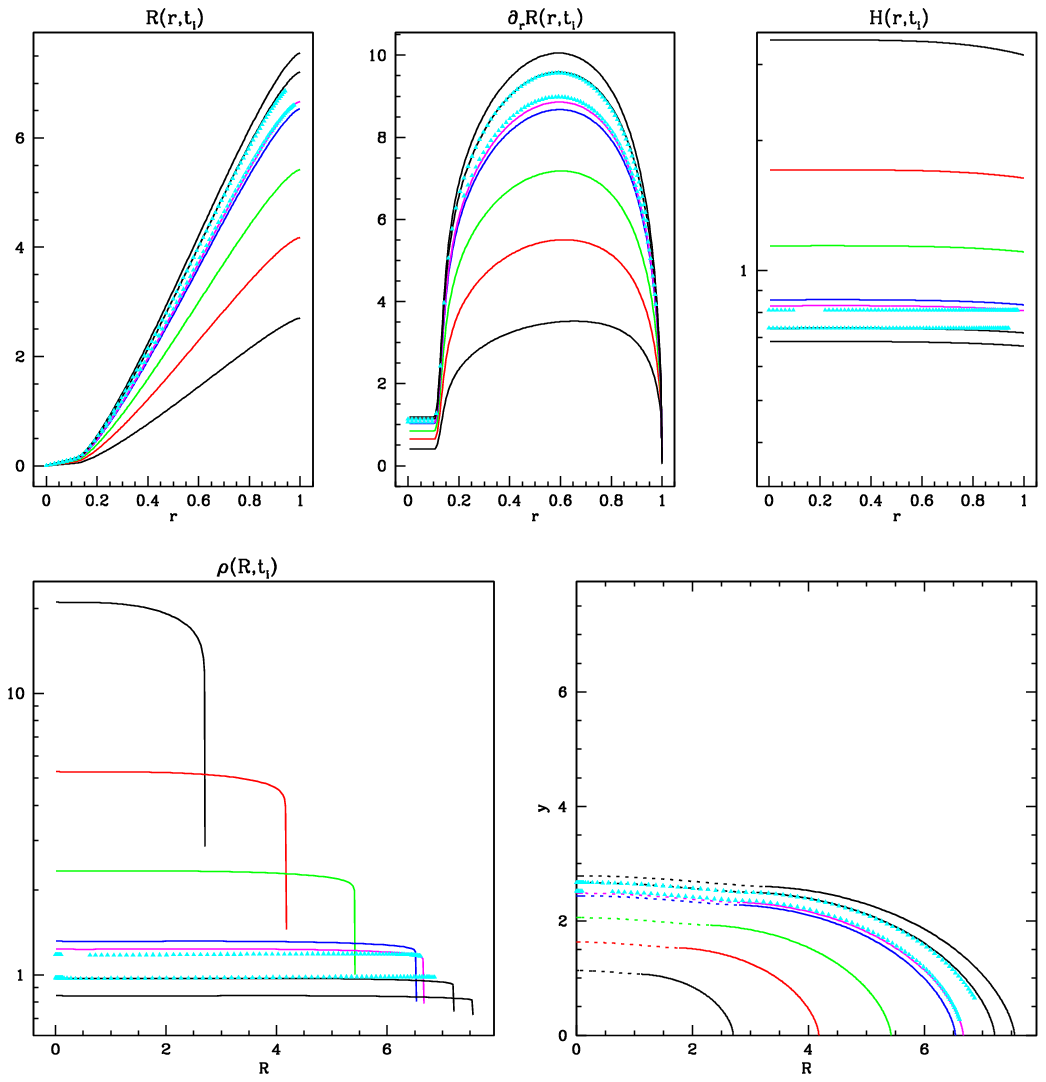}
      \caption{Similar figure as fig. \ref{fig_model1}, here for $\Lambda=0$, $\log(M)=2$, \mbox{$e_c=0$}, $\log(m_c/m_{c,\rm{sync}})=-2.4$ and $r_{e_0}=.5$~. The boundary parameters are \mbox{$\lambda=\nu=\frac12$} and $\mu=1$, hence $\delta=\frac12$. In the coding of fig. \ref{fig_paramscan}, this would be a magenta model. The curves are drawn for times $t_i= $ \{0.2 (black), 0.4 (red), 0.6 (green), 0.78 (blue), 0.8 (magenta), 0.9 (black), 0.97 (black)\}. The middle panel of the top row shows $\partial_rR(r,t_i)$. Here it is clearly visible that the model is synchronous (by design), since for a synchronous model $R\sim r$. At the boundary, the values range from $\partial_rR(r_b,t_1=.2)=.55$ to $\partial_rR(r_b,t_7=.97)=.04$ (not clearly visible). At later times, $\partial_rR(r_b,t)<0$. When $\partial_rR(r_b,t)=0$, there is a collision of shells at the boundary. Then new physics has to be introduced in order to leave that singular state. If the model can be continued up to maximum expansion, we may be in the condition explained in subsection~\ref{sect_mass_ejection} and \ref{sect_anom_red}. From the right panel on the top row we see that the values $H=H_{\rm o}-10\%$ are totally absent. In this case this comes about because they would require $t>0.97$, which is beyond the validity of the model. Note also that $\rho\to0$ at the boundary, because $\mu>\delta$ as explained in section~\ref{sect_behaviourX}. The embedding surfaces are oblate spheroids (in the regions where they are embedded in Euclidean space).
                    }
         \label{fig_model2}
   \end{figure}

\begin{figure}[h]
   \centering
   \includegraphics[width=110mm]{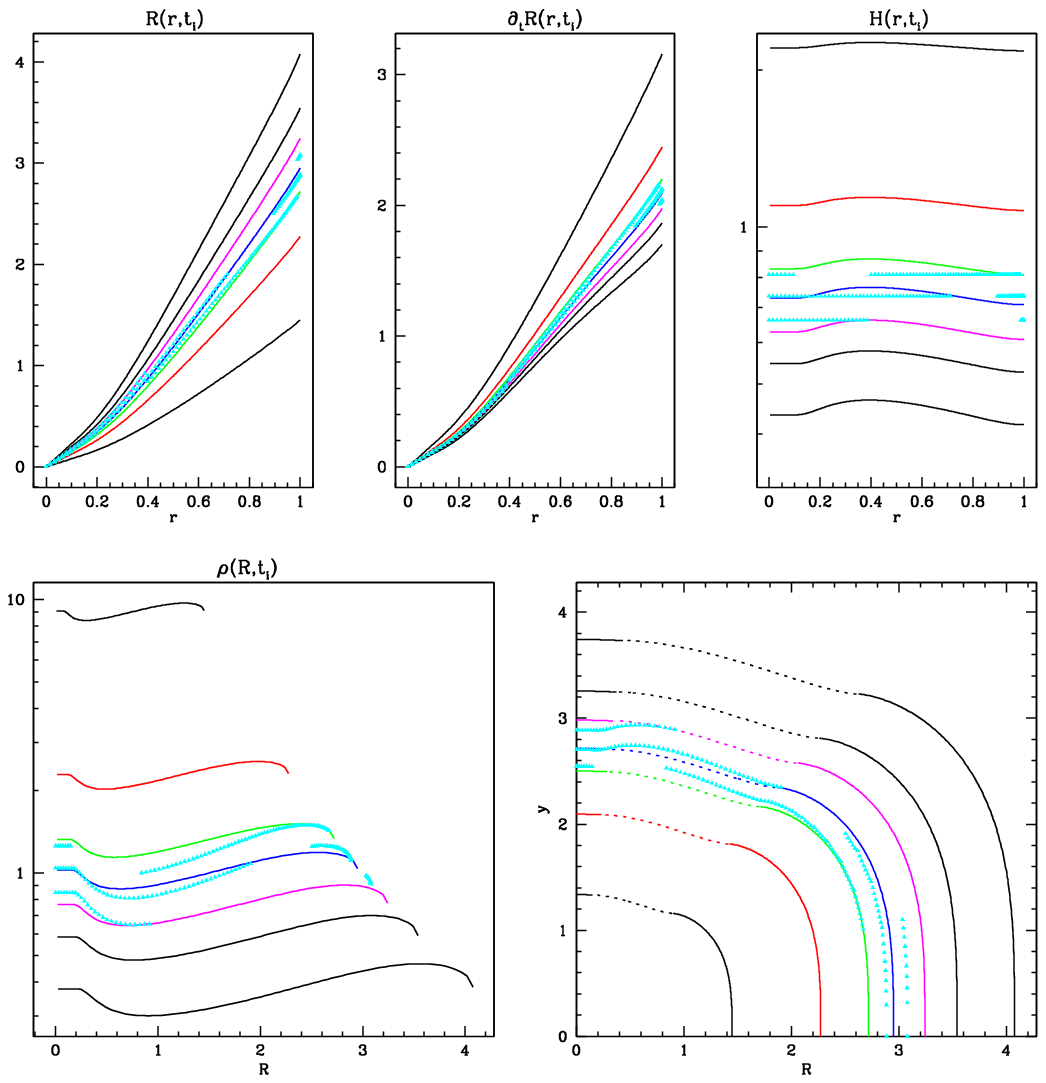}
      \caption{Similar figure as fig. \ref{fig_model1}, here for $\Lambda=0$, $\log(M)=0.9$, \mbox{$e_c=0.05$}, $\log(m_c/m_{c,\rm{sync}})=-0.8$ and $r_{e_0}=.7$~. The boundary parameters are \mbox{$\lambda=.7$}, $\mu=.3$ and $\gamma=0$, hence $\delta=.3$~. In the coding of \mbox{fig. \ref{fig_paramscan}}, this would be a cyan model. The curves are drawn for times $t_i= $ \{0.3 (black), 0.6 (red), 0.8 (green), 0.9 (blue), 1. (magenta), 1.2 (black), 1.5 (black)\}. The embedding surfaces are prolate spheroids (in the regions where they are embedded in Euclidean space).
                    }
         \label{fig_model3}
   \end{figure}

\section{Between a dust ball and a universe}

\subsection{An example}
\label{sect_limit}
We now discuss one particular class of models for which $\partial_rR(r_{b},t)=0$ at some time $t$, while $X(r_{b},t)$ is not singular nor zero, which is called a neck. Therefore the condition $\partial_rR(r_{b},t)=0$ is not visible in the metric (the radial term is non-zero and is a 'de l'H\^opital' type limit $0/0$). Out of this particular class we now consider the models for which $r_{\phi b}=r_{b}$, $\lambda>1$, $\mu>1$ and $\nu=1$, which are of a class we excluded in the previous section.

Using (\ref{drRasymp}) and (\ref{dercyc}) one finds that the metric approaches, for $r\to r_{b}$, the form
\begin{eqnarray}
\label{Schwarzlim}
ds^2 &=& dt^2 - {4\gamma^2(\lambda+1)^2\over\alpha}{1-\cyc+a\cyc^3\over\cyc}\,dr^2 - (2M\cyc)^2d\Omega^2,\nonumber\\
\end{eqnarray}
with $\cyc$ here shorthand for $\cyc(t/(2M))$, for clarity. Returning to the inner Schwarzschild-$\Lambda$ solution (\ref{InnerSchwarz}), we find that the time transformation
\begin{equation}
dt^2={dR^2\over\displaystyle{\Lambda R^2\over3}+{2M\over R}-1}
\end{equation}
happens to be the same as (\ref{soleqRtint}), and we obtain
\begin{equation}
\label{InnerSchwarz2}
ds^2 = dt^2 - {1-\cyc+a\cyc^3\over\cyc}d{\bar t}^2 - (2M)^2\cyc^2\,d\Omega^2,
\end{equation}
with again $\cyc=\cyc(t/(2M))$, which is identical to (\ref{Schwarzlim}) upon relabeling the radial coordinate $\bar t=[2\gamma(\lambda+1/)/\alpha]r$. Note also that the cylindrical geometry is quite obvious here, since both (\ref{InnerSchwarz}) and (\ref{InnerSchwarz2}) show that the angular components of the metric are independent of the radial coordinate $\bar t$.

\subsection{Mass ejection from a black hole}
\label{sect_mass_ejection}
Next to impossible as a neck (for which $X(r_{b},t)$ is not singular and not zero) may be in the spherical case globally, the condition $X(r_b,t)=0/0$ and finite may very well occur on a local scale with the spherical symmetry broken. From the asymptotic expression of $X(r,t)$ in (\ref{Xasymp}) it follows that
\begin{equation}
\lambda\ge1,\quad\mu\ge1,\quad\nu\ge1\quad {\rm and} \quad \min(\lambda,\mu,\nu)=1.
\end{equation}
When both $\partial_rR(r)$ and $\sqrt{1-2e(r)}$ are zero, i.e. when shells collide at the boundary $r_b$, the boundary could be crossed, and if this happens at maximum expansion,\footnote{In the other cases the crossing material would enter the space between the universe and the Schwarzschild radius, where only inward motion is possible and this material would therefore instantaneously enter the universe again with a locally Lorentzian outward velocity.} matter and radiation could leave the black hole. Clearly, it remains a tough act to climb the gravitational wall in order to reach an outside observer, but since the radial coefficient $X(r,t)$ is locally and temporarily regular, this is perfectly possible. We are then in the condition of a dust ball (section \ref{sect_dustball}), without Schwarzschild-$\Lambda$ radius. This condition may also be relevant for 'small' and rotating black holes (recall that the Kerr solution at the poles is locally very similar to the Schwarzschild solution), because, according to this mechanism, matter and radiation could escape from the poles of a rotating black hole, thereby causing periodic bipolar outbursts.

In order to make the structural constraints for this to happen somewhat more concrete, let us assume $\Lambda=0$, hence $a=0$. At the moment of maximum expansion, $\partial_\psi\cyc=0$, and therefore the expressions (\ref{Rasymp}), (\ref{drRasymp}), (\ref{Xasymp}) and (\ref{Xasympcoef}) reduce to 
\begin{equation}
X(u) = -4\sqrt{\alpha}\lambda M \,u^{\lambda-1} +2{\beta\over\sqrt{\alpha}}\,u^{\mu-1}
\end{equation}
with
\begin{equation}
\lambda\ge1,\quad\mu\ge1\quad {\rm and} \quad \min(\lambda,\mu)=1.
\end{equation}

\subsection{On the anomalous redshifts}
\label{sect_anom_red}

We can also make a connection with the so-called "anomalous redshifts". This phenomenon, which has been around for about 50 years now, comes down to the assertion that high redshift quasars seem physically associated with low redshift, relatively nearby and active galaxies. This interpretation has been lifelong advocated by H. Arp, though the vast majority of the astronomical community interprets these redshifts as cosmological, and puts the phenomena aside as merely chance alignments and projection effects. It does not help, of course, that there is no obvious physical explanation for Arp's interpretation. An accessible and short report on the phenomenon by Arp himself can be found in Pecker \& Narlikar (\cite{peck}). 

Since we argued in the previous subsection that, under specific conditions, material can leave a rotating black hole at the poles, we can see that the "quasars" can be explained as being such material. That material has to climb a large potential well, though, hence the redshifts observed are largely gravitational. A particularly nice example is the Seyfert NGC~3516 @ $z=0.009$, which has 6 X-ray sources @ $z=(2.1,1.4,.93,.69,.33,.089)$ aligned along the minor axis that are bipolarly distributed. Moreover, sources with higher redshifts are closer to the galaxy than those with the lower redshifts. This we can (qualitatively) explain by the overall mechanism of the ejection. At ejection, the radial coefficient $X(r,t)$ of the metric, which used to be $+\infty$, is finite but of variable magnitude depending on the details of the shell collision. The higher $X(r,t)$, the more space will have to be crossed in the radial direction, the higher the reddening will be and the more momentum will be lost. Hence the ejecta with the highest momentum at exit, that therefore get the farthest out from the galaxy, will be the lesser redshifted. In other words, the redshift is inversely proportional to the ease with which the material could escape. Of course, this assumes that the initial momentum prior to ejection is for all ejecta similar, which need not be the case. 

The case NGC~3516 also shows that the ejection can be recurrent, as demonstrated in the `ideal' case of section \ref{sect_mass_ejection}. Therefore NGC~3516 is actually also in this respect corroborating evidence for the physics proposed in this paper. 

In this context, the issue of quantized redshifts is never far away. It is conceivable that there is some relation, but in this paper we will not pursue this issue.  

\section{Orbits}
\label{sect_orbits}

\subsection{The equations of the motion}
\label{sect_orbits_eq_motion}

Due to the spherical symmetry, the motion is planar and we choose any meridional plane $\varphi$. Hence we can adopt the metric
\begin{equation}
\label{2metric}
ds^2 = dt^2 - X^2(r,t)\,dr^2 - R^2(r,t)\,d\vartheta^2,
\end{equation}
leading to the following set of equations for the geodesics:
\begin{equation}
\label{geod1}
\left\{\begin{array}{lcl}
\ddot r &=& \displaystyle -{\partial_tX^2\over X^2}{\dot r}{\dot t} - {\partial_rX^2\over2X^2}{\dot r}^2 + {\partial_rR^2\over2X^2}{\dot \vartheta}^2\nonumber\\[3mm]
\ddot t &=&\displaystyle -\frac12\partial_tX^2{\dot r}^2 - \frac12\partial_tR^2{\dot \vartheta}^2\nonumber\\[3mm]
\dot \vartheta &=& \displaystyle{h\over R^2},
\end{array}
\right.
\end{equation}
where $h$ is the specific relativistic angular momentum (a constant), and differentiation with respect to arc length $s$ is denoted by a dot.

Instead of $\dot r$, which is the shell label change rate with proper time along the orbit, we introduce the local Lorentzian radial velocity component
\begin{equation}
\label{def_wprime}
w' = X\dot r.
\end{equation}
This leaves us with the following system of 5 coupled first order differential equations:
\begin{equation}
\label{geod2}
\left\{\begin{array}{lcl}
\dot r &=& \displaystyle {w'\over X}\\[2mm]
\dot \vartheta &=& \displaystyle {h\over R^2}\\[2mm]
\dot t &=& \tilde E\\
\dot w' &=& \displaystyle -{\partial_tX\over X}\tilde E \,w' + \sqrt{1-2e}\,{h^2\over R^3}\\[2mm]
\dot {\tilde E} &=& \displaystyle -{\partial_tX\over X}{w'}^2 - {\partial_tR\over R}{h^2\over R^2}
\end{array}
\right.
\end{equation}
where we also introduced $\tilde E$. We note that, interestingly, the $\partial_rX$ has disappeared.

The initial conditions are given by the coordinates $(r_0,\vartheta_0)$ and the Lorentzian components of the velocity 
\begin{equation}
(X_0{\dot r}_0, R_0{\dot\vartheta}_0)=(w'_0,h/R_0). 
\end{equation}
Since we have an additional constant of the motion $\varepsilon$ following from the metric (\ref{2metric}):
\begin{equation}
\label{defu}
{\tilde E}^2 = \varepsilon + X^2{\dot r}^2+ R^2{\dot\vartheta}^2 = \varepsilon + {w'}^2+ R^2{\dot\vartheta}^2,
\end{equation}
where $\varepsilon=1$ for a massive particle and $\varepsilon=0$ for light, we have also
\begin{equation}
\label{defu0}
{\tilde E}^2_0= \varepsilon+{w_0'}^2+ R^2_0{\dot\vartheta}^2_0,
\end{equation}
which completes the necessary initial conditions. The quantity $\tilde E$ is therefore the relativistic energy $E$ relative to the rest mass energy $m_0c^2$ as measured by a comoving observer \mbox{$(w'_0=\dot\vartheta_0=0)$}:
\begin{equation}
\label{defu1}
\tilde E = {E\over m_0c^2} = {dt\over ds}.
\end{equation}

In the actual integrations we make no use of (\ref{defu}), since efficiency considerations are not critical, and (\ref{defu}) can rather be used to check the accuracy of the integrations. Also, we replace the derivatives $d\over ds$ on the left hand sides of (\ref{geod2}) and (\ref{geod3}) with $d\over dt$, which means that the right hand sides are divided by $\tilde E$. This has the advantage that the independent variable of the orbits is the cosmic time, on which all observers agree. We retain the information on the arc length by replacing $\dot t=\tilde E$ by 
\begin{equation}
{ds\over dt}= {\tilde E}^{-1}.
\end{equation}

\subsection{The orbits at the boundary \texorpdfstring{$r_{b}$}{rb}}
\label{sect_orbitbound}
\subsubsection{The equations}
\label{sect_orbitbound_equations}

The function $X(r_{b},t)$ is singular. In order to carry the integration over that limit in a stable way, we set
\begin{equation}
\label{defxiOP}
r = r_{b}(1-|\xi|^{\hat\delta}),\qquad -1\le\xi\le1,
\end{equation}
with 
\begin{equation}
\label{defdelta}
\hat\delta = {\delta}^{-1}\ge2
\end{equation}
and $\delta$ defined in (\ref{defdeltahat}). 
Inversion leads to
\begin{equation}
\label{defxiOP1}
u = r_{b}-r=r_{b}|\xi|^{\hat\delta}\quad,\qquad |\xi|=\left({u\over r_b}\right)^{1/\hat\delta}=\left(1-{r\over r_b}\right)^{\delta}.
\end{equation}
We refer for this choice to section~\ref{sect_alternuniv}, where it was shown that such a change in shell label transforms the singular metric at the boundary into a Lorentz 
metric. We denote
\begin{equation}
\label{defepsxi}
\epsilon_\xi = {\rm sign}(\xi).
\end{equation}
As a consequence of the introduction of this new shell label $\xi$, the disk $0\le r\le r_b$ is now covered by 2 sheets, one for $\xi\ge0$, or $\epsilon_\xi=+1$, and one for $\xi\le0$, or $\epsilon_\xi=-1$.

In order to verify that the coordinate transformation (\ref{defxiOP1}) is consistent with the radial velocity $w'$, we write $w'$ in terms of $\xi$ and $\dot\xi$:
\begin{equation}
\label{def_wprimexi}
w'=\epsilon_\xi w_\xi\dot\xi
\end{equation}
with
\begin{equation}
\label{def_wxi}
w_\xi=-r_{b}\hat\delta X|\xi|^{\hat\delta-1}=-r_{b}^{\delta}\hat\delta X u^{1-\delta}<0.
\end{equation}
When $r\to r_{b}$, $w_\xi$ remains finite if and only if, taking into account the asymptotic behavior of $X(r,t)$ in (\ref{Xasymp1}), definition (\ref{defdelta}) holds. Hence we find
\begin{equation}
\label{wxilim}
w_\xi=-r_{b}^{\delta}\hat\delta X_{\rm lim} \quad {\rm for} \quad r= r_b .
\end{equation}

With the expressions (\ref{def_wprimexi}) and (\ref{def_wxi}) we see that $w'$ is a step function at $r=r_{b}$. This follow also from (\ref{def_wprime}), since $\dot r$ changes sign at $r_b$ ($\dot r=0$ at $r_b$). The velocity $w'$ is thus the Lorentzian radial velocity on sheet $\epsilon_\xi=+1$ and the negative of it on sheet $\epsilon_\xi=-1$. We would thus need to stop and restart the integration upon arrival at that shell. We can avoid this by defining
\begin{equation}
\label{defwvel}
w = \epsilon_\xi w'=\epsilon_\xi X\dot r= {w_\xi\,\dot\xi},
\end{equation}
where we used (\ref{def_wprimexi}). This makes $w$ the true local Lorentzian radial velocity.

The equations involving $w$ and $\xi$ in (\ref{geod2}) now read
\begin{equation}
\label{geod3}
\left\{\begin{array}{lcl}
\dot \xi &=& w_\xi^{-1} w\\
\dot w &=& \displaystyle -{\partial_tX\over X}\tilde E\, w + \sqrt{1-2e}\,{h^2\over R^3}\epsilon_\xi\\[2mm]
\dot {\tilde E} &=& \displaystyle -{\partial_tX\over X}w^2 - {\partial_tR\over R}{h^2\over R^2}.
\end{array}
\right.
\end{equation}
These equations behave well at $r=r_{b}$.
The appearance of $\epsilon_\xi$ in (\ref{geod3}) is a consequence of the shift from $\dot r$ to the Lorentzian $w$, since we need to take account of the sign of $\xi$.

In the case that $e(r)=0$ throughout, we can keep the same formalism, except that we now define $\hat\delta=1$:
\begin{equation}
\label{defxiCS}
r = r_{b}(1-\xi),\quad {\rm with} \quad -\infty<    \xi\le1.
\end{equation}
Clearly, there is only one sheet.

Of course, we can arrive at the same conclusions with respect to the behaviour of $r(s)$ in the vicinity of the boundary solely from an analysis of the geodesics (\ref{geod1}). The first equation can be rewritten as
\begin{equation}
\ddot u + 2{\partial_tX\over X}\dot u \tilde E- {\partial_rX\over X}{\dot u}^2 + {1-2e\over\partial_rR}{h^2\over R^3} = 0.
\end{equation}
It reduces for $u\to0$ to
\begin{equation}
\ddot u + 2{\partial_tX_{\rm lim}\over X_{\rm lim}}\dot u \tilde E + {\delta-1\over u}{\dot u}^2 + X_{\rm lim}^{-1}{h^2\over R^3}u^{\lambda} = 0,
\end{equation}
taking into account (\ref{Xasymp1}) and (\ref{drRasymp1}). 
The fourth term tends to zero, and we retain
\begin{equation}
\label{orbit_at_boundary}
\ddot u + 2{\partial_tX_{\rm lim}\over X_{\rm lim}}\tilde E\dot u+{\delta-1\over u}{\dot u}^2 = 0.
\end{equation}
There is an exact solution for such an equation, in the assumption that 
\begin{equation}
A=2{\partial_tX_{\rm lim}\over X_{\rm lim}}\tilde E
\end{equation}
is slowly varying (thus assuming it to be constant). In that case the equation does not explicitly depend on $s$, and a substitution $f=\dot u$ yields a first order linear differential equation in $f(u)$. We find the following solution, including the condition $u(0)=0$:
\begin{equation}
u(s) = \left|C\bigl(1-e^{-As}\bigr)\right|^{\hat\delta},
\end{equation}
with $C$ an arbitrary constant.\footnote{In the general solution, the other arbitrary constant is a shift of the origin, since (\ref{orbit_at_boundary}) does not depend explicitly on $s$ (assuming $A$ to be a constant).}

For small $s$ and/or $A$,
\begin{equation}
u(s) = |CAs|^{\hat\delta}.
\end{equation}
This result means that the variable $\xi$ as defined in (\ref{defxiOP}) or (\ref{defxiOP1}) is indeed a linear function of $s$ in the vicinity of the boundary. If $\hat\delta=1$, we recover (\ref{defxiCS}).

Finally, we note that $u(s)=0$ is also a solution. From (\ref{geod2}) and/or (\ref{geod3}) it can be seen that this is the only 'circular' orbit to be found, be it an unstable one.

\subsubsection{The physical nature of the boundary \texorpdfstring{$r_{b}$}{rb}}
\label{sect_physnatrb}

The definition (\ref{defwvel}) has a very physical rationale: a traveler (not comoving) who arrives at the boundary will not experience a jolt of any kind, but will simply continue his/her journey 'outwards', though he/she will move 'inwards' because he/she revisits shells with decreasing label $r$. This must, as it is a consequence of the analysis in section~\ref{sect_infall}: if a particle would travel 'beyond' $r_b$, it would enter the space-time between $R(r_b,t)$ and the Schwarzschild radius $R_S$ where outward travel is impossible. The strange situation thus occurs that our outward traveler will at the boundary have to change $\epsilon_\xi$, which is only of relevance with respect to the shell labels in the universe, not locally. Put differently, an observer who sends a projectile outwards (e.g. on an inertial track), will get it back, but with opposite $\epsilon_\xi$.

All orbits that reach the boundary will thus reflect from the boundary in the $(r,t)$ picture: tangential velocity is conserved, shell label change rate flips sign. The boundary is therefore a gravitational mirror. If close enough to the boundary in order to allow for short enough light travel times, one could see 2 images of the same object, or actually see oneself in the mirror (but at an earlier epoch).\footnote{This is of course not different from a mirror as we know it in daily life, but there time differences are of the order of $10^{-8}$s.}  In a cosmological setting  these images would correspond to different ages, and different aspect angles, so these would be hard to identify. The only (theoretical) way an observer could ascertain his/her position as being on the boundary, and at the same time detect the direction towards the center, is by identifying 2 lines of sight, in opposite directions, that contain the same objects, as one would do with an ordinary mirror. The radial direction is the only one on which one would see the same objects, with the same aspect angle and the same age (but see next paragraph). One way to distinguish which direction is the outward one is to observe additional objects, which are to be found in the outward direction since these are falling in from the mother universe. Barring these, one could state that, in some sense, 'inward' and 'outward' do not exist there.

We now investigate what happens with the spin of an object when it arrives at the boundary. Since spin changes sign when reflected in a mirror, we can expect that the spin, as defined with reference to the universe, will change sign after touching the boundary. To see intuitively how this works, we first consider a comoving observer who sees a spinning traveler. Both are close to the boundary. Both will see the local universe direct and mirrored: the observer thus sees the traveler in a direct image and in a mirrored image (just like an ordinary mirror), and these images will have opposite spin. When the traveler approaches the boundary, the observer will see both images of the traveler coming closer together. When the traveler is  sufficiently close to the boundary, the light paths towards the observer will eventually coincide, and the observer cannot tell which image of the traveler is direct, and which image is reflected. After that, both images will again part. None of the images change spin, they 'only' exchange the adjectives 'direct' and 'reflected'. Of course, the traveler didn't physically change spin, as defined with reference to his/her own local reference frame. In other words: suppose that the traveler leaves the observer on a journey, and both have (thus) had the occasion to physically ascertain their relative spins. When, after recoiling at the boundary, the traveler returns to the observer, both will find the other with a spin opposite to the spin at departure. Stronger still: when you send off a right hand in the radial outward direction, you will get a left hand back, since it literally and physically flew into its mirrored world.

What is the point of view of the traveler with respect to the universe? As he/she approaches the boundary, the hemisphere of the local universe centered around the anti-center $C'$ will start to look like the hemisphere of the local universe centered around the center $C$, up to the moment that they are equal upon arrival at the boundary (suppose no infalling material or radiation from outside the universe). The 2 directions towards the 'center' are then on 2 diametrically opposite points. Suppose now that the observer's spin is tangential, i.e. perpendicular to the direction of the center $C$. The plane perpendicular to his/her spin we call the horizontal plane, and we call the zenith the direction of his/her spin vector. How could the traveler determine his/her spin with respect to the universe? He/she could look at 2 standard stars $A$ and $B$ on the horizon with an angular separation of, say, a few degrees, and observe which of the stars passes first his/her line-of-sight. Let $A$ be left from $C$, and $B$ right. Thus $B$ passes before $A$ (northern pole rotation). This is a clear procedure if far enough from the boundary, but when approaching the boundary, the region around the anticenter $C'$ increasingly starts to look like the region around the center $C$, as already indicated. At the boundary, the mirror image of $B$ is now left from the mirror image of $C$ in the direction $C'$, and the mirror image of $A$ is now to right. Hence $A$ passes before $B$,  causing the determination of the opposite spin. The same effect is also clear when looking perpendicular to $CC'$ where you see the mirroring right in the field of view. On the boundary, there is no way to tell 'direct' from 'mirrored'. Right after touching, the direct image becomes the mirror image and vice versa. After some time after touching, the mirror images disappear, and the determination of the spin will be permanently towards the nadir. Again, the traveler didn't change spin, the image of the universe did.

Clearly, physically, there is nothing special going on. Our 'outbound' traveler is continuing traveling in his/her view in an outward direction (not necessarily on a radial orbit of course): his/her orbit has no turning point, except in a coordinate system that has a singular radial metric coefficient, such as $(r,t)$ but unlike the local Lorentzian system. The effect is apparent in the sense that it is the universe that presents itself 'upside down'. In order to make this clear, we recall that in the standard cosmological picture of a closed universe as the surface of a complete sphere, the traveler simply would have passed from the 'upper hemisphere' to the 'lower hemisphere', and would continue exploring new horizons. Not so here, as explained in section~\ref{sect_embed}. Our outbound traveler flies towards and into the mirror image of his/her universe.

We now decompose the spin vector into its radial component and tangential component. The radial component will not flip when touching the boundary, the tangential will.  Suppose that the observer emits a photon radially outward. Because of helicity, it has a spin with only a radial component. He/she will get that photon back, but with the opposite helicity: parity is not conserved. The boundary could therefore be a model for spinor-like behaviour. We recall that at the boundary, space is as Lorentzian as anywhere else, since the singularity in the $(r,t)$ coordinate system can be transformed away (section~\ref{sect_alternuniv}). Hence the analogy of the sphere is locally valid.

This brings us to yet another characterization of the boundary. Instead of passing from the upper hemisphere to the lower hemisphere as in conventional cosmology, the outward traveler passes from the upper side of the (only) hemisphere, to the underside of the (same) hemisphere. Hence, when the traveler happens to return to the shells he/she has already visited on his/her outward journey, he/she will find everything left handed that used to be right handed, and vice versa! In the synchronous case \ref{sect_synchronous}, this translates into 'standing on' the hemisphere for $0\le\chi<\pi/2$ and 'hanging down from' the same hemisphere for $\pi/2<\chi\le\pi$. Spin and helicity are proportional to $\cos\chi$.

Finally, the boundary is a special place because there material objects and photons can appear 'out of nothing', i.e. they are falling into the universe. As already noted in the discussion connected with (\ref{entervelocity}), they can have any radial velocity in the local Lorentzian frame, including thus positive ones. 

\begin{figure}[ht]
   \centering
   \includegraphics[width=15cm]{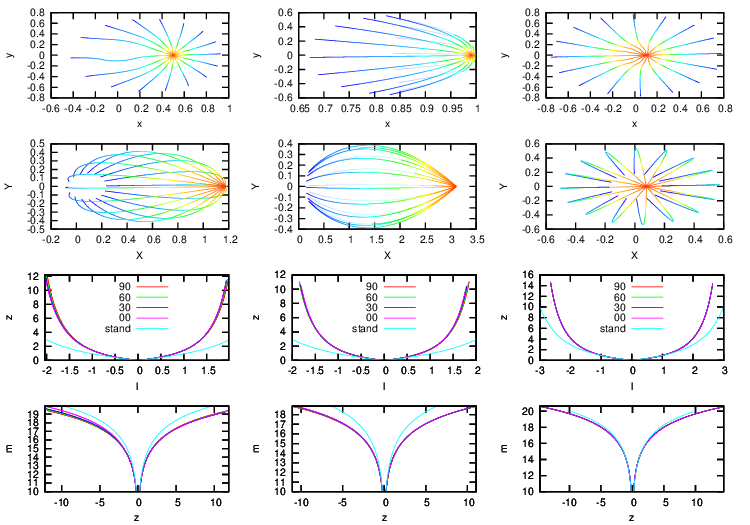}
      \caption{\textbf{Two top rows:} Light paths in the same model as the one in figure \ref{fig_model1}. The colours in the paths are indicative of the cosmic time. Light travels from the distant past (blue, $t_{\rm e}=200$ Ma in all 3 cases) to the present (red, defined as the cosmic time $t_{\rm o}$ when the Hubble parameter is the observed one, which is dependent on position). The top row shows paths in coordinate space $(x,y)=(r\cos\vartheta,r\sin\vartheta)$. The second row shows paths in physical space $({\cal X},{\cal Y})=(R\cos\vartheta,R\sin\vartheta)$. For clarity, the aspect ratio is not one-to-one.The right column shows light that arrives at a point on the border of the synchronous volume $(r_{\rm o}=.1)$ at $t_{\rm o}=12$ Ga, the arrival point in the middle column is close to the boundary ($r_{\rm o}=.99$, $t_{\rm o}=8.8$ Ga) and the arrival point in the left column is on shell $r_{\rm o}=.5$ and $t_{\rm o}=9.3$ Ga. In all cases, the arrival point has $\vartheta_{\rm o}=0$, and 16 light paths are shown that arrive from various directions, from $\theta_{\rm o}=2.5^\circ$, every $22.5^\circ$. Only in the middle column are there photons that have recoiled at the boundary. The part of their path on the other sheet than the one of the observer is drawn in lighter hues. On the blue side of the rays the photons were traveling when the universe was very young, hence they all tend to $({\cal X},{\cal Y})=(0,0)$, which is the stationary center of the universe, from which it evolved from 'a point' at $t=0$. The 3D impression which one may get from the figures on the second row is false. This is especially clear from the left and right panels: none of the light rays reach the boundary.\newline 
\textbf{Third row:} the Hubble law $z(\ell)$ for different directions (indicated in degrees) and compared with the standard model. Negative $\ell$ correspond to the indicated angle $+180^\circ$. The Hubble law is very similar in all directions, which is by design, since $H$ and $I$ do not differ by more that 10\%. From the middle column it is clear a recoil from the boundary has no effect on $z(\ell)$, showing again that the boundary is not a locus of singularity.\newline
\textbf{Bottom row:} the magnitude-redshift relation $m(z)$ for objects with an absolute magnitude of $M_{\rm bol}=-30$. Negative $z$ again correspond to the angle $+180^\circ$. The colours have the same meaning as on the third row. 
                    }
         \label{fig_model1_light}
   \end{figure}

\begin{figure}[ht]
   \centering
   \includegraphics[width=15cm]{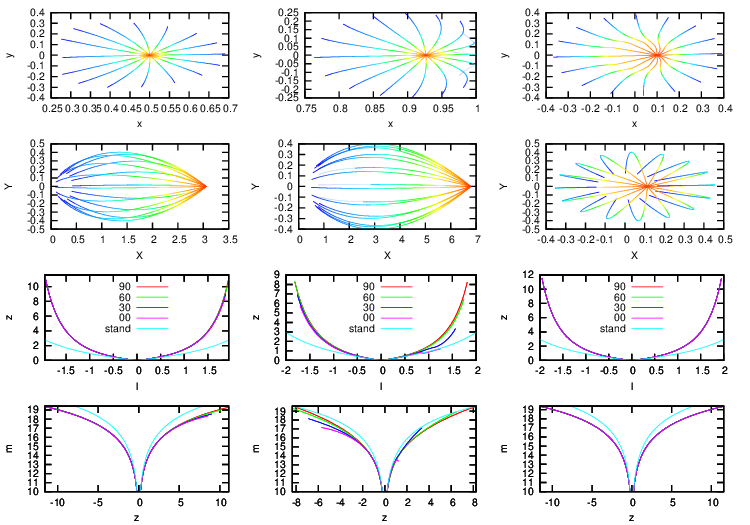}
      \caption{Same as fig. \ref{fig_model1_light}, here for the model of fig. \ref{fig_model2}.
The right column shows light that arrives at a point on the border of the synchronous volume $(r_{\rm o}=.1)$ at $t_{\rm o}=9$ Ga, the arrival point in the middle column is close to the boundary ($r_{\rm o}=.93$, $t_{\rm o}=8.9$ Ga) and the arrival point in the left column is on shell $r_{\rm o}=.5$ and $t_{\rm o}=9$ Ga. We note that the relations $z(\ell)$ and $m(z)$ at the observer close to the boundary (middle column)  start to deviate from each other at larger $z$. }
         \label{fig_model2_light}
   \end{figure}

\begin{figure}[ht]
   \centering
   \includegraphics[width=15cm]{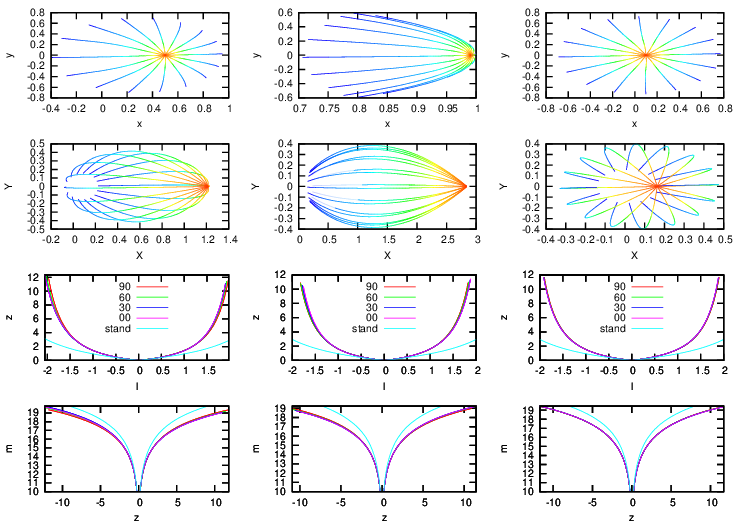}
      \caption{Same as figs. \ref{fig_model1_light} and \ref{fig_model2_light}, here for the model of fig. \ref{fig_model3}.
The right column shows light that arrives at a point on the border of the synchronous volume $(r_{\rm o}=.1)$ at $t_{\rm o}=9$ Ga, the arrival point in the middle column is close to the boundary ($r_{\rm o}=.99$, $t_{\rm o}=8.7$ Ga) and the arrival point in the left column is on shell $r_{\rm o}=.5$ and $t_{\rm o}=9.3$ Ga. In the spatial view in the middle column we see from the change of the color hue that the rays arriving at $2.5^\circ$, $25^\circ$ and $-20^\circ$ have recoiled at the boundary during their journey. This is hard to see in the coordinate view, because the prolate embedding diagram of fig. \ref{fig_model3} shows that there is a lot of radial distance to cross for little progress in shell label. 
       } 
         \label{fig_model3_light}
   \end{figure}

\section{Light}
\label{sect_light}
\subsection{Redshift}
\label{sect_redshift}

The redshift $z$ is calculated by considering 2 photons separated by 1 light period $\delta t$, on paths $t(p)$ and $t(p)+\delta t(p)$, as seen by co-moving observers. Since $\delta t$ equals the wavelength ($c=1$), $d_p\delta t(p)\,\Delta p$ is the wavelength change on the path over a $\Delta p$, and hence ${d_p\delta t\over\delta t} \Delta p= d_p z\,\Delta p$ the change in the redshift $z$ over that $\Delta p$. From (\ref{4metric}) we have $g_{00}=1$ and $g_{0,i}=0$, $i=1,2,3$, so we can use the theorem in appendix~\ref{sect_transfer}, more in particular (\ref{eqntau_LT}), which states that $\tilde E\,\delta t$ is a constant along the orbit, and thus
\begin{equation}
\label{cnstauu}
\tilde E\,\delta t=\tilde E_0\delta t_0=\delta t_0.
\end{equation}
The second equality expresses the assumption that the photon is emitted at $t_0$ in the rest frame. 

Since we have ${\tilde E}(p)$ available from the integration, we can calculate redshifts as follows. Suppose we integrate forward in time between 2 cosmic times $t_0<t_1$, and consider a photon with wavelengths $\lambda_0=c\delta t_0=\delta t_0$ and $\lambda_1=c\delta t_1=\delta t_1$. Then
\begin{equation}
\label{redshift_forward}
z = {\lambda_1-\lambda_0\over\lambda_0}={\delta t_1-\delta t_0\over\delta t_0} = {\delta t_1\over\delta t_0}-1= {1\over {\tilde E}(t_1)}-1.
\end{equation}
When we integrate backward in time between 2 cosmic times $t_0>t_1$, we obtain
\begin{equation}
\label{redshift_back}
z = {\lambda_0-\lambda_1\over\lambda_1}= {\tilde E}(t_1)-1.
\end{equation}
In the latter case, $t_1$ is the time at emission $t_{\rm e}$, and we recover of course the well-known
\begin{equation}
{\delta t_{\rm e}\over\delta t_0}= {1\over1+z}.
\end{equation}
In order to derive the Hubble law, we use (\ref{defu}) and (\ref{eqntau_app}), the latter of which reads
\begin{equation}
\label{eqntau}
{d_p\delta t\over\delta t} = d_p z =\partial_t{\tilde E}.
\end{equation}
We obtain the following differential equation for the redshift along a light path:
\begin{equation}
\dot z = {1\over2\tilde E}\partial_t\left(X^2{\dot r}^2+ R^2{\dot\vartheta}^2\right)={1\over\tilde E}\left(X\partial_tX\, {\dot r}^2+ R\partial_tR\,{\dot\vartheta}^2\right).
\end{equation}
For a light ray making an angle $\theta$ with the outward pointing radial direction, we can set
\begin{equation}
\label{def_lightangle}
X_0\dot r_0 = \cos\theta_0  \qquad {\rm and} \qquad R_0\dot\vartheta_0=\sin\theta_0.
\end{equation}
The Hubble law expresses redshift as a function of the geometrical distance $\ell$ from the emitter to the observer at some time $t_0$, at the state the universe is in at $t_0$. It is therefore only meaningful if the universe did not change appreciably during the travel time of the photons. If we insist on $z(\ell)$ without restrictions on the size of $\ell$, though there is no photon that could ever have traveled that $\ell$, this distance satisfies
\begin{equation}
\dot\ell=\sqrt{X_0^2{\dot r}^2+ R_0^2{\dot\vartheta}^2},
\end{equation}
which can be added to the set (\ref{geod2}). Locally, we have
\begin{equation}
\label{Hubble}
{dz\over d\ell} = {1\over {\tilde E}}{\displaystyle {\partial_tX\over X} \cos^2\theta+ {\partial_tR\over R}\sin^2\theta \over\sqrt{X_0^2{\dot r}^2+ R_0^2{\dot\vartheta}^2}}.
\end{equation}
Since in this paper $t\equiv c\,{\rm t}$, the replacement of the derivative with respect to $ct$ with a derivative with respect to t would result in multiplying the left hand side with $c$, yielding the familiar velocity ${\upsilon}$.

Equation (\ref{Hubble}) gives us occasion to define the radial Hubble parameter 
\begin{equation}
\label{radHubble}
I(r,t) = {\partial_tX(r,t)\over X(r,t)}={\partial_{tr}^2R\over\partial_rR}, 
\end{equation}
and thus
\begin{equation}
\label{Hubble1}
{dz\over d\ell} = {1\over {\tilde E}}{I(r,t) \cos^2\theta+ H(r,t)\sin^2\theta \over\sqrt{X_0^2{\dot r}^2+ R_0^2{\dot\vartheta}^2}}.
\end{equation}
In the limit $\ell\to0$ and for the current epoch $t_{\rm o}$, we obtain for (\ref{Hubble})
\begin{equation}
\label{Hubble_small}
z = \left[H_{\rm o} + (I_{\rm o}-H_{\rm o})\cos^2\theta_{\rm o}\right]\,\ell.
\end{equation}
We see that in principle we could recover the Hubble parameter in its standard definition by observing the velocity-distance relation in directions perpendicular to the center of the universe. The above equation provides also a means to detect the direction of the center of the universe in a perfectly spherical world.

Figures \ref{fig_model1_light}, \ref{fig_model2_light} and \ref{fig_model3_light} show some typical light paths. We refer to the captions for more information.

\subsection{On the relation between \texorpdfstring{$H(r,t)$}{H(r,t)} and \texorpdfstring{$I(r,t)$}{I(r,t)}}
\label{sect_HI}
In the general case, $I(r,t)$ is quite independent of the tangential Hubble parameter $H(r,t)$. In the synchronous case $I(t)=H(t)$ everywhere, though.

It is not hard to produce models for expanding universes, which have $\partial_tR/R>0$ (of course), while $\partial_tX/X<0$ in some parts! Since we can exclude these occurrences, such as others whereby $H$ and $I$ differ by more than, say $20\%$, this constraint is an important one in order to select regions in a model that suit the observational constraints. Note however that in the center $H(t)=I(t)$ and therefore this constraint is of no significance in the central regions of any model we consider in this paper.

It turns out that
\begin{equation}
\partial_rH = (I-H){\partial_rR\over R}.
\end{equation}
The sign of $\partial_rH$ is thus the same as the sign of $I-H$.

We can make this qualitative statement somewhat more quantitative when we express $\partial_rH$ as a function of $p(r)$ and ${\rm cyc}(a,e,\psi)$ as we know from (\ref{Rexplgen}). We find
\begin{equation}
\partial_rH={d_r\omega\over\omega}H-\omega\,d_r\psi\left(\frac12\cyc^{-3}-a+{H^2\over\omega^2}\right).
\end{equation}
The appearance of $d_r\psi=d_r[\omega(r)(t+\phi(r))]$ makes it clear that the phase function will be instrumental if it is desired to keep $I$ close enough to $H$.

\subsection{The observed surface brightness and flux}
\label{sect_obsflux}
Because isotropy is lost in the spherical case, an expanding light cone will not generate a constant flux on its surface. Therefore we have to analyze the transformation of surface brightnesses and fluxes in a more fundamental manner than what is sufficient in the isotropic case.

Consider 2 photons that are emitted in directions that differ infinitesimally by $d\theta$. They will separate from each other along the light path by an infinitesimal distance $dl$, perpendicular to the path.
In the Euclidean case we have
\begin{equation}
{dl\over d\theta} = \ell,
\end{equation}
with $\ell$ the geometrical distance, and this expression is valid independent of direction and place of emission. The above expression is the basis of a flux calculation. It is also known as the angular distance, often denoted by $D_A$. We will need to define $D_A$ at observer and emitter, $D_{\rm o}$ and $D_{\rm e}$ respectively (appendix~\ref{sect_angdistance}).
In our case, the above expression is only an approximation for small $\ell$, because here $dl/d\theta$ depends on the light ray, and must be calculated via an integration. Obviously, this complicates flux calculations considerably. To that end we derive in appendix~\ref{sect_angdistance} two additional differential equations (\ref{eqdeltal}) and (\ref{deltaw1}) that add to the equations for the light ray.

In appendix~\ref{sect_surface brightness} we calculate the surface brightness along the light ray (\ref{transintens1}), defined as the power of the light ray at a certain wavelength $\lambda$, within a wavelength range $d\lambda$ and an elementary solid angle $d\Omega$, and incident on an elementary surface $dS$. As is well known, and contrary to the Euclidean case, it is not a constant anymore. 

In appendix~\ref{sect_flux} we calculate the apparent magnitude of an object with a given absolute magnitude at emission time as the expression (\ref{transmagn}). 

\subsection{The light ray for small orbit parameter \texorpdfstring{$p$}{p}}

\subsubsection{The light ray}
We find in the vicinity of $(\xi,\vartheta)=(\xi_0,0)$ or $(r_0,0)$, with $r_0\ne r_b$, in first order in $p$
\begin{equation}
\label{xi_firstorder}
\xi(p) = \xi_0 + \xi_1\tilde E_0p + {\mathcal O}(p^2) \quad  {\rm with} \quad \xi_1 = w_{\xi,0}^{-1}\cos(\theta_0)
\end{equation}
or
\begin{equation}
\label{r_firstorder}
r(p) = r_0 + \epsilon_\xi{\cos(\theta_0)\over X_0}\tilde E_0p + {\mathcal O}(p^2).
\end{equation}
In these expressions, $w_{\xi,0}=w_\xi(r_0,t_0)$ and $X_0=X(r_0,t_0)$. When $\theta_0=\pi/2$, we have to go to second order. We find
\begin{equation}
\xi(p) = \xi_0 + \xi_2{\tilde E_0}^2p^2 \quad\!\! {\rm with} \quad\!\! \xi_2=\frac{\epsilon_\xi}2w_{\xi,0}^{-1}
\sqrt{1-2e_0}{\sin^2\theta_0\over R_0}.
\end{equation}
or
\begin{equation}
r(p) = r_0 + {\sqrt{1-2e_0}\over X_0}{\sin^2\theta_0\over R_0}{\tilde E_0}^2p^2,
\end{equation}
again, with the obvious notations $R_0=R(r_0,t_0)$ and $e_0=e(r_0)$.
On the boundary, expression (\ref{xi_firstorder}) remains valid, while for the shell label
\begin{equation}
r(p) = r_b - \left(\hat\delta X_{\rm lim}\right)^{-\delta}|\cos(\theta_0)\tilde E_0p|^{\hat\delta} + {\mathcal O}(p^{\hat\delta+1}).
\end{equation}
The case $r_0=r_b$ and $\theta_0=\pi/2$ yields the only 'circular' orbit, as remarked upon at the end of section~\ref{sect_orbitbound_equations}.

For $\vartheta$ we find easily
\begin{equation}
\vartheta(p) = {\sin{\theta_0}\over R_0}\tilde E_0p+ {\mathcal O}(p^2).
\end{equation}
For $\theta_0=0$ the light ray has $\vartheta(p)=0$.

Equally straightforward is the expansion 
\begin{equation}
\label{t_firstorder}
t(p) = t_0 + \tilde E_0p+ {\mathcal O}(p^2).
\end{equation}
The radial velocity reads
\begin{equation}
w(p) = \tilde E_0\cos\theta_0 + w_1\tilde E_0p+ {\mathcal O}(p^2)
\end{equation}
with
\begin{equation}
w_1 = -I_0\cos\theta_0 + {\sqrt{1-2e_0}\over R_0}\sin^2\theta_0\,\epsilon_\xi,
\end{equation}
and $I_0$ shorthand for $I(r_0,t_0)$ (and similarly later on for $H_0=H(r_0,t_0)$). The coefficient $w_1$ is never zero, except on the boundary when $\theta_0=\pi/2$ which is the case of the circular orbit $w(p)=0$.

As for the relative energy, we know from classical cosmology that, in order to dig out the deceleration parameter, we will need to go second order:
\begin{equation}
\label{tildeE_p}
{\tilde E}(p) = \tilde E_0  + \tilde E_1{\tilde E_0}^2p + \tilde E_2\tilde E_0^3 p^2 + {\mathcal O}(p^3).
\end{equation}
We find
\begin{eqnarray}
\label{tilde_E_1}
{\tilde E}_1&=& -{\partial_t X_0\over X_0}\cos^2\theta_0 - {\partial_t R_0\over R_0}\sin^2\theta_0 \nonumber\\
&=&-I_0\cos^2\theta_0 - H_0\sin^2\theta_0.
\end{eqnarray}
For the second order coefficient we introduce
\begin{equation}
\label{def_HtIt}
H^t = -{R\partial_t^2R\over \left(\partial_tR\right)^2}=-{\partial_t^2R\over RH^2}=q \,\,\,\, {\rm and} \,\,\,\, 
I^t = -{X\partial_t^2X\over \left(\partial_tX\right)^2}=-{\partial_t^2X\over XI^2}
\end{equation}
together with
\begin{equation}
\label{def_Hrt_Irt}
H^{rt} = {\partial_rR\over R}-{\partial_{rt}^2R\over \partial_tR}\quad{\rm and}\quad
I^{rt} = {\partial_rX\over X}-{\partial_{rt}^2X\over \partial_tX}. 
\end{equation}
We note that $H$ and $I$ have the dimension of inverse length, while $H^t$, $H^{rt}$, $I^t$ and $I^{rt}$ are dimensionless. We also recognize in $H^t$ the familiar deceleration parameter $q$ of classical cosmology.  

With all these newly defined functions, we obtain:
\begin{equation}
\label{redshift_small_p2}
\tilde E_2 = \tilde E_{2a} + \tilde E_{2b}
\end{equation}
with
\begin{equation}
\label{redshift_small_p2_a}
\tilde E_{2a} = I_0^2\cos^2\theta_0 + H_0^2\sin^2\theta_0
\end{equation}
and
\begin{eqnarray}
\label{redshift_small_p2_b}
\tilde E_{2b}&=& \cos\theta_0\sin^2\theta_0{\sqrt{1-2e_0}\over R_0}\epsilon_\xi\, (H_0-I_0) + \nonumber\\
&& \quad+\frac12\sin^2\theta_0\,H_0^2(1+H_0^t) + \frac12\cos^2\theta_0\,I_0^2(1+I_0^t)+\nonumber\\
&& \quad+ \frac12{\cos\theta_0\over X_0}\epsilon_\xi\left[\sin^2\theta_0\,H_0H^{rt}_0 + \cos^2\theta_0\,I_0I^{rt}_0\right].
\end{eqnarray}
The expressions for the redshift (\ref{redshift_forward}) and (\ref{redshift_back}) can also be written as
\begin{equation}
\label{redshift_small_p}
z = \left[ I_0\cos^2\theta_0 + H_0\sin^2\theta_0\right]|p|,
\end{equation}
showing that $z$ is independent of the sign of $p$.

\subsubsection{The photometry}
We start with the the expression for $(dl/d\theta)(p)$.
Despite the relative complexity of the relevant differential equations (\ref{eqdeltal}) and (\ref{deltaw1}), we obtain to second order in $p$, after quite some calculations
\begin{equation}
\label{dl_over_dtheta_p}
{dl\over d\theta} = |p| + {\mathcal O}(p^3)
\end{equation}
indicating that the second order term is identically zero. This absolute value is a consequence of $dl/d\theta\ge0$. This expression is valid at any point along the light ray. However, the expressions (\ref{transintens1}) for the surface brightness and (\ref{transmagn}) for the apparent magnitude feature $dl_{\rm e}/d\theta_{\rm o}$ and $dl_{\rm o}/d\theta_{\rm e}$. 

The expression for $dl_{\rm e}/d\theta_{\rm o}$ is the easier one. We need to integrate the light ray backwards (thus backwards in time). Hence the observed redshift is, with (\ref{redshift_back}) and (\ref{tildeE_p})
\begin{equation}
z = {\tilde E}_{\rm e} -1 = \tilde E(p_{\rm e}) - 1 = \tilde E_{1,\rm o}p + \tilde E_{2,\rm o} p^2 + {\mathcal O}(p^3).
\end{equation}
After inversion $p(z)$ and insertion in (\ref{dl_over_dtheta_p}) we obtain
\begin{equation}
\label{dele_over_dto_p}
{dl_{\rm e}\over d\theta_{\rm o}} = -{z\over\tilde E_{1,\rm o}}\left(1-\tilde E_{2,\rm o}\tilde E_{1,\rm o}^{-2}z\right).
\end{equation}
To calculate $dl_{\rm o}/d\theta_{\rm e}$, we must integrate a light path from the emitter to the observer, hence the parameter $p$ at emission time $p_{\rm e}=0$ and at the observer $p_{\rm o}=p$. The result needs to be expressed as a function of redshift $z=1/\tilde E_{\rm o}-1$ according to (\ref{redshift_forward}). This means that the expression (\ref{tildeE_p}) has to be expanded at the emittor, who has $\tilde E_{0,{\rm e}}=1$:
\begin{equation}
\tilde E_{\rm o} = 1 + \tilde E_{1,\rm e}p + \tilde E_{2,\rm e}p^2 + {\mathcal O}(p^3).
\end{equation}
For $\tilde E_{2,\rm e}$ we can in (\ref{redshift_small_p2}) simply replace the subscript 0 on the right hand side with subscript o since that term is second order:
\begin{equation}
\label{E_0_expanded_in_e}
\tilde E_{\rm o} = 1 + \tilde E_{1,\rm e}p + \tilde E_{2,\rm o}p^2 + {\mathcal O}(p^3).
\end{equation}
For $\tilde E_{1,\rm e}$ we need to write the right hand side of (\ref{tilde_E_1}) in terms of the observables with subscript o via a first order expansion. This proceeds by writing
\begin{equation}
{\tilde E}_{1,\rm e} = {\tilde E}_{1,\rm o}  + \partial_r{\tilde E}_{1,\rm o}(r_{\rm e}-r_{\rm o}) + 
\partial_t{\tilde E}_{1,\rm o}(t_{\rm e}-t_{\rm o}) + {\mathcal O}(p^2)
\end{equation}
which we transform with (\ref{r_firstorder}) and (\ref{t_firstorder}) in 
\begin{eqnarray}
\label{tildeE_e}
{\tilde E}_{1,\rm e}&=&{\tilde E}_{1,\rm o}  + \partial_r{\tilde E}_{1,\rm o}w\epsilon_\xi X_{\rm o}^{-1}(p_{\rm e}-p_{\rm o}) + 
\partial_t{\tilde E}_{1,\rm o}(p_{\rm e}-p_{\rm o}) + {\mathcal O}(p^2)\nonumber\\
&=&{\tilde E}_{1,\rm o}  - \partial_r{\tilde E}_{1,\rm o}w\epsilon_\xi X_{\rm o}^{-1}p 
-\partial_t{\tilde E}_{1,\rm o}p + {\mathcal O}(p^2)
\end{eqnarray}
After some calculations we obtain for (\ref{E_0_expanded_in_e})
\begin{equation}
\tilde E_{\rm o} = \tilde E_{1,\rm o}p + {\tilde E}'_{2,{\rm o}}p^2 + \tilde E_{11,\rm o}pz 
\end{equation}
with
\begin{equation}
\tilde E_{11,\rm o} = 2\sin^2\theta_{\rm o}(H_{\rm o}-I_{\rm o}).
\end{equation}
and
\begin{equation}
\label{redshift_small_p2_accent}
{\tilde E}'_2 = \tilde E_{2a} - \tilde E_{2b}
\end{equation}
with both terms defined in (\ref{redshift_small_p2_a})  and (\ref{redshift_small_p2_b}). 

Next we calculate
\begin{equation}
z = {1\over{\tilde E}_{\rm o}}-1= -\bigl(\tilde E_{1,\rm o}+\tilde E_{11,\rm o}z\bigr)p + \left({\tilde E}_{1,\rm o}^2-\tilde E_{2,\rm o}\right)p^2,
\end{equation}
which is to be inverted:
\begin{equation}
\label{delo_over_dte_p}
{dl_{\rm o}\over d\theta_{\rm e}}=p = -{z\over\tilde E_{1,\rm o}}\left[1-\left(1+\tilde E_{11,\rm o}\tilde E_{1,\rm o}^{-1}-{\tilde E}'_{2,\rm o}\tilde E_{1,\rm o}^{-2}\right)z\right].
\end{equation}
We will not explicitly write down the expression for the observed surface brightness (\ref{transintens1}) since it follows quite trivially from (\ref{transintens1}), (\ref{dele_over_dto_p}) and (\ref{delo_over_dte_p}). We note that the factors $-z{\tilde E}^{-1}_{1,\rm o}$ in front of the latter expressions cancel.

The magnitude-redshift relation (\ref{transmagn}) reads, with (\ref{delo_over_dte_p}) and some more manipulations:
\begin{eqnarray}
\label{transmagn_small}
 m_{X_{\rm o}}^{({\rm mag})}(t_{\rm o},\emph{\textbf{r}}_{\rm o},z) &=&  5\log_{10}(3.066)  + 40 + M_{X_{\rm e}}^{({\rm mag})}(t_{\rm e},\emph{\textbf{r}}_{\rm e}) - \nonumber\\
&&- 5\log_{10}\left[ I_{\rm o}\cos^2\theta_{\rm o}+H_{\rm o}\sin^2\theta_{\rm o}\right] +\nonumber\\
&&+5\log_{10}z
   + {5\over{\rm ln}10}\tilde E_{1,\rm o}^{-1}\left({\tilde E}'_{2,\rm o}\tilde E_{1,\rm o}^{-1}-\tilde E_{11,\rm o}\right)z.\nonumber\\
\end{eqnarray}

\subsection{The synchronous universe}
\label{sect_photometry_sync}

\subsubsection{Light paths and the horizon}
\label{sect_lightpaths_sync}
It is useful to relate the general analysis above to the well-known standard cases. The metric (\ref{isochmetrichi}) can be rewritten in the 2D case as
\begin{equation}
\label{isochmetric2}
ds^2 = dt^2 - \left(2M\,\cyc\right)^2\bigl[d\chi^2+\sin^2\chi\,d\vartheta^2\bigr].
\end{equation}
We will first consider a radial light ray $d\vartheta=0$. The last 2 equations of (\ref{geod3}) easily combine to the constants of the motion
\begin{equation}
\label{cyct_iso}
w=\pm {\tilde E} \qquad {\rm and} \qquad {\tilde E}\,\cyc = {\tilde E}_{\rm o}\cyc(t_{\rm o})=\cyc(t_{\rm o}),
\end{equation}
while the first equation of (\ref{geod3}) produces
\begin{equation}
\label{iso_chi_int}
|\chi-\chi_{\rm o}| = {1\over2M}\int^{t_{\rm o}}_t{dt'\over\cyc(t')}=\omega\!\!\int^{t_{\rm o}}_t{dt'\over\cyc(t')},
\end{equation}
which also follows, of course, directly from (\ref{isochmetric2}). We assumed a light ray that arrived at the observer at $\chi_{\rm o}$ and cosmic time $t_{\rm o}$ from being emitted at $\chi$ on an earlier time $t$.

We can be more explicit regarding (\ref{iso_chi_int}). We express it firstly as a function of the state $\psi$, initially defined in (\ref{def_psi}) as the state $\psi(r,t)$ of shell $r$ at $t$, but here it can be taken as the state of the universe at $t$. Clearly
\begin{equation}
\label{iso_chi_int_psi}
|\chi-\chi_{\rm o}| = \int^{\psi_{\rm o}}_\psi{d\psi'\over\cyc(\psi')}.
\end{equation}
Next we assume that $\psi_{\rm o}-\psi$ is not larger than 1 period, or, with the notations of (\ref{defpsiperiodic})
\begin{equation}
0\le\psi\le\psi_{\rm o}\le2\bar\psi_{\rm max}.
\end{equation}
Passing to the new variable $\cyc$, we obtain, with the notations of appendix \ref{sect_cyc},
\begin{eqnarray}
\label{deltachi_int_cyc}
|\chi-\chi_{\rm o}| &=& \int^{\cyc_{\rm o}}_\cyc{u^{-1/2}\,du\over\sqrt{1-\epsilon u+au^3}}\qquad0\le\psi\le\psi_{\rm o}\le\bar\psi_{\rm max}\nonumber\\
&=& \int_{\cyc_{\rm o}}^\cyc{u^{-1/2}\,du\over\sqrt{1-\epsilon u+au^3}}\qquad\bar\psi_{\rm max}\le\psi\le\psi_{\rm o}\le2\bar\psi_{\rm max}\nonumber\\ 
&=& \int^{w_0}_\cyc{u^{-1/2}\,du\over\sqrt{1-\epsilon u+au^3}}+
\int_{\cyc_{\rm o}}^{w_0}{u^{-1/2}\,du\over\sqrt{1-\epsilon u+au^3}}
\nonumber\\
&&\qquad\qquad\qquad\qquad\qquad
0\le\psi\le\bar\psi_{\rm max}\le\psi_{\rm o}\le2\bar\psi_{\rm max}
\nonumber\\
\end{eqnarray}
As was already pointed out in sections \ref{sect_embed} and \ref{sect_sync_closed}, this is only formally the same analysis as in the standard Robertson-Walker models: here the center and the boundary are special places globally (not locally). The case $\epsilon_\xi=+1$ corresponds to $0\le\chi\le\pi/2$ and the case $\epsilon_\xi=-1$ corresponds to $\pi/2\le\chi\le\pi$. Moreover, the circles with radii $\chi$ and $\pi-\chi$ are to be identified. A radial orbit is different from a general orbit, because the center is a special point. However, looking at these general orbits, they appear as `great circles' in an embedding diagram (see section \ref{sect_embed}), just as the light rays in the figures \ref{fig_model1_light}, \ref{fig_model2_light} and \ref{fig_model3_light} suggest paths that are similar to great circles. Hence one could, for every great circle, define a radial angle $\chi''$ similar to $\chi$. We will need this extended definition of $\chi$ in chapter \ref{sect_Mach}.

Finally we note that the expressions (\ref{deltachi_int_cyc}) can be used to calculate the horizon during a finite time lapse. We assume the start of time at $\cyc=0$. It thus suffices to substitute $\psi=\cyc=0$ in the first equation:
\begin{equation}
\chi_{\rm horizon} = \int^{\cyc_{\rm o}}_0{u^{-1/2}\,du\over\sqrt{1-\epsilon u+au^3}}.
\end{equation}
This integral is evaluated in equation (\ref{Delta_chi_max}). As an example, we consider the case of a closed universe with $a=0$. At maximum expansion $\cyc_{\rm o}=w_0=1$ and we obtain, with $w_1=-\infty$ and $w_2=+\infty$ (the cubic in (\ref{deltachi_int_cyc}) is reduced to linear) that $\chi_{\rm horizon}=\pi$. At that time, every object is visible twice: once in a direct image, and once in a reflected image. This is obvious for an observer at the center $C$, but holds for every observer if we consider the angle $\chi''$.

\subsubsection{Photometry}
As for the redshift, we use (\ref{redshift_back}) and (\ref{cyct_iso}) to get the familiar
\begin{equation}
\label{redshift_iso}
z+1 = {\tilde E} = {\cyc(t_{\rm o})\over\cyc}={{\cal R}_{\rm o}\over{\cal R}_e},
\end{equation}
with ${\cal R}_{\rm e}={\cal R}(t_{\rm e})$ and ${\cal R}_{\rm o}={\cal R}(t_{\rm o})$ the radii of the universe at the time of emission resp. the time of observation.

For the Hubble law, we insert (\ref{Rsync}) into (\ref{Hubble}):
\begin{equation}
z = {\partial_tR\over R}\ell = {d_t\cyc\over \cyc}\ell.
\end{equation}

For the parameters introduced in the previous subsection, we find (omitting all subscripts o)
\begin{equation}
H=I,\quad H^t=I^t=q, \quad H^{rt}=I^{rt}=0.
\end{equation}
and
\begin{equation}
\tilde E_1=-H,\quad \tilde E_{2a} = H^2,\quad \tilde E_{2b}=H^2{1+q\over2}.
\end{equation}
All this yields
\begin{equation}
{dl_{\rm e}\over d\theta_{\rm o}}={z\over H_{\rm o}}\left[1-{3+q_{\rm o}\over2}z\right]
\end{equation}
and
\begin{equation}
{dl_{\rm o}\over d\theta_{\rm e}}={z\over H_{\rm o}}\left[1-{1+q_{\rm o}\over2}z\right].
\end{equation}
In this case however, we can do better. From geometric considerations (the paradigm of the inflating balloon) we can see that
\begin{equation}
{dl_{\rm e}\over d\theta_{\rm o}} = {\cal R}_{\rm e}\tilde r \quad {\rm and} \quad {dl_{\rm o}\over d\theta_{\rm e}} = {\cal R}_{\rm o}\tilde r
\end{equation}
with $\tilde r$ the normalised radial shell label (\ref{def_rnorm}) of the observer if the emitter is placed at $\tilde r=0$ (which can always be arranged for). The rod is now placed perpendicular to the plane of the light ray (say a meridian) and hence a $dl$ is a small arc length on a small circle subtended by a $d\theta$ which is now actually a $d\varphi$. The radius of the small circle is ${\cal R}\tilde r$. Unfortunately, this nice trick works only if the embedding diagram is a circle (and thus the 2-D universe a sphere).

Turning to the surface brightness, we see from (\ref{transintens1}) that we need the product 
\begin{equation}
\label{dldth_sync_first_order}
\left({dl_{\rm e}\over d\theta_{\rm o}}{d\theta_{\rm e}\over dl_{\rm o}}\right)^2= 1-2z
\end{equation}
in first order in $z$. Apparently, $q_{\rm o}$ has disappeared. 

Hence
\begin{equation}
\label{syncdomdS}
{dS_{\!\rm e}\over d\Omega_{\rm o}}={dS_{\!\rm o}\over d\Omega_{\rm e}}\left({\cal R}_{\rm e}\over {\cal R}_{\rm o}\right)^2, 
\end{equation}
(actually also $d\Omega_{\rm o}=d\Omega_{\rm e}$), and we obtain, using (\ref{redshift_iso}), for the transformation of the surface brightness (\ref{transintens1}) the well-known relation
\begin{eqnarray}
{\cal I}_{\rm o}(t_{\rm o},\lambda_{\rm o},\Omega_{\rm o},\emph{\textbf{r}}_{\rm o})= {1\over(1+z)^5}
{\cal I}_{\rm e}(t_{\rm e},\lambda_{\rm e},\Omega_{\rm e},\emph{\textbf{r}}_{\rm e}),
\end{eqnarray}
Now we also see that (\ref{dldth_sync_first_order}) is the first order expansion in $z$ of $(1+z)^{-2}$.

As for the transformation of the fluxes (\ref{transmagn_small}), also that expression is considerably simplified. We recover the familiar
\begin{eqnarray}
\label{transmagnsync}
 m_{X_{\rm o}}^{({\rm mag})}(t_{\rm o},\emph{\textbf{r}}_{\rm o},z) &=&  5\log_{10}(3.06)  + 40 + M_{X_{\rm e}}^{({\rm mag})}(t_{\rm e},\emph{\textbf{r}}_{\rm e}) - \nonumber\\
&& -5\log_{10}H_{\rm o}+5\log_{10}z +{5\over2\,{\rm ln}10}(1-q_{\rm o})z.\nonumber\\
\end{eqnarray}
In this case, one can actually obtain an exact expression, for $\Lambda=0$ that is. It is called the expression of Mattig, and in order to derive it one uses the well-known parametric expression for cyc. For $\Lambda\ne0$ a much more complicated parametric expression for cyc can be derived, using $\chi$ as given in (\ref{iso_chi_int}) as the parameter. However, it involves elliptic functions and the roots of the cubic $1-\epsilon x + ax^3$ (see app.~\ref{sect_cyc}), and is therefore unlikely to lead to a closed expression.

One can easily prove that the (locally) synchronous models are the only ones for which $I=H$ in a finite region of space and therefore the only ones for which $m(z)$ is direction independent.

Finally, the relation between luminosity distance and angular distance at the observer (\ref{rellumangdist}) is, with (\ref{syncdomdS}), the familiar relation
\begin{equation}\label{rellumangdistiso}
    D_L = (1+z)^2D_{\rm o}.
\end{equation}

\subsection{Comparison with observations}
\label{sect_comp_obs}

The obvious observational effect of the spherical cosmological models we consider here is the directional dependence in the classical observational tests. The relation $z(\ell)$ shows a quadrupole effect, and the $m(z)$ relation has in addition a dipole term. Studies, such as the one by McClure \& Dyer \cite{clure}, that look for so-called anisotropies in the Hubble flow, focus on anisotropies that are caused by relatively nearby disturbances (such as the great attractor). It will be hard to separate these disturbances from the effects of a spherical cosmological model in the large, and this in the presence of important uncertainties in the interpretation of the raw observations.

We now work out the relevant quantities in the Hubble law $z(\ell)$ as given by (\ref{Hubble_small}) and the magnitude-redshift relation $m(z)$ as presented in (\ref{transmagn_small}) for the simplest possible non-trivial model for a locally flat universe: the one with $\Lambda=0$ given in appendix~\ref{model_e0Lambda0}. Since the shell label is arbitrary anyway, we can suffice with $p=r$ and thus
\begin{equation}
R=r\,\bigl[t+\phi(r)\bigr]^{2/3}. 
\end{equation}
Note that this particular choice for $r$ endows it with the dimension of a length to the power 1/3. Since $e(r)=0$ we have
\begin{equation}
X=\partial_rR=\bigl[t+\phi(r)\bigr]^{2/3}+\frac23r\phi'(r)\bigl[t+\phi(r)\bigr]^{-1/3}.
\end{equation}
A prime denotes derivative with respect to shell label $r$, and we will in the sequel omit the explicit notation for the shell label dependence of $\phi(r)$.
Since
\begin{equation}
H=\frac23{1\over t+\phi}
\end{equation}
we have directly an estimate for $t+\phi$, about 9 Ga, and an alternative form for
\begin{equation}
X = (t+\phi)^{2/3}(1+r\phi'H).
\end{equation}
Next we find
\begin{equation}
I = H{1-r\phi'H/2\over1+r\phi'H}.
\end{equation}
The 2 parameters $H$ and $I$ are sufficient to fit the Hubble law, including the quadrupole. $H$ and $I$ then
yield $r\phi'$. The deceleration parameter 
\begin{equation}
H^t=q=1/2,
\end{equation}
as also follows from (\ref{defq}), and
\begin{equation}
I^t = \frac12{(1+r\phi'H)(1-2r\phi'H)\over(1-r\phi'H/2)^2}.
\end{equation}
Both $H^t$ and $I^t$ therefore are fixed if the Hubble law is known to sufficient accuracy. The other parameters in the $m(z)$ relation are
\begin{equation}
H^{rt} = \frac1r\left(1+\frac32r\phi'H\right)
\end{equation}
featuring additionally the parameter $r$, and finally the rather messy looking
\begin{eqnarray}
I^{rt}H^{-1} &=& {2\phi'+r\phi''-r{\phi'}^2H/2\over1+r\phi'H} +\nonumber\\
&&+ \frac12{4\phi'+2r\phi''-r{\phi'}^2H-3r{\phi'}^2H^2\over 1-r\phi'H/2}
\end{eqnarray}
adding $\phi''$ to the parameter list. Clearly there is enough freedom to fit $z(\ell)$ and $m(z)$, even if the data are good enough to contain directional information. Finally, as already remarked upon in subsection~\ref{sect_HI}, the phase function comes to the fore as instrumental in the fitting process.

\section{Primordial black holes}
\label{sect_primBH}

One of the important discoveries in the field of dynamics of galaxies within the past 15 years is the strong evidence for super massive black holes in the center of galaxies. In addition, a reasonably well established correlation exists between the mass of the central black hole and the mass of the galaxy (including dark matter, see e.g. Baes {\it et al.} \cite{baes}). This points to a scenario in which at least some black holes were primordial, with the galaxies and the dark matter subsequently accreting around them.

We now investigate what happens in the vicinity of a black hole when the universe is in a collapse phase. We start with the standard equations of the motion around a Schwarzschild black hole, with the metric (\ref{Schwarz}) and $\Lambda=0$. There are 2 constants of the motion. On the one hand
\begin{equation}\label{u_Schwarz}
    \tilde E_\infty = \left(1-{2M_\bullet\over R}\right) {d\bar t\over ds} = {E_\infty\over m_0c^2} =  \left(1-{\boldsymbol{\upsilon}_\infty^2\over c^2}\right)^{-1/2}
\end{equation}
with '$\infty$' not te be taken literally, but standing for 'sufficiently far from the black hole where general relativistic effects are negligible', $m_0$ the rest mass of a test particle and 
\begin{equation}
\label{def_M_BH}
M_\bullet = {G{\cal M}_\bullet\over c^2},
\end{equation}
with ${\cal M}_\bullet$ the mass of the black hole.

\begin{figure}
   \centering
   \includegraphics[width=130mm]{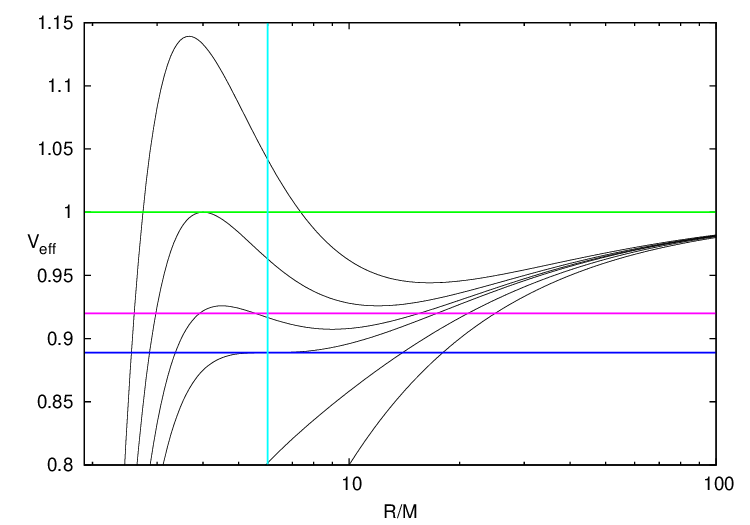}
      \caption{A few $V_{\rm eff}(R/M)=V_{\tilde h}(\tilde R)$ curves. The axes are truncated for clarity: the curves pass through $(\tilde R,V_{\tilde h})=(2,0)$ and tend asymptotically to $V_{\tilde h}=1$ for $R\to+\infty$. The curve with the blue tangent drawn has $\tilde h={\tilde h}_{\rm lim}=2\sqrt3$. The tangent line is at $\tilde E^2_\infty=8/9=V_{\rm max}({\tilde h}_{\rm lim})$, and touches $V_{\tilde h}$ at $\tilde  R_{\rm lim}=6$, marked by the vertical cyan line. All $V_{\tilde h}(\tilde R)$ curves have in common that their second derivative is zero there. All orbits with $\tilde h\le 2\sqrt3$ will plunge into the black hole. Two such orbits are shown: the 'lowest' one is the radial orbit $\tilde h=0$. The 3 curves 'above' the orbit $\tilde h=2\sqrt3$ have angular momenta $\tilde h>2\sqrt3$ and 2 extrema. The lowest curve of these 3 together with the magenta ${\tilde E^2_\infty}$ corresponds to an orbit with apocenter about $3.5M_\bullet$ that plunges into the black hole or to an orbit with pericenter somewhat smaller than $6M_\bullet$ and apocenter about $15 M_\bullet$. The second one has $\tilde h=4$ and has its maximum at ${\tilde E^2_\infty}=1$ for $\tilde R=4$ (the minimum is at $\tilde R=12$). For that energy (the green line) it is a marginally bound orbit that either barely escapes the plunge by skimming the critical pericenter $\tilde R=4$, or that has barely insufficient radial momentum to make the escape. The top curve represents the effective potential for some higher angular momentum $\tilde h$.}
         \label{fig_infall}
   \end{figure}

On the other hand
\begin{equation}\label{def_h}
    h = R^2{d\vartheta\over d(s/c)}=R^2{d\vartheta\over d\tau}
\end{equation}
with $\tau$ the proper time on the geodesic. The expression (\ref{def_h}) is identical (up to factor $c$) to the third equation of the cosmological geodesics (\ref{geod1}). This is relevant, since on small scales the universe can be considered synchronous, and the black hole could therefore be taken as the center of the universe, and thus in the center of a spherically symmetric field. The metric yields
\begin{equation}
\label{V_eff_schwarzacc}
    \left(dR\over ds\right)^2 = {\tilde E_\infty}^2-V_{\tilde h}(\tilde R)\ge0
\end{equation}
with
\begin{equation}\label{V_eff_schwarz}
    V_{\tilde h}(\tilde R) = \left(1-{2\over \tilde R}\right)\left(1+{{\tilde h^2}\over {\tilde R}^2}\right).
\end{equation}
the effective potential. In this expression 
\begin{equation}
\label{def_tildeR_tildeh}
\tilde R = \frac R{M_\bullet}\qquad {\rm and} \qquad \tilde h = {h\over M_\bullet c}={h\over G{\cal M}_\bullet/c}.
\end{equation}
We now analyze these effective potential curves (\ref{V_eff_schwarz}), and refer to fig. \ref{fig_infall}. 

The derivative of $V_{\tilde h}(\tilde R)$ equals, apart from a factor $2{\tilde R}^{-2}$,
\begin{equation}
\label{der_V_eff_schwarz}
1-{\tilde h^2\over\tilde R^2}(\tilde R - 3).
\end{equation}
The effective potential has 2 extrema for $\tilde h>2\sqrt3$ at the radii $\tilde R_1(\tilde h)<\tilde R_2(\tilde h)$:
\begin{equation}
\label{schwarz_extrema}
    \tilde R_{1,2}(\tilde h) = {\tilde h\over2}\left(\tilde h+\epsilon\sqrt{{\tilde h}^2-12}\right)
\end{equation}
with $\epsilon=-1$ for the maximum at $\tilde R_1$ and $\epsilon=+1$ for the minimum at $\tilde R_2$. We find the following expressions for the maximum $V_{\rm max}(\tilde h)=V_{\tilde h}\bigl(\tilde R_1(\tilde h)\bigr)$ and the minimum $V_{\rm min}(\tilde h)=V_{\tilde h}\bigl(\tilde R_2(\tilde h)\bigr)$:
\begin{equation}\label{def_Vlim}
    V_{\rm max,min}(\tilde h) = {(36+{\tilde h}^2){\tilde h} -\epsilon ({\tilde h}^2-12)^{3/2}\over54{\tilde h}}.
\end{equation}
A particle with negative radial velocity will fall into the black hole if one of the following three conditions is satisfied: 
\begin{enumerate}
\item $\tilde h<2\sqrt3={\tilde h}_{\rm lim}$. The parameter ${\tilde h}_{\rm lim}$ is therefore the smallest value for $\tilde h$ that allows non-plunging orbits.
\item $\tilde R$ is smaller than the smallest radius of intersection of ${\tilde E_\infty}^2$ with ${\tilde V_{\tilde h}}$.
\item ${\tilde E_\infty}^2\ge V_{\rm max}(\tilde h)$. The parameter $V_{\rm max}(\tilde h)$ is therefore the largest value of the energy for a given $\tilde h$ that allows non-plunging orbits if the radial velocity is negative.
\end{enumerate}
It follows that 
\renewcommand{\labelenumi}{\roman{enumi}}
\begin{enumerate}
\item
a particle that has negative radial velocity and finds itself closer to the black hole than the minimum of all $\tilde R_1(\tilde h)$ will plunge into the black hole. This minimum occurs in the limit $\tilde R_1(\tilde h)$ for $\tilde h\to+\infty$ and equals $\tilde R=3$
\item a particle that is farther away than the maximum of all $\tilde R_1(\tilde h)$ will never be in condition (2). That radius is $\tilde R_{\rm lim}=6$. 
\end{enumerate}

From (\ref{u_Schwarz}) we see that in a collapsing synchronous universe $\tilde E$ will increase, since $|\boldsymbol{\upsilon}_\infty|\cal R$ is a constant and therefore the magnitude of the velocity far from the black hole $|\boldsymbol{\upsilon}_\infty|$ will increase. On the other hand, the angular momentum is a constant both in the Schwarzschild metric and in the cosmological metric. Hence the effective potential does not change during the collapse. It follows from the form of $V_{\tilde h}(\tilde R)$ that material that was originally in a stable orbit around the black hole will find itself on an increasingly more radial orbit and, at some moment in time (during the part of the orbit with negative radial velocity) finally on a plunge orbit. The higher $\tilde E$, the more rapidly the fall will be. Hence the black hole will start to accrete matter it would never have accreted in a static universe. Therefore, the stronger the collapse phase, the more massive the black holes in it would grow. In the limit, a black hole in the center of a synchronous universe would accrete all the matter in the universe! But as already mentioned before, this is not the scenario that is being advanced in this paper as a likely one,\footnote{Note that no fundamental changes would have to be made in this paper in order to include a black hole at the center of the universe. Instead of $m(0)=0$ we would simply have to assign a non-zero value to $m(0)$.} though that scenario is in line with the concept of a universe inside a black hole. Instead, the collapse would probably halt much sooner, at various places and at various times,\footnote{Recall that clusters of galaxies occur in spongelike structures.} and the material that was lucky enough not to fall into the black hole, would simply be a galaxy in the newly started expansion phase of the universe.

In the sequel we will also need the inverse of (\ref{def_Vlim}) with $\epsilon=-1$, yielding the angular momentum $\tilde h$ that corresponds to an effective potential curve that has a given $\tilde E_\infty^2$ at its maximum. Somewhat surprisingly, that inversion can be written down explicitly:
\begin{equation}
\label{tildehEinfty}
\frac1{12}\tilde h^2(\tilde E_\infty)=W{\sqrt{W+4}+\sqrt{W}\over\sqrt{W+4}+3\sqrt{W}}+1,\qquad 
2W=9\tilde E_\infty^2-8\ge0.
\end{equation}
As a bonus, essentially the same calculation yields the angular momentum $\tilde h$ that corresponds to an effective potential curve that has a given $\tilde E_\infty^2$ at its minimum, or, more physically phrased, the angular momentum $\tilde h$ that corresponds to the circular orbit with given $\tilde E_\infty$:
\begin{equation}
\label{tildehEinfty_circ}
\frac1{12}\tilde h^2_{\rm circ}(\tilde E_\infty)=W{\sqrt{W+4}-\sqrt{W}\over\sqrt{W+4}-3\sqrt{W}}+1,\qquad 
1>2W=9\tilde E_\infty^2-8\ge0.
\end{equation}
This result can be linked with the elementary result for circular orbits in the Newtonian limit, here for large $R$, which is
\begin{equation}
E_{\infty,{\rm Newt}}= -\frac{m_0}2{G{\cal M}_\bullet\over R}=-\frac{m_0}2{c^2\over\tilde R},
\end{equation}
hence $\tilde E_\infty=1-1/(2\tilde R)$ or $\tilde E_\infty^2=1-1/\tilde R$. On the other hand, if we set $2W=1-\epsilon$, hence $\tilde E_\infty^2=1-\epsilon/9$, we obtain from (\ref{tildehEinfty_circ}) that $\tilde h^2_{\rm circ}(\tilde E_\infty)=9/\epsilon$ and therefore $\tilde h^2_{\rm circ}=\tilde R$, as should, since $h_{\rm circ}^2=R^2\upsilon^2=R^2(G{\cal M}_\bullet/R)=RG{\cal M}_\bullet=\tilde R(G{\cal M}_\bullet/c)^2$. . 

\section{Dark matter}
\label{sect_darkmatter}
\subsection{Introduction}
\label{sect_darkmatter_intro}

The inclusion of dark matter must be part of any cosmology. The proof of its existence rests essentially on the dynamics of galaxies and clusters of galaxies, and is most obvious in an analysis of the rotation curves of spiral galaxies.

Neutrino's are not considered to be viable candidates for dark matter in the standard cosmology, because, in order to be effective, their masses and/or their densities would need to be too high to allow for an open universe. In the model we consider here, the universe is closed by design, and this argument disappears. Moreover, in this paper we cannot invoke much a priori information about the latest compact phase of our universe, so from that side no useful constraint is to be expected. In this section we therefore investigate more closely the possibility that dark matter is constituted 'simply' of neutrino's. Clearly, this has been looked into earlier (see e.g. Paganini {\it et al.} \cite{pag} for the case of galaxies and Nieuwenhuizen \cite{nieuw} for the case of the galaxy cluster Abell 1689). We will discuss this now again, in the context of the universes we consider here.

In the same vein as we argued that in our paradigm the CMB is isotropic of necessity, we can probably say the same about the cosmic neutrino background (CNB), at least for those neutrino's that have been around for multiple expansion and contraction phases.\footnote{The CNB is not to be confused with the neutrino flux at higher energies, which is, just as radiation, highly anisotropic. For a speculative picture of the early stages that we refer to, see section~\ref{sect_early}.} We will consider Fermi-Dirac or Maxwell-Boltzmann distributions. It must be stressed that our analyses are equally valid for any dark matter candidate, as long as it satisfies (approximately) a Maxwell-Boltzmann distribution, which is what nature tells us that it must (see section~\ref{sect_darkmatter_gals}).
 
\subsection{The field equations}
\label{sect_field}
We will assume that the mass distribution $\rho_\nu(R)$ has selfconsistently settled down in a steady state around a mass concentration. We will continue to use the notations employed in chapter~\ref{sect_primBH}. 
For a mass of ${\cal M}_\bullet=10^6M_\odot$ we find
$M_\bullet =  4.805\times10^{-8}\,{\rm pc}$, 
and we define $\tilde M_\bullet$ as $M_\bullet$ in this unit:
\begin{equation}
\label{def_tilde_M}
M_\bullet\equiv\tilde M_\bullet\times 4.805\times10^{-8}\,{\rm pc}.
\end{equation}

In the general relativistic case, we assume the following metric:
\begin{equation}
\label{metric_static}
ds^2=e^{\nu(R)}{d\bar t}^2 - e^{\lambda(R)}dR^2 - R^2\,d\Omega^2.
\end{equation}
The Einstein field equations determine $\lambda(R)$ and $\nu(R)$ (see Landau \& Lifshitz,  \cite{landau}). The function $\lambda(R)$ is determined by the first field equation 
\begin{equation}
\label{Einstein1}
e^{-\lambda(R)}\left({d_R\lambda(R)\over R}-{1\over R^2}\right)+{1\over R^2}={8\pi G\over c^4}T_0^0(R)={8\pi G\over c^2}\rho_\nu(R).
\end{equation}
In analogy with the Schwarzschild case, we denote
\begin{equation}
\label{genrelBH_defm}
e^{-\lambda(R)} = e^{-\lambda(\tilde R)} = 1-{2m(R)\over R}\equiv1-{2\tilde m(\tilde R)\over\tilde R}.
\end{equation}
and find
\begin{equation}\label{genrelBH1a}
d_Rm(R) = {4\pi G\over c^2}\rho_\nu(R)\,R^2=\kappa\rho_\nu(R)\,R^2,
\end{equation}
We note that (\ref{genrelBH1a}) is essentially identical to (\ref{defrho}), and we therefore know that $m(R)$ is the gravitating mass function (with unit length). Passing to dimensionless quantities:
\begin{equation}\label{genrelBH1}
d_{\tilde R}{\tilde m}(\tilde R) = 4\pi\,\tilde\rho_\nu(\tilde R)\,{\tilde R}^2,
\end{equation}
with $\tilde\rho_\nu(\tilde R)$ the dimensionless mass density expressed in ${\cal M}_\bullet/M_\bullet^3$ and $\tilde m(\tilde R)=m(R)/M_\bullet\ge1$. If we denote by $\tilde R_{\rm in}$ the radius inside which there is no mass, then if follows from section \ref{sect_primBH} that $\tilde R_{\rm in}\ge3$ and ${\tilde m}(\tilde R_{\rm in})=1$.

The mass function clearly follows from the metric (\ref{metric_static}) and (\ref{genrelBH_defm}):
\begin{equation}
\label{defcalMR}
{\cal M}(\tilde R) = 4\pi{\cal M}_\bullet\int_{\tilde R_{\rm in}}^{\tilde R}{\tilde\rho_\nu(\tilde R')\,\tilde R'^2d\tilde R'\over\sqrt{1-{2\tilde m(\tilde R')/\tilde R'}}}
\end{equation}
In the Newtonian case
\begin{equation}
{\cal M}_{\rm Newt}(R) = {c^2\over G}m(R)
\end{equation}
and we recover of course the familiar
\begin{equation}\label{genrelBH1_class}
d_R{\cal M}_{\rm Newt}(R) = 4\pi R^2\rho_\nu(R).
\end{equation}

The function $\nu(R)$ is connected by the second field equation
\begin{equation}
\label{genrelBH2a}
e^{-\lambda(R)}\left({d_R\nu(R)\over R}+{1\over R^2}\right)-{1\over R^2}=-{8\pi G\over c^4}T_1^1(R)={8\pi G\over c^4}{\cal P}_R(R),
\end{equation}
with the radial pressure ${\cal P}_R(R)$. Again, we write the function $\nu(R)$ as
\begin{equation}
\label{genrel_def_Xi}
e^{\nu(R)} = e^{\nu(\tilde R)} = 1-{2\,\Xi(R)\over R}\equiv1-{2\,\tilde\Xi(\tilde R)\over\tilde R}.
\end{equation}
The function $\Xi(R)/R$ plays the role of a general relativistic gravitational potential of the distributed matter around the black hole, and $\tilde \Xi(\tilde R)=\Xi(R)/M_\bullet$. We obtain
\begin{equation}
\label{genrelBH2}
\bigl(\tilde R-2\tilde m\bigr)\, d_{\tilde R}\tilde\Xi = \tilde\Xi-\tilde m -4\pi\,\tilde{\cal P}_R\,{\tilde R}^2\bigl(\tilde R - 2\,\tilde\Xi\bigr)
\end{equation}
with the dimensionless radial pressure $\tilde{\cal P}_R(\tilde R)$ expressed in ${\cal M}_\bullet c^2/M_\bullet^3$. We omitted the explicit dependence on $\tilde R$ of $\tilde\Xi(\tilde R)$, $\tilde m(\tilde R)$ en $\tilde{\cal P}_R(\tilde R)$ for clarity. Denoting
\begin{equation}
\label{def_Psitilde}
\tilde\Psi(\tilde R)\equiv{\tilde\Xi(\tilde R)\over \tilde R}={\Xi(R)\over R}
\end{equation} 
we can cast (\ref{genrelBH2}) into the form
\begin{equation}
\label{genrelBH3}
\bigl(\tilde R-2\tilde m\bigr)\tilde R\, d_{\tilde R}\tilde\Psi = (2\,\tilde\Psi-1)(\tilde m +4\pi{\tilde R}^3\tilde{\cal P}_R),
\end{equation}
or, returning to $\nu(\tilde R)$,
\begin{equation}
\label{genrelBH4}
\bigl({\tilde R\over2}-\tilde m\bigr)\tilde R\, d_{\tilde R}\nu = \tilde m +4\pi{\tilde R}^3\tilde{\cal P}_R.
\end{equation}
Because of the signature of the metric, $d_{\tilde R}\tilde\Psi\le0$ or $d_{\tilde R}\nu\ge0$ with the equality only possible if $\tilde m +4\pi{\tilde R}^3\tilde{\cal P}_R=0$.

In the Newtonian limit ($\tilde{\cal P}_R\sim \tilde\rho\langle{v^2_R\over c^2}\rangle\to0$) and for large $R$ (compared to $M_\bullet$), equation (\ref{genrelBH3}) reduces to
\begin{equation}
\label{genrelBH2lim}
d_{\tilde R}\tilde\Psi(\tilde R)=-{\tilde m(\tilde R)\over\tilde R^2}.
\end{equation}
Turning on the dimensions again, we find for (\ref{genrelBH2lim}) the well-known
\begin{equation}
d_R\Psi(R)= -{G{\cal M}(R)\over R^2}
\end{equation}
if
\begin{equation}
\label{def_Psi}
\Psi(R)=c^2\tilde\Psi(\tilde R),
\end{equation}
which is therefore the Newtonian binding potential. 

Alternatively, in the same limit, (\ref{genrelBH2}) becomes
\begin{equation}
\label{dRXi}
d_{\tilde R}\tilde\Xi(\tilde R) = {\tilde\Xi(\tilde R)-\tilde m(\tilde R)\over\tilde R}
\end{equation}
which is also a known Newtonian result, since classical potential theory learns
\begin{equation}
\label{Xi_newton1}
R\Psi(r) = G{\cal M}(R) + 4\pi G\,R\!\int_{R}^{R_b}\!\!\!R'\,\rho(R')\,dR'
\end{equation}
for a mass distribution with finite total mass and boundary \mbox{$R_b\le+\infty$}, and thus
\begin{equation}
\label{Xi_newton2}
d_{R}[R\Psi(R)] = 4\pi G\!\int_{R}^{R_b}\!\!\!R'\,\rho(R')\,dR'={R\Psi(R)-G{\cal M}(R)\over R}\ge0.
\end{equation}
Passing to dimensionless quantities, we find
\begin{equation}
c^2d_{\tilde R}[{\tilde R}\tilde\Psi({\tilde R})]= {c^2{\tilde R}\tilde\Psi({\tilde R})\over {\tilde R}} - {c^2\tilde m({\tilde R})\over {\tilde R}},
\end{equation}
identical to (\ref{dRXi}). Hence, in the Newtonian limit,
\begin{equation}
\Xi(R) = {R\Psi(R)\over c^2}.
\end{equation}

We don't need the last 2 field equations 
\begin{eqnarray}
\label{Einstein3}
{e^{-\lambda(R)}\over2}\left[d^2_R\nu(R)+
\frac12d_R[\nu(R)-\lambda(R)]\bigl[d_R\nu(R)+{2\over R}\bigr]\right]\!\!&=&\!\!
-{8\pi G\over c^4}T_2^2(R)\nonumber\\
\!\!&=&\!\!-{8\pi G\over c^4}T_3^3(R)\nonumber\\
\end{eqnarray}
that involve the tangential pressures $T_2^2=T_3^3={\cal P}_\vartheta={\cal P}_\varphi\equiv{\cal P}_T/2$ in order to determine $\tilde m(\tilde R)$ and $\tilde\Xi(\tilde R)$, since these equations follow from the first 2 by virtue of the fact that the divergence of the $({}_1^1)$ tensor $T_\mu^\nu$ is zero ($T_{\mu\,;\sigma}^\sigma=0$). This property of a covariant 4-vector field yields again 4 equations. It turns out that 3 of them are trivial $0=0\,$'s. The remaining one gives rise to the equation of hydrostatic equilibrium:
\begin{equation}
d_R{\cal P}_R-{d_R\Psi\over1-\displaystyle{2\Psi\over c^2}} \left({{\cal P}_R\over c^2}+\rho_\nu\right) +
{2\over R}\left({\cal P}_R-\frac12{\cal P}_T\right)=0.
\end{equation}
The limit $c\to+\infty$ recovers the well-known spherical Jeans equation of stellar dynamics.

Finally we note that if we didn't have a distribution function to derive $\rho_\nu$ and ${\cal P}_R$ from, or an equation of state connecting the mass density and the pressures, the above equations would not suffice to determine them. This is, of course, a well-known feature shared with the Newtonian case.

Given expressions for $\tilde\rho_\nu(\tilde R)$ and $\tilde{\cal P}_R(\tilde R)$, the functions $\tilde m(\tilde R)$ and $\tilde \Xi(\tilde R)$ or $\tilde \Psi(\tilde R)$ or $\nu(\tilde R)$ are determined by integrating the equations (\ref{genrelBH1}) and (\ref{genrelBH2}) or (\ref{genrelBH3}) or (\ref{genrelBH4}) from the inside out, starting from $\tilde R_{\rm in}$, with 
\begin{equation}
\rho(\tilde R)=0,\,\, 2\le\tilde R<\tilde R_{\rm in}\quad{\rm and\,\, thus}\quad\tilde m(\tilde R_{\rm in})=1, 
\end{equation}
and a free parameter
\begin{equation}
\tilde\Xi_{\rm in}=\tilde\Xi(\tilde R_{\rm in}) \quad{\rm or}\quad \tilde\Psi_{\rm in}=\tilde\Psi(\tilde R_{\rm in})\quad{\rm or}\quad \nu_{\rm in}=\nu(\tilde R_{\rm in}).
\end{equation}
However, $\tilde\Xi_{\rm in}\ge1$ since $\tilde\Xi(\tilde R)=1$ is the solution of the naked black hole. The physical interpretation of this parameter is best grasped from $\Psi_{\rm in}$, since that is the general relativistic generalization of the depth of the potential well caused by the black hole and the matter around it.  

Leaving the solution for $\tilde R\ge\tilde R_{\rm in}$ for later, we still need to determine the solution inside $\tilde R_{\rm in}$. The solution of the `homogeneous' equation (\ref{genrelBH2}) with $\tilde m(\tilde R)=1$ and $\tilde{\cal P}_R(\tilde R)=0$ is 
\begin{equation}
\tilde\Xi(\tilde R)= 1 + C(\tilde R-2).
\end{equation}
The linear term in $\tilde R$ also affects the constant in $g_{00}(\tilde R)=e^{\nu(\tilde R)}$. With $\tilde\Xi_{\rm in}=\tilde\Xi(\tilde R_{\rm in})$ we find
\begin{equation}
\label{g00Rin}
g_{00}(\tilde R)={\tilde R_{\rm in}-2\,\tilde\Xi_{\rm in}\over\tilde R_{\rm in}-2}\left(1-{2\over\tilde R}\right),\qquad2\le\tilde R\le\tilde R_{\rm in}.
\end{equation}
Clearly a $\tilde\Xi_{\rm in}>1$ causes clocks to run slower than the $1-2/\tilde R$ rate of the naked black hole, as a consequence of the additional matter around it.

We will now consider two regimes where the neutrino's manifest themselves quite differently.

\subsection{Neutrino's in the vicinity of a quiescent black hole}
\label{sect_neut_BH}

In this section we follow up on the formation of primordial black holes (chapter~\ref{sect_primBH}), and investigate what could remain around such a black hole as a result of a collapse annex accretion process. We need not consider $R<3M$ since between $2M$ and $3M$ all matter falls into the black hole anyway (as explained in chapter~\ref{sect_primBH}), in the additional assumption  that matter in that zone has entered that zone from the outside and was not launched in it from the inside. Note that this (mildly) invalidates the assumption of steady state, since the black hole will grow, but we assume that this growth, after the formation process, is slow. 

\subsubsection{The orbital dynamics}
\label{sect_orbit_dyn}

When integrating over the dark matter distribution, we should only include those that are not on plunging orbits, since we can assume that all material on them has crashed into the black hole. Therefore we reconsider the analysis that let to the infall conditions in section~\ref{sect_primBH}, but now for the metric
\begin{equation}
\label{metric_static1}
ds^2=g_{00}(\tilde R){d\bar t}^2 - {dR^2\over\displaystyle1-{2\tilde m(\tilde R)\over\tilde R}} - R^2\,d\Omega^2
\end{equation}
with
\begin{equation}
\label{def_g00}
g_{00}(\tilde R)=1-{2\,\tilde\Xi(\tilde R)\over\tilde R}\equiv\gamma_{00}^2.
\end{equation}

We find
\begin{equation}
\label{V_eff_ineq}
    \left(dR\over ds\right)^2 = {\displaystyle1-{2\tilde m(\tilde R)\over\tilde R}\over g_{00}(\tilde R)}
    \left({\tilde E_\infty}^2-V_{\tilde h}(\tilde R)\right)\ge0
\end{equation}
with
\begin{equation}
\label{V_eff}
    V_{\tilde h}(\tilde R) = g_{00}(\tilde R)\left(1+{{\tilde h^2}\over {\tilde R}^2}\right).
\end{equation}
These expressions are the obvious generalizations of (\ref{V_eff_schwarzacc}) and (\ref{V_eff_schwarz}). 

From fig. \ref{fig_infall} we can surmise that, also in this case, the effective potential curves $V_{\tilde h}(\tilde R)$ have 0, 1 or 2 extrema for finite $\tilde R$. In the case of 2 extrema, occurring at $\tilde R_1(\tilde h)<\tilde R_2(\tilde h)$, the smallest is a maximum. The extrema can be found rather explicitly using the second field equation (\ref{genrelBH2a}), which we first recast, with the help of (\ref{genrelBH_defm}), in the form
\begin{equation}
\tilde R\bigl[\tilde R-2\tilde m(\tilde R)\bigr] {d_{\tilde R}\gamma_{00}(\tilde R)\over\gamma_{00}(\tilde R)}=
\tilde m(\tilde R) + 4\pi \tilde R^3 \tilde{\cal P}_R(\tilde R).
\end{equation}
This enables us to write the extrema of the effective potential (\ref{V_eff}) 
\begin{equation}
\frac12{d_{\tilde R}g_{00}(\tilde R)\over g_{00}(\tilde R)}\left(1+{\tilde h^2\over\tilde R^2}\right)-{\tilde h^2\over\tilde R^3}=\frac12{d_{\tilde R}V_{\tilde h}(\tilde R)\over g_{00}(\tilde R)}
\end{equation}
as the solutions for $\tilde R$ of
\begin{equation}
\label{condition_k}
1-{\tilde h^2\over\tilde R^2}{\tilde R-3\tilde m(\tilde R)-4\pi \tilde R^3 \tilde{\cal P}_R(\tilde R)\over\tilde m(\tilde R)+4\pi \tilde R^3 \tilde{\cal P}_R(\tilde R)} ={d_{\tilde R}V_{\tilde h}(\tilde R)\over d_{\tilde R}g_{00}(\tilde R)}=0,
\end{equation}
somewhat remarkably in a rather closed form with no explicit differentiation involved.
For the Schwarzschild case $\tilde m(\tilde R)=1$ and $\tilde{\cal P}_R(\tilde R)=0$, and we recover (\ref{der_V_eff_schwarz}). Clearly we cannot make the solutions of (\ref{condition_k}) more explicit without knowledge of $\tilde m(\tilde R)$ and $\tilde{\cal P}_R(\tilde R)$, but we will assume that the conditions for infall will be qualitatively identical to the ones stated in section~\ref{sect_primBH}, on which the figure is based.  

Since for physical reasons $d_{\tilde R}g_{00}(\tilde R)>0$, the left hand side of expression (\ref{condition_k}) has the same sign as $d_{\tilde R}V_{\tilde h}(\tilde R)$, and thus has maxima $\tilde R_1$ and minima $\tilde R_2$ in the same positions. That left hand side therefore is
\begin{enumerate}
\item positive if $\tilde R$ is smaller than the maximum of $V_{\tilde h}(\tilde R)$
\item negative if $\tilde R$ is between the maximum and the minimum of $V_{\tilde h}(\tilde R)$ \item positive if $\tilde R$ is larger than the minimum of $V_{\tilde h}(\tilde R)$
\end{enumerate}
if these extrema are present (this depends on $\tilde h$). In the first case we are on a plunging orbit, in the second and third we may, depending on $\tilde E_\infty$. 

\subsubsection{The dark matter response density}

In order to obtain the response dark matter density, we first rewrite (\ref{def_h}) as
\begin{equation}
h = R^2{d\vartheta\over dt_R}{dt_R\over d\tau}\equiv R\upsilon_T\tilde E_R
\end{equation}
with 
\begin{equation}
\label{def_upsilonT}
\upsilon_T=R{d\vartheta\over dt_R}
\end{equation}
the tangential component of the velocity, $t_R$ the time measured by a stationary observer at $R$ and 
\begin{equation}
\label{def_tilde_ER}
\tilde E_R = {dt_R\over d\tau}
\end{equation}
the relativistic energy, normalised to the rest energy, measured by the same observer of a moving object with proper time $\tau$. In terms of the normalised quantities we introduced so far, we obtain for the tangential component of the linear momentum
\begin{equation}
\label{def_tildepT}
p_T = {m_0\upsilon_T\over\sqrt{1-\boldsymbol\upsilon^2/c^2}}= {E_R\over c^2}\upsilon_T=m_0\tilde E_R\upsilon_T\equiv m_0c{\tilde p_T}
\end{equation}
which also defines the dimensionless $\tilde p_T=p_T/(m_0c)$.

All this prepares us for the use of (\ref{aniso}), after passing to dimensionless quantities:
\begin{eqnarray}
\label{neut_response1}
\rho(\tilde R) &=& 4\pi m^3c^3\!\!\!\int_{\tilde E_{R,\rm min}}^{\gamma_{00}^{-1}}\!\!\! f \bigl(\gamma_{00}\tilde E_R\bigl)\,\tilde E_R\,d\tilde E_R\!\!\int_{\tilde p_{T,\rm min}}^{\sqrt{\tilde E_R^2-1}}\!\!\!\!{{\tilde p_T}\,d{\tilde p_T}\over\sqrt{\tilde E_R^2-1-{\tilde p_T}^2}},\nonumber\\
&&
\end{eqnarray}
where $f(\tilde E_\infty)$ a distribution function normalised so as to produce a mass when integrating over a volume in phase space.

As to the integration limits, the upper bound in $\tilde E_R$ follows from (\ref{rel_Einfty_ER}) and the fact that we consider only dark matter particles that are permanently trapped around the black hole, hence $\tilde E_\infty<1$. We recall that the upper bound in $\tilde p_T$ can be understood by the requirement (\ref{V_eff_ineq}) of positive squared radial velocity 
\begin{equation}
\label{V_eff_inequality1}
\tilde E_\infty^2=g_{00}(R)\tilde E_R^2\ge V_{\tilde h}(R)=g_{00}(R)(1+\tilde p_T^2). 
\end{equation}
We note that this inequality, or alternatively (\ref{defeR}), is consistent with the expression
\begin{equation}
\label{def_tilde_ER1}
\tilde E_R = \sqrt{1+\tilde p_R^2 + \tilde p_T^2}
\end{equation}
with $\tilde p_R$ the dimensionless radial momentum, defined in analogy with $\tilde p_T$. The upper bound in the integration of $\tilde p_T$ is therefore $\sqrt{\tilde E_R^2-1}$.

The lower bounds in (\ref{neut_response1}) are not specified, because we have yet to exclude the plunging orbits, and this needs a more careful analysis. If we were to include them, the lower bound in $\tilde p_T$ would be $\tilde p_{T,\rm min}=0$, corresponding to a radial orbit with no angular momentum. The lower bound in $\tilde E_R$ would then be $\tilde E_{R,\rm min}=1$, corresponding to a radial orbit at its apocentrum, hence with zero velocity and $\tilde E_R$ therefore equal to the rest mass energy 1 (note that $\tilde E_R$ is not a constant of the motion).

\subsubsection{The integral space}

Unlike the Newtonian case of a spherically distributed mass distribution, a point $(\tilde E_\infty^2,\tilde h^2)$ in integral space does not necessarily determine the geometry of an orbit as a shell inside a pericenter radius and an apocenter radius. It could also be a plunging orbit, depending on the radius, needed to decide whether we are in the first case (i) mentioned at the end of section \ref{sect_orbit_dyn} or not. Therefore the interpretation of integral space is radius-dependent. Only if we exclude plunging orbits, is there a unique correspondence between a pair $(\tilde E_\infty^2,\tilde h^2)$ and a shell with a pericenter and an apocenter radius.

\begin{figure}[ht]
   \centering
   \includegraphics[width=88mm]{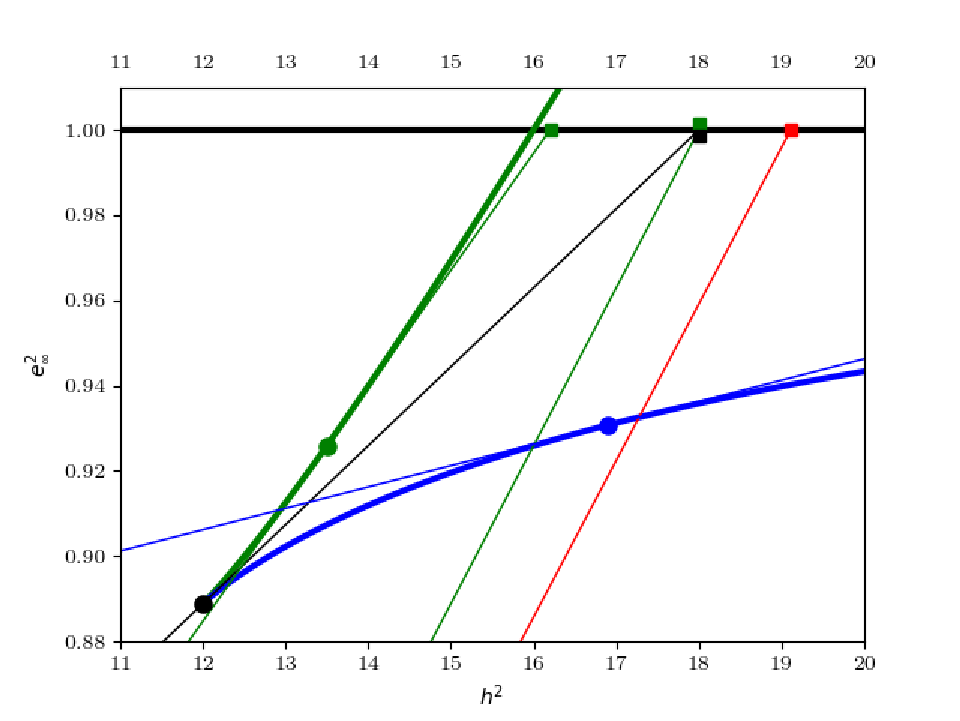}
      \caption{The integral space. Explanation in the text.  }
         \label{fig_intspace}
   \end{figure}

Figure \ref{fig_intspace} shows integral space for a naked black hole. When plunging orbits are allowed, the inequality (\ref{V_eff_ineq}):
\begin{equation}
\label{V_eff_inequality}
    \tilde E_\infty^2 \ge g_{00}(\tilde R)\left(1+{{\tilde h^2}\over {\tilde R}^2}\right).
\end{equation}
forms in integral space $(\tilde E_\infty^2,\tilde h^2)$ a right-angled triangular region bounded by
\begin{enumerate}
\item
the line $\tilde h^2=0$, or $\tilde p_{T,\rm min}=0$ in the integral (\ref{neut_response1}) (not shown for clarity)
\item
the line $E_\infty^2=1$, or $\tilde E_{R,\rm min}^2=g_{00}(\tilde R)^{-1}$ in the integral (\ref{neut_response1}). We do not consider the unbound orbits $E_\infty^2>1$.
\item
the line (\ref{V_eff_inequality}), with the equality sign. This is the hypotenuse of the triangle,  representing the orbits in $\tilde R$ that have zero radial velocity. We call $\tilde R$ the parameter of the hypotenuse. Five of them are shown: 2 green ones, one red, one black and one blue.  
\begin{itemize}
\item
These hypotenuses intersect the vertical  $\tilde h^2=0$ in $\tilde E_\infty^2=g_{00}(\tilde R)$, or $\tilde E_{R,\rm min}=1$ in the integral (\ref{neut_response1}). As already indicated, these are orbits with zero angular momentum, thus radial orbits, with $\tilde R$ as apocenter. The test particle is momentarily at rest. The intersection is a monotonically increasing function of $\tilde R$.
\item
The intersection of the hypothenuse with the line $\tilde E_\infty^2=1$ at $\tilde h^2=\tilde R^2(g_{00}(\tilde R)^{-1}-1)$ is the marginally bound orbit for which $\tilde R$ is the pericenter. It has the largest local energy $\tilde E_R$ of all orbits on the hypotenuse. It is indicated by a square. In the Schwarzschild case, which is the case shown, the squares are at $\tilde h^2=2\tilde R^2/(\tilde R-2)$. The blue square is not shown for clarity (it would be at $\tilde h^2\sim30$). 
\end{itemize}
In general, the local energy $\tilde E_R$ of the orbits on a hypotenuse increases 'from left to right'. 
\end{enumerate} 

\subsubsection{The integral subspace of non-plunging orbits}

\paragraph{The two branches}

Equation (\ref{V_eff_inequality}) defines 3 families of hypotenuses in integral space, depending on whether one chooses radii in an interval in which the effective potential curves $V_{\tilde h}(\tilde R)$ have a minimum, a maximum, or no extremum. Equation (\ref{V_eff_inequality}) together with (\ref{condition_k}), which determines the extremum of $V_{\tilde h}(\tilde R)$, form the equations of an envelope that can be split in 2 branches. The dots in figure \ref{fig_intspace} show the location where the depicted family member touches the envelope.  

\begin{itemize}
\item
The blue family, for $\tilde R_{\rm lim}\le\tilde R$, generates the concave blue branch, corresponding to the minima. The position of the squares increasing monotonously with $\tilde R$.
\item
The green family, for $3\le\tilde R\le\tilde R_{\rm lim}$, generates the (slightly) convex green branch corresponding to the maxima. The green family shows a more complicated behavior. In the Schwarzschild case the position of the green squares $\tilde h^2=2\tilde R^2/(\tilde R-2)$ decreases from $\tilde R=6$ at $\tilde h^2=18$ till $\tilde R=4$ at $\tilde h^2=16$, which happens to be also the intersection of the line $\tilde E_\infty^2=1$ with the green envelope. In the sequel we will meet that radius again. For $4\ge\tilde R\ge3$ the position of the green square increases, thus reversing trend, to end up again at $\tilde h^2=18$ for $\tilde R=3$. That family member is shown in the picture and has no dot, because it touches the green envelope at infinity, being the asymptote of the green branch.
\item
The red family, for $2\le\tilde R<3$, does not form branches for $\tilde h^2>0$. The position of the red squares continues the trend set by the green squares, and increases dramatically with decreasing $\tilde R$, up to the limit towards infinity for $\tilde R\to2$, where the limit hypotenuse of the red family of zero velocity curves is $\tilde E_\infty=0$. This also means a dramatic increase in accessible integral space. We have already seen that all orbits inside the sphere $\tilde R=3$ are plunging. Therefore the red family does not concern us here. A point on a red line of course is a zero radial velocity orbit at that radius, but they all have their apocenter at that radius. They differ by the number of windings needed to complete the plunge. This number increases with increasing $\tilde h$.
\end{itemize}
The radius $\tilde R_{\rm lim}$ that is common to both families defines $\tilde R_{\rm lim}$ and corresponds to that unique $\tilde h$ for which the maximum and the minimum of $V_{\tilde h}(\tilde R)$ coincide. In figure \ref{fig_intspace} (naked black hole) $\tilde R_{\rm lim}=6$. The zero radial velocity line and the cusp corresponding to the common member $\tilde R=\tilde R_{\rm lim}$ are shown in black. The corresponding angular momentum we denote by $\tilde h_{\rm lim}$. In the Schwarzschild case $\tilde h_{\rm lim}^2=12$ and $\tilde R_{\rm lim}=6$.

\paragraph{Characterization of the branches}
We first notice that, because of (\ref{V_eff_ineq}) and (\ref{V_eff_inequality1}), that
\begin{equation}
\left(dR\over ds\right)^2=\left(1-{2\tilde m(\tilde R)\over\tilde R}\right)\tilde p_R^2
\end{equation}
and therefore no perturbation $(\delta\tilde p_R,\delta\tilde p_T)$ for fixed $\tilde R$ will violate (\ref{V_eff_inequality}).

The blue branch is, just as in the Newtonian case, the locus of all (stable) circular orbits. We recall from the Newtonian case that the circular orbit with radius $\tilde R$ divides the tangent line with parameter $\tilde R$ in `the lower half', which represents orbits for which $\tilde R$ is the apocenter, and `the upper half', representing the orbits for which $\tilde R$ is a pericenter. 

In order that an orbit is a non-plunging one, we learn from figure \ref{fig_infall} that $\tilde E_\infty^2$ must be smaller than the maximum of $V_{\tilde h}(\tilde R)$ for finite $\tilde R$, that therefore must be present. We denote that maximum by $V_{\rm max}(\tilde h)$, as we did in the Schwarzschild case in section~\ref{sect_primBH}. The equation $\tilde E_\infty^2=V_{\rm max}(\tilde h)$ is the equation of the green envelope in figure~\ref{fig_intspace}. Hence 
\begin{equation}
\label{infall_cond}
V_{\tilde h}(\tilde R)\le\tilde E_\infty^2\le V_{\rm max}(\tilde h).
\end{equation}
At radii where the equality $E_\infty^2=V_{\tilde h}(\tilde R)$ holds, we are in a pericenter or apocenter of the orbit, since, according to (\ref{V_eff_ineq}), the radial velocity is zero there. For the specific combination $(\tilde E_\infty^2,\tilde h^2)$ for which \mbox{$\tilde E_\infty^2= V_{\rm max}(\tilde h)$}, such as indicated in figure \ref{fig_intspace} by the green dot, we have an orbit that has just enough angular momentum to be a non-plunging orbit. Referring to figure \ref{fig_infall}, the radius where the effective potential curve $V_{\tilde h}(\tilde R)$ touches the line $\tilde E_\infty$ is the pericenter of that orbit. We call such an orbit one with an unstable pericenter. For somewhat different energy or angular momentum, this orbit could flip into an orbit with that radius as (stable) pericenter or a plunging orbit with that radius as apocenter.\footnote{These orbits are also called unstable circular orbits. For all practical purposes, this is a misnomer, since an additional force would be required to keep the orbit at fixed $R$.}

All non-plunging orbits therefore are located within the wedge formed by the blue and green branches. In topological terms, this wedge is a set that is neither open nor closed. The green branch forms an open part of the boundary, the blue branch is the closed part of the boundary.

\paragraph{Characterization of the orbits between the branches}
Out of a given point $(\tilde E_\infty^2,\tilde h^2)$ representing an non-plunging orbit, one can always draw a tangent (the `upper one') to the blue branch. The parameter of that line is the apocenter of the orbit. 

As for the pericenter, there are 2 possibilities. 
\begin{enumerate}
\item
The point has another tangent (the `lower one') to the blue branch, the parameter of which is the pericenter of the orbit. For this to be the case, the orbit must lie 'to the right' of the black hypotenuse. 
\item
The point has a 'lower' tangent to the green branch, the parameter of which is again the pericenter of the orbit. For this to be the case, the orbit must lie 'to the left' of the black hypotenuse. The black hypotenuse thus divides the 2 cases.
Orbits 'to the left' of the black hypotenuse have no 'upper' tangent to the green branch. 
\end{enumerate}

The black dot is the unstable circular orbit. All non-plunging orbits are located between the 2 branches.

\subsubsection{The integration over $\tilde p_T$}

The condition (\ref{infall_cond}) on $\tilde E_\infty$ in order to prevent a particle from falling into the black hole assumes $\tilde R$ and $\tilde h$ as a given. However, what we need in the integration (\ref{neut_response1}) is the condition on $\tilde h$ given $\tilde R$ and $\tilde E_\infty$. To that end, we note that the function $V_{\rm max}(\tilde h)$ is an increasing function of $\tilde h$, which follows from the form (\ref{V_eff}) of $V_{\tilde h}(\tilde R)$. Hence we can calculate the minimum $\tilde h$ for a given $\tilde E_\infty$, in order to have a non-plunging orbit, being an orbit with unstable pericenter, and we denote that minimum by $\tilde h(\tilde E_\infty)$. The specific combination mentioned above can thus be denoted by $(\tilde E_\infty^2,\tilde h^2(\tilde E_\infty^2))$ and is again the equation of the green envelope. The condition we impose is therefore in essence a lower limit on the angular momentum, and thus, with given $\tilde R$, we deal with a lower limit on the tangential component of the linear momentum:
\begin{equation}
\label{pTlim1}
\tilde p_T\ge\tilde p_{T,{\rm min}}(\tilde R,\tilde E_\infty)\equiv\tilde p_{T,{\rm min}}(\tilde R,\tilde E_R)={\tilde h(\tilde E_\infty)\over\tilde R}.
\end{equation}
In the Schwarzschild case, it is implicitly given by the inversion for $\tilde h$ of (\ref{def_Vlim}) for $V_{\rm max}(\tilde h)=\tilde E_\infty^2$, and explicitly given by expression (\ref{tildehEinfty}). In the present case, we need to solve (\ref{condition_k}), which we write as 
\begin{equation}
\tilde h^2=\tilde R^2{\tilde m(\tilde R)+4\pi \tilde R^3 \tilde{\cal P}_R(\tilde R)
\over
\tilde R-3\tilde m(\tilde R)-4\pi \tilde R^3 \tilde{\cal P}_R(\tilde R)} 
\end{equation}
and (\ref{V_eff}) 
\begin{equation}
\tilde E_\infty^2=  g_{00}(\tilde R)\left(1+{{\tilde h^2}\over {\tilde R}^2}\right)=g_{00}(\tilde R)\left.\tilde E_R^2\right|_{\tilde p_R=0}
\end{equation}
for $\tilde R$ and $\tilde h$. The function $\tilde E_\infty^2(\tilde h^2)$ appears here as an explicit parameter representation (in function of the parameter $\tilde R$), if only we knew $\tilde m(\tilde R)$, $\tilde{\cal P}_R(\tilde R)$ and $g_{00}(\tilde R)$. Fortunately, the radius $\tilde R_1[\tilde h(\tilde E_\infty)]$ at which the curve $V_{\tilde h(\tilde E_\infty)}(\tilde R)$ attains it maximum is always smaller or equal than the $\tilde R$ for which we are calculating the response, lest we be on a plunging orbit. Since we will calculate the response via the system of differential equations (\ref{genrelBH1}) and (\ref{genrelBH2}) or (\ref{genrelBH3}) from the inside out, we can tabulate $\tilde h(\tilde E_\infty)$ as we integrate along (see also later). 

\subsubsection{The integration over $\tilde E_R$}

For the sequel we will need to ascertain whether $\tilde R$ is smaller or larger that $\tilde R_{\rm lim}$. In the numerical calculations, we will therefore check the sign of the derivative of  (\ref{condition_k}), which can be done by checking the (sign of the) derivative of the function
\begin{equation}
\label{condition_k1}
-{1\over\tilde R^2}{\tilde R-3\tilde m(\tilde R)-4\pi \tilde R^3 \tilde{\cal P}_R(\tilde R)\over\tilde m(\tilde R)+ 4\pi \tilde R^3 \tilde{\cal P}_R(\tilde R)}
\end{equation}
which has, obviously, the same sign as the derivative of (\ref{condition_k}). It serves the same function as (\ref{der_V_eff_schwarz}). The zero of the derivative of (\ref{condition_k1}) will thus mark the radius where the extrema of $V_{\tilde h_{\rm lim}}(\tilde R)$ coincide. Hence that radius is also $\tilde R_{\rm lim}$.

We now place ourselves at some $\tilde R_0\ge\tilde R_{\rm lim}$, and visualize the curves $V_{\tilde h}(\tilde R)$ for increasing $\tilde h$, starting from $\tilde h_{\rm lim}$. Initially, the maxima $V_{\rm max}(\tilde h)$ will be smaller than $V_{\tilde h}(\tilde R_0)=\tilde E_{\infty,\rm min}^2$ giving rise to plunging orbits. The first marginally non-plunging orbit, i.e. the orbit that has just enough angular momentum to realize a non-zero pericenter, is the one for which $V_{\rm max}(\tilde h)=V_{\tilde h}(\tilde R_0)$, and for $\tilde h$ larger than this limit value we can realize an integration interval in $\tilde p_T$ of non-plunging orbits since 
\begin{equation}
\label{condition_Enonplung}
V_{\rm max}(\tilde h)\ge\tilde E_\infty^2\ge V_{\tilde h}(\tilde R_0)=\tilde E_{\infty,\rm min}^2.
\end{equation}
We recognize the same condition as (\ref{infall_cond}), on which we based our analysis of the $\tilde p_T$ interval. Hence the lower limit for $\tilde E_R$, which we denoted by $\tilde E_{R,\rm min}$ in (\ref{neut_response1}), follows from the solution for $\tilde h$ of 
\begin{equation}
\label{ERmin1}
\tilde E_{\infty,\rm min}^2\equiv V_{\rm max}(\tilde h)=V_{\tilde h}(\tilde R_0).
\end{equation}
This condition depends only on $\tilde R$, $\tilde m(\tilde R)$ and $\tilde{\cal P}_R(\tilde R)$, since (\ref{ERmin1}) solves for $\tilde h$. Denoting the solution $\tilde h_{\rm min}$, we find
\begin{equation}
\label{ERmin1a}
\tilde E_{R,\rm min}^2=1+\left(\tilde h_{\rm min}(\tilde R_0)\over\tilde R_0\right)^2 \quad {\rm for} \quad \tilde R_0>\tilde R_{\rm lim}.
\end{equation}
The orbit with $(\tilde E_{R,\rm min},\tilde h_{\rm min})$ is in figure \ref{fig_intspace} the intersection of a blue zero radial velocity line with parameter $\tilde R_0$ and the green envelope, and also the lower left corner of the region of integral space accessible to $\tilde R_0$ (non-plunging orbits). For that orbit the radius $\tilde R_0$ is an apocenter. The pericenter is the unstable radius $\tilde R_1(\tilde h_{\rm min}(\tilde R_0))<\tilde R_{\rm lim}$. It is the same orbit that constitutes the lower bound of the integration in $\tilde p_T$ for that energy $\tilde E_{R,\rm min}$. In the figure, $\tilde R_1$ could be constructed by determining the green zero radial velocity curve that is tangent at $(\tilde E_{R,\rm min},\tilde h_{\rm min})$ and ascertaining its radius. For $\tilde R_0\xrightarrow[>]{}\tilde R_{\rm lim}$, $\tilde R_1\xrightarrow[>]{}\tilde R_{\rm lim}$ and for $\tilde R_0=\tilde R_{\rm lim}$ we have the unstable circular orbit at $\tilde R_{\rm lim}$, which is actually the unstable limit of a plunging orbit with apocenter $\tilde R_0$.

Our second case is $\tilde R_0<\tilde R_{\rm lim}$. For a non-plunging orbit $\tilde R_0>\tilde R_1(\tilde h)$ and the expression (\ref{condition_k}) for $d_{\tilde R}V_{\tilde h}(\tilde R)$ is negative. Hence we need the condition
\begin{equation}
\label{ERmin2}
{\tilde p_T}^2>{\tilde m(\tilde R_0)+ 4\pi \tilde R_0^3 \tilde{\cal P}_R(\tilde R_0)\over \tilde R_0-3\tilde m(\tilde R_0)-4\pi \tilde R_0^3 \tilde{\cal P}_R(\tilde R_0)}\equiv{\tilde p_T}^2(\tilde R_0), 
\end{equation}
which means that we are on the 'downward' branch of $V_{\tilde h}(\tilde R)$. This condition only depends on $\tilde R$,  $\tilde m(\tilde R)$ and $\tilde{\cal P}_R(\tilde R)$, and not on $\tilde E_R$. The tangential momentum ${\tilde p_T}^2(\tilde R_0)$ is thus the minimum tangential momentum the orbit must have. The corresponding $\tilde h=\tilde R_0{\tilde p_T}(\tilde R_0)$ is such that $V_{\tilde h}(\tilde R)$ has a maximum at $\tilde R_0$. Also in this case, it implies a minimum energy. This energy is \mbox{$\tilde E_\infty^2=V_{\tilde h}(\tilde R_1(\tilde h))$}, with $\tilde R_1(\tilde h)=\tilde R_0$, by virtue of (\ref{ERmin2}). Hence the horizontal line with that $\tilde E_\infty$ is the tangent at the maximum of $V_{\tilde h}(\tilde R)$, which is at $\tilde R_0$; smaller $\tilde p_T$ would give rise to plunging orbits, smaller energies preclude motion. This unique orbit has ${\tilde p}_T^2 ={\tilde p_T}^2(\tilde R_0)$ and $\tilde p_R(\tilde R_0)=0$, yielding thus a minimum energy $\tilde E_{R,\rm min}^2=1+{\tilde p_T}^2(\tilde R_0)$. It is an orbit with an unstable pericenter $\tilde R_0$.  
In figure \ref{fig_intspace} it is the energy of the green dot on the green zero radial velocity curve with parameter $\tilde R_0$. Angular momenta that are smaller than the angular momentum of the green dot give rise to plunging orbits.  Like the blue dots, the green dots divide the tangent into 2 parts. The upper part are orbits with pericenter $\tilde R_0$ and an apocenter that increases if we approach $\tilde E_\infty=1$. The lower part (not very well visible in the figure because the curvature of the green envelope is small) do not represent orbits \underline{at $\tilde R_0$}, because they are below the tangent that is the zero radial velocity curve. When $\tilde R_0\xrightarrow[<]{}\tilde R_{\rm lim}$, the apocenter of the orbit of which $\tilde R_0$ is the pericenter tends towards $R_{\rm lim}$, yielding in the limit the unstable circular orbit.

During the calculation of the solution from the inside out, the tracing of ${\tilde p_T}(\tilde R)$ is thus a similar process as the tracing of $\tilde h(\tilde E_\infty)$.

\subsubsection{The moments of the distribution function}
Having established the integration limits in (\ref{neut_response1}), we can now perform the integration over $\tilde p_T$:
\begin{eqnarray}
\label{neut_response2}
\rho(\tilde R) &\!=\!& 4\pi m^3c^3\!\!\!\int_{\tilde E_{R,\rm min}}^{\gamma_{00}^{-1}}\!\!\!\!f\bigl(\gamma_{00}\tilde E_R\bigl)\,\tilde E_R\!
\left[\tilde E_R^2-1-\tilde p_{T,{\rm min}}^2(\tilde R,\tilde E_R)\right]^\frac12\!d\tilde E_R . \nonumber\\
\end{eqnarray}
with
\begin{equation}\label{expr_E_R,min}
\tilde E_{R,\rm min}^2=\left\{
\begin{array}{ll}
\displaystyle 1+{\tilde p_T}^2(\tilde R)            &\quad {\rm for}\quad \tilde R\le\tilde R_{\rm lim} \\[1mm]
\displaystyle 1+\left(\tilde h_{\rm min}/\tilde R\right)^2          &\quad {\rm for}\quad \tilde R\ge\tilde R_{\rm lim}\\
\end{array}
\right.
\end{equation}
and $\tilde p_{T,{\rm min}}^2(\tilde R,\tilde E_R)$ given by (\ref{pTlim1}).

The mean velocity $\langle\beta\rangle(\tilde R)=\langle|\boldsymbol\upsilon|/c\rangle(\tilde R)$ is obtained by multiplying the expression (\ref{neut_response2}) for $\tilde\rho(\tilde R)$ by 
\begin{equation}\label{def_betamean}
\sqrt{1-\tilde E_R^{-2}}
\end{equation}
to yield $\tilde\rho(\tilde R)\langle\beta\rangle(\tilde R)$.

The radial pressure is obtained via the expression:
\begin{equation}
\label{expr_PR_gen}
{\cal P}_R = n\langle\upsilon_R p_R\rangle
\end{equation}
with $n$ the number density. Taking into account that
\begin{equation}
\upsilon_R = {p_R\over m_0}\sqrt{1-\boldsymbol\upsilon^2/c^2}={p_R\over m_0\tilde E_R}
\end{equation}
we have
\begin{equation}
\upsilon_Rp_R={m_0^2c^2({\tilde E_R}^2-1)-p_T^2\over m_0\tilde E_R} =
m_0 c^2{{\tilde E_R}^2-1-\tilde p_T^2\over \tilde E_R}.
\end{equation}
Thus we can calculate the radial pressure by multiplying the inner integrand in (\ref{neut_response1}) by $c^2(\tilde E_R^2-1-{\tilde p_T}^2)/\tilde E_R$
or, after integration, the remaining integrand in (\ref{neut_response2}) by
\begin{equation}
\frac{c^2}3{\tilde E_R^2-1-{\tilde p_{T,{\rm min}}}^2\over\tilde E_R},
\end{equation}
yielding
\begin{eqnarray}
\label{pres_response}
{\cal P}_R(\tilde R) &\!=\!& {4\pi\over3} m^3c^5\!\!\!\int_{\tilde E_{R,\rm min}}^{\gamma_{00}^{-1}}\!\!\!\!f\bigl(\gamma_{00}\tilde E_R\bigl) \left[\tilde E_R^2-1-\tilde p_{T,{\rm min}}^2(\tilde R,\tilde E_R)\right]^\frac32\!d\tilde E_R.
\nonumber\\
\end{eqnarray}

The tangential pressures are the same in any tangential direction, and we have
\begin{equation}
\label{expr_PT_gen}
{\cal P}_\varphi={\cal P}_\vartheta=\frac12{\cal P}_T = \frac12n\langle\upsilon_T p_T\rangle=\frac12m_0c^2\langle{\tilde p_T^2\over\tilde E_R}\rangle.
\end{equation}
A similar calculation as above yields
\begin{eqnarray}
\label{presT_response}
\frac12{\cal P}_T(\tilde R) &\!=\!& {4\pi\over3} m^3c^5\!\!\!\int_{\tilde E_{R,\rm min}}^{\gamma_{00}^{-1}}\!\!\!\!f\bigl(\gamma_{00}\tilde E_R\bigl)d\tilde E_R \!\sqrt{\tilde E_R^2-1-\tilde p_{T,{\rm min}}^2(\tilde R,\tilde E_R)}\times\nonumber\\
&&\qquad\qquad \times\left[\tilde E_R^2-1+\frac12\tilde p_{T,{\rm min}}^2(\tilde R,\tilde E_R)\right].
\nonumber\\
\end{eqnarray}
This is not the same expression as for the radial pressure, since we eliminated plunging orbits. If we would not do that, $\tilde p_{T,{\rm min}}^2(\tilde R,\tilde E_R)=0$, and the 3 pressures would be equal, as should for an isotropic distribution function. Clearly
$\frac12{\cal P}_T(\tilde R) >{\cal P}_R(\tilde R)$.

\subsubsection{The central regions}
\paragraph{The inner boundary}
Obviously all the above integrals will be zero if \mbox{$\gamma_{00}^{-1}\le\tilde E_{R,\rm min}$}. Therefore $\tilde R$ must satisfy 
\begin{equation}\label{def_tilde_R_in}
g_{00}^{-1}(\tilde R) \ge 1+{\tilde p_T}^2(\tilde R). 
\end{equation}
Since inside that radius, there are no bound orbiting masses, we can use the Schwarzschild case ${\tilde p_T}^2(\tilde R)=1/(\tilde R - 3)$, as follows from (\ref{der_V_eff_schwarz}) or (\ref{ERmin2}) with $\tilde m(\tilde R)=1$. We recall from section~\ref{sect_field} that, when integrating equations (\ref{genrelBH1}) and (\ref{genrelBH2}) from the inside out, we start with two free parameters which could be taken as $\tilde R_{\rm in}$ and $\tilde\Xi_{\rm in}=\tilde\Xi(\tilde R_{\rm in})$. Here our insistence on bound orbits expressed by (\ref{def_tilde_R_in}) and the explicit expression for $g_{00}$ as given by (\ref{g00Rin}) produces a relation between $\tilde R_{\rm in}$ and $\tilde\Xi_{\rm in}$ (or $\tilde\Psi_{\rm in}$):
\begin{equation}
\label{expr_tilde_R_in}
\tilde R\ge\tilde R_{\rm in} = {4\,\tilde\Xi_{\rm in}\over2\,\tilde\Xi_{\rm in}-1}={4\tilde\Psi_{\rm in}+1\over2\tilde\Psi_{\rm in}},\qquad
2\,\tilde\Xi_{\rm in}=4\tilde\Psi_{\rm in}+1.
\end{equation}
As we noted following (\ref{genrelBH3}), $\tilde\Xi_{\rm in}\ge1$ and thus $\tilde\Psi_{\rm in}\ge\frac14$. The metric coefficient $g_{00}$, already calculated in (\ref{g00Rin}) can be written in terms of the only parameter that remains:
\begin{equation}
\label{g00Xiin}
g_{00}(\tilde R)=\tilde\Xi_{\rm in}(3-2\,\tilde\Xi_{\rm in})\left(1-{2\over\tilde R}\right).
\end{equation}

We now determine the characteristics of that particular orbit with pericenter $\tilde R_{\rm in}$. The tangential component of the linear momentum there equals
\begin{equation}\label{pT_Xi_in}
\tilde p_T^2(\tilde R_{\rm in})={2\,\tilde\Xi_{\rm in}-1\over3-2\,\tilde\Xi_{\rm in}}={2\tilde\Psi_{\rm in}\over1-2\tilde\Psi_{\rm in}}, 
\end{equation}
and the angular momentum equals
\begin{equation}\label{h_Xi_in}
\tilde h_{\rm in}={4\,\tilde\Xi_{\rm in}\over\sqrt{(3-2\,\tilde\Xi_{\rm in})(2\,\tilde\Xi_{\rm in}-1)}}={4\tilde\Psi_{\rm in}+1\over\sqrt{2\tilde\Psi_{\rm in}(1-2\tilde\Psi_{\rm in})}}. 
\end{equation}
By construction, $V_{\rm max}(\tilde h_{\rm in})=1$, since the integration interval collapses at $\tilde R=\tilde R_{\rm in}$ for $\gamma_{00}\tilde E_R=\tilde E_\infty=1$. Hence the apocenter of the orbit is $+\infty$, where the velocity is zero. We find
\begin{equation}\label{E_R_Xi_in}
\tilde E_R^2(\tilde R_{\rm in})={2\over3-2\,\tilde\Xi_{\rm in}}={1\over1-2\tilde\Psi_{\rm in}}.
\end{equation}
The (tangential) velocity at pericenter equals, with the expressions (\ref{def_tildepT}),
\begin{equation}\label{vT_Xi_in}
\tilde\upsilon_T(\tilde R_{\rm in})=c\sqrt{\tilde\Xi_{\rm in}-\frac12}=c\sqrt{2\tilde\Psi_{\rm in}}.
\end{equation}
All the above equations and considerations involving $\tilde\Xi_{\rm in}$ impose the limits
\begin{equation}
1\le\tilde\Xi_{\rm in}<\frac32 \quad {\rm or} \quad \frac14\le\tilde\Psi_{\rm in}<\frac12\quad {\rm and} \quad 3<\tilde R_{\rm in}\le4.
\end{equation}
The equality signs pertain to the naked black hole.

Since the maximum of the effective potential $V_{\tilde h_{\rm in}}(\tilde R)$ touches $\tilde E_\infty=1$,
$\tilde h_{\rm in}$ is also the minimum angular momentum a marginally bound orbit should have in order to remain in orbit around the black hole.  In the Schwarzschild case, this normalized specific angular momentum equals $\tilde h_{\rm in}=4$ and the critical pericenter $\tilde R_{\rm in}=4$ (see also fig. \ref{fig_infall}). If mass distributions are added, $\tilde h_{\rm in}$ increases, indicating that the added mass 'sucks' test orbits into the abyss. Interestingly, when the depth of the potential well $\tilde\Psi_{\rm in}\to\frac12$, which is its maximum value, $\tilde h_{\rm in}\to+\infty$.

\paragraph{The central response}
It is instructive to establish the behaviour of  $\rho(\tilde R)$ and ${\cal P}_R(\tilde R)$ for \mbox{$\tilde R\to\tilde R_{\rm in}$}, which can be done analytically since the integration interval collapses there. The analysis is a bit more complicated than meets the eye in (\ref{neut_response2}) and (\ref{pres_response}), though.

We start by noting that all energies $\tilde E_\infty$ that are permissible for non-plunging orbits are the maximum of a unique effective potential curve $V_{\tilde h}(\tilde R)$. We can therefore consider the transformation from $\tilde E_\infty$ to the radius $y$ which is the radius at which the effective potential curve $V_{\tilde h}(\tilde R)$ attains its maximum. For the radii $\tilde R\to\tilde R_{\rm in}$ we are in the Schwarzschild case, and thus, with (\ref{ERmin2})
\begin{equation}
\tilde p^2_T(y) = {1\over y-3},\qquad \tilde h^2(y)={y^2\over y -3}=\tilde h^2(\tilde E_\infty).
\end{equation}
From these expressions it is reasonable to expect that the transformation from $\tilde E_\infty$ to $y$ may result in a simplification. The maximum of $V_{\tilde h}(\tilde R)$ is its value in $y$:
\begin{equation}
\tilde E^2_\infty(y)=g_{00}(y)\left[1+\tilde p^2_T(y)\right]=g_{00}(y){y-2\over y-3}.
\end{equation}
This equation defines the transformation from $y$ to $\tilde E_\infty$. The transformation $\tilde E^2_\infty(y)$ is a decreasing one. The inverse transformation can be made explicit with (\ref{g00Xiin}) and results in a quadratic equation with solution
\begin{equation}
\label{trans_E_Infty_to_y}
y = {4\over2-\frac32u+\frac12\sqrt{(9u-8)u}},\quad u= {\tilde E_\infty^2\over\tilde\Xi_{\rm in}(3-2\,\tilde\Xi_{\rm in})}.
\end{equation}
For the local energy we find
\begin{equation}
\label{ERfxy}
\tilde E_R^2(\tilde R,y)={y-2\over y-3}\,\,{g_{00}(y)\over g_{00}(\tilde R)}={\tilde R\over \tilde R-2}\,\,{(y-2)^2\over y(y-3)}.
\end{equation}
The lower limit in the integral over $\tilde E_R$ equals 
\begin{equation}
\tilde E^2_{R,{\rm min}}=1+\tilde p^2_T(\tilde R)= {\tilde R-2\over \tilde R-3}.
\end{equation}
If we compare this expression with (\ref{ERfxy}) we see that the upper limit in the integral for $y$ equals $\tilde R$. This can be verified using (\ref{trans_E_Infty_to_y}) if we observe that
\begin{equation}
\tilde E^2_{\infty,{\rm min}}=g_{00}(\tilde R)\tilde E^2_{R,{\rm min}}=g_{00}(\tilde R){\tilde R-2\over \tilde R-3}=\tilde\Xi_{\rm in}(3-2\,\tilde\Xi_{\rm in}){(\tilde R-2)^2\over \tilde R(\tilde R-3)}
\end{equation}
and hence
\begin{equation}
u={(\tilde R-2)^2\over \tilde R(\tilde R-3)}.
\end{equation}
The upper limit in the integral over $\tilde E_\infty$ equals 1, or 
\begin{equation}
u={1\over\tilde\Xi_{\rm in}(3-2\,\tilde\Xi_{\rm in})}.
\end{equation}
Insertion into (\ref{trans_E_Infty_to_y}) yields, quite elegantly, for the lower limit of the integral over $y$ the value $y=\tilde R_{\rm in}$.

The radicand in the integrals such as (\ref{neut_response2}) transforms, after some rather lucky factorization, into
\begin{equation}
E_R^2-1-{\tilde h^2(\tilde E_\infty)\over \tilde R^2}={(2-\tilde R)y+4\tilde R\over \tilde R^2(\tilde R-2)y(y-3)}(\tilde R-y)^2.
\end{equation}
The term in $(\tilde R-y)^2$ reflects the fact that when $\tilde E_\infty\to\tilde E_{\infty,{\rm min}}$ then $y\to \tilde R$. The transformation $\tilde E^2_\infty(y)$ being a decreasing one, hence $y\le \tilde R$, and
\begin{equation}
4\tilde R-(\tilde R-2)y\ge4\tilde R-(\tilde R-2)\tilde R=\tilde R(6-\tilde R)\ge0.
\end{equation}
This allows us to write down the square root:
\begin{equation}
\sqrt{\tilde E_R^2-1-\tilde p_{T,{\rm min}}^2(\tilde R,\tilde E_R)}={\sqrt{(2-\tilde R)y+4\tilde R}\over \tilde R\sqrt{(\tilde R-2)y(y-3)}}(\tilde R-y).
\end{equation}
Lastly, the Jacobian of the transformation from $\tilde E_R$ to $y$ equals
\begin{equation}
d\tilde E^2_R= -{\tilde R\over \tilde R-2}\,\,{(y-2)(6-y)\over y^2(y-3)^2}\,\,dy.
\end{equation}
Collecting all the elements so far, we obtain finally for (\ref{neut_response2}): 
\begin{eqnarray}
\rho(\tilde R) &=& 4\pi m^3c^3\,\frac12{1\over(\tilde R-2)^{3/2}}\int_{\tilde R_{\rm in}}^{\tilde R} f(\tilde E_\infty(y))\,dy\times\nonumber\\
&&\quad \times{\sqrt{(2-\tilde R)y+4\tilde R}\,(\tilde R-y)(y-2)(6-y)\over y^{5/2}(y-3)^{5/2}}.
\end{eqnarray}
The transformation that yields the above form is valid for \mbox{$\tilde R\le\tilde R_{\rm lim}$}, as long as $g_{00}(\tilde R)$ does not differ significantly from the expression for empty space (\ref{g00Xiin}).

In a similar way we obtain for the radial pressure
\begin{eqnarray}
{\cal P}_R(\tilde R) &=& {4\pi\over3} m^3c^5\,\frac12{1\over\tilde R^{5/2}(\tilde R-2)^2}\int_{\tilde R_{\rm in}}^{\tilde R} f(\tilde E_\infty(y))\,dy\times\nonumber\\
&&\quad \times\,\,{\left[(2-\tilde R)y+4\tilde R\right]^{3/2}\!(\tilde R-y)^3(6-y)\over y^3(y-3)^3}.
\end{eqnarray}

The analysis can be generalized for general $g_{00}(\tilde R)$ and for \mbox{$\tilde R>\tilde R_{\rm lim}$},  but we will not pursue this any further. Our main purpose is to study the limit $\tilde R\to\tilde R_{\rm in}$. We find
\begin{equation}
\label{rho_to_Rin}
\rho(\tilde R)\to\rho_{\rm in}{\sqrt{(2-\tilde R)\tilde R_{\rm in}+4\tilde R}\over(\tilde R-2)^{3/2}}(\tilde R-\tilde R_{\rm in})^2
\end{equation}
with
\begin{equation}
\label{rho_in}
\rho_{\rm in}={\pi\over\sqrt2} m^3c^3f(1){(2\,\tilde\Xi_{\rm in}-1)^3(4\,\tilde\Xi_{\rm in}-3)\over(2\,\Xi_{\rm in})^{5/2}(3-2\,\tilde\Xi_{\rm in})^{5/2}}.
\end{equation}
Similarly
\begin{equation}
{\cal P}_R(\tilde R)\to{\cal P}_{R,{\rm in}}{\left[(2-\tilde R)\tilde R_{\rm in}+4\tilde R\right]^{3/2}\over\tilde R^{5/2}(\tilde R-2)^2}(\tilde R-\tilde R_{\rm in})^4
\end{equation}
with
\begin{equation}
\label{PR_in}
{\cal P}_{R,{\rm in}}={\pi\over48} m^3c^5f(1){(2\,\tilde\Xi_{\rm in}-1)^5(4\,\tilde\Xi_{\rm in}-3)\over(2\,\Xi_{\rm in})^3(3-2\,\tilde\Xi_{\rm in})^3}.
\end{equation}
For the mean velocity we find
\begin{equation}
\langle\beta\rangle(\tilde R_{\rm in})=\sqrt{\tilde\Xi_{\rm in}-\frac12},
\end{equation}
which we already knew from the characteristic (\ref{vT_Xi_in}) of the unique orbit at $\tilde R_{\rm in}$ for which $\beta=\beta_T$. 

Since both $\tilde\rho(\tilde R)$ and $\tilde{\cal P}_R(\tilde R)$ tend to 0 for $\tilde R\to\tilde R_{\rm in}$, we can integrate the system (\ref{genrelBH1}) and (\ref{genrelBH2}) there with a series expansion. We find 
\begin{equation}
\label{tilde_m_small}
\tilde m(\tilde R)\sim1+{4\pi\over3}\tilde R_{\rm in}^2\tilde\rho_{\rm in}(\tilde R-\tilde R_{\rm in})^3\equiv1+\tilde m_3(\tilde R-\tilde R_{\rm in})^3
\end{equation}
with $\rho_{\rm in}=\tilde\rho_{\rm in}{\cal M}/M^3$ and
\begin{equation}
\label{Xi_small}
\tilde\Xi(\tilde R) \sim \tilde\Xi_{\rm in}+(\tilde\Xi_{\rm in}-1)(\tilde\Xi_{\rm in}-\frac12)(\tilde R-\tilde R_{\rm in}) -\tilde\Xi_4(\tilde R-\tilde R_{\rm in})^4
\end{equation}
with
\begin{equation}
\tilde\Xi_4=\frac18\tilde m_3\,\tilde\Xi_{\rm in}(3-2\,\tilde\Xi_{\rm in})(2\,\tilde\Xi_{\rm in}-1).
\end{equation}
In the expressions above it is, of course, understood that the nonlinear terms in $(\tilde R-\tilde R_{\rm in})$ are zero if $\tilde R\le\tilde R_{\rm in}$. The linear terms in (\ref{Xi_small}) are compliant with (\ref{g00Xiin}), since we obtain here 
\begin{equation}
g_{00}(\tilde R)\sim\tilde\Xi_{\rm in}(3-2\,\tilde\Xi_{\rm in})(1-{2\over\tilde R}) + {2\over\tilde R}\tilde\Xi_4(\tilde R-\tilde R_{\rm in})^4.
\end{equation}


For larger $\tilde R$, the upper bound in the expressions for (\ref{neut_response2}) and (\ref{pres_response}) for $\rho(\tilde R)$ and ${\cal P}_R(\tilde R)$ tends to a constant, as does the lower bound. Hence there will be a range of radii where $\rho(\tilde R)$ and ${\cal P}_R(\tilde R)$ are fairly constant. 


\subsubsection{Application}
\label{sect_application}
We will now apply the formalism to dark matter particles that obey a Fermi-Dirac distribution. The rationale for this is that we know, from the flat rotation curves, that at least some dark matter obeys such a distribution (or a Maxwell-Boltzmann distribution).

We will express our results in terms of a reference mass $m_s=1$ eV (which is to be considered as an effective mean of the masses of all the dark matter species and their antiparticles) and a reference temperature of $T_s=2K$. The Fermi-Dirac and Boltzmann distributions also feature $m_0/(kT)$. Assuming $\upsilon_s=10^3$ km/s as a unit of velocity, this yields 
\begin{equation}
a_s\equiv{m_s\over kT_s}=6.441\times10^{-2}\upsilon_s^{-2}.
\end{equation}
The actual value of this parameter we will denote by 
\begin{equation}\label{defaprime}
a \equiv a_s\tilde a, \qquad \tilde a= {m_0/m_s\over T/T_s}\equiv{\tilde m\over\tilde T}
\end{equation}
with $\tilde a$ reflecting our ignorance about the appropriate value of the quotient $m/T$. We define
\begin{equation}
\label{def_ac}
a_c \equiv ac^2,
\end{equation}
and note that
\begin{equation}\label{def_as}
a_sc^2={m_s c^2\over kT_s}=5797.
\end{equation}
With these definitions, we obtain
\begin{equation}
\label{F-D}
f(\tilde E_\infty) = {2g\over {\rm h}^3}{m_0\over1+\exp\left[a_c(\tilde E_\infty-\tilde E_{\infty,N})\right]}
\end{equation}
featuring Planck's constant h and the number of dark matter species $g$. The normalisation respects the Pauli principle, i.e. $\tilde E_{\infty,N}$ should be smaller than the minimum $\tilde E_\infty$ for non-plunging orbits. In the case of a naked Schwarzschild black hole that energy equals $\tilde E_\infty^2=8/9$ (see fig. \ref{fig_infall}). In the general case, the minimum energy possible for a non-plunging orbit is $V_{\rm max}(\tilde h_{\rm lim})$, which is realized by the marginally bound orbit with the unstable pericenter $\tilde R_{\rm lim}$. Hence
\begin{equation}
-\infty\le\tilde E_{\infty,N}\le V_{\rm max}(\tilde h_{\rm lim}).
\end{equation}
The equality $\tilde E_{\infty,N}=V_{\rm max}(\tilde h_{\rm lim})$ would mean that the dark matter distribution in phase space is totally degenerate on the unstable circular orbit (with radius $\tilde R_{\rm lim}$). 

We obtain for $\tilde\rho$: 
\begin{eqnarray}
\label{neut_response3}
\tilde\rho(\tilde R)^{\left({\cal M}_\bullet\over M_\bullet^3\right)} &=& 3.802\times10^{-26}g\,\tilde m^4\tilde M_\bullet^2\,\times\nonumber\\
&&\times\int_{\tilde E_{R,\rm min}}^{\gamma_{00}^{-1}}\!\!{ \tilde E_R\,d \tilde E_R\over1+e^{a_c(\gamma_{00}\tilde E_R-\tilde E_{\infty,N})}}
\sqrt{\tilde E_R^2-1-\tilde p_{T,{\rm min}}^2}\,\,.\nonumber\\
\end{eqnarray}
The number density is of course similar:
\begin{eqnarray}
\label{neut_response4}
n(\tilde R)^{(m^{-3})} \!\!&=& 1.311\times10^{19}g\,\tilde m^3\,\times\nonumber\\
&&\times\int_{\tilde E_{R,\rm min}}^{\gamma_{00}^{-1}}\!\!{ \tilde E_R\,d \tilde E_R\over1+e^{a_c(\gamma_{00}\tilde E_R-\tilde E_{\infty,N})}}
\sqrt{\tilde E_R^2-1-\tilde p_{T,{\rm min}}^2}\,\,.\nonumber\\
\end{eqnarray}
The radial pressure reads:
\begin{eqnarray}
\label{press_response}
\tilde{\cal P}_R(\tilde R)^{\left({\cal M}_\bullet c^2\over M_\bullet^3\right)} \!\!\!\!\!&&= 1.267\times10^{-26}g\,\tilde m^4\tilde M_\bullet^2\,\times\nonumber\\
&&\times\int_{\tilde E_{R,\rm min}}^{\gamma_{00}^{-1}}\!\!{ d \tilde E_R\over1+e^{a_c(\gamma_{00}\tilde E_R-\tilde E_{\infty,N})}}
\left[\tilde E_R^2-1-\tilde p_{T,{\rm min}}^2\right]^{3/2}\!\!.\nonumber\\
\end{eqnarray}
The coefficient in front of (\ref{press_response}) simply equals the coefficient in front of (\ref{neut_response3}) divided by 3. The expression for the tangential pressures follows from (\ref{presT_response}) and (\ref{press_response}), mutatis mutandis in the integrand.

Close to $R_{\rm in}$ we obtain for (\ref{rho_in})
\begin{equation}
\tilde\rho_{\rm in}^{\left({\cal M}_\bullet\over M_\bullet^3\right)}=6.721\times10^{-27}{g\,\tilde m^4\tilde M_\bullet^2 \over1+e^{a_c(1-\tilde E_{\infty,N})}}{(2\,\tilde\Xi_{\rm in}-1)^3(4\,\tilde\Xi_{\rm in}-3)\over(2\,\Xi_{\rm in})^{5/2}(3-2\,\tilde\Xi_{\rm in})^{5/2}}
\end{equation}
and mutatis mutandis for the number density
\begin{equation}
n_{\rm in}^{\left({\cal M}\over M^3\right)}=2.318\times10^{18}{g\,\tilde m^3 \over1+e^{a_c(1-\tilde E_{\infty,N})}}{(2\,\tilde\Xi_{\rm in}-1)^3(4\,\tilde\Xi_{\rm in}-3)\over(2\,\Xi_{\rm in})^{5/2}(3-2\,\tilde\Xi_{\rm in})^{5/2}}.
\end{equation}
For (\ref{PR_in}) we get
\begin{equation}
\tilde {\cal P}_{R,{\rm in}}^{\left({\cal M}_\bullet c^2\over M_\bullet^3\right)}=3.960\times10^{-28}{g\,\tilde m^4\tilde M_\bullet^2 \over1+e^{a_c(1-\tilde E_{\infty,N})}}{(2\,\tilde\Xi_{\rm in}-1)^5(4\,\tilde\Xi_{\rm in}-3)\over(2\,\Xi_{\rm in})^3(3-2\,\tilde\Xi_{\rm in})^3}.
\end{equation}

\subsubsection{Some examples and discussion}
As already indicated below equation (\ref{genrelBH1_class}), the density (\ref{neut_response3}) and the pressure (\ref{press_response}) are determined by integrating the system (\ref{genrelBH1}) and (\ref{genrelBH2}) over $\tilde R$ with the initial conditions $\tilde m(\tilde R_{\rm in})=1$ and $1\le\tilde \Xi_{\rm in}<\frac32$. Since $\tilde\rho>0$ everywhere, there is no clear recipe for determining the outer boundary $\tilde R_b$. It will depend on the mass of the black hole, the formation process, the amount of dark matter available and temperature of the dark matter particles.\footnote{The outward integration we use here is of course not a reflection of what would have happened physically. The collapse and accretion processes would have run their courses, and in the end $\tilde\Xi_{\rm in}$ would have followed from that, much like, classically, the mass density $\rho(R)$ comes first, and the central potential follows from
\begin{equation}
\lim_{R\to0}{\Xi(R)\over R} = 4\pi G\!\int_0^{R_b}\!\!\!x\,\rho(x)\,dx.
\end{equation}}

The solution clearly depends on $2\times3$ parameters. There are the 3 scaling parameters: average mass of the dark matter particle $m$, number of degrees of freedom $g$ and geometrical mass of the black hole $M$. The other 3 are the temperature $T$, normalisation $\tilde E_{\infty,N}$ and the depth of the potential well $\tilde\Psi_{\rm in}$. The outer boundary $\tilde R_b$ is to some extent also a free parameter. 

In the assumption that $\tilde\Xi_{\rm in}$ is largely determined by the mass of the surrounding galaxy, we can obtain a relation involving $\tilde\Xi_{\rm in}$ and the mass of the galaxy. Assuming a simple Hernquist potential
\begin{equation}
\label{Hern_pot}
\tilde\Psi_{\rm H}(R)={G({\cal M_{\rm H}+{\cal M})}\over c^2}{1\over R+R_{\rm H}}={{\cal M}_{\rm H}+{\cal M}\over{\cal M}}{1\over\tilde R+\tilde R_{\rm H}}
\end{equation}
we find
\begin{equation}
\tilde\Xi(\tilde R)={{\cal M}_{\rm H}+{\cal M}\over{\cal M}}{\tilde R\over\tilde R+\tilde R_{\rm H}}\sim{{\cal M}_{\rm H}+{\cal M}\over{\cal M}} {\tilde R\over\tilde R_{\rm H}}
\end{equation}
since $\tilde R_{\rm H}$ is much larger than $\tilde R_{\rm in}$ near the black hole. This approximation is probably valid for many more potentials for which it makes sense to define a `mean' radius $R_{\rm H}$. If the dynamical effects of the stellar mass in the close vicinity of the black hole is negligible compared to these of the mass of the black hole, we can match the linear terms in (\ref{Xi_small}) with the above expression to obtain at the match point
\begin{equation}
\label{sol_Rmatch}
\tilde R_{\rm match} = {\tilde\Xi_{\rm in}(3-2\,\tilde\Xi_{\rm in})\over {{\cal M}_{\rm H}+{\cal M}\over{\cal M}\tilde R_{\rm H}}-(\tilde\Xi_{\rm in}-1)(\tilde\Xi_{\rm in}-\frac12)}.
\end{equation}
This radius should be larger than $\tilde R_{\rm in}$, leading to
\begin{equation}
{2G({\cal M}_{\rm H}+{\cal M})\over c^2}{1\over R_{\rm H}}=2{{\cal M}_{\rm H}+{\cal M}\over{\cal M}\tilde R_{\rm H}}<\tilde\Xi_{\rm in}-\frac12
\end{equation}
which is hardly a constraint since the above expression translates into the requirement that the scaling radius $R_{\rm H}$ of the galaxy, expressed in Schwarzschild radii of the galaxy, should be larger than $1/(\tilde\Xi_{\rm in}-\frac12)<2$. 

\begin{figure}[ht]
   \centering
   \includegraphics[width=88mm]{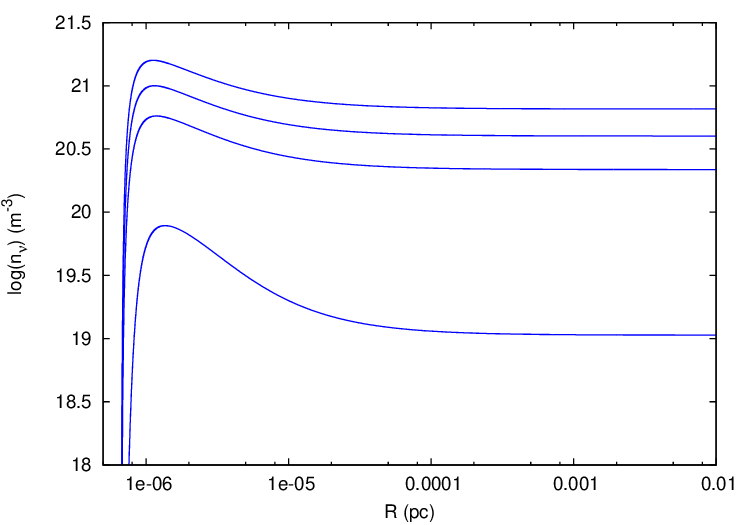}
      \caption{The logarithm of the dark matter number density, expressed in ${\rm m}^{-3}$, as a function of $\tilde R$, expressed in pc. We adopt $\tilde m=4$, $g=6$, $\tilde \Xi_{\rm in}=1.3$, $\tilde E_{\infty,N}=.7$ and $\tilde M=4.3$, which is the putative mass of the black hole in the center of our galaxy (Genzel {\it et al.} \cite{genzel}). From the lower curve up are the temperatures: $T=(10^3,5\times10^3,10^4,10^5)$ K, or $\bar t/\tilde m=(100,500,10^3,10^4)$. Therefore the part of the Fermi-Dirac distributions that we consider is largely degenerate. We see that, apart from the very center, the density is essentially a constant. Note also that the density rises extremely steeply in the center, since $n=0$ at $R_{\rm in}=6.7\times10^{-7}{\rm pc}$.
                   }
         \label{fig_neutndens}
   \end{figure}

\begin{figure}[ht]
   \centering
   \includegraphics[width=80mm]{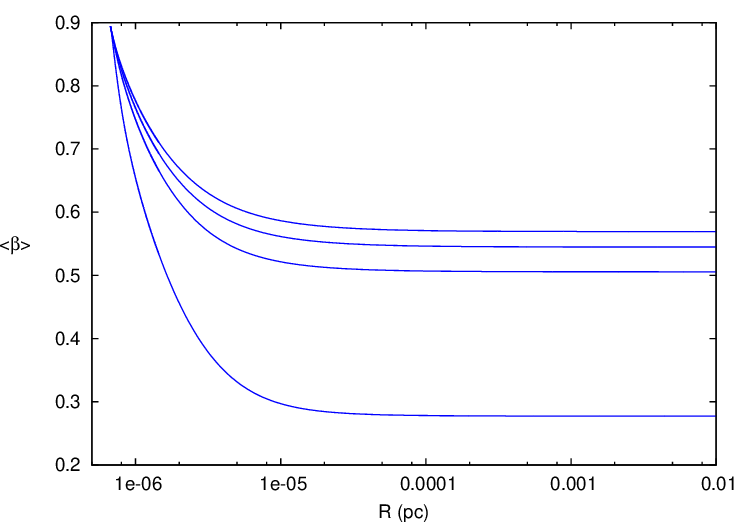}
      \caption{The mean dark matter velocity $\langle\beta\rangle$, with the same parameters as fig. \ref{fig_neutndens}}
         \label{fig_neutbeta}
   \end{figure}

\begin{figure}[ht]
   \centering
   \includegraphics[width=80mm]{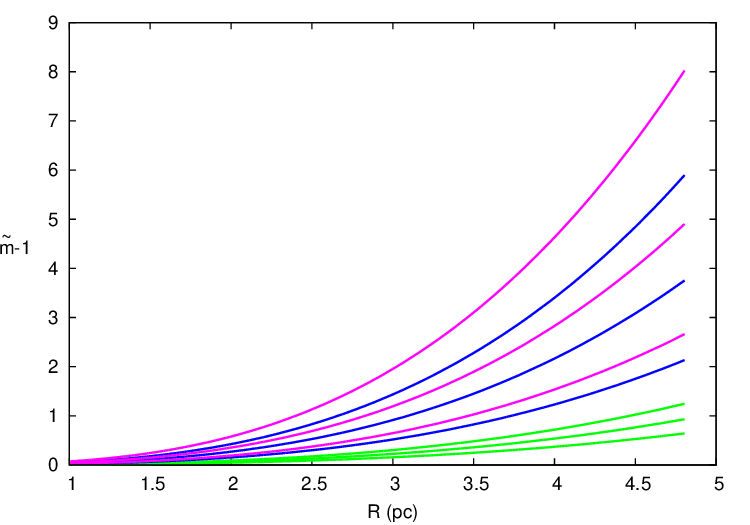}
      \caption{The total dark matter mass, expressed in $M$, as a function of $R$. As explained in the text, the curves $\tilde m(R)-1$ are essentially cubic polynomials in $R$. The parameters are the same as fig. \ref{fig_neutndens}. Particle masses are $m=2.5$ eV (green), $m=3.7$ eV (blue) and $\tilde m=4$ eV (magenta). For each particle mass, 3 temperatures are shown: \mbox{$T=(5\times10^3,10^4,10^5)$ K}, the lower curves corresponding to the lower temperatures. According to Genzel {\it et al.} (\cite{genzel}), there is in a sphere of 5 pc around the black hole at the center of our Galaxy about as much dark matter as the mass of the black hole itself, hence $\tilde m(5)-1\sim1$. If, instead of $\tilde E_{\infty,N}=.7$ we would have taken the maximum possible $\tilde E_{\infty,N}=.88$ (close to the Schwarzschild value), the 3 curves corresponding to the 3 temperatures would almost coincide with the one for the highest temperature ($10^5$ K).                   }
         \label{fig_neutcentralmass}
   \end{figure}

If ${\cal M}_{\rm H}\to0$, equation (\ref{sol_Rmatch}) has the solution $\tilde\Xi_{\rm in}=1$ and $\tilde R_{\rm match}=\tilde R_{\rm H}(=4)$. For very deep potential wells, the expression (\ref{Hern_pot}) should be replaced by a general relativistic one. Assuming that in reality it is never really necessary to recur to general relativity for galaxy dynamics, the (unattainable) limit $\tilde\Xi_{\rm in}\to\frac32$ would lead to \mbox{$2{{\cal M}_{\rm H}+{\cal M}\over{\cal M}\tilde R_{\rm H}}\to1$} which means that $R_H$ would be equal to the Schwarzschild radius of the mass of the galaxy as a whole. Physically this makes sense, but it is quite impossible in the present state of the universe. It might be considered as a possibility though in the dense phases of the history of our universe.

We don't see the usual dependence on $T^3$ for the density or $T^4$ for the pressure. This is a result from the integration, which is only over a small range in $\tilde E_R$, whereas usually one integrates from $1$ to $+\infty$.\footnote{In the latter case, a substitution $u=a_c\gamma_{00}\tilde E_R$ would be in order, leading to the temperature dependence just mentioned.} As a consequence, for values of $\tilde m_\nu$ and $\tilde t$ such that \mbox{$\tilde t/\tilde m_\nu>10^3$}, the part of the neutrino distribution that we integrate over is completely degenerate. 


The figures are models for the dark matter envelope of the black hole at the center of the Milky Way. For a discussion, we refer to the captions.

\begin{figure}[ht]
   \centering
   \includegraphics[width=80mm]{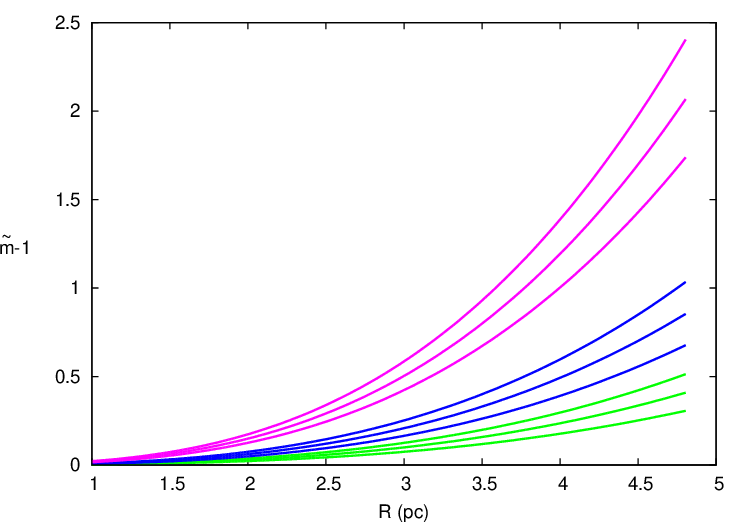}
      \caption{The total dark matter mass, expressed in $M$, as a function of $R$. The setup and the parameters are the same as fig. \ref{fig_neutcentralmass}, but here the particle mass is fixed at  $m=2$ eV, and $\tilde\Xi_{\rm in}=1.3$ (green), $\tilde\Xi_{\rm in}=1.35$  (blue) and $\tilde\Xi_{\rm in}=1.4$ (magenta). From this figure follows that  2 eV may be a lower limit for the particle mass, since $\tilde\Xi_{\rm in}=1.4$ may be considered as representative for a rather extreme collapse/accretion. 
                   }
         \label{fig_neutcentralmass_psi}
   \end{figure}

\subsection{Neutrino dark matter distributions in galaxies}
\label{sect_darkmatter_gals}
\subsubsection{On the formation of the neutrino halo}
\label{sect_intro_neutrinohalo} 

Any model for dark matter must be able to produce flat rotation curves. A necessary and sufficient condition in a spherical geometry is that the total mass enclosed within a certain radius increases roughly linearly with that radius. If that mass obeys collisionless dynamics, there are many equilibrium anisotropic distribution functions (functions of energy and the magnitude of the angular momentum) that would qualify. However, if the full freedom of anisotropic distribution functions is to be allowed, it seems unlikely that in all cases they would 'end up' producing flat rotation curves. For the subset of isotropic distribution functions on the contrary, a necessary and sufficient configuration is the (approximately) selfconsistent isothermal sphere, which, if fully selfconsistent without other mass distributions, depends on only one parameter, the temperature. 

There is at least one more inference that must be made in connection with flat rotation curves: if the dark matter is isothermal, it must also be one phase, i.e. it is not a superposition of isothermal distributions with different temperatures. Of course, it cannot be excluded that neutrino's exist in different phases with different temperatures. Since the temperature drops with the expansion of the universe, at a certain epoch $t_{\rm lock}$ a treshold must have been attained at which the coldest of the neutrino phases that are abundant enough to yield a dynamically significant dark matter component and that happened to be in the neighbourhood of a galaxy became 'trapped' or 'locked' into the gravitational field of the galaxy in a relatively short period of time (compared with the timescale of the expansion phase, of course).  Note also that the neutrino's have always been in equilibrium around the galaxy, only at $t_{\rm lock}$ (and later) the coldest dynamically significant phase becomes for the most part bound.  

We note also that, because isothermal distributions are the only isotropic distribution functions that produce flat rotation curves,\footnote{Assuming sphericity. Any departure of sphericity is not likely to change the qualitative arguments in this section.} the functional form of the distribution clearly must remain approximately constant during halo formation. This means that, in virtue of Jeans's theorem, the halo formation must have been sufficiently fast, lest the galaxy's gravity impinge too much on the form of the distribution function.

This scenario yields a constraint. A flat rotation curve can only be obtained if the mass density is close to an exponential of the potential. If the distribution is to be bound, the expression of the mass density (\ref{isothermdist4}) shows that the argument $a\Psi$ must be large enough in order that the normalized incomplete gamma function $\gamma_N(\frac32,a\Psi)$ is close to 1. This is the case if
\begin{equation}
a{GM_{\rm tot}(R)\over R}>f_1\sim5.
\end{equation}
with $M_{\rm tot}$ the mass of the galaxy and the neutrino's combined inside the radii $R$ where the rotation curve is flat. In terms of $\tilde a$, as defined in (\ref{defaprime}), we obtain
\begin{equation}\label{condaprime}
\tilde a>359 f_1 {R^{({\rm kpc})}\over M_{\rm tot}^{(10^{10}M_\odot)}}. 
\end{equation}
For a selfconsistent isothermal distribution the constant circular velocity $V_c$ equals
\begin{equation}\label{vcirc_isotherm}
V_c=\sqrt{\frac2a}
\end{equation}
and thus
\begin{equation}\label{cond_vcirc_isotherm}
V_c^{({\rm km/s})}<294 \left({M_{\rm tot}^{(10^{10}M_\odot)}\over f_1R^{({\rm kpc})}}\right)^{1/2}. 
\end{equation}

All these inequalities seem to be consistent with the observations.
If we adopt $V_c=200\,{\rm km/s}$ as a representative value we obtain $\tilde a=776$.  This may seem like a large number (hence large average neutrino mass or very cold neutrino distribution, or a combination of both), but it is simply an observational fact, and it satisfies the constraint (\ref{condaprime}) for, say, a Milky Way like galaxy. Therefore this order of magnitude for $\tilde a$  is not a consequence of the paradigm we adopted for this paper, but must be explained by any dark matter candidate. If $m_\nu\sim2$ eV as suggested in fig. \ref{fig_neutcentralmass_psi}, then $T\sim5.16\times10^{-3}$ K.

Condition (\ref{cond_vcirc_isotherm}) assures that, if the mass in neutrino's dominates the stellar mass, the neutrino mass density essentially follows the well-known isothermal sphere law
\begin{equation}
\label{def_rho_isoth}
\rho_{\rm isoth}(R) = (2\pi Ga)^{-1}R^{-2},
\end{equation}
clearly showing the direct relation between $a$ and the amount of dark matter. Hence neutrino's in circular orbits will have the same velocities as stars and gas in circular orbits.

On the other hand, inspection of figure \ref{fig_model1} or \ref{fig_model2} reveals that between 'lock-in' time and the current epoch, the radius of our universe, as defined in (\ref{def_radius_universe}), has increased by a rather modest one digit factor, presumably larger than 3. Assuming the standard relation
\begin{equation}
{T\over T_0}=\left({\cal R}\over{\cal R}_0\right)^2
\end{equation}
the neutrino temperature now can be estimated as a factor of 10-50 smaller than at 'lock-in' time.

At $t_{\rm lock}$, in addition, the cosmic deceleration should be less than the gravitational attraction of the galaxy, so that the neutrino's, which were at temperature $T_{\rm lock}$, would not continue to cool appreciably, lest a temperature gradient occur which we don't see. As a consequence, the cosmic evolution stops being important for these neutrinos, and they become a gravitationally bound distribution, in equilibrium around the galaxy.

In order to treat this condition semi-quantitatively, we note that for any material particle that freely moves (i.e. on a geodesic, but not necessarily comoving and in the absence of an appreciable gravitational field) with velocity $\boldsymbol{\upsilon}$ with respect to a comoving observer at the same location in a synchronous universe (on the scales of galaxy clusters this is a good approximation) $|\boldsymbol{\upsilon}|{\cal R}$ is a constant. This means that
\begin{equation}
\dot{\boldsymbol{\upsilon}} = -\boldsymbol{\upsilon}{\dot{\cal R}\over{\cal R}} = -\boldsymbol{\upsilon}H(t).
\end{equation}
The cosmic deceleration therefore acts as a kind of friction that is proportional to the velocity $|\boldsymbol{\upsilon}|$.
The attraction due to the galaxy must be considerably larger than this frictional force, and hence
\begin{equation}
{GM_{\rm tot}\over \langle R\rangle^2} = f_2\langle\upsilon\rangle H
\end{equation}
with $\langle R\rangle$ a characteristic size, $\langle\upsilon\rangle$ a characteristic velocity of the neutrino distribution and $f_2$ a dimensionless fudge factor. This factor stands for all the unknowns, is presumably somewhere in the one digit range, but probably larger than 5, meaning that the cosmic deceleration is only between $20\%$ and $10\%$ of the galactic attraction. For an isothermal distribution 
\begin{equation}
\langle\upsilon\rangle={2\over\sqrt\pi}V_c
\end{equation}
and thus
\begin{equation}
H= {1\over f_2}{\sqrt\pi\over2 V_c}{GM_{\rm tot}\over\langle R\rangle^2}
\end{equation}
or, with $V_c^2=GM_{\rm tot}/\langle R\rangle$,
\begin{equation}
H ={\sqrt\pi\over2f_2}{V_c\over\langle R\rangle}.
\end{equation}
If we insert $V_c=200\,{\rm km/s}$, $\langle R\rangle=15\,{\rm kpc}$ and $f_2=10$ we obtain ${\rm H}\sim{\rm1181\,km/s/Mpc}$, or in our units $H\sim11$. Referring to the figures \ref{fig_model1} and \ref{fig_model2} that show models for which $H_{\rm o}$ occurs at $t_{\rm o}\sim12-13\,{\rm Ga}$ after the beginning of the latest expansion phase, we see that the 'lock-in' time must have occurred before the universe was $1\,{\rm Ga}$ far in that latest expansion phase. Given the great uncertainties in this qualitative argument, it is also possible that the neutrino's were already locked up when the latest expansion phase commenced.  

Any more detailed analysis of this 'trapping' of neutrino's by galaxies would require a large cosmological scale simulation with many clusters, galaxies and neutrino clouds, thereby forgoing the fact that, in our paradigm, the initial conditions for such a simulation are basically unknown anyway. 
 
\subsubsection{A simple model for a Milky Way type galaxy}
\label{sect_simple_model}

The neutrino density at infinity will be expressed as a factor of the standard closure density (see equation (\ref{rhounit}) and the comment there) and we write
\begin{equation}\label{def_rhoprime}
\rho_{\nu,\infty}=\rho_{\rm crit}\,\tilde \rho_{\rm crit}=\tilde\rho_{\rm crit}10^{-26}\,{\rm kg}\,{\rm m}^{-3}
\end{equation}
according to (\ref{def_rhoc}), with $\tilde \rho_{\rm crit}$ reflecting our uncertainty as to the value of the background neutrino density at 'lock-in' time.

As argued above, we assume that the neutrino's are bound to the galaxy, and therefore we adopt the halo density (\ref{isothermdist4}).
The essential ingredient for flat rotation curves is a central mass concentration. The reason is that the signature of the exponential in  (\ref{isothermdist4}) must be present, which is only the case if the potential well is deep enough in order that a sufficient number of neutrino's are bound. Therefore a bulge-like structure must be present if we are to obtain a flat rotation curve. This is consistent with the bulges being old. They may be formed, together with the globulars, in an era preceding the latest expansion phase. Denoting the mass density of the galaxy, excluding the dark matter, by $\rho_L(R)$, we look for an equilibrium configuration that can be found by solving the spherical Poisson equation. It is clear that, because of the assumption of sphericity, we do not need to aim for much detail, and the model therefore is not much more than a proof of concept.  

In this subsection we adopt 1 Mpc as the unit of distance, express masses in the unit $10^{11}M_\odot$ and keep $\upsilon_S=10^3$ km/s as the unit of velocity. The neutrino density at infinity is then
\begin{equation}
\rho_{\nu,\infty}=1.469\,\tilde \rho_{\rm crit}.
\end{equation}
Thus we obtain the form of the Poisson equation that we are to integrate:
\begin{eqnarray}
\label{Poisson}
d^2_R\tilde\Psi +{2\over R}d_R\tilde\Psi &=& -3.500\times10^{-4}\tilde a\left[\rho_L + \rho_{\nu,\infty} e^{\tilde\Psi}\gamma_N(\textstyle{\frac32},\tilde\Psi)\right].\nonumber\\
\end{eqnarray}
with
\begin{equation}\label{def_tildePsi}
\tilde \Psi(R) = a\Psi(R),
\end{equation}
$\Psi(R)$ the Newtonian binding potential, and $\rho_L$ the mass in stars and dust, including the central mass concentration.

We will treat (\ref{Poisson}) as an initial value problem and we integrate outwards, and thus we need $\tilde\Psi(0)$ and $d_R\tilde\Psi(0)$.  From the previous subsection, we know that the mass density is essentially constant around the central black hole. Therefore, we can start the integration at about 5 pc ($5\times10^{-6}$ in our units), and an enclosed mass of  $10^6M_\odot$ ($10^{-5}$ in our units), which is presumably an underestimate. The potential is quadratic, the coefficient of $R^2$ is fixed by the choices above, but we can still choose $\tilde\Psi(0)$. It turns out that $\tilde\Psi(0)=.44$ is not a bad choice.

For $\rho_L$ we adopt a Hernquist bulge and an exponential envelope
\begin{equation}
\label{rho_L}
\rho_L = {M_H R_H\over 2\pi R(R_H+R)^3}+ {M_e\over8\pi R_e}e^{-R/R_e}
\end{equation}
with $M_H=0.15$, $R_H=8\times10^{-4}$ (Widrow \& Dubinsky \cite{widrow}), $M_e=.1$ (rather arbitrary) and $R_e=2.1\times10^{-3}$ (from fig. 15 in Fathi et.al. \cite{fathi}).
The integration is stopped when the total mass in neutrino's equals the total mass in a constant density sphere with density  $\rho_{\nu,\infty}$, indicating that the neutrino's have settled in a selfconsistent distribution. 

\begin{figure}[ht]
   \centering
   \includegraphics[width=130mm]{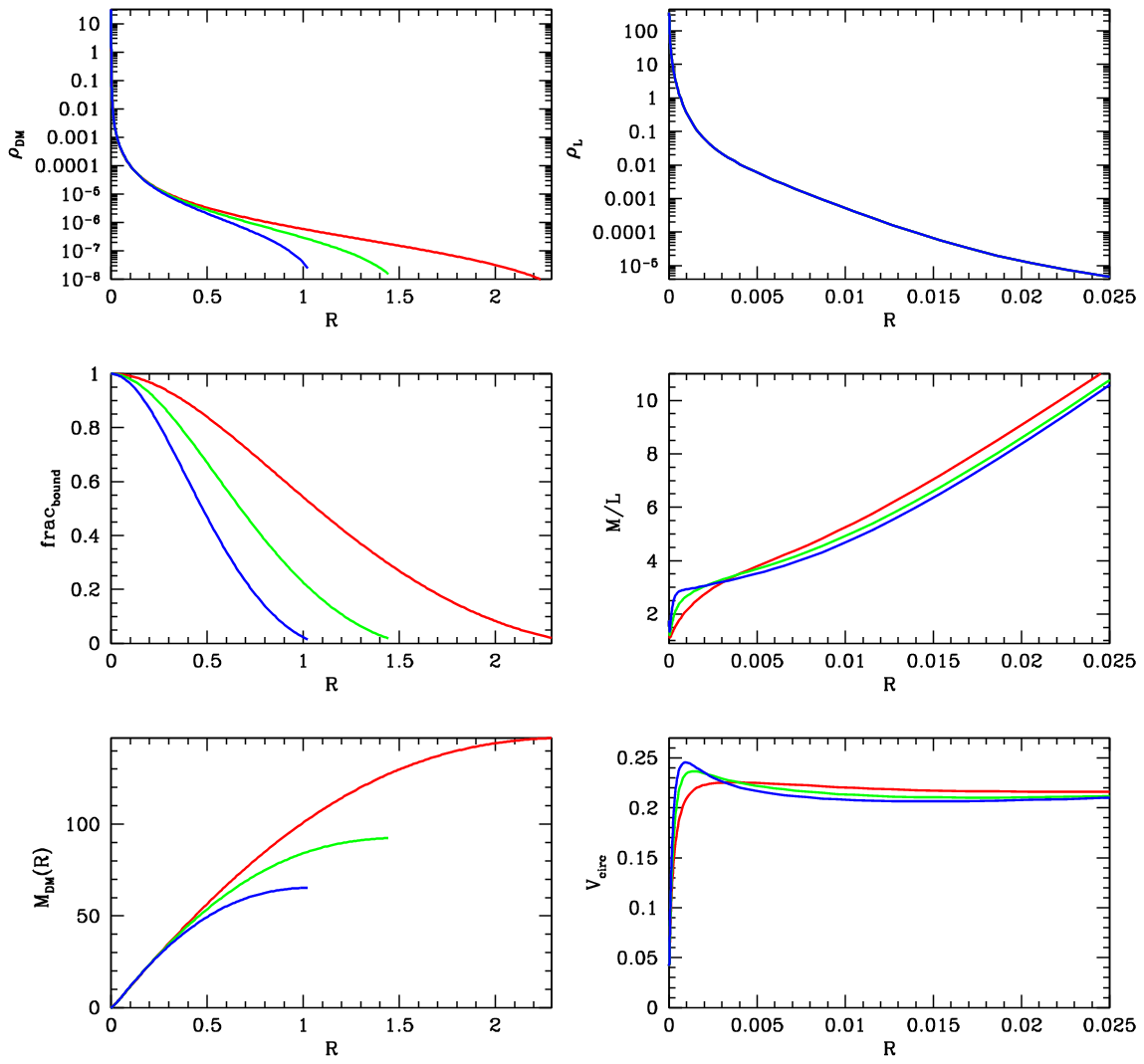}
      \caption{Models for a Milky Way type galaxy. Upper left panel: dark matter density. Upper right panel: luminous matter density. Middle left panel: fraction of neutrino's that is bound to the galaxy. Middle right panel: cumulative mass to light ratio. Lower left panel: cumulative mass. Lower right panel: rotation curve. Three models are shown: $\tilde\rho_{\rm crit}=2$ (red), $\tilde\rho_{\rm crit}=5$ (green) and $\tilde\rho_{\rm crit}=10$ (blue).                  }
         \label{fig_milkyway}
   \end{figure}

\section{Dark energy}
\label{sect_darkenergy}
The case for dark energy rests essentially on 3 arguments: the occurrence of a big bang and the consequences that come with it (such as the nucleosynthesis of light elements), the magnitude-redshift relation and the analysis of the CMB. In our paradigm, the first argument is powerless, since there is no such thing as a big bang 'of creation' (section~\ref{sect_early}). As to the second argument, it is very plausible from subsection~\ref{sect_comp_obs} that the $z(\ell)$ and $m(z)$ relations can be satisfied by our models, even with $\Lambda=0$. If needed, we can invoke $\Lambda\ne0$ which has the same effect as the dark energy, but that choice is then to be a feature of the 'mother universe' that our universe contains, and of which all knowledge is futile anyway. In general, one is not used to consider black holes with a $\Lambda\ne0$. The main characteristics of them are touched upon in appendix~\ref{sect_Novrad}. Lastly, the universes we consider have a CMB that is isotropic by design, and if it contains fluctuations, they cannot learn us very much because they are the remnants of details of history that do not change the grand picture.

\section{Mach's principle and Newton's second law}
\label{sect_Mach}

Mach's principle is rooted in a discussion that dates back to Newton and Leibniz. To the latter, space was simply a construct which was useful to stage the introduction of coordinates, and he correctly realized that the actual choice of the coordinates is quite arbitrary. Hence, any physical theory should be independent of the choice of a coordinate system, and therefore he was an advocate of covariance {\it avant la lettre}. Newton on the other hand quite convincingly proved Leibniz 'wrong' with his famous pail experiment, showing that one must endow the concept of 'absolute space' with physical reality. 

Here, just as in conventional cosmology, it is quite natural to promote the comoving frame to the status of coordinated absolute space, and since comoving observers do not experience any accelerations, the special significance of their coordinate systems does right to the special significance of absolute space.

The above statement does not explain, however, what the physical mechanisms are behind the 'punishments' that inhabitants of coordinate frames that are not comoving must endure, such as  the so called 'apparent forces' (linear acceleration, centrifugal and Coriolis forces). These forces are proportional to mass, but they have nothing to do with gravitation. Hence the need for inertial mass $m_I$ as distinct from the gravitational mass $m_G$. Mach's idea is simple and attractive: it is the whole body of the universe that defines the absolute frame. Yet, a physical explanation to back up the idea is not part of Mach's principle, and as such Mach's principle is therefore 'nothing more' than the assertion in the previous paragraph. It does suggest, however, a link between $m_I$ and $m_G$.

General relativity evades the issue, in a sense. It replaces Newton's second law $F=m_I \times a$ with the geodetic equations. These deal with accelerations, not with forces. Hence there is no need to introduce $m_I$. This also means that we cannot use general relativity as a tool to investigate Mach's principle.

We now revisit Mach's principle in the framework of our universes. We invoke a cosmic (dark) matter background $\rho(t,\mathbf{r})$. It has been argued in section~\ref{sect_darkmatter_intro} that it is very likely to be highly isotropic, and thus $\rho(t)$. The attraction of that isotropic background on a mass $m_G$ causes a net force that is zero, as was already known to Newton, be it that in his case the net force was the subtraction of 2 infinities. Here, however, the universes are finite in spacetime, and no singularities are to be confronted. Therefore, we can actually calculate something.

The zero net attraction causes a tension, however. A similar situation occurs if a (massless) particle is attached to two very long springs that pull at it in opposite directions with equal force, so that there is equilibrium. Because the springs are very long, Hooke's law (typically $F=kx$, with $x$ the distance from the equilibrium point and $k$ the spring constant) does not apply, and the force is independent of $x$. If we apply a force on the particle, this will cause an acceleration. The more the springs are tautened however, the smaller the acceleration will be for the same applied force. The particle thus acquires a stiffness due to the tension in the springs.

In order to obtain the tension $F$ that a massive test particle experiences, we now calculate the acceleration due to the mass density $\rho(t)$ in one hemisphere. We can work in the weak field limit, and thus the $r^{-2}$ law applies. However, our universes are not static. The 'distance' between 2 massive points is set by the light travel time, since it follows from the mathematical structure of the Einstein equations that gravitation propagates along null geodesics.

In a static universe, we would obtain the tension $F$ by integration of the $r^{-2}$ gravitational attraction law over one hemisphere: 
\begin{equation}
\label{F_Mach1}
F=m_G\int {(4\pi G)\rho\over (4\pi r^2)}r^2\,dr\int_{-{\pi\over2}}^{\pi\over2}\!\!|\sin\vartheta|\cos\vartheta\,d\vartheta\int_0^\pi\!\!d\varphi.
\end{equation}
Here we turn to the formalism proper to the synchronous universes of section~\ref{sect_synchronous}.
The metric (\ref{isochmetrichi}) gives rise to the volume element at emission time $t'$
of an attractor
\begin{equation}
dV = {\cal R}^3(t')\,\sin^2\chi'\,|\sin\vartheta'|\,d\chi'\,d\vartheta'\,d\varphi',
\end{equation}
which should replace the volume element $r^2\,dr|\sin\vartheta|d\vartheta d\varphi$ in (\ref{F_Mach1}). The factor $4\pi r^2$ in the denominator of (\ref{F_Mach1}) is the Euclidean surface of the sphere centered on the attractor with coordinates $(t',\chi',\vartheta',\varphi')$ on which the test particle with coordinates $(t,\chi,\vartheta,\varphi)$ is located. This should be replaced by the surface area of the sphere in curved spacetime
\[
2\pi{\cal R}^2(t)\sin^2[\chi''(\chi,\vartheta,\chi',\vartheta')]
\]
with $\chi''$ the angle between attractor and test particle on the great circle that connects both and that was referred to at the end of section \ref{sect_lightpaths_sync}. In this expression, we changed coordinate systems to put the attractor at the center and integrated over a suitable defined $\varphi''$ (the factor $\pi$) and $\vartheta''$ (the factor 2).
Collecting these changes and integrating over $\varphi'$, we obtain
\begin{eqnarray}
F(t,\chi,\vartheta,\varphi)&=&{Gm_G\over2{\cal R}^2(t)}\!\int\!\!\sin^2\chi'\,d\chi' \!\!\int_{-{\pi\over2}}^{\pi\over2}\!d\vartheta'
|\sin\vartheta'|\cos\vartheta'\times\nonumber\\
&&\qquad\times\,{\rho[t'(\chi'')]\,{\cal R}^3[t'(\chi'')])\over \sin^2[\chi''(\chi,\vartheta,\chi',\vartheta')]}.
\end{eqnarray}
For the case we consider here, the synchronous universes, this expression simplifies a lot. We use the explicit expressions (\ref{rhosync}) for $\rho(t)\,\,(= \rho_{c,{\rm sync}}/ \cyc^3)$ and (\ref{radius_sync}) for ${\cal R}(t)\,\,(=\cyc/\omega_{\rm sync})$, and obtain
\begin{equation}
\label{F_Mach2}
F(t,\chi,\vartheta,\varphi)={Gm_G\over2{\cal R}^2(t)}{\rho_{c,{\rm sync}}\over \omega_{\rm sync}^3}\!\!\int\!\!d\chi' \!\!\int_{-{\pi\over2}}^{\pi\over2}\!\!\!d\vartheta'\,
{\sin^2\chi'\,|\sin\vartheta'|\cos\vartheta'\over \sin^2[\chi''(\chi,\vartheta,\chi',\vartheta')]}.
\end{equation}
This is still a complicated integral, as it stands, but it is a purely geometrical one. We can now use the embedding paradigm which transforms the 2D universe to the surface of a sphere, be it that `the lower hemisphere' is to be identified with `the upper hemisphere'. This allows us to see the lines of force as great circles on the surface of that sphere. Hence we can again change the coordinate system $(\chi',\vartheta')$ to a new one $(\chi'',\vartheta'')$ in which the test particle is at $\chi''=0$. This makes the integrand trivial, since $\chi''=\chi'$. We can integrate over the angles $\vartheta''$ 

\begin{small}
\[\int_{-\pi/2}^{\pi/2}\!\!|\sin\vartheta''|\cos\vartheta''\,d\vartheta''=1
\]
\end{small}

to obtain
\begin{equation}
\label{F_Mach3}
F(t)={Gm_G\pi\over2{\cal R}^2(t)}{\rho_{c,{\rm sync}}\over \omega_{\rm sync}^3}\int \!\!d\chi''.
\end{equation}
In order to calculate the integral over $\chi''$, we recall from section \ref {sect_lightpaths_sync} that for a closed synchronous universe with $\Lambda=0$, $\chi''_{\rm horizon}=\pi$ for every observer during a full expansion or a full contraction. Two images (a direct one and a reflected one, the latter being in the opposite direction of the direct one in the case of perfect symmetry) of every object are produced during one full expansion or one full contraction. These images will not appear at the same time, of course, but within the period $t_b$. The same is true for the lines of force, which are basically the same thing: there will be one direct line of force, and one reflected line of force. The integral over $\chi''$ measures the amount of universe that contributes to the attraction. 

According to (\ref{deftb}) we denoted the time-lapse of a full expansion or contraction with $t_b$ (in our units), with $\phi=0$ for simplicity. Hence, 
\begin{equation}
\label{Mach_tb}
{\rm t}_b=\pi M/c,
\end{equation}
where we divided by $c$ in order to get explicitly a time-lapse. Thus, the integral in (\ref{F_Mach3}) increases linearly with time, and if we wait for ${\rm t}_b$, the integral equals $\pi$:
\begin{equation}
\label{F_Mach4}
F({\rm t})={Gm_G\pi^2\over2{\cal R}^2(t)}{\rho_{c,{\rm sync}}\over \omega_{\rm sync}^3}{{\rm t}\over {\rm t}_b}, \qquad 0\le{\rm t}\le{\rm t}_b.
\end{equation}

Substitution of (\ref{Mach_tb}), (\ref{rho_0_iso}):
\begin{equation}
\rho_{c,{\rm sync}}={3\over2\kappa}{1\over(2M)^2}={3c^2\over8\pi G}{1\over(2M)^2},
\end{equation}
and (\ref{omega_0_M}): $\omega_{\rm sync}=1/(2M)$ into (\ref{F_Mach3}), yields
\begin{equation}
\label{inforce_theor1}
F({\rm t})= m_G\frac38{c^3\over{\cal R}^2(t)}{\rm t}=m_G\frac38{c^2\over{\cal R}^2(t)}t,\qquad 0\le t\le\pi M.
\end{equation}
With (\ref{radius_sync}): ${\cal R}=1/\omega_{\rm sync}\cyc(t)=(2M)\cyc(t)$ we end up with
\begin{equation}
\label{inforce_theor2}
F(t)= m_G\times \frac38{c^2 \over(2M)^2}{1\over\cyc^2(t)}t,\qquad 0\le t\le\pi M.
\end{equation}
Isotropy of $F({\rm t})$ is assured by the isotropy of the universe, which in our paradigm is a consequence of the very many cycles our universe has gone through. We argued that in that case a given mass should influence our test particle through (many) different lines of force. Hence we don't need the restriction to one cycle anymore, and we can state
\begin{equation}
\label{inforce_theor3}
F(t)= m_G\times \frac38{c^2 \over(2M)^2}{1\over\cyc^2(t)}t.
\end{equation}
If we add to this the logically consistent hypothesis that our universe once was born as a (regular) black hole in our mother universe, of course very much longer ago than 13.6 billion years, and that it got its mass by accretion, we must also try to take these accretion events into account.

The mass of a closed synchronous universe is given by  (\ref{tot_mass_sync}):
\begin{equation}
\label{Mach_mtot}
{\cal M}_{\rm tot}=3\pi^2{M\over\kappa}
\end{equation}
in which we substituted (\ref{rho_0_iso}). Hence, with (\ref{Mach_tb})
\begin{equation}
{\rm t}_b={\kappa\over3\pi c}{\cal M}_{\rm tot}.
\end{equation}
Suppose we have a number of $N$ accretion events, separated in time by intervals $A_n$, $n=1,\ldots,N$. The universe is born with an initial mass ${\cal M}_0$ at event $n=0$. At accretion event $n$ a mass ${\cal M}_n$ is added. Of course $\sum_{n=0}^N{\cal M}_n={\cal M}_{\rm tot}$, the total mass of the universe.  As for the full age ${\cal A}$ of the universe,
\begin{equation}
{\cal A}=\sum_{n=1}^NA_n+A_{N+1}
\end{equation}
with $A=A_{N+1}$ the time-lapse since the last accretion event, which might be of the order of what we probably would call the age of the universe.

In time-lapse $A_1$ between creation and the first accretion event, there were $A_1/{\rm t}_{b,0}$ periods,\footnote{It doesn't matter that this number is in general not an integer, since the horizon increases linearly with time.} with ${\rm t}_{b,0}=\kappa{\cal M}_0/(3\pi c)$, each producing the number of images that are included in (\ref{inforce_theor3}), and these are images of a fraction ${\cal M}_0/{\cal M}_{\rm tot}$ of the current total mass. In time $A_2$ between the first and the second accretion event, there were $A_2/{\rm t}_{b,1}$ periods, with ${\rm t}_{b,1}=\kappa({\cal M}_0+{\cal M}_1)/(3\pi c)$, each producing the number of images that are included in (\ref{inforce_theor3}), and these are images of a fraction $({\cal M}_0+{\cal M}_1)/{\cal M}_{\rm tot}$ of the current total mass. Continuing that way, we find for the number of images
\[
\sum_{n=1}^N{A_n\over\kappa\big(\sum_{i=0}^{n-1}{\cal M}_i\big)/(3\pi c)}{\sum_{i=0}^{n-1}{\cal M}_i\over {\cal M}_{\rm tot}}+ {A_{N+1}\over\kappa{\cal M}_{\rm tot}/(3\pi c)}
\]
or
\begin{equation}
{3\pi c\over\kappa}{\sum_{n=1}^{N+1}{A_n}\over{\cal M}_{\rm tot}}
={3\pi c\over\kappa}{{\cal A}\over{\cal M}_{\rm tot}}.
\end{equation}
This is again a linear increase with time, and (\ref{inforce_theor3}) therefore remains valid, irrespective of the accretion history:
\begin{eqnarray}
\label{inforce_theor4}
F(t)&=& m_G\times \frac38{c^2 \over[2M(t)\cyc(t)]^2}{\cal A}\nonumber\\
&=& m_G\times 3.57\times10^{-10}{{\cal A}^{(10\,{\rm Ga})}\over\left[{\cal R}(t)^{(3\,{\rm Gpc})}\right]^2}\,\,\,\,{\rm m/s^2}\nonumber\\
&=&m_G\times a(t)=m_I\times a(t).
\end{eqnarray}
The factor $a(t)$ that $m_G$ is multiplied with should be one if $m_G$ is to be equal to $m_I$. Hence we need a truly very long age $\cal A$ to make this work, though we should add that $\cal A$ is relevant for our mother universe, of which we know nothing anyway.

The only constraint we have on ${\cal R}(t_{\rm o})=2M(t_{\rm o})\cyc(t_{\rm o})$, with $t_{\rm o}$ the current epoch, is the (tangential) Hubble parameter $H_{\rm o}$. In section \ref{sect_solH0}, equation (\ref{solcycH0a}), we obtained an expression for $\cyc(t)$, given the parameter $b=[(2M)H_{\rm o}]^{-2}$, here assuming $\Lambda=0$ and $\epsilon=1$. For small $M$ (this seems imperative if (\ref{inforce_theor4}) is to work) and thus large $b$, we obtain 
\begin{equation}
\label{cyc_Mach}
\cyc=9\times 2^{-4/3}b^{-2/3}=9(MH_{\rm o})^{4/3},
\end{equation}
and hence we get for (\ref{inforce_theor3})
\begin{eqnarray}
\label{a_Mach}
a(t)&=& {1\over6^3}\left[[2M(t)]{H_{\rm o}\over2}\right]^{-8/3}\!\!\!{c^2\over[2M(t)]^2}{\cal A}\nonumber\\
&=&6.33\times10^{-11}\left[2M(t)^{(3\,{\rm Gpc})}\right]^{-14/3}\!\!\!
{\cal A}^{(10\,{\rm Ga})}\,\,\,{\rm m/s^2}.
\end{eqnarray}
It thus seems imperative to consider the possibility that our universe is smaller than we currently think. The phenomenon that a universe has a boundary, functioning as a gravitational mirror (section \ref{sect_physnatrb}), which is a novel concept introduced in this paper, could certainly cause us to interpret certain objects as being far away that are in reality images in reflection off the boundary. There would be no way to ascertain the presence of the boundary, unless both images would appear to collide. It would be non-trivial however to identify objects `in direct view' and `in reflection', since these views would be at different aspect angles and (much) different ages. Be it as it may,  for $.01<M(t)<.05$, which is still larger than the size of typical galaxy clusters,  it is possible to obtain $a(t)=1$ for `reasonable' ages $\cal A$, certainly because we have no clue what the latter should be anyway.

Another consequence of (\ref{inforce_theor3}) is that for very small $M(t)$, which may still be huge since $M(t)$ is expressed in $3.066\,{\rm Gpc}$, the inertial acceleration  can become huge. That means that no known force except gravitation would be able to throw a massive particle off its geodesic course, not only preventing the formation of atoms, but presumably also acting as a driver in synchronizing a universe and/or keeping it that way. This could explain why our universe appears to be highly synchronized, which thus would be a relic of a very distant past, during which our universe was completely ionized but thermally cold because of the complete absence of random motions.
 
\section{Summary}
In this paper, we consider spherically symmetrical cosmological models that are the interior of black holes, for our purposes defined as the matter in the spacetime inside a Schwarzschild horizon. 

By virtue of the spherical symmetry, the universe can be ordered in spherical shells, which are given a label which is not a radius, but these labels preserve order with the surface area of the shells. The labels are time independent, in contrast with the surface area  (subsection~\ref{sect_defgen}).

We employ the well-known Lema\^\i tre-Tolman models  (section~\ref{chapt_LT}), but recast them in a form that suits our purposes. More in particular, we introduce  in 
subsection~\ref{sect_definitions} the function 'cyc', which is a generalization of the well known cycloidal solution of the Robertson-Walker (RW) models. We study it extensively in appendix~\ref{sect_cyc}. This function not only packs all technical mathematical details of the time evolution equation into one mathematical object, thereby formally separating all mathematical details from the essence of the physics, but its introduction is also almost necessary in order to deal elegantly with the complications that arise from the spherical symmetry. In subsection~\ref{sect_conn_standard} we briefly show how  'cyc' connects with the standard practices in cosmology and the parameters of the 'concordance' $\Lambda$CDM model.

We present the view that, in order to understand the current state of our universe, it is essential to consider states in the very distant past in which the universe was not in expansion. In section~\ref{chapt_early} we muse about these phases, and argue that the isotropy of the Cosmic Microwave Background can be understood as a consequence of them, in a qualitative sense that is. A big bang is still possible, but (a) how 'big' that 'bang' was we cannot know, (b) it is possible that there were many 'bangs' (in space and/or time), and (c) there is no need to invoke 'bangs of creation' or (d) to consider 'bangs' that originate out of a 'point' in space. 

Shells can collide (subsections \ref{sect_phasefunction} and \ref{sect_collapse}), or put differently, a universe not necessarily evolves in a synchronous way. This possibility adds an important twist to cosmological models, since RW models (these belong to a class which we call synchronous models) do not allow for collisions. Colliding shells provide a natural mechanism to make filamentary structures\footnote{In this paper we remain in the spherical symmetry. The structures are thus spherical shells.}. Collisions also mark singularities in the metric. These do not invalidate the geometrical theory of gravitation, of course, but their occurrence shows the danger in assuming that one metric can be employed, without limits in space and time, to describe the universe.  We argue that there is an important difference between the validity period of a (geometrical) cosmological model and the real age of the universe. We must acknowledge that the latter may be impossible to determine, in contrast with the age of the structures in it, which can be dated. If we consider the consequences of collisions together with the ideas of the previous paragraph, one can envisage a universe as comprised of many smaller entities, somewhat like a raspberry or a pomegranate.

We consider a rather general class of specific cosmological models, but must admit, of course, that we cannot be exhaustive (subsection~\ref{sect_asynchronous}). Also models that have infinite extent or that are not the interiors of black holes are part of this class, but are not the subject of this paper. Barring these models, our models have an outermost shell, the boundary, beyond which there is no matter. The RW models are 'almost everywhere' identical to the synchronous models (subsection~\ref{sect_synchronous}), the exception being that they differ from these on the boundary, which the RW models simply don't have. 

The Hubble parameter is dependent on the viewing direction, and we define $H_{\rm o}$ to be the current value as measured in the direction  perpendicular to the center of the universe. This definition is consistent with the original definition in the standard cosmological models. We consider only models that realise $H_{\rm o}$ somewhere and sometime (subsection~\ref{sect_H0}), and classify them in some loose sense accordingly (subsection~\ref{sect_asynchronous}). In subsection~\ref{sect_behaviourX} we also consider another, rather technical, classification that has to do with the behaviour of the radial metric coefficient at the boundary. It is important in the calculation of orbits.

Contrary to the standard models, we have to cope with planar orbits instead of radial orbits, which is the subject of section~\ref{sect_orbits}. We devote a special section (section~\ref{sect_light}) to light, since it can be argued that photons are the only particles that are essentially not comoving and thus can be considered to be traveling on a cosmological scale. In subsection~\ref{sect_orbitbound} we examine the behaviour of the orbits in the vicinity of the boundary. We show that no particle can cross the boundary. Upon arrival they are reflected back into the universe. The boundary therefore effectively acts as a gravitational mirror. Unlike an ordinary mirror, however, one effectively flies into the mirror image upon arrival at the boundary, causing a spinor-like effect. This mirror-boundary could have far-reaching consequences. To pick one, it could be that our universe looks larger than it is. Indeed, we may find ourselves as if we were inside a spherical mirror. For that mirror to have effective observational consequences one needs (light travel) time, however (and transparent space of course). We discuss this and other rather intriguing aspects of the boundary in subsection~\ref{sect_physnatrb}.

If our universe is confined inside a Schwarzschild horizon, we must also consider 'the mother universe'. In particular, we must consider material that falls into our universe from there.
The connection between the Schwarzschild-$\Lambda$ metric of our mother universe and the metric of our universe is elucidated in appendix~\ref{sect_Novikov}. We employ Novikov coordinates, which are the coordinates of an observer who moves in a swarm of radially infalling shells, and who determines his radial position by the shell that passes by. These coordinates give rise to a diagonal metric, but are nevertheless largely unused because of the metric coefficients which are fairly complicated in their original form. The function 'cyc' however hides the mathematical details, and the resulting metric turns out, in hindsight not surprisingly, to be of the form of the metric of our universe, but it is one that is, in contrast, collapsing. Additional great features of the Novikov coordinates are that they can be easily interpreted and there is no horizon singularity. Hence it is possible to treat the infalling particle problem, since spacetime is not treated differently inside or outside the Schwarzschild radius. This is outlined in section~\ref{sect_beyond}, with the aid of the formalism developed in subsection~\ref{sect_NoviSchwarz}.

In section~\ref{sect_beyond} we also discriminate between the Schwarzschild sphere of the universe and the boundary of the universe. The picture emerges, at least in the ideal spherical and ordered case without colliding shells, of a kind of 'breathing' universe in the spacetime that does not belong to the universe and that is 'interior' to the Schwarzschild radius.\footnote{Strictly spoken, there should be nothing left between the edge of the universe and its Schwarzschild radius, but we keep the possibility open that, with the help of non-spherical evolutions in the past, there are some leftovers from a previous expansion phase. In any case, whatever there is, it continuously falls back into the universe.} The maximum expansion of the universe occurs at the very moment when also the universe fills its Schwarzschild sphere. At all other times, it does not, and there is (empty) space-time  between the boundary of the universe and the Schwarzschild horizon, space-time that can be described with a Novikov metric.  In this picture, it therefore is not correct to ascertain that an expanding universe 'creates' space along with its expansion. Our universes simply expand into the preexisting space-time interior to their Schwarzschild spheres.

The above picture also points to an intriguing possibility that can occur when a black hole with a universe inside is in a state in which its universe reaches the Schwarzschild radius.
In section~\ref{sect_mass_ejection} we show that, if at that very moment of maximum expansion the shells at the boundary also collide, it is possible that the boundary, and hence the Schwarzschild shell, is crossed from the inside to the outside. Unlikely as this may be globally, it could be realized locally (thereby breaking the spherical symmetry of course). This could be relevant for 'ordinary' black holes, which are most likely rotating and described by a Kerr metric. This metric however is locally, at the poles, similar to the Schwarzschild metric, and therefore emission at the poles could be a possibility, thereby enabling bipolar outflows. 

In section~\ref{sect_anom_red} we link this mechanism also to H. Arp's anomalous redshifts, since material that manages to escape from a black hole must climb out of a rather formidable gravitational well. Barring the cases of chance alignments, the anomalous redshifts are therefore not cosmological but gravitational according to this mechanism.\footnote{We do not advocate that all Arp's cases are of this nature. Surely there can be chance alignments. Hence statistics on Arp's catalog is not very relevant to the point made in this paper.}

In standard cosmologies, any point can be the center of the universe, any light ray can therefore be a radial one, and flux calculations are straightforward by considering the surface of a sphere centered at the origin. Not so in this case, and in appendices \ref{sect_angdistance} and \ref {sect_surface brightness_flux} we set up the necessary machinery to calculate the magnitude as a function of redshift (section~\ref{sect_light}), which is needed to fit models with observations. We find, of course, that the relation depends on the direction of the line-of-sight. Though a fit to observations is way beyond the scope of this paper, we show in section~\ref{sect_comp_obs} that the presence of a phase function suffices to fit current observations without having to invoke dark energy. We also compare some of our models with the consensus model, and find that we can reproduce it, of course, but also with $\Lambda=0$.\footnote{Since this paper includes the case $\Lambda\ne0$ and RW models are part of the class we consider (almost everywhere), a satisfying fit to observations is a trivial matter anyway.} 

In section~\ref{sect_darkenergy} we briefly comment on dark energy. Since dark energy and $\Lambda>0$ are essentially the same thing observationally, in our models a $\Lambda\ne0$ is a property of the mother universe, of which we never will know anything except via whatever falls from there into our universe.

In section~\ref{sect_primBH} we speculate about the formation of galaxies. Not much more can be done, since our models have a history before the latest expansion phase that we are in now. We argue that in our paradigm one should seriously consider that at least some black holes are primordial to our expansion phase, either by infall from our mother universe, or by growth in a collapse phase. Indeed, we show that black holes can easily gobble up a lot of material if the collapse around it is strong enough. The well-established relation between central black holes and the mass of the halo for spiral galaxies suggests that at least spirals are basically giant accretion disks around massive black holes, formed in a (local) collapse phase. The picture of our universe as a black hole in a mother universe, together with our universe comprising numerous super massive black holes that are the seeds of new universes, is logically consistent and therefore rather appealing.

As to the dark matter, in section~\ref{sect_darkmatter} we pick up the old idea that it is made up of neutrino's, but stress that the analysis is valid for any particle that obeys a Fermi-Dirac or Boltzmann distribution. For this to work, the dark matter must be massive enough, cold enough, and dense enough, all of which are problematic properties in standard cosmology because they would close the universe. Not here, on the contrary: our models are closed by design. We show in section~\ref{sect_neut_BH} that the remnant of a gravitational collapse that forms or grows a primordial black hole is a nearly constant density sphere of trapped dark matter particles. For a Milky Way type black hole, that sphere easily may contain as much mass as the black hole itself within about 5 pc, which is not contrary to observations of our Milky Way black hole environment. For larger black holes, the dark matter envelope can be very massive depending on its extent, and therefore the inferred black hole masses can include a very important contribution from the dark matter envelope.

On a larger scale, the flat rotation curves of spirals show that the dark matter distribution must have been isothermal at the moment in the expansion when they were 'locked in' into the galaxies (section~\ref{sect_intro_neutrinohalo}). Just like the isotropy of the CMB, this points to the scenario that our universe must be much older than the latest expansion phase, such that it was able to also produce a isotropic and isothermal cosmic dark matter background. It should be emphasized, though, that the structures we now see need not nearly be as old. In section~\ref{sect_simple_model} we produce a simple spherical model which includes the essential ingredients of a Milky Way type galaxy with a flat rotation curve. 

In the final section, we revisit Mach's principle in the light of our universes. We show that inertial mass can be explained as a kind of stiffness that every massive particle acquires as a consequence of the attraction by the cosmic dark matter background. 

\acknowledgments
      Much of this work has been done in my \emph{free time}. I would like to thank my wife Tine who has given me all the freedom to work on this paper. Comments by J. Binney on an advanced version of the paper are gratefully acknowledged. I also thank the geometer H. Van Maldeghem for his comments on the geometrical aspects of this paper.

\appendix

\section{The function \texorpdfstring{$\cyc$}{cyc} and its associate functions}
\label{sect_cyc}
Not all material in this appendix is new. This appendix is intended to be functional for the paper, self-contained and rather complete, and it presents the mathematics of the solution of the evolution equation in a consistent manner.

\subsection{Some general formulae}

Equations (\ref{soleqRtint}) and (\ref{Rexplgen}) lead us to define the function $\cyc(a,\epsilon,\psi)$ through the following implicit expression:
\begin{equation}
\label{defcyc}
\int_0^{\cyc(a,\epsilon,\bar\psi)}{\sqrt{u}\,\,du\over\sqrt{1-\epsilon u+au^3}} = \bar\psi.
\end{equation}
In the case that the radicand in the denominator has (a) positive root(s), we denote the (smallest) root with $w_0(a,\epsilon)$ and (\ref{defcyc}) defines the principal branch of a periodic function. Still in that case, we denote
\begin{equation}
\label{defpsimax}
\bar\psi_{\rm max}(a,\epsilon)=\int_0^{w_0(a,\epsilon)}{\sqrt{u}\,\,du\over\sqrt{1-\epsilon u+au^3}}.
\end{equation}
and define
\begin{eqnarray}
\label{defpsiperiodic}
\bar\psi &=& \left\{
\begin{array}{lll}
\psi & \quad0\le\psi\le\bar\psi_{\rm max} &{\rm (upward\, branch)} \\[2mm]
2\bar\psi_{\rm max}-\psi & \quad\bar\psi_{\rm max}\le\psi\le2\bar\psi_{\rm max}
& {\rm (downward\, branch)}\\
\end{array}
\right.\nonumber\\
\end{eqnarray}
with $\psi$ given by (\ref{soleqRtint}), and endless repeats for $\psi>2\bar\psi_{\rm max}$. The function $\cyc$ resembles a cycloidal function.

In the case that the radicand in the denominator of (\ref{defcyc}) has no positive root, $\bar\psi_{\rm max}=+\infty$, the function $\cyc$ is monotonically increasing and $\bar\psi=\psi$.

In contrast to conventional cosmology (the synchronous cases of sections \ref{sect_e0} and \ref{sect_synchronous}), where $\cyc$ only depends on $t$, and, moreover, $\cyc$ only enters in the time dependency of the radius of the universe, we are here faced with a function that is in addition dependent on $r$ through $\psi$, $a$ end $\epsilon$. It therefore also enters into the spatial part of the metric in a non trivial way. Hence, instead of satisfying oneself with a numerical solution of (\ref{defcyc}), some more analysis is needed for efficiency reasons. Since $\cyc(r,t)=\cyc(a(r),\epsilon(r),\psi(r,t))$, we need to calculate partial derivatives of $\cyc(a,\epsilon,\psi)$ with respect to all three arguments, and subsequently apply the chain rule. The latter operations are cumbersome but trivial, the former are also cumbersome but less trivial.

The easy derivatives we need are those with respect to $\psi$. We find
\begin{equation}
\label{dercyc}
\partial_\psi\cyc = \pm {\sqrt{1-\epsilon\,\cyc+a\,\cyc^3}\over\sqrt{\cyc}}
\end{equation}
The minus sign in (\ref{dercyc}) applies to the downward branch of the cycloidal case. This is compatible with the left hand side in (\ref{eqRt}), which allows both signs for $\partial_t R$.
We further need
\begin{equation}
\label{der2cyc}
\partial^2_\psi\cyc = -{1\over2\,\cyc^2}+a\,\cyc
\end{equation}
and also
\begin{equation}
\partial^3_\psi\cyc = -({1\over\cyc^3}+a)\,\partial_\psi\cyc.
\end{equation}
As to $\partial_\epsilon\cyc$, we find after manipulation of (\ref{defcyc}),
\begin{equation}
\partial_\epsilon\cyc = -\frac12{\rm decyc},
\end{equation}
with
\begin{equation}
\label{defdecyc}
{\rm decyc} = {\sqrt{1\!-\!\epsilon\,\cyc\!+\!a\,\cyc^3}\over\sqrt{\cyc}}
\int_0^\cyc\!\!\!{u^{3/2}\,du\over(1\!-\!\epsilon u\!+\!au^3)^{3/2}}.
\end{equation}
This function is finite for all $\bar\psi\le w_0$.

The expression $\partial_a\cyc$ needs some further manipulations. In fact, in turns out convenient to consider $a\partial_a\cyc$:
\begin{eqnarray}
a\partial_a\cyc &=& {\sqrt{1\!-\!\epsilon\,\cyc\!+\!a\,\cyc^3}\over\sqrt{\cyc}}\Bigl[(1\mp1) \,a\partial_a\psi_{\rm max}-\nonumber\\
&&\qquad-\int_0^\cyc
\!a\partial_a\left(1\!-\!\epsilon u\!+\!au^3\right)^{-1/2}\sqrt u\,du\Bigr].
\end{eqnarray}
The term in $a\partial_a\psi_{\rm max}$ is needed should the phase function $\phi$ depend on $a$ (as is the case in appendix~\ref{sect_Novikov}).

The above expression transforms, after performing a differentiation with respect to $1/a$:
\begin{eqnarray}
\label{adacycfin}
2a\partial_a\cyc = {\sqrt{1\!-\!\epsilon\,\cyc\!+\!a\,\cyc^3}\over\sqrt{\cyc}}
\bigl[\bar\psi&+&2(1\mp1)a\partial_a\psi_{\rm max}\bigr]-\nonumber\\
&-&\dacyc(a,\cyc)
\end{eqnarray}
with
\begin{eqnarray}
\label{defdacyc}
\dacyc(a,\cyc) &=& {\sqrt{1\!-\!\epsilon\,\cyc\!+\!a\,\cyc^3}\over\sqrt{\cyc}}
\int_0^\cyc\!\!\!{\sqrt u\,(1-\epsilon u)\,du\over(1\!-\!\epsilon u\!+\!au^3)^{3/2}}.\nonumber\\
\end{eqnarray}
Also this function is finite for all $\bar\psi\le w_0$.

The derivatives $\partial_\psi\cyc$, $\partial^2_\psi\cyc$, $\partial_\epsilon\cyc$ and $a\partial_a\cyc$ are all we need to integrate orbits. For the photometry however, we need to dig deeper into the second and third derivatives.

As for the second derivatives, these are the easy ones:
\begin{equation}
\partial^2_{\epsilon\psi}\cyc= {1\over2\partial_\psi\cyc}\left[-1 + 2(\partial_\psi^2\cyc)\,\partial_\epsilon\cyc\right]
\end{equation}
and
\begin{equation}
a\partial^2_{a\psi}\cyc= {1\over2\partial_\psi\cyc}\left[a\,\cyc^2 + 2(\partial_\psi^2\cyc)\,a\partial_a\cyc\right].
\end{equation}
Those with respect to $\epsilon$ and $a$ are more involved:
\begin{equation}
\partial^2_\epsilon\cyc = {\partial_\epsilon\cyc\over|\partial_\psi\cyc|}\left[\partial^2_{\epsilon\psi}\cyc-{1\over2|\partial_\psi\cyc|}\right] -\frac34\,{\rm d2ecyc}
\end{equation}
with
\begin{equation}
\label{defd2ecyc}
{\rm d2ecyc} = {\sqrt{1\!-\!\epsilon\,\cyc\!+\!a\,\cyc^3}\over\sqrt{\cyc}}
\int_0^\cyc\!\!\!{u^{5/2}\,du\over(1\!-\!\epsilon u\!+\!au^3)^{5/2}},
\end{equation}
\begin{eqnarray}
  \partial_\epsilon(a\partial_a\cyc)= &\displaystyle{\partial^2_{\epsilon\psi}\cyc\over\partial_\psi\cyc}&\!\!(a\partial_a\cyc) -\frac34\,{\rm d2eacyc} -\nonumber\\
  && -{\frac32-\frac32\epsilon\,\cyc+a\,\cyc^3\over1-\epsilon\,\cyc+a\,\cyc^3}\partial_\epsilon\cyc
\end{eqnarray}
with
\begin{equation}
\label{defd2eacyc}
{\rm d2eacyc} = {\sqrt{1\!-\!\epsilon\,\cyc\!+\!a\,\cyc^3}\over\sqrt{\cyc}}
\int_0^\cyc{u^{3/2}(1-\epsilon u)\,du\over(1\!-\!\epsilon u\!+\!au^3)^{5/2}},
\end{equation}
and
\begin{eqnarray}
  a\partial_a(a\partial_a\cyc)=&\displaystyle{a\partial^2_{a\psi}\cyc\over\partial_\psi\cyc}&\!\!(a\partial_a\cyc)
  + |\partial_\psi\cyc|\left(\frac32a\partial_a\bar\psi\pm a^2\partial^2_a\psi\right)-\nonumber\\
  && -\frac12{1-\epsilon\,\cyc\over1-\epsilon\,\cyc+a\,\cyc^3}a\partial_a\cyc +\nonumber\\
  &&+\frac34\,{\rm dacyc} -\frac34\,{\rm d2acyc}
\end{eqnarray}
with
\begin{equation}
\label{defd2acyc}
{\rm d2acyc} = {\sqrt{1\!-\!\epsilon\,\cyc\!+\!a\,\cyc^3}\over\sqrt{\cyc}}
\int_0^\cyc\!\!\!{\sqrt{u}(1-\epsilon u)^2\,du\over(1\!-\!\epsilon u\!+\!au^3)^{5/2}}.
\end{equation}
The other third derivatives we need are $\partial_\epsilon\partial^2_\psi\cyc$, $a\partial_a\partial^2_\psi\cyc$, $\partial^2_\epsilon\partial_\psi\cyc$, $a\partial_a\partial_\epsilon\partial_\psi\cyc$, $(a\partial_a)^2\partial_\psi\cyc$. These are rather trivially obtained from the above expressions, and do not generate any additional irreducible integrals.

\subsection{On the numerical calculation of \texorpdfstring{$\rm cyc$}{cyc}}
In order to calculate $\cyc$, we need to numerically invert (\ref{defcyc}). Efficiency can be gained if we have an estimate. For $\psi$ in the vicinity of 0, we obtain
\begin{eqnarray}
\label{smallcyc}
\cyc(a,\epsilon,\psi) = x + \sum_{i=1}^6c_ix^{i+1} +{\mathcal O}(x^8),
\end{eqnarray}
with $x=(3\psi/2)^{2/3}\ge0$ and
\begin{eqnarray}
c_1 &=& -\frac\epsilon5\nonumber\\
c_2 &=& -{3\epsilon^2\over5^2\times7}\nonumber\\
c_3 &=& {1\over3^2}\left(a-{23\over5^3\times7}\epsilon^3\right) \nonumber\\
c_4 &=& -{2\epsilon\over3^2\times5\times11}\left(2a+{947\over5^3\times7^2}\epsilon^3\right) \nonumber\\
c_5 &=& -{\epsilon^2\over5\times7\times11\times13}\left(2a-{37\times89\over5^4\times7}\epsilon^3\right) \nonumber\\
c_6 &=& {1\over3^4\times5}\left(a^2 + {2^2\times127\over5\times7\times11\times13}a\epsilon^3-
{2^2\times19\times31817\over5^5\times7^3\times11\times13}\epsilon^6\right). \nonumber\\
\end{eqnarray}
This expression allows one to calculate $\cyc(a,\epsilon,\psi)$ up to and including order $\psi^{4+2/3}$.  

When $\cyc$ is bounded, it is bounded by the smallest positive root $w_0(a,\epsilon)$ of $1-\epsilon\,\cyc+a\,\cyc^3$ (as defined in this section, see also later). In the vicinity of
$\cyc(\bar\psi_{\rm max})=w_0$, with $\psi\le\bar\psi_{\rm max}$, we find
\begin{equation}
\cyc(a,\epsilon,\psi) = w_0 - \sum_{i=1}^4a_{2i}u^{2i}-{\cal O}(u^{10}),
\end{equation}
with $u=(\bar\psi_{\rm max}-\psi)/2$ and
\begin{eqnarray}
a_2 &=& w_0\left(-2a+{1\over w_0^3}\right)=\epsilon w_0^{-1}-3w_0a \nonumber\\
a_4 &=& {a_2\over3}\left(a+{1\over w_0^3}\right)={a_2\over3}{\epsilon\over w_0^2} \nonumber\\
a_6 &=& {a_2\over45}\left(2a^2-14{a\over w_0^3}+{11\over w_0^6}\right) \nonumber\\
a_8 &=& {a_2\over315}\left(a^3 + 129{a^2\over w_0^3}-204{a\over w_0^6}+73{1\over w_0^9}\right).
\end{eqnarray}
In the next subsections, we will study the integrals (\ref{defcyc}), (\ref{defdecyc}), (\ref{defdacyc}), (\ref{defd2ecyc}), (\ref{defd2eacyc}) and (\ref{defd2acyc}) in detail. We will consider the 3 cases $\epsilon<0$, $\epsilon=0$ and $\epsilon>0$ separately, though all the integrals clearly indicate that the limit $e(r)\to0$ is infinitely smooth.

\subsection{The case \texorpdfstring{$e(r)=0$}{e(r)=0} over some range in \texorpdfstring{$r$}{r}}
\label{sect_e0}

We discuss this case first, and in some detail, because the standard $\Lambda$CDM cosmological model belongs to this class. This case is actually a special case of the more general case \ref{cyccase2}, but it is special because all the functions we need can be made explicit in terms of elementary functions. The solutions of (\ref{soleqRtint}) are particularly simple indeed, and they all represent Euclidean flat models at any given time. We consider only expansion.

\subsubsection{\texorpdfstring{$\Lambda\ne0$}{Lambda<>0} and  \texorpdfstring{$m(r)\ne0$}{m(r)<>0}}
We find
\begin{equation}
\label{defpe0om0}
p(r) = \sqrt[3]{6m(r)},\qquad\omega={1\over\sqrt3}\qquad{\rm and}\qquad a=\Lambda.
\end{equation}
There are 2 subcases.
\begin{equation}
\cyc(\psi)=\left[{1\over\sqrt{\Lambda}}\sinh({3\sqrt{\Lambda}\over2}\psi)\right]^{2/3},
\end{equation}
for $\Lambda>0$, and
\begin{equation}
\cyc(\psi)=\left[{1\over\sqrt{|\Lambda|}}\sin({3\sqrt{|\Lambda|}\over2}\psi)\right]^{2/3},
\end{equation}
for $\Lambda<0$. The factorization in $p(r)$ and $\rm cyc$ is one that yields a regular limit for cyc if $\Lambda\to0$. In the bound case, \mbox{$\bar\psi_{\rm max}=\pi/(3\sqrt{|\Lambda|})$} and $w_0=\cyc(\bar\psi_{\rm max})=|\Lambda|^{-1/3}=(-a)^{-1/3}$. We note in passing that $\partial_a(w_0)(a)>0$.

In these expressions, there is an implicit unit of length. This can be avoided by using the definition (\ref{defepsilon2}) of representation (b), but then $\omega(r)$ and also $a(r)$ are functions of $r$, which is unnecessary.

The classification in section~\ref{sect_classH0} for the solutions for $H_{\rm o}$ is applicable.
In order to realize $H_{\rm o}$, there is therefore the condition
\begin{equation}
\label{condH0e0}
\Lambda\le3H_{\rm o}^2\sim1.64
\end{equation}
if $\Lambda\ge0$. The function $t_0(r)$, discussed in \ref{sect_solH0}, can be calculated explicitly. We obtain
\begin{equation}
\label{t0e0Lgt0}
\left[\sinh\bigl({\sqrt{3\Lambda}\over2}(t_0+\phi(r))\bigr)\right]^2 = {\Lambda\over3H_{\rm o}^2-\Lambda}
\end{equation}
for $\Lambda>0$, and
\begin{equation}
\label{t0e0Llt0}
\left[\sin\bigl({\sqrt{3|\Lambda|}\over2}(t_0+\phi(r))\bigr)\right]^2 = {|\Lambda|\over
|\Lambda|+3H_{\rm o}^2}
\end{equation}
for $\Lambda<0$. Whether this also corresponds to a positive $t_0(r)$ depends on the phase function $\phi(r)$.

\subsubsection{\texorpdfstring{$\Lambda=0$}{Lambda=0} and \texorpdfstring{$m(r)\ne0$}{m(r)<>0}}
\label{model_e0Lambda0}
We find
\begin{equation}
R(r,t) = \left({9m(r)\over2}\right)^{1/3}\!\!\!(t+\phi(r))^{2/3}.
\end{equation}
This is a spherically anisotropic generalisation of the familiar Einstein - de Sitter universe. This means in particular that, though space is Euclidean at any time, expansion is not uniform, and can be described with a scale function $m(r)$ and a phase function $\phi(r)$. This solution can also be brought in the form (\ref{Rexplgen}), adhering to the definition of $\omega$ in (\ref{defomega}), and we find
\begin{equation}
\label{cyc0Lambda0}
\cyc(\psi)=({3\over2}\psi)^{2/3},
\end{equation}
with $p(r)$ arbitrary. Clearly this case is the limit $\Lambda\to0$ of the previous cases $\Lambda\ne0$, and therefore we can as well adopt (\ref{defpe0om0}).

The Hubble parameter simplifies to
\begin{equation}
H(r,t) = \frac23\,\,\,{1\over t+\phi(r)}.
\end{equation}
and therefore the current epoch is given by a generalisation of a well known elementary expression
\begin{equation}
t_{\rm o}(r) = \frac23\frac1{H_{\rm o}}-\phi(r)
\end{equation}
which is the limit of the expressions (\ref{t0e0Lgt0}) and (\ref{t0e0Llt0}).

\subsubsection{\texorpdfstring{$m(r)=0$}{m(r)=0}}
In empty space $\Lambda$ must be positive or zero, and one obtains the anisotropic generalisation of the de Sitter universe, which is described, after an appropriate shell label transformation, as
\begin{equation}
p(r) = P, \qquad \omega = \sqrt{\Lambda\over3}, \qquad \cyc(\psi)=\exp(\psi),
\end{equation}
with $P$ an arbitrary scale factor with the dimension of a length. We have $H=\sqrt{\Lambda/3}$ and $q=-1$. Euclidean space has $\Lambda=0$.

\subsubsection{Discussion}
We see that, in all these models, $R(r,t)$ is a separable function of $r$ and $t$ if the phase function is a constant. Hence we can relabel $r$ such that $p(r')\sim r'$ (see also section~\ref{sect_synchronous} for the meaning of this choice). The non-trivial dependence of these models on $r$ therefore solely resides in the phase function $\phi(r)$. Shells will not intersect if shells with smaller $r$ start their expansion later, which means that $d_r\phi(r)\ge0$.

\subsection{The case \texorpdfstring{$\Lambda=0$}{Lambda=0} and \texorpdfstring{$e(r)\ne0$}{e(r)<>0}}

This case demands a separate treatment, because there is only 1 root in the denominator of (\ref{defcyc}).

\subsubsection{The case \texorpdfstring{$\epsilon>0$}{epsilon>0}}
In this case the integration in (\ref{defcyc}) is bounded by the root \mbox{$\cyc=1$} of the denominator.  We obtain the well known
\begin{equation}
\label{defcyc0}
\epsilon^{3/2}\psi = \arcsin\sqrt{\epsilon\cyc} - \sqrt{\epsilon\cyc}\sqrt{1-\epsilon\,\cyc}.
\end{equation}
It defines the familiar cycloidal function which rises from 0 to $w_0=1/\epsilon$ in the interval $\psi\in[0,\pi/2]\epsilon^{-3/2}$, and declines from $w_0=1/\epsilon$ to 0 in the interval $\psi\in[\pi/2,\pi]\epsilon^{-3/2}$. Clearly $\bar\psi_{\rm max}=(\pi/2)\epsilon^{-3/2}$. 
We recover (\ref{cyc0Lambda0}) in the limit $\epsilon\to0$. The familiar parametric representation of the cycloid is recovered by setting
\begin{equation}
\sqrt{\epsilon\cyc}=\sin\frac u2  \quad {\rm or} \quad R=\frac p\epsilon(1-\cos u)
\end{equation}
leading to
\begin{equation}
2\epsilon^{3/2}\psi = u - \sin u.
\end{equation}
We also need
\begin{eqnarray}
\label{decyc0}
\partial_\epsilon\cyc = -{1\over2\epsilon^2}\left(3-\epsilon\,\cyc-3{\sqrt{1-\epsilon\,\cyc}\over\sqrt{\epsilon\,\cyc}}\arctan{{\sqrt{\epsilon\,\cyc}}\over\sqrt{1-\epsilon\,\cyc}}\right)\nonumber\\
\end{eqnarray}
and
\begin{equation}
\label{de2pcyc0}
\partial^2_\epsilon\cyc = -{2\over\epsilon}\partial_\epsilon\cyc -{F_+\over2\epsilon^2}\left(\cyc+\epsilon\partial_\epsilon\cyc\right)
\end{equation}
with
\begin{equation}
F_+=-1+\frac3{2\epsilon\,\cyc}\left(-1 +{(\epsilon\,\cyc)^{-1/2}\over\sqrt{1-\epsilon\,\cyc}}\arctan{{\sqrt{\epsilon\,\cyc}}\over\sqrt{1-\epsilon\,\cyc}}\right)
\end{equation}
For small $\epsilon$ the above expressions need power expansions in $\epsilon$ for numerical stability.

\subsubsection{The case \texorpdfstring{$\epsilon<0$}{epsilon<0}}
The function $\cyc$ is unbounded, and is implicitly given by the elementary integral
\begin{equation}
|\epsilon|^{3/2}\psi = \sqrt{|\epsilon|\cyc}\sqrt{1-\epsilon\,\cyc}-\ln\bigl(\sqrt{|\epsilon|\,\cyc}+\sqrt{1-\epsilon\,\cyc}\bigr).
\end{equation}
Again, a perhaps more familiar parametric form is
\begin{equation}
\sqrt{|\epsilon|\cyc}=\sinh\frac u2  \quad {\rm or} \quad R=\frac p{|\epsilon|}(\cosh u - 1)
\end{equation}
leading to
\begin{equation}
2|\epsilon|^{3/2}\psi = \sinh u - u.
\end{equation}
As for $\partial_\epsilon\cyc$ and $\partial^2_\epsilon\cyc$, we obtain quite analogously as in the previous case
\begin{eqnarray}
\partial_\epsilon\cyc &=& -{1\over2\epsilon^2}\Biggl(3-\epsilon\,\cyc-\nonumber\\
&&\quad-3{\sqrt{1-\epsilon\,\cyc}\over\sqrt{|\epsilon|\,\cyc}}\ln\bigl(\sqrt{|\epsilon|\,\cyc}+\sqrt{1-\epsilon\,\cyc}\bigr)\Biggr),
\end{eqnarray}
\begin{equation}
\label{de2mcyc0}
\partial^2_\epsilon\cyc = -{2\over\epsilon}\partial_\epsilon\cyc -{F_-\over2\epsilon^2}\left(\cyc+\epsilon\partial_\epsilon\cyc\right)
\end{equation}
with
\begin{eqnarray}
F_-=-1&+&\frac3{2\epsilon\,\cyc}\Biggl(-1 + \nonumber \\
&&\quad+{(|\epsilon|\,\cyc)^{-1/2}\over\sqrt{1-\epsilon\,\cyc}}\ln\bigl(\sqrt{|\epsilon|\,\cyc}+\sqrt{1-\epsilon\,\cyc}\bigr)\Biggr).
\end{eqnarray}

\subsection{The case \texorpdfstring{$0<\Lambda\le{4\over9}\epsilon^3\omega^2$}{0<Lambda<=4/9epsilon**3omega**2} or \texorpdfstring{$0<a\le\frac4{27}\epsilon^3$}{0<a<=4/27epsilon**3} with \texorpdfstring{$\epsilon>0$}{epsilon>0}}
\label{cyccase1}
\subsection{The roots of the cubic}
The roots of the denominator in (\ref{defcyc}) are real. One of them is negative, which we denote by $w_1$, the other 2 are positive and are denoted by $w_0$ and $w_2$, with $w_0<w_2$. Hence, $\cyc$ is bounded by $w_0$. Since the cubic is already in the Cardano form, we know that $w_0=-w_1-w_2=|w_1|-w_2$. We find
\begin{equation}
\label{equ_w}
w_0 = 2\sqrt{\epsilon\over3a}\cos\left(\frac13\arctan\sqrt{{4\epsilon^3/(27a)-1}} +\frac\pi3\right)
\end{equation}
and
\begin{equation}
w_n = 2\sqrt{\epsilon\over3a}\cos\left(\frac13\arctan\sqrt{{4\epsilon^3/(27a)-1}} +{(2n+1)\pi\over3}\right)
\end{equation}
for $n=1,2$.

It is instructive to investigate the structure of the roots in both limits. When $0^+\leftarrow a$, we write
\begin{equation}
\frac13\arctan\sqrt{{4\epsilon^3/(27a)-1}} = \frac\pi6 - \delta,
\end{equation}
with
\begin{equation}
\delta = \frac13\sum_{k\ge0} {(-1)^k\over2k+1}\left({a\over(4\epsilon^3/27-a)}\right)^{k+1/2}\to0,
\end{equation}
we find
\begin{equation}
w_0 = 2\sqrt{\epsilon\over3a}\sin\delta \to 1,
\end{equation}
\begin{equation}
w_2 = \sqrt{\epsilon\over3a}(\sqrt3\cos\delta-\sin\delta) \to +\infty
\end{equation}
and
\begin{equation}
w_1 = \sqrt{\epsilon\over3a}(-\sqrt3\cos\delta-\sin\delta) \to -w_2 \to -\infty.
\end{equation}
In the other limit $a\to{\frac4{27}\epsilon^3}^-$ we find $w_0=w_2=\frac3{2\epsilon}$ and $w_1=-\frac3\epsilon$. One can also prove that $\partial_aw_0(a)>0$ in this case,\footnote{
For any root
\begin{equation}
\partial_aw ={w^4\over3-2\epsilon w}.
\end{equation}
For $a=0$ this expression is positive for $w_0=1$. Hence $w_0(a)$ is increasing and $\partial_aw$ remains positive up to the limit $a\to\frac4{27}\epsilon^3$ where $w_0=\frac32$.}
and therefore $w_0$ is to be found in the interval $]1,\frac3{2\epsilon}]$.

\subsection{The integral for \texorpdfstring{$\bar\psi$}{psi}}
Turning to the integral (\ref{defcyc}), we find, with the transformation
\begin{equation}
\label{Carlsontrans}
z=(\cyc-u)/u
\end{equation}
that
\begin{eqnarray}
\bar\psi &=& \cyc^{3/2}\!\!\!\int_0^{+\infty}\!\!\!\!\!{dz\over(1+z)\sqrt{z-z_0}\sqrt{z-z_1}\sqrt{z-z_2}} \nonumber \\
\end{eqnarray}
with
\begin{equation}
z_0 = {\cyc\over w_0} -1  \qquad  z_{1,2} = {\cyc\over w_{1,2}} -1.
\end{equation}
or, in Carlson's (\cite{carlson}) notation,
\begin{equation}
\label{cyccarlson}
\bar\psi = \frac23\cyc^{3/2}R_J(-z_0,-z_1,-z_2,1).
\end{equation}
In this expression, $R_J$ denotes Carlson's elliptic integral of the third kind. It is possible to write (\ref{cyccarlson}) in terms of the Legendre elliptic functions (see e.g. http://dlmf.nist.gov/19.25.iii), but there seems to be little advantage in doing so, especially because the elliptic function of the third kind is involved.

The maximum value of $\cyc$ equals $w_0$. At that point
\begin{equation}
\label{psimax}
\bar\psi_{\rm max} = \frac\pi2w_0^{3/2}F_1(\frac12,\frac12,\frac12,2;-\frac {w_0}{w_1},-\frac {w_0}{w_2})
\end{equation}
featuring the hypergeometric function of 2 variables of the first kind.

As in the case $a=0$, the function $\cyc$ shows a cycloidal behaviour. It rises from 0 to $w_0$ in the interval $\psi\in[0,\bar\psi_{\rm max}]$, and declines from $w_0$ to 0 in the interval $\psi\in[\bar\psi_{\rm max},2\bar\psi_{\rm max}]$.

\subsection{The integral for dacyc}
We now turn to $\dacyc(a,\psi)$. Decomposition into partial fractions of $(1-u)/(1-u+au^3)$ leads to
\begin{eqnarray}
\label{dacyc1}
\frac32{a\over\cyc^{3/2}}&\!\!\!\!\displaystyle\int_0^{\cyc}&\!\!\!\!{\sqrt u(1-u)\over(1-\epsilon u+au^3)^{3/2}}\,du= \nonumber\\
&&{\epsilon-1/w_2\over (w_2-w_0)(w_2-w_1)}R_D(-z_0,-z_1;-z_2)\nonumber\\
&+&{\epsilon-1/w_0\over (w_0-w_2)(w_0-w_1)}R_D(-z_2,-z_1;-z_0)\nonumber\\
&+&{\epsilon - 1/w_1\over (w_2-w_1)(w_0-w_1)}R_D(-z_0,-z_2;-z_1),
\end{eqnarray}
with
\begin{equation}
R_D(x,y;z)= R_D(z)=\frac32\int_0^{+\infty}{dt\over\sqrt{t+x}\sqrt{t+y}(t+z)^{3/2}}
\end{equation}
Carlson's elliptic integral of the second kind. The second notation is shorthand and omits explicit reference to the two other roots $-x$ and $-y$ that are symmetrical. The second term on the right hand side of (\ref{dacyc1}) is singular for $\cyc\to w_0$. Partial integration resolves this singularity, and results in the general expression
\begin{equation}
\label{relRDs}
R_D(-z_0) + R_D(-z_1) + R_D(-z_2) = {3\over\sqrt{1-\epsilon\,\cyc+a\,\cyc^3}} ,
\end{equation}
leading to
\begin{eqnarray}
\dacyc(a,\psi) &=& w_0\,\cyc\Biggl\{ -2A + \frac23\sqrt{1-\cyc+a\,\cyc^3}\times\nonumber\\
&&\qquad\qquad\Bigl[A_1R_D(-z_1)+A_2R_D(-z_2)\Bigr]\Biggl\},
\end{eqnarray}
with
\begin{eqnarray}
A &=&\displaystyle(\epsilon-\frac1{w_0})\displaystyle(1-\frac {w_0}{w_2})^{-1}(1-\frac {w_0}{w_1})^{-1} \nonumber\\
A_1 &=&\displaystyle(\epsilon-\frac1{w_1})\displaystyle(1-\frac {w_1}{w_2})^{-1}(1-\frac{w_0}{w_1})^{-1}+A \nonumber\\
A_2 &=&\displaystyle(\epsilon-\frac1{w_2})\displaystyle(1-\frac {w_0}{w_2})^{-1}(1-\frac{w_2}{w_1})^{-1}+A.
\end{eqnarray}
Finally, the limit $\psi\to\bar\psi_{\rm max}$ in expressions (\ref{adacycfin}) en (\ref{defdacyc}) yields
\begin{eqnarray}
a\partial_a\bar\psi_{\rm max} = &\displaystyle{w_0^{5/2}\over3}&\Bigl[A_2R_D(0,1-\frac {w_0}{w_1},1-\frac {w_0}{w_2})\Bigr.+\nonumber\\
&&\quad\,\,+\Bigl.A_1R_D(0,1-\frac {w_0}{w_2},1-\frac {w_0}{w_1})\Bigr] -\frac12\bar\psi_{\rm max},\nonumber\\
\end{eqnarray}
which will be needed in appendix~\ref{sect_Novikov}.

All other integrals can be calculated in a similar manner. They are of the form
\begin{equation}
\int_0^\cyc{\sqrt{u}f(a,\epsilon,u)\over(1-\epsilon u+au^3)^{1/2+n}}\,dx, \qquad n=1,2
\end{equation}
and $f(a,\epsilon,u)$ a polynomial in $u$. The integrand can be transformed with the transformation (\ref{Carlsontrans}), and decomposed into partial fractions to yield a sum of integrals of the form
\begin{equation}
A_{i,n}\!\int_0^{+\infty}\!\!\!{dz\over\sqrt{(z-z_0)(z-z_1)(z-z_2)}}{1\over(z-z_i)^n}, \quad i=0,1,2,
\end{equation}
which can, after another partial integration ($n=2$), a partial fraction decomposition and relation (\ref{relRDs}) be reduced to a sum involving $R_D(-z_1)$ and $R_D(-z_2)$. These calculations are elementary but cumbersome, are not likely to deliver enlightening insights and are best left to some calculating device.

\subsection{The integral for decyc}
The expression for $\partial_\epsilon\cyc$ completes the expressions for the first derivatives:
\begin{eqnarray}
\partial_\epsilon\cyc(a,\psi) &=& w_0\,\cyc\Biggl\{ -E + \frac13\sqrt{1-\cyc+a\,\cyc^3}\times\nonumber\\
&&\qquad\qquad\Bigl[E_2R_D(-z_2)+E_1R_D(-z_1)\Bigr]\Biggl\},
\end{eqnarray}
with
\begin{eqnarray}
E &=&\displaystyle(1-\frac {w_0}{w_2})^{-1}(1-\frac {w_0}{w_1})^{-1} \nonumber\\
E_1 &=&\displaystyle(1-\frac {w_1}{w_2})^{-1}(1-\frac{w_0}{w_1})^{-1}+E \nonumber\\
E_2 &=&\displaystyle(1-\frac {w_0}{w_2})^{-1}(1-\frac{w_2}{w_1})^{-1}+E.
\end{eqnarray}

\subsection{The integral for the light path in the synchronous case} 
In section \ref{sect_photometry_sync}, equation (\ref{deltachi_int_cyc})  we considered integrals of the kind 
\begin{equation}
\Delta\chi = \int_0^{\cyc}{u^{-1/2}\,du\over\sqrt{1-\epsilon u+au^3}}.
\end{equation}
This one is actually the most elementary of them all. With the same transformations as above we obtain
\begin{eqnarray}
\Delta\chi &=& \sqrt{\cyc}\int_0^{+\infty}{dz\over\sqrt{z-z_0}\sqrt{z-z_1}\sqrt{z-z_2}}
\nonumber\\
&=&2\sqrt{\cyc}\,R_F(-z_0,-z_1,-z_2)
\end{eqnarray}
with $R_F$ Carlson's elliptic integral of the first kind.

For the special case $\cyc=w_0$ we find
\begin{equation}
\Delta\chi = 2\sqrt{w_0}R_F(0,1-{w_0\over w_1},1-{w_0\over w_2}).
\end{equation}
or, with the multipication theorem for $R_F$,
\begin{equation}
\Delta\chi = 2R_F(0,{1\over w_0}-{1\over w_2},{1\over w_0}-{1\over w_1}).
\end{equation}
In this case, there is an advantage to pass to the Legendre elliptic integral. We use the identity $(z\ge y)$
\begin{equation}
R_F(0,y,z)={1\over\sqrt{z}}K\left(\sqrt{1-\frac yz}\right)
\end{equation}
and obtain\footnote{One may notice that the lhs of this equation is symmetric in $y$ and $z$, while the rhs is not. If we interchange $y$ and $z$, one obtains the well-known identity $K(k)=k^{-1}K(ik/k')$, in the usual notations.}  
\begin{equation}
\label{Delta_chi_max}
\Delta\chi={2\over\sqrt{1-w_0/w_1}}K\left(\sqrt{1/w_2-1/w_1\over1/w_0-1/w_1}\right).
\end{equation}

\subsection{The case \texorpdfstring{$\Lambda<0$}{Lambda>0} or \texorpdfstring{$a<0$}{a<0} and the case \texorpdfstring{$\Lambda>{4\over9}\epsilon^3\omega^2$}{Lambda>4/9epsilon**3omega**2} or \texorpdfstring{$a>\frac4{27}\epsilon^3$}{a>4/27epsilon**3} with \texorpdfstring{$\epsilon>0$}{epsilon>0}}
\label{cyccase2}
There is 1 real root, and 2 complex conjugate ones. The real root reads
\begin{equation}
(2a)^{1/3}w_0 = \left(-1+\sqrt{1-{4\epsilon^3\over27a}}\right)^{1/3} \!\!\!+ \left(-1-\sqrt{1-{4\epsilon^3\over27a}}\right)^{1/3}\!\!\!.
\end{equation}
The complex conjugate roots are
\begin{equation}
w_{1,2} = -w_0/2\pm ic,
\end{equation}
with
\begin{eqnarray}
\sqrt{3\over|a|}Dc^{-1} &=& \left({\epsilon^3\over27} - {a\over2}+\sqrt{|a|}D\right)^{1/3} + \nonumber\\
&& \qquad\qquad+\left({\epsilon^3\over27}-{a\over2}-\sqrt{|a|}D\right)^{1/3} - \frac\epsilon3
\end{eqnarray}
and
\begin{equation}
D = \sqrt{{|a|\over4}-{{\rm sign}(a)\epsilon^3\over27}}.
\end{equation}

When $\Lambda<0$, the real root is positive, and $\cyc$ is bounded by $w_0$. Again one can prove\footnote{
The proof follows trivially from the expression
\begin{equation}
\partial_aw ={w^4\over1-2aw^3}.
\end{equation}} 
that in this case $\partial_aw_0(a)>0$ and $w_0$ can be found in the interval $]0,1[$.

Again, the limit $a\to0^-$ is not very transparent. After resolving the indeterminacy $\infty-\infty$ we find
\begin{equation}
w_0^{-1} = \left({\epsilon^3\over27}-{a\over2}+\sqrt{|a|}D\right)^{1/3} \!\!\!+ \left({\epsilon^3\over27}-{a\over2}-\sqrt{|a|}D\right)^{1/3}\!\!\! + \frac\epsilon3,
\end{equation}
leading to $w_0\to\epsilon$. For the complex roots we find $c\to\sqrt\epsilon/|a|\to+\infty$.
In the limit $a\to-\infty$ we find $w_0\to0^+$ and $c\to0$.

When $\Lambda>{4\over9}\epsilon^3\omega^2$ the real root is negative, and $\cyc\ge0$ can be arbitrarily large. For $a\to+\infty$ we find $w_0\to0^-$ and $c\to0$. In the limit ${4\epsilon^3\over27}^+\leftarrow a$ we find $c=0$ and the 2 complex roots co\"incide
in the real $w_{1,2}=-w_0=(4/a)^{1/3}=3/\epsilon$.

The elaboration of the integrals is essentially the same as in the previous case, since the Carlson elliptic integrals can, in this case, be extended to complex arguments.

For completeness, $\epsilon=0$ is the (trivial) limit of this case. We recover the 3 roots of $-1/a$:
the real root
\begin{equation}
w_0=(-a)^{-1/3}
\end{equation}
and the 2 complex conjugate ones
\begin{equation}
w_{1,2}=\left(-\frac12\pm{\sqrt3\over2}i\right)w_0.
\end{equation}

\subsection{The case \texorpdfstring{${4\over9}\epsilon^3\omega^2\le\Lambda<0$}{4/9epsilon**3omega**2<=Lambda<0} or \texorpdfstring{$\frac4{27}\epsilon^3\le a<0$}{4/27epsilon**3<=a<0} with \texorpdfstring{$\epsilon<0$}{epsilon<0}}
We can reduce this case to the one discussed in subsection~\ref{cyccase1}, because the case $\epsilon<0$ and $a<0$ follows immediately from that analysis if we change the sign of $u$, and thus the sign of the roots, and change also the sign of $\Lambda$. There are 3 real roots, 2 of which are negative. The positive root is $-w_1$ in the notations of subsection~\ref{cyccase1}, which thus takes on the function of $w_0$. The function $\cyc$ has a maximum and shows a cycloidal behaviour. For ${\frac4{27}\epsilon^3}^+\leftarrow a$ we obtain $-w_1=3/\epsilon$, while for $a\to0^-$ the root $-w_1\to+\infty$. Clearly $\partial_a(-w_1)(a)>0$.

\subsection{The case \texorpdfstring{$\Lambda>0$}{Lambda>0} or \texorpdfstring{$a>0$}{a>0} and the case \texorpdfstring{$\Lambda<{4\over9}\epsilon^3\omega^2$}{Lambda<4/9epsilon**3omega**2} or \texorpdfstring{$a<\frac4{27}\epsilon^3$}{a<4/27epsilon**3} with \texorpdfstring{$\epsilon<0$}{epsilon<0}}
We can reduce this case to the one discussed in subsection~\ref{cyccase2}. There is 1 real root.
For $\Lambda>0$ the root $-w_0$ is negative, and the function $\cyc$ is unbounded. 
In the limit $a\to+\infty$ we find $-w_0\to0^-$.

For $\Lambda\le-{4\over9}\epsilon^3\omega^2$ the root $-w_0$ is positive, the function $\cyc$ has a maximum and shows a cycloidal behaviour. For $-\infty\leftarrow a$ we find $0^+\leftarrow-w_0$. In the limit $a\to{4\epsilon^3\over27}^-$ we find \mbox{$w_0=(-4/a)^{1/3}=-3/\epsilon$}, and thus $w_0$ can be found in the interval $]0,\frac3\epsilon]$.
Again $\partial_a(w_0)(a)>0$.

\begin{table}
\caption{Summary of the behaviour of $\cyc$}
\label{tab_cyc}
\renewcommand{\arraystretch}{1.5}
\begin{center}
\begin{tabular}{|l|l|l|}
  \hline
  $\epsilon=+1$ &  $\Lambda>{4\over9}\epsilon^3\omega^2$  & unbound \\ \cline{2-3}
  $\epsilon(r)>0$ &  $-\infty<\Lambda\le0\le\Lambda\le{4\over9}\epsilon^3\omega^2$ & $0<w_0\le1\le w_0\le\frac3{2\epsilon}$ \\
\hline
\hline
  $\epsilon=-1$ &  $\Lambda\ge0$  & unbound \\ \cline{2-3}
  $\epsilon(r)<0$ &  $-\infty<\Lambda\le{4\over9}\epsilon^3\omega^2\le\Lambda<0$ & $0<w_0\le-\frac3\epsilon\le w_0<+\infty$ \\
  \hline
\hline
  $\epsilon=0$ &  $\Lambda\ge0$  & unbound \\ \cline{2-3}
  $\epsilon(r)=0$ &  $-\infty<\Lambda<0$ & $0<w_0<|\Lambda|^{-1/3}$ \\
\hline
\end{tabular}
\end{center}
\renewcommand{\arraystretch}{1.0}
\end{table}

\section{Novikov coordinates}
\label{sect_Novikov}
\subsection{Radial orbits for material particles}
\label{sect_Novrad}

We consider radial motion for particles with mass in the metric (\ref{Schwarz}):
\begin{equation}
ds^2 = \left(1\!-\!{2M\over R}\!-\!{\Lambda R^2\over3}\right)d{\bar t}^2 - {dR^2\over\displaystyle\left(1\!-\!{2M\over R}\!-\!{\Lambda R^2\over3}\right)} - R^2\,d\Omega^2
\end{equation}
with $M>0$. The equation for the time-like geodesics reads
\begin{equation}
\label{novgeod}
\left(1-{2M\over R}-{\Lambda R^2\over3}\right){d{\bar t}\over dt} = \tilde E_\infty \equiv V_0(R){d{\bar t}\over dt},
\end{equation}
with $t$ the proper time (or the arc length, which is the same because of the convention (\ref{deft})) of the particle. The relativistic energy $\tilde E_\infty$ relative to the rest mass energy appears as a constant of the motion.\footnote{We continue to use the symbol $\tilde E_\infty$ though that is only a proper notation for $\Lambda=0$.} The notation $V_0(R)$ is consistent with (\ref{V_eff_schwarz}) for $\tilde h=0$ and upon inclusion of $\Lambda$.  Insertion into the metric yields
\begin{equation}
\label{novdRdt}
d_tR = \pm \sqrt{{\tilde E_\infty}^2- V_0(R)}.
\end{equation}
Clearly $V_0(R)$ is the effective potential: motion is only possible if \mbox{${\tilde E_\infty}^2\ge V_0(R)$}. 

The indefinite integral reads
\begin{equation}
\label{nov_intorbitR}
\int {\sqrt{R}\,dR \over \sqrt{2M+({\tilde E_\infty}^2-1)R+\frac13\Lambda R^3}} = \pm\int dt.
\end{equation}
The solutions of this equation have been studied in appendix~\ref{sect_cyc}, and are
\begin{equation}
\label{novRt}
R = P \,\cyc\left(a_P,\epsilon_N,\sqrt{2M\over P^3}t + \bar\psi_{\rm init}\right), \qquad a_P = \frac\Lambda3{P^3\over2M}.
\end{equation}
In this expression
\begin{equation}
\epsilon_N={\rm sign}({\tilde E_\infty}^2 -1), 
\end{equation}
the shell parameter $P>0$  is defined by
\begin{equation}
\label{novconst}
{\tilde E_\infty}^2 = 1-\epsilon_N{2M\over P},
\end{equation}
and $\bar\psi_{\rm init}$ is some initial state.

We now consider radial orbits that have zero radial velocity at $t=0$ for some $R=R_0$ and that therefore are bound. This also defines $\bar\psi_{\rm init}$. We denote them by $R_P(t)$, which is a sufficient notation because $M$ and $\Lambda$ are overall constants. Because of the spherical symmetry, we will call these henceforth 'shells' with parameter $P$. 

The Schwarzschild radius $R_S$ is the smallest positive root of $V_0=0$, which translates into 
\begin{equation}
\label{def_a_S}
1-\epsilon_Nu+a_Su^3=0  \!\!\quad {\rm with}\! \quad a_S=\frac43\Lambda M^2 \!\!\quad {\rm and}\! \quad R_S=2Mu.
\end{equation}
Thus, with the notations of appendix~\ref{sect_cyc}, $R_S=2Mw_0(a_S)$, hence $V_0(R_S)=0$. From examination of $V_0(R)$ it appears that $d_RV_0(R)>0$ for $R\le R_S$, which means that there is exactly one orbit that has $R\le R_S$ and that just reaches $R=R_S$ where it has zero radial velocity. From the denominator of the integral in (\ref{nov_intorbitR}), with $R=Pu$, we find that we recover (\ref{def_a_S}) if $P=2M$, which is therefore the unique orbit that `starts off' with zero radial velocity at the Schwarzschild radius. 

We now consider the 3 cases $\Lambda=0$, $\Lambda<0$ and $\Lambda>0$ separately.

\subsubsection{\texorpdfstring{$\Lambda=0$}{Lambda=0}}

We find $0\le\tilde E_\infty<1$ because of boundedness, $\epsilon_N=+1$ by virtue of (\ref{novconst}) and thus $P\ge2M$. Shells that start at rest at $t=0$ at some radius $R_0\ge R_S$ fall inwards, all the way to $R=0$:
\begin{equation}
\label{novradlamda0}
R_P(t) = P \,\cyc\left(0,+1,\sqrt{2M\over P^3}t + {\pi\over2}\right).
\end{equation}

\begin{figure}[ht]
   \centering
   \includegraphics[width=130mm]{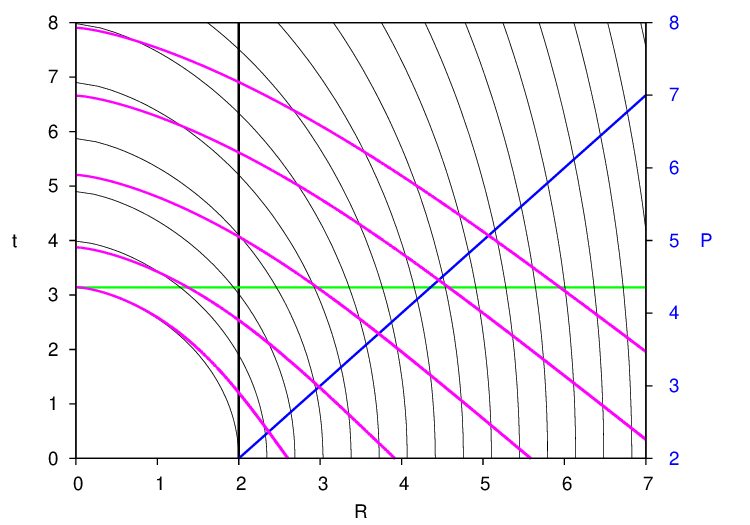}
      \caption{Infalling shells $R_P(t)$ for the case $\Lambda=0$ and $M=1$ (black thin lines). The shells start from rest at $t=0$ at a radius $R_0\ge R_S=2$, and time is indicated on the left axis. The Schwarzschild radius $R_S$ is indicated by the black vertical line. The blue curve is $P(R_0)$, which is labeled on the right axis. In this case, $P(R_0)=R_0$ because $a=0$. After a time indicated by the green horizontal line (in this particular case $t=\pi$), the 'first' shell (i.e. the one with $R_0=R_S$) has reached $R=0$, and from then on all of spacetime is covered. The magenta lines are infalling light shells. Every orbit (including light) in spacetime will find itself instantaneously intersecting some shell $R_P(t)$ if it is located 'above' the light shell that crashes into $R=0$ together with the radial shell that started off at $t=0$ on the Schwarzschild radius. That light shell is the one most to the lower left in the figure.
                   }
         \label{fig_novi0}
\end{figure}
This is the easiest case to interpret, since all the shells have a Newtonian negative energy. The marginally bound shell with $\tilde E_\infty=1$ and $P=+\infty$ starts off at infinity, while the shell with $\tilde E_\infty=0$ and $P=2M$ starts off at the Schwarzschild radius.

\subsubsection{\texorpdfstring{$\Lambda<0$}{Lambda<0}}

The effective potential $V_0(R)$ is a monotonically increasing function and thus all lines $V_0(R)=\tilde E_\infty$ with $0\le\tilde E_\infty<+\infty$ intersect $V_0(R)$ at some $R$. No particle is unbound, also not those with $\tilde E_\infty\ge1$. This means that the cosmological constant 'always wins' in attracting radial shells to the center. 

If $\tilde E_\infty<1$ and thus $\epsilon_N=+1$ and $P\ge2M$ we find similar solutions as (\ref{novradlamda0}):
\begin{equation}
\label{novradlamdalt0a}
R_P(t) = P \,\cyc\left(a_P,+1,\sqrt{2M\over P^3}t + \bar\psi_{\rm max}(a_P)\right),
\end{equation}
with $\bar\psi_{\rm max}(a_P)$ as calculated in appendix~\ref{sect_cyc}.
These shells start at rest at $t=0$ at some radius $R_0$ with $R_S\le R_0<\sqrt[3]{-6M/\Lambda}$: the upper limit follows from $V_0\le\tilde E_\infty^2<1$ and thus $2M/R+\Lambda R^2/3>0$.

For all other start radii $R_0>\sqrt[3]{-6M/\Lambda}$ the relations $\tilde E_\infty>1$, $\epsilon_N=-1$ and $P>0$ hold, together with the solution
\begin{equation}
\label{novradlamdalt0b}
R_P(t) = P \,\cyc\left(a_P,-1,\sqrt{2M\over P^3}t + \bar\psi_{\rm max}(a_P)\right).
\end{equation}

In both cases, the shell that starts at $R_0=\sqrt[3]{-6M/\Lambda}$ has $\tilde E_\infty=1$ and hence shell parameter $P=+\infty$. The shell that is marginally bound, starting off at $R=+\infty$ has $\tilde E_\infty=+\infty$ and $P=0$.

\begin{figure}[ht]
   \centering
   \includegraphics[width=130mm]{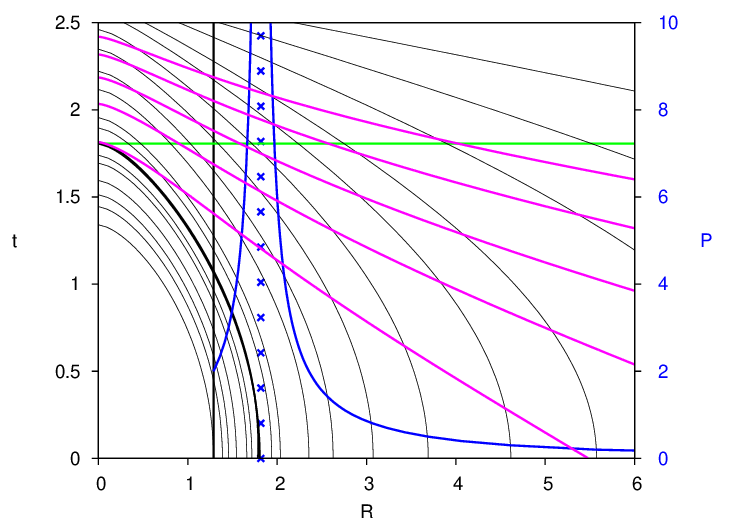}
   \caption{Infalling shells $R_P(t)$ for the case $\Lambda=-1.0$ and $M=1$ (black thin lines). The shells start from rest at $t=0$ at a radius $R_0\ge R_S$, and time is indicated on the left axis. The Schwarzschild radius $R_S$ is the bold vertical line. The 2 families of shells are separated by the start radius $\sqrt[3]{-6M/\Lambda}$, indicated by the vertical line with the crosses. The shell with this start radius is drawn in bold. The family that starts at radii smaller than $\sqrt[3]{-6M/\Lambda}$ has $\epsilon_N=+1$, the other family has $\epsilon_N=-1$. The blue curves are $P(R_0)$ for both families, and share the right axis. After a time indicated by the green horizontal line, all shells of the family with $\epsilon_N=+1$ have crashed into $R=0$, and from then on all of spacetime is covered by the family $\epsilon_N=-1$. The magenta lines are infalling light shells. Every orbit (including light) in spacetime will find itself instantaneously intersecting some shell $R_P(t)$ of family $\epsilon_N=-1$ if it is located 'above' the light shell that crashes into $R=0$ together with the radial shell that started off at $t=0$ on the radius $\sqrt[3]{-6M/\Lambda}$. That light shell is the one most to the lower left in the figure.
                    }
         \label{fig_novin}
   \end{figure}

\subsubsection{\texorpdfstring{$\Lambda>0$}{Lambda>0}}

In this case, the effective potential $V_0$ has a maximum at $\sqrt[3]{3M/\Lambda}$, where it attains the value $\sqrt[3]{9\Lambda M^2}$. We need $V_0(R)>0$
in order to have a metric with the required signature. Hence $9\Lambda M^2<1$. If $M=0$ (case that is not relevant in the context of this paper), there is universal expansion. The presence of the black hole makes bound shells possible in a finite part of space. Space is confined to the inner radius $R_S$ and the outer radius $R_2=2Mw_2(a_S)$, with $w_0$ and $w_2$ as defined in appendix~\ref{sect_cyc}.

Radial shells that have zero radial velocity at some R between $R_S$ and $R_2$ have therefore $0\le\tilde E_\infty<1-\sqrt[3]{9\Lambda M^2}$, hence, with (\ref{novconst}), $\epsilon_N=+1$ and $2M\le P<\sqrt[3]{(8M)/(9\Lambda)}$.

Because $V_0(R)$ has a maximum at $\sqrt[3]{3M/\Lambda}$, points at rest at $t=0$ at some radius $R_S\le R_0<\sqrt[3]{3M/\Lambda}$ will fall into the black hole according to (\ref{novradlamdalt0a}). On the contrary, points at rest at $t=0$ at some radius $\sqrt[3]{3M/\Lambda}<R_0\le R_2$ will be driven towards $R_2$, and beyond toward infinity. The shell $R_2$ has the same role as the Schwarzschild radius, but instead of a black hole horizon it is a 'black sky' horizon. In this latter case the motion is not defined by (\ref{defcyc}), which uses $0\le\cyc=R/P\le w_0$. Instead, these shells are given by
\begin{equation}
\label{novradlamdagt0}
\int_{w_2}^{\cyc'(a,\psi)}{\sqrt{u}\,\,du\over\sqrt{1-u+au^3}} = \psi.
\end{equation}
We will not discuss them further.

A point at rest at $R=\sqrt[3]{3M/\Lambda}$ will remain at rest, on an unstable shell though. For that shell $\tilde E_\infty=1-\sqrt[3]{9\Lambda M^2}$
\begin{figure}[ht]
   \centering
   \includegraphics[width=130mm]{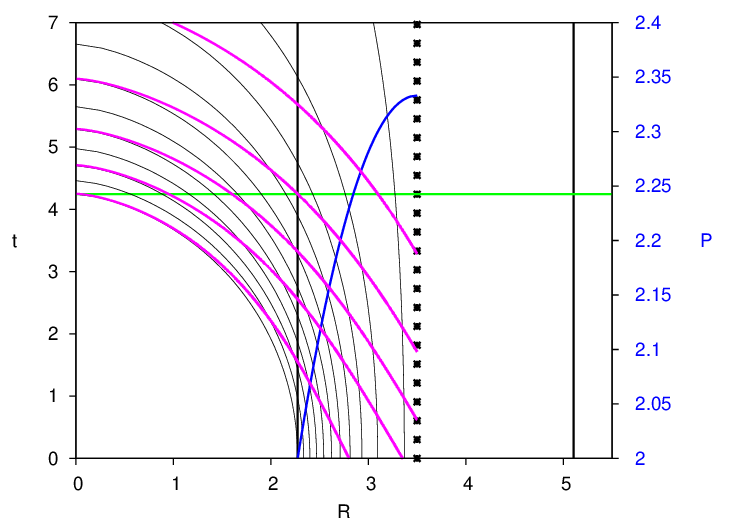}
      \caption{Infalling shells $R_P(t)$ for the case $\Lambda=.7$ and $M=1$ (black thin lines).
      Only the shells are drawn that start from rest at $t=0$ at a radius $R_S\le R_0\le\sqrt[3]{3M/\Lambda}$ that can be expressed by the function $\cyc$, and time is indicated on the left axis. The Schwarzschild radius $R_S$ and the radius $R_2$ are the bold vertical lines. The limit radius $\sqrt[3]{3M/\Lambda}$ is indicated by the vertical line with the crosses. The blue curve is $P(R_0)$, and is labeled on the right axis. After a time indicated by the green horizontal line, all of space inside $\sqrt[3]{3M/\Lambda}$ is covered. The magenta lines are infalling light shells. Every orbit (including light) in spacetime will find itself instantaneously intersecting some shell $R_P(t)$ if it is located 'above' the light shell that crashes into $R=0$ together with the radial shell that started off at $t=0$ on the Schwarzschild radius. That light shell is the one most to the lower left in the figure.
}

         \label{fig_novip}
   \end{figure}

\subsection{The Novikov metric}
\label{sect_Novi_metric}

In this subsection we will transform the Schwarzschild-$\Lambda$ metric into a representation in comoving coordinates. One such a frame can be obtained by considering a swarm of shells of the kind we have just come to consider. The essential feature of the swarm is that they all have zero radial velocity at the same \mbox{Schwarzschild-$\Lambda$} coordinate time $\bar t=0$. We arrange the swarm of shells such that at that time all allowed radii $R\ge R_S$ are uniquely occupied. We denote with $t$ the proper time on a shell (i.e. the arc length), and since it is a proper time, it can be used as a coordinate time that all shells can agree on. We set $t=0$ at the start of their journey, and thus all shells have zero \mbox{Schwarzschild-$\Lambda$} radial velocity at coordinate time $\bar t=0$ and proper time in their shell $t=0$. At any later coordinate time $\bar t$, any allowed radius $R\ge R_S$ will be covered by a shell, however with increasing radial Schwarzschild-$\Lambda$ velocity as $\bar t$ progresses, since coordinate shells crash into the black hole or disappear into the black sky, but are replaced by others with non-zero velocity.

We now consider the coordinates $(t,P)$. These coordinates were first introduced by Novikov (see \cite{MTW}). We call $P$ the (Novikov) shell label. We consider only the coordinate shells with label $P$ that fall into the black hole. We again must distinguish 3 cases.

For $\Lambda=0$, we see that after a warm-up time that is equal to the free fall (proper) time of shell $P=2M$ all radii $R\ge0$ are covered, as indicated in figure \ref{fig_novi0}. From that time on, the minimum shell label, which was initially $P=2M$, will steadily increase. All shells have $\epsilon_N=+1$

For $\Lambda<0$ we define the warm-up time as the time needed to crash the outermost shell of the first family with $\epsilon_N=+1$ into $R=0$ (see figure \ref{fig_novin}). From that time on, all of space is covered by shells with $\epsilon_N=-1$.

For $\Lambda>0$ the warm-up time is equal to the crash time of shell $P=2M$ (see figure \ref{fig_novip}), be it that we only consider the region of space that allows infalling shells. All these shells have $\epsilon_N=+1$.

Note that, as has been remarked in the captions of the figures \ref{fig_novi0}, \ref{fig_novin} and \ref{fig_novip}, these warm-up times can be relaxed somewhat by considering the light shells, but this is of no particular importance.

From the above analysis, we conclude that 
\begin{equation}
\label{def_epsilonN_metric}
\epsilon_N={\rm sign}(\Lambda),
\end{equation}
with the convention that ${\rm sign}(0)=+1$.

For the transformation of the Schwarzschild-$\Lambda$ metric (\ref{Schwarz}) into $(t,P)$, we make use of (\ref{novgeod}) and (\ref{novconst})
\begin{equation}
\label{novttrans}
\partial_t{\bar t}_P=d_t{\bar t}_P = {\displaystyle\sqrt{1-\epsilon_N{2M\over P}}\over{V_0(R_P)}}
\end{equation}
together with (\ref{novdRdt}) and (\ref{novconst}) which can be written as
\begin{equation}
\label{novRtrans}
\partial_tR_P=d_tR_P = -\sqrt{1-V_0(R_P)-\epsilon_N{2M\over P}}.
\end{equation}
The first equality in these equations follows from the fact that the 'Novikov swarm' evolves at constant $P$ and hence
\begin{equation}
d_t\bar t_P = \partial_t\bar t_P + \partial_P\bar t_P\,d_tP = \partial_t\bar t_P  
\end{equation}
and similarly for $d_tR_P$. Note that we do not seek (and need) an explicit expression for $\bar t_P$.

The metric (\ref{Schwarz}) transforms into
\begin{equation}
\label{Schwarznovi}
ds^2 = dt^2 - {\bigl[\partial_PR_P(t)\bigr]^2\over\displaystyle1-\epsilon_N{2M\over P}}\,dP^2 - R^2_P(t)\,d\Omega^2,
\end{equation}
after some rather lengthy calculations, especially to verify algebraically that the cross term
\begin{eqnarray}
2\Bigg[ & &\!\!\!\!\!\!\sqrt{1-\epsilon_N{2M\over P}} \partial_P{\bar t}_P+\nonumber\\
&&\qquad+{1\over V_0(R_P)}\sqrt{1-V_0(R_P)-\epsilon_N{2M\over P}}\partial_PR_P\Bigg]\,\, 
dt\,dP
\end{eqnarray}
is zero. This must, since the shells start at rest at $t=0$, therefore they are orthogonal to 3-space at that time and hence, (\ref{Schwarznovi}) is the Gaussian extension into proper time of a 3-metric. In order to prove this, one can consider its time derivative and show that it is zero. We are, of course, free to redefine the origin of time at any time, as long as the shells are such that they once were all together at rest at the same moment.

The metric (\ref{Schwarznovi}) is of the form (\ref{4metric}). In fact, it is a special case of (\ref{4metric}), since substitution of $r$ by $P$ and the choice $e(P)=\epsilon_N M/P$ transforms (\ref{4metric}) in (\ref{Schwarznovi}). Interestingly, just as is the case for metric (\ref{4metric}), $\Lambda$ does not appear explicitly as a number in (\ref{Schwarznovi}), and that information is fully contained in $R_P(t)$. The metric is valid for $P>2M$ if $\epsilon_N=+1$, or $P>0$ if $\epsilon_N=-1$.

Finally, we note that at any time $|\partial_PR_P(t)|>0$ since none of the shells cross (as is also obvious from the figures \ref{fig_novi0}, \ref{fig_novin} and \ref{fig_novip}), and hence $P$ and $R_P$ preserve order.

\subsection{Orbits}
\label{sect_Novi_orbit}

Since the Novikov metric has the same form of (\ref{4metric}), the analysis developed in section~\ref{sect_orbits} is valid, with the additional simplification that there is no singularity in the metric. All radii, also these inside $R_S$, can be treated equally.

As an example, we will now show that inside $R_S$ only inward motion is possible. Therefore we consider light that is sent radially in $(-)$ or out $(+)$:
\begin{equation}
dt = \pm{\partial_PR_P\over\sqrt{1-\epsilon_N\displaystyle{2M\over P}}}dP.
\end{equation}
The increment in $R$ equals
\begin{equation}
dR = \partial_PR_P\,dP+\partial_tR_P\,dt=\left(\partial_tR_P\pm\sqrt{1-\epsilon_N\displaystyle{2M\over P}}\,\right)dt
\end{equation}
which reads, taking into account (\ref{novRtrans})
\begin{equation}
dR = \left(-\sqrt{1-V_0(R_P)-\epsilon_N\displaystyle{2M\over P}}\pm\sqrt{1-\epsilon_N\displaystyle{2M\over P}}\,\right)dt.
\end{equation}
Clearly, if $V_0(R)<0$, i.e. inside $R_S$, the right hand side is always negative.

In the following figures, the structure of the orbits is elucidated by showing the light cones at various 'places' in the Novikov representation.

\begin{figure}[ht]
   \centering
   \includegraphics[width=130mm]{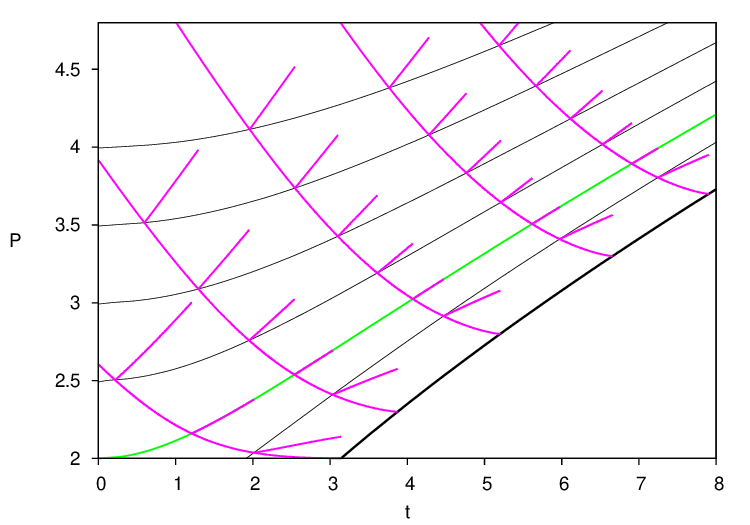}
      \caption{Light cones for the case $\Lambda=0$ and $M=1$ (as in figure \ref{fig_novi0}) in the Novikov representation $(t,P)$. The black lines are the loci of constant $R$. Two of them are highlighted. The thick black line is the locus $R=0$, the green one is the Schwarzschild radius $R=R_S$. Radii increase from lower right to upper left. Novikov shells are horizontal lines. The slope of the black lines can be understood by noting that the larger $P$, the larger $R_0$ and thus the longer it takes (in proper time $t$) to reach a certain $R$. Below $R=0$ there are no points in the Novikov metric. The magenta lines that run from upper left to lower right are infalling light rays (the same ones as in figure \ref{fig_novi0}). Where they cross the black lines, they form one of the 'legs' of the local light cone, and also the outgoing light ray at that point is shown. Outside the Schwarzschild radius, outgoing rays travel towards larger radii. At the Schwarzschild radius, light rays follow the Schwarzschild radius, indicating that an outgoing light ray there 'stays put'. Inside the Schwarzschild radius, even outgoing rays travel towards smaller radii, and will eventually meet the thick black line $R=0$.
}\label{fig_novilight0}
\end{figure}

\begin{figure}[ht]
   \centering
   \includegraphics[width=130mm]{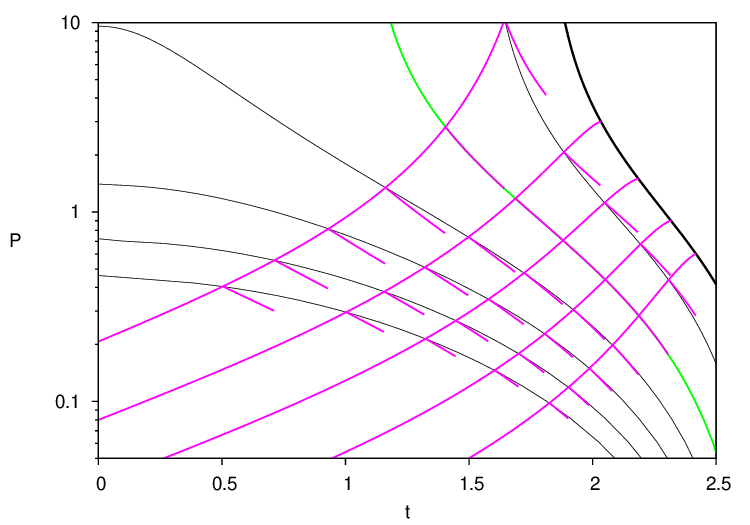}
      \caption{Light cones for the case $\Lambda=-1.0$ and $M=1$ (as in figure \ref{fig_novin}, same setup as figure \ref{fig_novilight0}). In this case $\epsilon_N=-1$ and therefore $\partial_PR_P<0$. Radii increase from upper right to lower left. The slope can be understood by noting that the smaller $P$, the larger $R_0$ and thus the longer it takes to reach a given radius. the origin. Above the locus $R=0$, there are no points in the Novikov metric. The magenta lines that run from lower left to upper right are infalling light rays (the same ones as in figure \ref{fig_novin}). Where they cross the black lines, they form one of the 'legs' of the local light cone, and also the outgoing light ray at that point is shown. Outside the Schwarzschild radius, outgoing rays travel towards larger radii. At the Schwarzschild radius, light rays follow the Schwarzschild radius, indicating that an outgoing light ray there 'stays put'. Inside the Schwarzschild radius, even outgoing rays travel towards smaller radii, and will eventually meet the thick black line $R=0$.
}\label{fig_novilightn}
\end{figure}

\begin{figure}[ht]
   \centering
   \includegraphics[width=130mm]{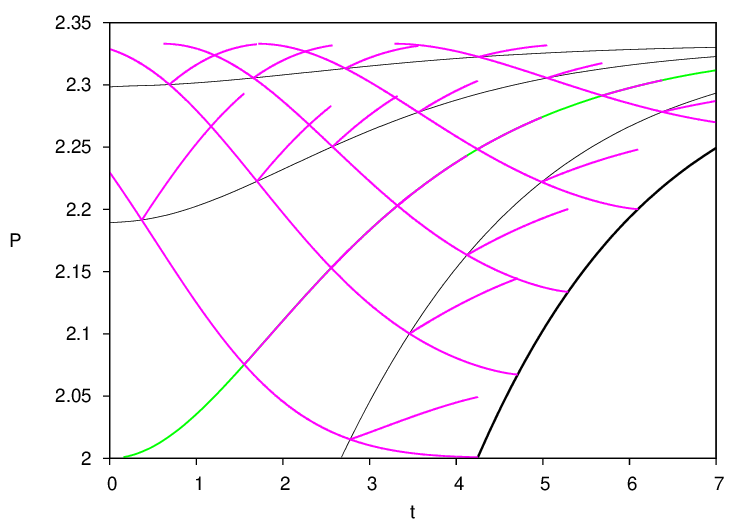}
      \caption{Light cones for the case $\Lambda=.7$ and $M=1$ (as in figure \ref{fig_novip}, same setup as figures \ref{fig_novilight0} and \ref{fig_novilightn}). This case is qualitatively similar to the case $\Lambda=0$. Only loci of constant $R$ and light rays inside $\sqrt[3]{3M/\Lambda}$ are shown.
}\label{fig_novilightp}
\end{figure}

\subsection{Transition from orbit representation between Schwarzschild-\texorpdfstring{$\Lambda$}{Lambda} and Novikov coordinates}
\label{sect_NoviSchwarz}

In the Schwarzschild-$\Lambda$ frame a stationary observer will assign to an infalling Novikov-shell the local Lorentzian radial velocity
\begin{equation}
\label{novschwarzupsilon0}
\upsilon_N(\bar t,R) = {dR_P/\sqrt{V_0(R)}\over\sqrt{V_0(R)}\,d\bar t}={d_{\bar t}R_P\over V_0(R)}<0.
\end{equation}
The numerator in the first fraction stands for a length measurement, the denominator for a time measurement. 
In a Novikov frame (note that there are an infinite number of them, since one can choose $t=0$), on the other hand, a stationary observer will find the Schwarzschild-$\Lambda$ shells of constant $R$ moving radially outward. The Novikov observer's local Lorentzian radial length measurement equals
\begin{equation}
|\partial_P R_P\,dP|/\sqrt{1-\epsilon_N(2M/P)}
\end{equation}
which happens at constant $t$. Hence we can rewrite $\partial_P R_P\,dP$ as
\begin{equation}
dR_P = \partial_P R_P\,dP + \partial_tR_P\,dt=\partial_PR_P\,dP
\end{equation}
and the Novikov stationary observer assigns to the outgoing Schwarzschild-$\Lambda$ shell the local Lorentzian radial velocity
\begin{eqnarray}
\label{novschwarzupsilon}
\upsilon_N(t,P)&=&{|dR_P|/\sqrt{1-\epsilon_N(2M/P)}\over dt}\nonumber\\
&=&{|d_tR_P|\over\sqrt{1-\epsilon_N(2M/P)}}>0.
\end{eqnarray}
Both velocities are equal in absolute value, as also follows from (\ref{novttrans}). Note that, using (\ref{novdRdt}) and (\ref{novconst}),
\begin{equation}
\label{novschwarzupsilon1}
|\upsilon_N|={\sqrt{\tilde{E}_\infty^2-V_0}\over\tilde E_\infty}= \sqrt{1-{V_0\over\tilde{E}_\infty^2}}\le1
\end{equation}
in the region of space where the  Schwarzschild-$\Lambda$ metric is valid ($V_0(R)>0$).

An orbit in the Schwarzschild-$\Lambda$ metric transforms into an orbit in the Novikov metric as follows. As above, we assign to both the stationary Schwarzschild-$\Lambda$ observer and the stationary Novikov observer their own local Lorentzian coordinate frame with the $\cal X$-axis in the outward $R$ direction and the $\cal Y$ axis in a tangential direction. In the local Lorentzian Schwarzschild-$\Lambda$ frame, a particle with coordinates $(R,\Omega)$ and coordinate velocity $(d_{\bar t}R,d_{\bar t}\Omega)$ has a velocity
\begin{equation}
\label{expr_VXS_VYS}
(V_{{\cal X},S},V_{{\cal Y},S}) = \left({d_{\bar t}R\over V_0(R)},{Rd_{\bar t}\Omega\over\sqrt{V_0(R)}})\right).
\end{equation}
This velocity transforms in a velocity $(V_{{\cal X},N},V_{{\cal Y},N})$ in the local Lorentzian Novikov frame with the Lorentz transformation
\begin{equation}
\label{Lorentz_transfo}
V_{{\cal X},N}= {V_{{\cal X},S}+\upsilon_N\over1+\upsilon_N V_{{\cal X},S}}\qquad V_{{\cal Y},N}=V_{{\cal Y},S}{\sqrt{1-\upsilon_N^2}\over1+\upsilon_N V_{{\cal X},S}}.
\end{equation}
Hence, the coordinate velocities in the Novikov frame are
\begin{equation}
(d_tP,d_t\Omega) = \left({\sqrt{1-\epsilon_N(2M/P)}\over|\partial_PR(P,t)|}V_{{\cal X},N},{1\over R}V_{{\cal Y},N}\right)
\end{equation}
and the coordinates are $(P,\Omega)$ with $P(R,t)$ depending on the particular Novikov frame (using the black lines in figures \ref{fig_novilight0}, \ref{fig_novilightn} or \ref{fig_novilightp}).

The Lorentz transformation causes a kind of refraction. Denoting
\begin{equation}
\label{def_VS_VN}
V_S=\sqrt{V_{{\cal X},S}^2+V_{{\cal Y},S}^2} \qquad  V_N=\sqrt{V_{{\cal X},N}^2+V_{{\cal Y},N}^2}
\end{equation}
and the respective angles with the outward radial direction by $\theta_S$ and $\theta_N$, we can calculate these angles with
\begin{equation}
\label{def_thetaS_thetaN}
\left\{
\begin{array}{rcl}
V_{{\cal X},S}  & =& V_S\cos\theta_S \\
V_{{\cal Y},S}  & =& V_S\sin\theta_S \\
\end{array}
\right.
\qquad
\left\{
\begin{array}{rcl}
V_{{\cal X},N}  & =& V_N\cos\theta_N \\
V_{{\cal Y},N}  & =& V_N\sin\theta_N .\\
\end{array}
\right.
\end{equation}
For light, the above relations simplify to the well-known
\begin{equation}
\cos\theta_N={\cos\theta_S+\upsilon_N\over1+\upsilon_N\cos\theta_S}.
\end{equation}
Wavelength is transformed with
\begin{equation}
\label{lambda_transfo}
\lambda_N={\sqrt{1-\upsilon_N^2}\over1+\upsilon_N\cos\theta_S}\lambda_S.
\end{equation}
The inverse refraction transformations are, of course, obtained by simply replacing $\upsilon_N$ with $-\upsilon_N$ and interchanging $N$ and $S$.

\section{A transport theorem for coordinate intervals}
\label{sect_transfer}

We consider 2 particles (material particles or photons) that are launched on the same orbit but separated by a coordinate interval $\delta x^\lambda$. We set out to study how $\delta x^\lambda(p)$ is transported along a geodesic, where, for a material particle, $p$ is the arc length along the geodesic or, for a photon, $p$ is a parameter such that the familiar geodesic equations hold:
\begin{equation}
\label{geodlight}
d_p\left(g_{\lambda\sigma}d_px^\sigma\right)=\frac12\partial_\lambda g_{\sigma\tau}\,d_px^\sigma d_px^\tau, \quad
\lambda,\sigma,\tau=0,1,2,3,
\end{equation}
with $\sigma$ and $\tau$ dummy indices. The evolution of the $x^\lambda$ coordinate along the 2 orbits therefore can be denoted by $x^\lambda(p)$ and $x^\lambda(p)+\delta x^\lambda(p)$. If we denote 
\begin{equation}
d_px^\lambda(p)=u(x)
\end{equation}
for the first particle, then 
\begin{equation}
d_px^\lambda(p)+d_p{\delta x^\lambda}(p)=u(x+\delta x^\lambda)=u(x)+\partial_{x^\lambda}u\,\delta x^\lambda
\end{equation}
for the second particle. Hence the transport differential equation
\begin{equation}
\label{eqntau_app}
{d_p{\delta x^\lambda}\over \delta x^\lambda} =\partial_{x^\lambda} u.
\end{equation}
We consider a metric for which $g_{\lambda\lambda}$ is independent of $x^\lambda$ and $g_{\lambda\mu}=0$, $\mu\ne\lambda$. Hence, from the $d p^{\,2}$ we get
\begin{equation}
g_{\lambda\lambda}u^2=\epsilon - g_{\sigma\tau}\,d_px^\sigma d_px^\tau,\qquad \sigma,\tau\ne\lambda
\end{equation}
with $\epsilon=1$ for a material particle and $\epsilon=0$ for light, and thus, upon partial differentiation with respect to $x^\lambda$,
\begin{equation}
\partial_{x^\lambda} u = -{1\over2ug_{\lambda\lambda}}\partial_\lambda g_{\sigma\tau}\,d_px^\sigma d_px^\tau.
\end{equation}
If we compare this expression with the $\lambda$-geodesic we find
\begin{equation}
\partial_{x^\lambda} u = -{1\over ug_{\lambda\lambda}}d_p(ug_{\lambda\lambda}),
\end{equation}
and thus, with (\ref{eqntau_app})
\begin{equation}
{d_p\delta x^\lambda\over\delta x^\lambda} = -{d_p(ug_{\lambda\lambda})\over ug_{\lambda\lambda}}.
\end{equation}
Hence the product
\begin{equation}
g_{\lambda\lambda}\,d_px^\lambda\,\delta x^\lambda
\end{equation}
is a constant along any orbit.

For the 3 metrics we consider in this paper it suffices to consider orbits in the meridional plane for which $\varphi$ is a constant. We find that $g_{\vartheta\vartheta}\,d_p\vartheta\,\delta\vartheta$ is a constant. Since the $\vartheta$-geodesic yields in addition that $g_{\vartheta\vartheta}\,d_p\vartheta=R^2d_p\vartheta$ is a constant (third equation of the system (\ref{geod1})), we find the obvious constant $\delta\vartheta$, indicating that there is rotational symmetry.

Somewhat less obvious is the application of this theorem to the $\bar t$ coordinate in the Schwarzschild-$\Lambda$ metric, for which the same argument as above yields that $(1-2M/R-\Lambda R^2/3)\,d_p\bar t\,\delta\bar t$ is a constant along a geodesic. From an analogous equation as (\ref{u_Schwarz}) we find that $(1-2M/R-\Lambda R^2/3)\,d_p\bar t$ is a constant of the motion, and hence also $\delta\bar t$. It expresses the fact that orbits calculated at different times but with the same initial conditions are identical, except for the time translation. This is instrumental in the derivation of the formula for the gravitational redshift.

For the L-T metric we obtain, with (\ref{defu1}), 
\begin{equation}\label{eqntau_LT}
d_pt\,\delta t=\tilde E\,\delta t
\end{equation}
as a constant along any geodesic, but here $\tilde E$ is not a constant.

\section{The angular distance}
\label{sect_angdistance}
\subsection{Definitions}
\label{sect_angdistance_defs}
The angular distance at the observer $D_{\rm o}$ is defined as the ratio of the linear size $dl_{\rm e}$ of an infinitesimal light emitting rod over the angle $d\theta_{\rm o}$ it subtends at the observer:
\begin{equation}
\label{defangdisto}
    D_{\rm o} = {dl_{\rm e}\over d\theta_{\rm o}},
\end{equation}
with the rod placed in the plane of the light ray that connects emitter and observer. It is at all times perpendicular to the tangent of the light ray. Therefore, $dl_{\rm e}$ is a length measurement at the emitter at the time of emission $t_{\rm e}$, while $d\theta_{\rm o}$ is an angle measurement at the observer at the time of observation $t_{\rm o}$. Because isotropy is lost in the spherical case, $D_{\rm o}$ will depend on the viewing direction.

Similarly, we will need to consider the angular distance at the emitter
\begin{equation}
\label{defangdiste}
 D_{\rm e} = {dl_{\rm o}\over d\theta_{\rm e}},
\end{equation}
which is the ratio of the linear size $dl_{\rm o}$ of a similar rod at the observer over the angle $d\theta_{\rm e}$ it subtends at the emitter. We will now derive expressions for these ratios.

We placed the rod in the plane of the light ray, but it need not be so. If we simplify the situation in a non-essential way by replacing the light ray by a line in Euclidean space, the rod could be on any plane through that line. So we can also state that the rod fills a circle with surface $\pi (dl)^2$, subtended by a solid angle 
\begin{equation}
\int_0^{d\theta}\sin\theta'd\theta'\int_0^{2\pi}d\varphi = \pi(d\theta)^2.
\end{equation}
A finite circle at the emitter need not be seen as a circle by the observer. In the limits $dl\to0$ and $d\theta\to0$ which we consider here, this will be the case, though. We can therefore easily pass from length and angle to surface and solid angle by squaring the ratios (\ref{defangdisto}) and (\ref{defangdiste}). Hence it also suffices to consider $dl$ and $d\theta$ in the plane of the light ray.  

We need to calculate the infinitesimal divergence $dl$ which is at every point along the orbit perpendicular to the tangent, and which is the consequence of a small deviation $d\theta$ from the initial direction $\theta_0$. If the integration along the orbit is executed backwards (starting from the observer, with initial condition $dl_{\rm e}=0$), we obtain $dl_{\rm e}/d\theta_{\rm o}$, which is the angular distance at observer (\ref{defangdisto}). If executed forward (starting from the emitter, with initial condition $dl_{\rm o}=0$) we obtain $dl_{\rm o}/d\theta_{\rm e}$. 

\subsection{The equations}
\label{sect_angdistance_eqs}
In the following analysis, we do not need to make the above distinction and we will consider the generic $dl/d\theta$. The problem at hand is known as the equations for the geodetic deviation. It can be shown quite generally that these are second order differential equation involving covariant differentiation and the Riemann tensor. We will follow a more intuitive (be it more elementary) approach.

Along the orbit, a small deviation $(\Delta r,\Delta\vartheta)$ from $(r,\vartheta)$ will generate
\begin{equation}
\label{deltal}
    \left(\Delta l\right)^2 = \left(X\Delta r\right)^2+\left(R\Delta\vartheta\right)^2.
\end{equation}
From the expression of $\dot r$ in (\ref{geod2}), we obtain
\begin{equation}\label{Xdeltar}
    {d\over dp}(X[\Delta r]) = \Delta w + {\partial_tX\over X}\tilde E \,(X[\Delta r])
\end{equation}
where we introduced the shorthand notation
\begin{equation}
\label{def_[Deltar]}
[\Delta r] = \epsilon_\xi \Delta r.
\end{equation}
The second equation of (\ref{geod2}) yields
\begin{eqnarray}\label{Rdeltafi}
    {d\over dp}(R\Delta\vartheta) &=& {\sqrt{1-2e}\over R}\epsilon_\xi\left[w(R\Delta\vartheta)-{2h\over r}(X[\Delta r])\right] + \nonumber \\
    &&\qquad+{\partial_tR\over R}{\tilde E}\,(R\Delta\vartheta)+{\Delta h\over R}.
\end{eqnarray}
Finally, the second equation of (\ref{geod3}) yields
\begin{eqnarray}\label{deltaw}
  {d\over dp}(\Delta w) &=& -\partial_r\left({\partial_tX\over X}\right){{\tilde E}w\over X}\epsilon_\xi\,(X[\Delta r]) -  {\partial_tX\over X}{\tilde E}\,\Delta w - \nonumber\\
  && \quad - {h^2\over R^3}\left({d_re\over\partial_r R}\right)X[\Delta r] + 2\sqrt{1-2e}\,\epsilon_\xi{h\Delta h\over R^3} - \nonumber\\
  && \qquad -3(1-2e){h^2\over R^4}(X[\Delta r]).
\end{eqnarray}
As for the initial conditions, we denote
\begin{equation}
    (w,\upsilon)=(X\dot r\epsilon_\xi,R\dot\theta)=(\cos\theta,\sin\theta)\,{\tilde E}
\end{equation}
which are the components of a vector that is proportional to the local Lorentzian velocity vector (it is identical at the start of the light ray, when $w^2+\upsilon^2=1=\tilde E_0$). We can assume \mbox{$\upsilon=h/R\ge0$}. Initially, the light ray makes an angle $0\le\theta_0\le\pi$ with the outgoing normal, given by $\tan\theta_0=\upsilon_0/w_0$. A small deviation $\Delta\theta_0$ leads to $(\Delta w_0,\Delta\upsilon_0)$. These components are not independent because $\Delta\tilde E_0=0$. We obtain
\begin{equation}
\Delta\theta_0=\Delta\upsilon_0/w_0
\end{equation}
and
\begin{equation}\label{deltah}
    {\Delta h\over\Delta\theta_0} = w_0R_0.
\end{equation}
From this equation and the linear structure of equations (\ref{Xdeltar}), (\ref{Rdeltafi}) and (\ref{deltaw}) we see that we can divide all the infinitesimals by $\Delta\theta_0$ in order to obtain equations involving finite quantities.

It is possible to combine all the above equations using the expression (\ref{deltal}).
The orthogonality of $(X[\Delta r],R\Delta\vartheta)$ with $(w,\upsilon)$ implies
\begin{equation}
    (X[\Delta r],R\Delta\vartheta)=(-\sin\theta,\cos\theta)\,\Delta l,
\end{equation}
using geometric considerations and $\Delta\vartheta>0$. We obtain, after some calculations:
\begin{eqnarray}\label{eqdeltal}
  {d\over dp}\left(\Delta l\over\Delta\theta_0\right)  &=& -\sin\theta\left(\Delta w\over\Delta\theta_0\right) + \cos\theta{w_0R_0\over R} +\nonumber\\
  &&\quad +\epsilon_\xi{\sqrt{1-2e}\over R}\cos\theta\,(1+\sin^2\theta)\,{\tilde E} \left(\Delta l\over\Delta\theta_0\right) + \nonumber\\
  && \quad+\left[\sin^2\theta{\partial_tX\over X} + \cos^2\theta{\partial_tR\over R}\right]{\tilde E}\left(\Delta l\over\Delta\theta_0\right).
\end{eqnarray}
Similarly, we can transform equation (\ref{deltaw}):
\begin{eqnarray}\label{deltaw1}
  {d\over dp}\left(\Delta w\over\Delta\theta_0\right) &=& -{\partial_tX\over X}{\tilde E}\left(\Delta w\over\Delta\theta_0\right) + \partial_r\left({\partial_tX\over X}\right){{\tilde E}w\over X}\epsilon_\xi\sin\theta\left(\Delta l\over\Delta\theta_0\right)  + \nonumber\\
  && + {\upsilon^2\over R}\left({d_re\over\partial_r R}\right)\sin\theta\left(\Delta l\over\Delta\theta_0\right)  + 2\sqrt{1-2e}{\upsilon\over R^2}\epsilon_\xi w_0R_0 + \nonumber\\
  && \qquad +3(1-2e)\left(\upsilon\over R\right)^2\sin\theta\left(\Delta l\over\Delta\theta_0\right).
\end{eqnarray}
The differential equations (\ref{eqdeltal}) and (\ref{deltaw1}) are sufficient to calculate $\Delta l/\Delta\theta_0$, and can be added to the equations of the light ray.
As to the initial conditions, we start with $\Delta l_0/\Delta\theta_0=0$. The orthogonality condition yields $\Delta w_0/\Delta\theta_0=-\upsilon_0$. 

\subsection{A theorem}
\label{sect_angdistance_theorem}
A further simplification, based on $w^2+\upsilon^2={\tilde E}^2$, and thus $\Delta w = -\Delta\upsilon\tan\theta$, can eliminate $\Delta w$ altogether.
Therefore we write
\begin{eqnarray}
\Delta\upsilon &=& {\Delta h\over R} - {h\over R^2}\partial_rR\,\Delta r={\Delta h\over R} - {h\over R}{\sqrt{1-2e}\over R}\epsilon_\xi(X[\Delta r])\nonumber\\
&=& {w_0R_0\over R}\Delta\theta_0 + \sin^2\theta{\sqrt{1-2e}\over R}\epsilon_\xi\, {\tilde E}\Delta l,
\end{eqnarray}
which transforms (\ref{eqdeltal}) into
\begin{eqnarray}
\label{dloverdthshort}
  {d\over dp}\left(\Delta l\over\Delta\theta_0\right)  &=& {R_0\cos\theta_0\over R\cos\theta} +
  \epsilon_\xi{\sqrt{1-2e}\over R\cos\theta}\,{\tilde E}\left(\Delta l\over\Delta\theta_0\right) + \nonumber\\
  && \quad+\left[\sin^2\theta{\partial_tX\over X} + \cos^2\theta{\partial_tR\over R}\right]{\tilde E}\left(\Delta l\over\Delta\theta_0\right).
\end{eqnarray}
This equation could at first sight replace one of the equations (\ref{deltaw}) and (\ref{eqdeltal}), were it not that it needs the resolution of a $0/0$ at the turning points of the light ray ($\theta=\pi/2$), which implies a further derivative. Therefore the above equation is only useful as a further relation involving $\Delta l/\Delta\theta_0$.

We now apply/verify equation (\ref{dloverdthshort}) in 3 cases. Firstly, for small $t$, we obtain
\begin{equation}
\label{dloverdthsmall}
{\Delta l\over\Delta\theta_0}= p
\end{equation}
which is the Euclidean limit.

Secondly, in the Euclidean case $e=0$, we have ${\Delta l/\Delta\theta_0}=p$ exactly, and (\ref{dloverdthshort}) simplifies to
\begin{equation}
R\cos\theta = R_0\cos\theta_0 + p,
\end{equation}
which is a rather obvious trigonometric relation valid for a straight line. This example also shows that (\ref{dloverdthshort}) is not sufficient to calculate $\Delta l/\Delta\theta$, since information on the orbit must also be included, in this case the fact that the orbit is a straight line.

Thirdly, we can also verify equation (\ref{dloverdthshort}) in the synchronous case. From the Euclidean case, we know that hidden in (\ref{dloverdthshort}) is a geometrical relation of the orbit. We adopt the notations of section~\ref{sect_photometry_sync}.
From geometrical considerations we now that
\begin{equation}
{\Delta l\over\Delta\theta_0}=R(\chi,t)={\cal R}(t)\sin\chi={\cyc(t)\over\omega}\sin\chi,
\end{equation}
where we made use of (\ref{sync_R_chi}) and (\ref{radius_sync}). Geometry learns us also that we can simplify the orbit to a radial one starting from the origin, and thus $\theta_0=0$ and $R_0=0$. 
Equation (\ref{dloverdthshort}) then reduces to
\begin{equation}
{d\over dt}\left(\Delta l\over\Delta\theta_0\right)  = \left(
  \epsilon_\xi{\sqrt{1-2e}\over R} + {\partial_tR\over R}\right)\left(\Delta l\over\Delta\theta_0\right),
\end{equation}
yielding
\begin{eqnarray}
{d_t\,\cyc\over\omega}\sin\chi+{\cyc\over\omega}\cos\chi\, d_t\chi&=&\left({\epsilon_\xi|\cos\chi|\over{\cal R}\sin(\chi)}+{d_t\,\cyc\over\cyc}\right){\cal R}\sin\chi.\nonumber\\
\end{eqnarray}
With the aid of (\ref{iso_chi_int}), this expression can be seen to yield an identity. This special case is also an occasion to see the sheet parameter $\epsilon_\xi$ `at work', since in the left hand side we find $\cos\chi$, while in the right hand side we find $\epsilon_\xi|\cos\chi|=\cos\chi$.

\section{The surface brightness and flux}
\label{sect_surface brightness_flux}

\subsection{The surface brightness}
\label{sect_surface brightness}

An elementary surface $dS_{\!\rm e}$ at the emitter sends photons to an elementary surface $dS_{\!\rm o}$ of the observer. Therefore the emitter sends his photons within his elementary solid angle $d\Omega_{\rm e}$ and the observer receives them within her elementary solid angle $d\Omega_{\rm o}$. The light paths that materialize this relation are not straight lines, nor can they be taken to be the simple radial geodesics as in the synchronous case, but have to be calculated. Conservation of photons requires
\begin{eqnarray}
n_{\rm o}(t_{\rm o},\lambda_{\rm o},\Omega_{\rm o},\emph{\textbf{r}}_{\rm o})\!&dt_{\rm o}&\!d\lambda_{\rm o}\,d\Omega_{\rm o}\,dS_{\!\rm o} = \nonumber\\
&&n_{\rm e}(t_{\rm e},\lambda_{\rm e},\Omega_{\rm e},\emph{\textbf{r}}_{\rm e})\,dt_{\rm e}\,d\lambda_{\rm e}\,d\Omega_{\rm e}\,dS_{\!\rm e},
\end{eqnarray}
with $n_{\rm e}$ (resp. $n_{\rm o}$) the number of photons emitted (resp. observed) at time $t$, location $\emph{\textbf{r}}$, with wavelength $\lambda$ and direction $\Omega$, during a time interval $dt$, on an elementary surface $dS$, within an elementary solid angle $d\Omega$ and in a wavelength range $d\lambda$. Defining the surface brightness as 
\begin{equation} \label{def_surf_bright}
{\cal I}(t,\lambda,\Omega,\emph{\textbf{r}})=E\,n(t,\lambda,\Omega,\emph{\textbf{r}})={hc\over\lambda}n(t,\lambda,\Omega,\emph{\textbf{r}}),
\end{equation}
at a specific wavelength and per unit of time, we obtain
\begin{equation}
\label{transintens}
{\cal I}_{\rm o}(t_{\rm o},\lambda_{\rm o},\Omega_{\rm o},\emph{\textbf{r}}_{\rm o})= {1\over(1+z)^3}{dS_{\!\rm e}\over d\Omega_{\rm o}}{d\Omega_{\rm e}\over dS_{\!\rm o}}
{\cal}I_{\rm e}(t_{\rm e},\lambda_{\rm e},\Omega_{\rm e},\emph{\textbf{r}}_{\rm e}),
\end{equation}
where we used
\begin{equation}
{E_{\rm o}\over E_{\rm e}}={dt_{\rm e}\over dt_{\rm o}}={\lambda_{\rm e}\over\lambda_{\rm o}}={1\over1+z}.
\end{equation}
We of course recover the local Lorentzian invariant ${\cal I}\nu^{-3}$, but spacetime contributes a factor due to the effect that the infinitesimal lightcone with top at the emitter is not equal to the one with top at the observer.

In the notations of appendix~\ref{sect_angdistance}, we obtain
\begin{equation}\label{relSang1}
    dl_{\rm e} = \sqrt{dS_{\!\rm e}} \quad {\rm and} \quad d\theta_{\rm o} = \sqrt{d\Omega_{\rm o}}
\end{equation}
and similarly
\begin{equation}\label{relSang2}
    dS_{\!\rm o} = dl_{\rm o}^2 \quad {\rm and} \quad d\Omega_{\rm e}= d\theta_{\rm e}^2.
\end{equation}
Using (\ref{relSang1}) and (\ref{relSang2}) the expression (\ref{transintens}) can be written as
\begin{equation}
\label{transintens1}
{\cal I}_{\rm o}(t_{\rm o},\lambda_{\rm o},\Omega_{\rm o},\emph{\textbf{r}}_{\rm o})= {1\over(1+z)^3}\left({dl_{\rm e}\over d\theta_{\rm o}}{d\theta_{\rm e}\over dl_{\rm o}}\right)^2
{\cal I}_{\rm e}(t_{\rm e},\lambda_{\rm e},\Omega_{\rm e},\emph{\textbf{r}}_{\rm e}),
\end{equation}
or arguably the more useful
\begin{eqnarray}
\label{transintens2}
\Delta\lambda_{\rm o}{\cal I}_{\rm o}(t_{\rm o},\lambda_{\rm o},\Omega_{\rm o},\emph{\textbf{r}}_{\rm o})= {\Delta\lambda_{\rm e}\over(1+z)^2}\left({dl_{\rm e}\over d\theta_{\rm o}}{d\theta_{\rm e}\over dl_{\rm o}}\right)^2\!\!
{\cal I}_{\rm e}(t_{\rm e},\lambda_{\rm e},\Omega_{\rm e},\emph{\textbf{r}}_{\rm e}).
\end{eqnarray}
Clearly, in Euclidean space, 
${\cal I}_{\rm o}(t_{\rm o},\lambda_{\rm o},\Omega_{\rm o},\emph{\textbf{r}}_{\rm o})=
{\cal I}_{\rm e}(t_{\rm e},\lambda_{\rm e},\Omega_{\rm e},\emph{\textbf{r}}_{\rm e})$.

\subsection{The flux}
\label{sect_flux}
As for the fluxes, we start from (\ref{transintens}). We integrate the left hand side at the observer over the total solid angle subtended by the emitter, and we integrate the right hand side over the full visible surface of the emitter, thereby capturing all photons. Hence,
\begin{eqnarray}
\left[\int {\cal I}_{\rm o}(t_{\rm o},\lambda_{\rm o},\Omega_{\rm o},\emph{\textbf{r}}_{\rm o})\,d\Omega_{\rm o}\right]&&= \nonumber\\ {1\over(1+z)^3}{d\Omega_{\rm e}\over dS_{\!\rm o}}&&\left[\int
{\cal I}_{\rm e}(t_{\rm e},\lambda_{\rm e},\Omega_{\rm e},\emph{\textbf{r}}_{\rm e})dS_{\!\rm e}\right]\!.
\end{eqnarray}
The integration on the right hand side equals, for an isotropic emitter, the luminosity $L$ at $\lambda_{\rm e}$ (which is a power) divided by $4\pi$, while the integration on the left hand side equals the flux ${\cal F}$ (which is power per unit surface) observed at $\lambda_{\rm o}$. Hence
\begin{eqnarray}
\label{transflux}
{\cal F}(t_{\rm o},\lambda_{\rm o},\emph{\textbf{r}}_{\rm o})= {1\over4\pi}{1\over(1+z)^3}{d\Omega_{\rm e}\over dS_{\!\rm o}}
L(t_{\rm e},\lambda_{\rm e},\emph{\textbf{r}}_{\rm e}).
\end{eqnarray}
The observed flux in band $X_{\rm o}$, that is the redshift transformed band from the band $X_{\rm e}$ at the emitter, equals
\begin{equation}
\label{transbolflux}
{\cal F}_{X_{\rm o}}(t_{\rm o},\emph{\textbf{r}}_{\rm o})= {1\over4\pi}{1\over(1+z)^2}{d\Omega_{\rm e}\over dS_{\!\rm o}}
L_{X_{\rm e}}(t_{\rm e},\emph{\textbf{r}}_{\rm e}).
\end{equation}
Turning to magnitudes, we recall that an absolute magnitude $M_{X_{\rm e}}$ is a flux at 10 pc. Hence, we rewrite (\ref{transbolflux}) as
\begin{equation}
{\cal F}_{X_{\rm o}}(t_{\rm o},\emph{\textbf{r}}_{\rm o})= {1\over(1+z)^2}{d\Omega_{\rm e}\over dS_{\!\rm o}}
\left[L_{X_{\rm e}}(t_{\rm e},\emph{\textbf{r}}_{\rm e})\over4\pi\,(10\,{\rm pc})^2\right]
(10\,{\rm pc})^2.
\end{equation}
in which we recognize inside the big brackets the flux of the emitter if it were placed at \mbox{$10\,{\rm pc}$}.
Passing to magnitudes, we obtain
\begin{equation}
 m_{X_{\rm o}}^{({\rm mag})}(t_{\rm o},\emph{\textbf{r}}_{\rm o})= -2.5\log_{10}\left({(10{\rm pc})^2\over(1+z)^2}{d\Omega_{\rm e}\over dS_{\!\rm o}}\right)
+ M_{X_{\rm e}}^{({\rm mag})}(t_{\rm e},\emph{\textbf{r}}_{\rm e}).
\end{equation}
In our unit of length
\begin{eqnarray}
\label{transmagn}
 m_{X_{\rm o}}^{({\rm mag})}(t_{\rm o},\emph{\textbf{r}}_{\rm o})&=& 5\log_{10}\left((1+z){dl_{\rm o}\over d\theta_{\rm e}}\right) +5\log_{10}(3.066)\nonumber\\
&&\qquad+ M_{X_{\rm e}}^{({\rm mag})}(t_{\rm e},\emph{\textbf{r}}_{\rm e}) + 40,
\end{eqnarray}
where we made again use of (\ref{relSang1}) and (\ref{relSang2}).

\subsection{The luminosity distance}
The luminosity distance is defined as
\begin{equation}
\label{deflumdist}
{\cal F} = {L\over4\pi D_L^2}.
\end{equation}
We find, with (\ref{transbolflux}),
\begin{equation}
\label{lumdist}
D_L^2 = (1+z)^2 {dS_{\!\rm o}\over d\Omega_{\rm e}}.
\end{equation}
Hence, with (\ref{defangdisto}), (\ref{relSang1}) and (\ref{relSang2})
\begin{equation}
\label{rellumangdist}
    D_L = (1+z){dl_{\rm o}\over d\theta_{\rm e}}=(1+z){dl_{\rm o}\over d\theta_{\rm e}}{d\theta_{\rm o}\over dl_{\rm e}}D_{\rm o}.
\end{equation}

\section{Distribution functions}
\label{app_isothermal}
In the presence of a static spherical mass distribution centered at $R=0$, geodetic motion has the constant of the motion (\ref{u_Schwarz}):
\begin{equation}
\label{def_tildeEpot}
\tilde E_\infty={E_\infty\over m_0c^2} = g_{00}{d\bar t\over ds}.
\end{equation}
In this expression $m_0$ denotes the mass of the particle. At the distance $R$ an observer measures time with a 'shell time' $\bar t_R$, and the metric yields
\begin{equation}\label{def_gam00}
    d\bar t_R = \sqrt{g_{00}}\,d\bar t \equiv\gamma_{00}\,d\bar t\equiv\sqrt{1-2\Psi(R)/c^2}\,d\bar t.
\end{equation}
Hence
\begin{equation}
\label{rel_Einfty_ER}
E_\infty= \gamma_{00}{d\bar t_R\over ds}m_0c^2=\gamma_{00}E_R
\end{equation}
with
\begin{equation}
\label{defeR}
    E_R=\sqrt{p^2c^2+m_0^2c^4}={m_0c^2\over\sqrt{1-\upsilon^2/c^2}}\equiv{E_0\over\sqrt{1-\upsilon^2/c^2}}
\end{equation}
the relativistic energy as measured by the shell observer at $R$, and $p=|\boldsymbol{p}|$ momentum and $\upsilon=|\boldsymbol{\upsilon}|$ velocity observed by the same observer.

Because of the spherical symmetry, the angular momentum $\boldsymbol{R}\times\boldsymbol{p}$ is also conserved. Denoting by $p_R$ the radial component and by $p_T$ the absolute value of the component perpendicular to the radial component (and hence $p^2=p_R^2+p_T^2$), we have $|\boldsymbol{R}\times\boldsymbol{p}|=Rp_T$. Any isotropic distribution $f(E_\infty)\,d^3\!\boldsymbol{R}\,d^3\!\boldsymbol{p}$ far from the mass distribution therefore transforms to
\begin{equation}
d^3\!\boldsymbol{R}\,2\pi f\bigl(\gamma_{00}E_R\bigl)\,dp_R\,p_T\,dp_T
\end{equation}
at distance $R$. In this equation the $2\pi$ already accounts for the integration over the position angle of the momentum vectors perpendicular to the radial component, since the distribution does not depend on it. For small velocities and a weak field, we of course recover
\begin{equation}
    E_\infty = E_0-m_0\Psi+\frac12m_0\upsilon^2=E_R-m_0\Psi,
\end{equation}
expressing the fact that at infinity, a particle has climbed the potential well. 

Passing from $p_R$ to $E_R$ using (\ref{defeR}), we obtain 
\begin{equation}
\label{aniso}
d^3\!\boldsymbol{R}\,2\pi(\pm) c^{-2}f\bigl(\gamma_{00}E_R\bigl){E_R\,dE_R\,p_T\,dp_T\over\sqrt{(E_R^2-m_0^2c^4)c^{-2}-p_T^2}}
\end{equation}
where the $(\pm)$ accounts for the sign of $p_R$.

When all $p_T$ that keep the radicand positive are allowed, it is not too hard to work out the expression of all the velocity moments
\begin{eqnarray}\label{velmoments}
    \int\!\! {f(E_\infty)\upsilon^{2n}\over\left(1-{\upsilon^2\over c^2}\right)^n}d^3\!\boldsymbol{p} &=& {4\pi c^{2n-3}\over E_0^{2n}}\!\!\int_{E_0}^{+\infty}\!\!\!f(E_\infty)\left(E_R^2-E_0^2\right)^{{n+1\over2}}\!\!E_R\,dE_R.\nonumber\\
    &&
\end{eqnarray}
The Boltzmann distribution
\begin{equation}
\label{Boltzmann}
    f(E_\infty)=f_0\exp\left(-{E_\infty\over kT}\right)
\end{equation}
can be integrated with (\ref{velmoments}). We obtain the density
\begin{eqnarray}
    \rho &=& 4\pi c^{-3}f_0\!\!\int_{E_0}^{+\infty}\!\!\!\exp\left(-{E_\infty\over kT}\right)\sqrt{E_R^2-E_0^2}\,E_R\,dE_R.\nonumber\\
    &&
\end{eqnarray}
which yields
\begin{equation}
\label{isothermdist0}
    \rho=4\pi f_0m_0^2c{kT\over \gamma_{00}}K_2\left({m_0c^2\over kT}\gamma_{00}\right),
\end{equation}
with $K_2$ the Bessel function of the third kind in the imaginary argument. 

If the argument of $K_2$ is small, as is likely the case in a compact phase, then the approximation $K_2(z)\to2z^{-2}$ is valid. If it is large, as is likely the case in the current epoch, we have $K_2(z)\to[\pi/(2z)]^{1/2}e^{-z}$. Hence
\begin{eqnarray}
\label{Boltz_temp}
\rho&\to& \left\{
\begin{array}{lcl}
\displaystyle f_0\,8\pi c^{-3}\left(kT\over\gamma_{00}\right)^3& {\rm if} \quad& \displaystyle{m_0c^2\over kT}<<1 \\[3mm]
\displaystyle f_0\left(2\pi m_0kT\over\gamma_{00}\right)^{3/2}\exp\left(-{m_0c^2\over kT}\gamma_{00}\right)&\, {\rm if} \quad& \displaystyle{m_0c^2\over kT}>>1.
\end{array}
\right.\nonumber\\
\end{eqnarray}
Another normalisation of (\ref{isothermdist0}) is
\begin{equation}
    \rho(\gamma_{00})={\rho_\infty\over \gamma_{00}}{K_2\left(\displaystyle{m_0c^2\over kT}\gamma_{00}\right)\over K_2\left(\displaystyle{m_0c^2\over kT}\right)}
\end{equation}
with $\rho_\infty$ the density of the isothermal distribution in the absence of a mass distribution. It is computationally convenient to consider the function
\begin{equation}
    K'_n(z) = e^zK_n(z)
\end{equation}
yielding
\begin{equation}\label{isothermdist1}
    \rho(\gamma_{00})={\rho_\infty\over \gamma_{00}}\exp\left[(1-\gamma_{00}){m_0c^2\over kT}\right]{K'_2\left(\displaystyle{m_0c^2\over kT}\gamma_{00}\right)\over K'_2\left(\displaystyle{m_0c^2\over kT}\right)}.
\end{equation}

If the quantity $\Psi/c^2$ is small, which is the case even for very massive and very compact mass concentrations on distances larger than a parsec, we can further simplify to
\begin{equation}\label{isothermdist2}
    \rho(\Psi)=\rho_\infty\exp\left({m_0\Psi\over kT}\right){K'_2\left(\displaystyle{m_0c^2\over kT}-{m_0\Psi\over kT}\right)\over K'_2\left(\displaystyle{m_0c^2\over kT}\right)}.
\end{equation}
If ${m_0c^2/(kT)}$ is large, then
\begin{equation}\label{isothermdist3}
    \rho(\Psi) = \rho_\infty\exp\left({m_0\Psi\over kT}\right).
\end{equation}
For the isotropic pressure $\cal P$, defined as one third of the moment of order 1 in the expression (\ref{velmoments}), we obtain
\begin{equation}\label{def_pressure}
{\cal P}(\Psi)={\rho_\infty\over m_0\gamma_{00}}{kT\over\gamma_{00}}{K_3\left(\displaystyle{m_0c^2\over kT}\gamma_{00}\right)\over K_2\left(\displaystyle{m_0c^2\over kT}\right)}.
\end{equation}

In the presence of a mass concentration, these distributions apply to the ensemble of all particles, whether they are bound to the structure or not. If we restrict us to the bound particles, the same analysis would apply but the integration would extend to $m_0\Psi$ rather than $+\infty$. Here we will consider only the analogue of (\ref{isothermdist3}), which is the well-known expression
\begin{equation}\label{isothermdist4}
    \rho(\Psi) = \rho_\infty e^{m_0\Psi/(kT)}\gamma_N(\textstyle{\frac32},{m_0\Psi\over kT}).
\end{equation}
with
\begin{equation}
    \gamma_N(\alpha,x)={1\over\Gamma(\alpha)}\int_0^xt^{\alpha-1}e^{-t}dt
\end{equation}
the normalized incomplete gamma function.

\onecolumn

\section{Definition of symbols}
\begin{table}[h]
\caption{Definition of upper case symbols}             
\centering                          
\begin{tabular}{l l l l}        
\hline\hline                 
Symbol & Equation & section &  Name \\    
\hline                        
$C$ & & \ref{sect_nat_curv} &\footnotesize  center of the universe \\
$D_L$ & \ref{deflumdist}  &  &\footnotesize   luminosity distance\\
$D_{\rm o}$, $D_{\rm e}$ & \ref{defangdisto}, \ref{defangdiste} & &\footnotesize  angular distance at observer, emitter \\
$E_R$, $E_0$ & \ref{defeR} & &\footnotesize  rel. en. meas. by stat. obs., rest mass en. \\
$\tilde E_\infty$, $E_\infty$  & \ref{u_Schwarz}, \ref{novgeod}, \ref{def_tildeEpot} & &\footnotesize  (norm.) rel. en. measured at infinity \\
$\tilde E$ & \ref{geod2} & \ref{sect_redshift} &\footnotesize  norm. rel. en. measured by comoving obs.\\
$\tilde E_R$ & \ref{def_tilde_ER}, \ref{def_tilde_ER1} & &\footnotesize  norm. rel. en. measured by a stat. obs. at $R$\\
$\tilde E_{R,\rm min}$ & \ref{neut_response1}, \ref{expr_E_R,min} & &\footnotesize  min. $\tilde E_R$ of non-plunging orbit around a BH\\
$\tilde E_{\infty,\rm min}$ &\footnotesize  \ref{ERmin1} & &\footnotesize  min. $\tilde E_\infty$ of non-plunging orbit around a BH\\
$\tilde E_{\infty,N}$ & \ref{F-D} & &\footnotesize  normalisation in Fermi-Dirac distr.\\
${\cal F}$, ${\cal F}_X$ & \ref{transflux}, \ref{transbolflux}& &\footnotesize  obs. flux: at specific wavelength, in band $X$  \\
$G$ & & &\footnotesize  gravitational constant \\
${\rm H}(r,t)$, $H(r,t)$ & \ref{defHtime}, \ref{defH} & &\footnotesize  tangential Hubble parameter \\
$H_0$ & \ref{defH0} & &\footnotesize  $H(r_0,t_0)$\\
${\rm H}_{\rm o}$&\ref{def_Ho}&&\footnotesize observed value for ${\rm H}(r_{\rm o},t_{\rm o})$\\
$H^t(r,t)$ & \ref{def_HtIt} & &\footnotesize  tangential deceleration parameter \\
$H^{rt}(r,t)$ & \ref{def_Hrt_Irt} & &\footnotesize  mixed deceleration parameter \\
$I(r,t)$ & \ref{radHubble} & &\footnotesize  radial Hubble parameter \\
$I^t(r,t)$ & \ref{def_HtIt} & &\footnotesize  radial deceleration parameter \\
$I^{rt}(r,t)$ & \ref{def_Hrt_Irt} & &\footnotesize  mixed radial deceleration parameter \\
${\cal I}_{\rm e}$, ${\cal I}_{\rm o}$ &  \ref{def_surf_bright}, \ref{transintens}& &\footnotesize  surface brightness at emitter, observer\\
$L$, $L_X$ & &\ref{sect_flux} &\footnotesize  rad. power at spec. wavelength, in band $X$ \\
$M$ & \ref{defM}, \ref{Schwarz}, \ref{defmr} & &\footnotesize  total eff. gravit. mass with dim. of length\\
$M_e$ & \ref{rho_L} & &\footnotesize  total mass in an exponential envelope \\
$M_\bullet$  & \ref{def_M_BH}  &   &\footnotesize    $G{\cal M}_ \bullet/c^2$ \\
$\tilde M$ & \ref{def_tilde_M} & &\footnotesize  $M$ in units of $10^6M_\odot$\\
$M_H$ & \ref{rho_L} & &\footnotesize  total mass in a Hernquist bulge \\
$\tilde M_\bullet$ & \ref{def_tilde_M} & &\footnotesize  geom. mass of BH hole in units $10^6M_\odot$ \\
$M_{X_{\rm e}}^{({\rm mag})}(t_{\rm e},\emph{\textbf{r}}_{\rm e})$ & \ref{transmagn} & &\footnotesize  absolute magnitude of emitter in band $X_{\rm e}$\\
$M_{\rm tot}(R)$ & & \ref{sect_intro_neutrinohalo} &\footnotesize  total mass of a (spiral) galaxy  \\
${\cal M}(r)$ & \ref{defcalM} &  &\footnotesize  cumulative mass inside shell $r$ \\
${\cal M}(R)$ & \ref{defcalMR} &  &\footnotesize  cumulative mass inside radius $R$ \\
$\Delta{\cal M}_i$ & & \ref{sect_collapse} &\footnotesize  mass of shell $i$ \\
${\cal M}_{\rm tot}$ & \ref{def_calMtot}, \ref{tot_mass_sync} &  &\footnotesize  total mass of the universe \\
${\cal M}_\bullet$ & \ref{def_M_BH} & &\footnotesize  mass of black hole\\
$O$ & & \ref{sect_nat_curv} &\footnotesize  position of the observer \\
$P$ & \ref{defP}, \ref{novRt} & &\footnotesize  max. shell param., Novikov shell label \\
$\tilde{\cal P}_R(R)$, ${\cal P}_R(R)$ &\ref{genrelBH2a}, \ref{expr_PR_gen} &&\footnotesize  (dimensionless) radial pressure \\
$\tilde{\cal P}_T$, ${\cal P}_T(R)$ & &\ref{sect_field}&\footnotesize  (dimensionless) tang. pressure \\
${\cal P}_\varphi(R)$, ${\cal P}_T(R)$ & &\ref{sect_field}&\footnotesize  tangential pressures \\
${\cal P}_{(\nu)}$ & \ref{def_pressure} & &\footnotesize  (neutrino) pressure\\
\hline\hline
\end{tabular}
\end{table}

\begin{table}[h]
\caption{Definition of upper case symbols (continued)}             
\centering                          
\begin{tabular}{l l l l}        
\hline\hline                 
Symbol & Equation & Section &  Name \\    
\hline                        
$R(r,t)$ & \ref{Rexplgen} & \ref{sect_defgen} &\footnotesize  radius\\
${}^4R(r,t)$, ${}^3R(r,t)$ & \ref{4Riemgen}, \ref{3Riemgen} &  &\footnotesize  4D, 3D Riemann curvature scalar\\
$R_{\rm max}(r)$ & \ref{defpmax} &  &\footnotesize  maximum radius of shell $r$ \\
$R_i(t)$ & & \ref{sect_collapse} &\footnotesize  radius of thin shell $i$\\
$R_{\rm max}(t)$ & \ref{defrmaxt} &  &\footnotesize  max. radius of a dust ball / universe @ $t$ \\
$R_{\rm max}$ & \ref{defrmax} &  &\footnotesize  maximum radius of the universe \\
$R_P(t)$ &  & \ref{sect_Novrad} &\footnotesize  Schw-$\Lambda$ radius of a Novikov shell\\
$R_e$, $R_H$ & \ref{rho_L} & &\footnotesize  scale length of exp., Hernquist bulge \\
$R_S$ & \ref{defschwarzRS} &  &\footnotesize  Schwarzschild radius\\
$R_2$ &  & \ref{sect_beyond}, \ref{sect_Novrad} &\footnotesize  black sky radius\\
$R_{\{\lambda,\mu,\nu\}}(t)$ & \ref{drRasymp}& &\footnotesize  coefficients of behaviour of $\partial_rR$ @ $r_b$\\
$\tilde R$, $\tilde R_0$ & \ref{V_eff_schwarz} & \ref{sect_neut_BH}&\footnotesize  norm. radius around a black hole, specific\\
$\tilde R_b$ & \ref{Xi_newton1} & &\footnotesize  outer boundary of mass distr. around BH\\
$\tilde R_{\rm in}$ & \ref{def_tilde_R_in}, \ref{expr_tilde_R_in} & &\footnotesize  inner boundary of mass distr. around BH\\
$\tilde R_{\rm lim}$ &  & \ref{sect_primBH}, \ref{sect_neut_BH}&\footnotesize  $\tilde R$ of circ. orbit with smallest $\tilde E_\infty$ and $\tilde h$\\
$\tilde R_1(\tilde h)$, $\tilde R_2(\tilde h)$ & \ref{schwarz_extrema} &  &\footnotesize   extrema in the eff. potential around BH\\
${\cal R}$  & \ref{radius_sync} & &\footnotesize  radius of a synchronous universe\\
${\cal R}_{\rm e}$, ${\cal R}_{\rm o}$  &  & \ref{sect_photometry_sync}&\footnotesize  $\cal R$ @ epoch of emission, observation\\
$dS_{\rm e}$, $dS_{\rm o}$ &  & \ref{sect_surface brightness} &\footnotesize  elementary surface at emitter, observer\\
$S^T\!(r,t)$, $S^T\!(R)$, $S^T_i(t)$ & \ref{def_ST}, \ref{Lorentz_dx} &\footnotesize  \ref{sect_collapse} &\footnotesize  tangential stretch rate, inside shell $i$\\
$T, \tilde T$ & \ref{defaprime}&  &\footnotesize  absolute temperature, in units of 2 K\\
$T_s$ & &   \ref{sect_application} &\footnotesize  2 Kelvin\\
$T_\mu^\nu$ &\ref{Einstein1}, \ref{genrelBH2a} &   &\footnotesize    energy-momentum tensor \\
$dV$ & \ref{defvolelem} & &\footnotesize  elementary volume element \\
$V_{\tilde h}(R)$ & \ref{V_eff_schwarz}, \ref{V_eff}& &\footnotesize  effective potential of an orbit with given $\tilde h$\\
$V_0(R)$ & \ref{deffR}, \ref{novdRdt}& &\footnotesize  effective potential of a radial orbit $\tilde h=0$\\
$V_{\rm max}(\tilde h)$ & \ref{def_Vlim}& &\footnotesize  max. $V_{\tilde h}(R)$ closest to the black hole\\
$V_c$ & & \ref{sect_intro_neutrinohalo} &\footnotesize  circular velocity of a flat rotation curve \\
$V_{{\cal X},N},V_{{\cal Y},N}$, $V_N$ & \ref{expr_VXN_VYN}, \ref{Lorentz_transfo}, \ref{def_VS_VN} & &\footnotesize  Lorentzian velocities in Novikov frame \\
$V_{{\cal X},S},V_{{\cal Y},S}$, $V_S$ & \ref{expr_VXS_VYS0}, \ref{expr_VXS_VYS}, \ref{def_VS_VN} & &\footnotesize  Lorentzian velocities in Schw-$\Lambda$ frame \\
$X(r,t)$ & \ref{defX}&  &\footnotesize  radial scale factor \\
$X_{\{\lambda,\mu,\nu\}}(t)$ & \ref{Xasympcoef} & &\footnotesize  coefficients of behaviour of $X$ at $r_b$ \\
$X_{\rm lim}(t)$ & \ref{defXlim} &  &\footnotesize  coef. of the singularity of $X$ at $r_b$ \\
$\cal X$ &   & \ref{sect_defgen} &\footnotesize  cartesian coordinate\\
$\cal Y$ &   & \ref{sect_defgen} &\footnotesize  cartesian coordinate\\
${\cal Y}_t(R)$ & \ref{defembedmet}, \ref{defembed} & &\footnotesize  embed. surf. isom. repres. 2D universe\\
$\cal Z$ &   & \ref{sect_defgen} &\footnotesize  cartesian coordinate\\
\hline\hline
\end{tabular}
\end{table}

\begin{table}[h]
\caption{Definition of lower case symbols }             
\centering                          
\begin{tabular}{l l l l}        
\hline\hline                 
Symbol & Equation & Section &  Name \\    
\hline                        
$a$, $\tilde a$ & \ref{defaprime} & &\footnotesize  parameter in isoth. distr., in units of $a_s$\\
$a_c$ & \ref{def_ac} & &\footnotesize  $ac^2$\\
$a_s$ & \ref{def_as} & &\footnotesize  $a$ for 1 eV and 2 K\\
$a_i$ & \ref{shell_evol}  &  &\footnotesize   $a(r)$ for shell $i$\\
$a_P$ & \ref{novRt} & &\footnotesize  $a$ in function cyc for a Novikov shell\\
$a_S$  & \ref{def_a_S}  &  &\footnotesize   $a_P$ of Novikov shell that starts at $R_S$\\
$a(r)$ & \ref{defa} & &\footnotesize  $\Lambda/[3\omega^2(r)]$\\
$b$ & \ref{defb} & &\footnotesize  $\omega^2(r)/(H_0^2-\Lambda/3)$\\
c, $c$ &&&\footnotesize  velocity of light\\
$\cyc(r,t)$ & \ref{Rexplgen} & &\footnotesize solution of the evolution equation\\
$\cyc(a,\epsilon,\bar\psi)$ & \ref{defcyc} & &\footnotesize  cyc as function of parameters and state\\
${\rm cc}_0(b)$ & \ref{solcycH0a}, \ref{solcycH0b}, \ref{solcc01}, \ref{solcc02} & &\footnotesize  cyc for $H(r,t)=H_0$\\
$e(r)$ & \ref{defX}, \ref{defer} &  &\footnotesize  energy function\\
$e_c$ & \ref{ecenterc}, \ref{defer} &  &\footnotesize   central scaling of $e(r)$\\
$e_i$ &  \ref{epacket} &   &\footnotesize    binding energy of shell $i$ \\
$e_{\rm sync}(\tilde r)$ & \ref{e_iso} & &\footnotesize  $e(\tilde r)$ for synchronous universe\\
$g_\nu$ & \ref{F-D} &&\footnotesize  degr. of freedom in Fermi-Dirac distr.\\
$g_{\lambda\mu}$  &  & \ref{sect_transfer} &\footnotesize  metric tensor \\
h & \ref{F-D} &&\footnotesize  Planck's constant \\
$h$ & \ref{geod1}, \ref{def_h} & &\footnotesize  specific relativistic angular momentum\\
$\tilde h$ & \ref{def_tildeR_tildeh}, \ref{V_eff} & &\footnotesize  normalized $h$\\
$\tilde h_{\rm in}$ &  \ref{h_Xi_in} & &\footnotesize  min. $\tilde h$ of marg. bnd non-plung. orbit\\
$\tilde h_{\rm lim}$ &  & \ref{sect_primBH}, \ref{sect_neut_BH}&\footnotesize  smallest $\tilde h$ for non-plunging orbits\\
$\tilde h_{\rm min}$ &  \ref{ERmin1} &&\footnotesize  $\tilde h$ of a marginally non-plunging orbit \\
$\tilde h(\tilde E_\infty)$ & \ref{pTlim1} & &\footnotesize  min. $\tilde h$ of non-plung. orbit given $\tilde E_\infty$\\
$\ell$ &  & \ref{sect_redshift} &\footnotesize  geometrical distance \\
$dl$, $\Delta l$ &  \ref{deltal} &  &\footnotesize  elem. dist. perp. to a light ray\\
$dl_{\rm o}$, $dl_{\rm e}$ &  & \ref{sect_angdistance_defs} &\footnotesize  $dl$ at observer, emitter\\
$m(r)$ & \ref{defrho}, \ref{relmM}, \ref{eqRt}, \ref{defmr}&  &\footnotesize  eff. grav. mass function\\
$m(R)$ & \ref{genrelBH_defm}&  &\footnotesize  eff. gravit. mass function around BH\\
$\tilde m(r)$ & \ref{def_tildem}, \ref{defmr} &  &\footnotesize   normalized $m(r)$\\
$\tilde m(R)$ & \ref{genrelBH_defm}, \ref{genrelBH1} &  &\footnotesize   normalized $m(R)$\\
$m(z)$ &&\ref{sect_obsflux}, \ref{sect_photometry_sync}, \ref{sect_flux}&\footnotesize  magnitude-redshift relation\\
$m_0$ & \ref{u_Schwarz} &\ref{sect_orbits_eq_motion}, \ref{app_isothermal}&\footnotesize  rest mass\\
$m_c$ & \ref{mcenter1}, \ref{defmr} &  &\footnotesize   central scaling of $m(r)$\\
$m_{c,{\rm sync}}$ & \ref{eff_mass_sync} &  &\footnotesize  $m_c$ for synchronous model $m(r)$\\
$m_i$  & \ref{mpacket}  &   &\footnotesize   gravitating mass @ shell i\\
$\Delta m_i$  & \ref{calMpacket}   &  &\footnotesize   $m_i-m_{i-1}$ \\
$\tilde m_\nu$ & \ref{defaprime}&  &\footnotesize  neutrino mass in units of 1 eV\\
$m_{\nu,s}$ & &  \ref{sect_application} &\footnotesize  mass equivalent of 1 eV\\
$m_{X_{\rm o}}^{({\rm mag})}(t_{\rm o},\emph{\textbf{r}}_{\rm o})$ & \ref{transmagn} & &\footnotesize  observed magnitude in band $X_{\rm o}$\\
$p$ & &\ref{sect_redshift}, \ref{sect_transfer}, \ref{app_isothermal}  &\footnotesize  param. of light path, momentum\\
$p(r)$ & \ref{defp} & &\footnotesize  shell parameter of shell with label $r$\\
$\tilde p(r)$ & \ref{def_tildep} & &\footnotesize  normalized shell parameter\\
\hline\hline
\end{tabular}
\end{table}

\begin{table}[h]
\caption{Definition of lower case symbols (continued)}             
\centering                          
\begin{tabular}{l l l l}        
\hline\hline                 
Symbol & Equation & Section &  Name \\    
\hline                        
$p_c$ & \ref{pcenter} & &\footnotesize  scaling for $p(r)$ \\
$p_{c,{\rm sync}}$ & \ref{p_c_iso} & &\footnotesize  $p_c$ for synchronous universe\\
$p_i$  & \ref{shell_evol}  &   &\footnotesize  $p(r)$ for shell $i$ \\
$\tilde p_T$, $p_T$ &\ref{def_tildepT} & \ref{app_isothermal} &\footnotesize  (norm.) tang. comp. of linear momentum \\
$\tilde p_R$, $p_R$ &\ref{def_tilde_ER1} & \ref{app_isothermal}&\footnotesize  (norm.) rad. comp. of linear momentum \\
$\tilde p_{T,\rm min}(\tilde R,\tilde E_R)$ & \ref{neut_response1}, \ref{pTlim1} & &\footnotesize  min. $\tilde p_T$ for a non-plunging orbit around a BH \\ 
$\tilde p_T(\tilde R)$ &  \ref{ERmin2} &  &\footnotesize   tang. comp. of $\tilde p$ of unstable pericenter orbit  \\  
$q(r,t)$ & \ref{defq} & &\footnotesize  classical deceleration parameter \\
$r$ & & \ref{sect_defgen} &\footnotesize  shell label \\
$\tilde r$ & \ref{def_rnorm}&  &\footnotesize  normalized shell label \\
$r_0$ & &  &\footnotesize  a specific shell label\\
$r_e$  & \ref{def_re}  &  &\footnotesize  label of shell inside which $e(r)=0$ \\
$r_{e_0}$  & \ref{def_re0}  &  &\footnotesize  label of isolated shell on which $e(r_{e_0})=0$ \\
$\emph{\textbf{r}}_{\rm o}$, $\emph{\textbf{r}}_{\rm e}$ &&\ref{sect_surface brightness}&\footnotesize  position vector of observer, emitter\\
$r_{ma}$ & \ref{defmr} & &\footnotesize  $r$ inside which $m(r)$ has a synch. behaviour \\
$r_{b}$ & \ref{defrb} &  &\footnotesize  $r$ of comoving matter outer boundary \\
$r_{ea}$& \ref{defer}& &\footnotesize  $r$ inside which $e(r)$ has a synch. behaviour \\
$r_{\rm sync}$ & \ref{defrsync} & &\footnotesize  $r$ inside which the universe is synchronous\\
$r_{\phi a}$, $r_{\phi b}$   & \ref{defphir} & &\footnotesize  $r$ at the interior/exterior of which $\omega(r)\phi(r)$ is const. \\
$s$ & &\ref{sect_defgen}, \ref{sect_orbits} &\footnotesize  arc length \\
$t$, t& \ref{deft} & &\footnotesize  (c $\times$) cosmic time \\
$t_c$ &  & \ref{sect_infall}  &\footnotesize   c $\times$ cosmic time in relation with the Schw-$\Lambda$ metric\\
$t_0$ & &  &\footnotesize  a specific cosmic time\\
$t_0(r)$ & \ref{deftbrb} & &\footnotesize  $t$ at which a given $H_0$ is realized @ shell $r$\\
$t_b(r)$ & \ref{deftb} & &\footnotesize  cosmic time of max. expansion of shell $r$, if bound\\
$t_{\rm e}$, $t_{\rm o}$ & & \ref{sect_redshift} &\footnotesize  cosmic time of emission, observation\\
$t_M$ & \ref{deftM} & &\footnotesize  validity time limit of the metric \\
$t_{\rm lock}$ & & \ref{sect_intro_neutrinohalo} &\footnotesize  cosmic time oat which neutrino's bind to a galaxy\\
$\bar t$, $\bar{\rm t}$ & \ref{Schwarz}, \ref{def_bar_t}&  &\footnotesize  (c $\times$) Schwarzschild-$\Lambda$ coordinate time\\
$\bar t_P$ &  & \ref{sect_Novi_metric}&\footnotesize  Schwarzschild-$\Lambda$ time $\bar t$ of a Novikov shell\\
$\bar {\rm t}_R$ &  & \ref{sect_neut_BH}&\footnotesize  time of stationary observer at $R$ in a Schw-$\Lambda$ metric\\
$u$ &  \ref{defx} & &\footnotesize  $r_b-r$\\
$\upsilon$  &   &  \ref{app_isothermal}  &\footnotesize   velocity measured by stationary observer\\
$\upsilon_N$ &  \ref{novschwarzupsilon0}, \ref{novschwarzupsilon}& &\footnotesize  rad. vel. diff. between Schw-$\Lambda$ and Novikov frame \\
$\upsilon_{iS}$ & \ref{vradinschwarz} & &\footnotesize  Lorentzian rad. vel. in the inner Schw-$\Lambda$ metric\\ 
$\upsilon_s$ & & \ref{sect_application}&\footnotesize  unit of velocity\\
$\upsilon_T$ & \ref{def_upsilonT} &&\footnotesize  tang. component of $\upsilon$ measured by stat. observer\\
$\boldsymbol{\upsilon}$ & & \ref{sect_intro_neutrinohalo} &\footnotesize  velocity vector with respect to comoving observer\\
$w_\xi$ & \ref{def_wxi} &  &\footnotesize  coef. in the regularized rad. equ. of an orbit\\
$w$, $w'$ & \ref{defwvel}, \ref{def_wprime}, \ref{def_wprimexi}& &\footnotesize  Lorentzian radial velocities \\
$w_0$ &  \ref{defpmax} & \ref{sect_cyc} &\footnotesize  smallest real root for cyc of $1-\epsilon\, \cyc +a\, \cyc^3=0$\\
$w_1$, $w_2$ &  & \ref{sect_cyc} &\footnotesize  the 2 roots of $1-\epsilon\, \cyc +a\, \cyc^3=0$ that are not $w_0$\\
$z$, $z(p)$, $z(\ell)$ & & \ref{sect_redshift} &\footnotesize  redshift, as function of orbit parameter, Hubble law\\
$z(t)$ & \ref{cosmozt} & &\footnotesize  standard cosmological redshift \\

\hline\hline
\end{tabular}
\end{table}

\begin{table}[h]
\caption{Definition of Greek symbols}             
\centering                          
\begin{tabular}{l l l l}        
\hline\hline                 
Symbol & Equation & Section &  Name \\    
\hline                        
$\Lambda$ & & & \footnotesize cosmological constant\\
$\Xi(R)$ & \ref{genrel_def_Xi} &&\footnotesize  $R\,\Psi(R)$ \\
$\tilde\Xi(\tilde R)$ &\ref{genrel_def_Xi} &&\footnotesize normalised $\Xi(R)$\\
$\tilde\Xi_{\rm in}$ &\ref{expr_tilde_R_in} &&\footnotesize   $\tilde\Xi(\tilde R_{\rm in})$\\
$\Phi$  & \ref{defphir} & &\footnotesize $\phi(r_{\phi b})$\\
$\Psi(R)$  & \ref{def_Psi}, \ref{def_gam00} & &\footnotesize gen. relat. analog of Newt. binding potential \\
$\tilde\Psi(R)$  & \ref{def_Psitilde}, \ref{def_tildePsi} & &\footnotesize normalized $\Psi(R)$ \\
$d\Omega$  & \ref{def_Omega} & &\footnotesize elementary solid angle\\
$\Omega_k$, $\Omega_M$, $\Omega_\Lambda$ & \ref{defOmegas} &  &\footnotesize standard cosmological $\Omega$'s\\
$d\Omega_{\rm e}$, $d\Omega_{\rm o}$ &  & \ref{sect_surface brightness} &\footnotesize elementary solid angle at emitter, observer\\
$\alpha$ & \ref{deferlim} &&\footnotesize shorthand for $\alpha_0$\\
$\alpha_i,\,i\ge0$ & \ref{defer} &&\footnotesize coefficients of the expansion of $e(x)$ at $r_b$\\
$\beta$ & \ref{defmrlim} &&\footnotesize shorthand for $\beta_0$\\
$\beta_i,\,i\ge0$ & \ref{defmr} &&\footnotesize coefficients of the expansion of $m(x)$ at $r_b$\\
$\langle\beta\rangle(\tilde R)$ & \ref{def_betamean}  &&\footnotesize mean $\upsilon/c$ of a mass distribution around a black hole \\
$\gamma$ & \ref{defgamma} &&\footnotesize shorthand for $\gamma_0$\\
$\gamma_i,\,i\ge0$ & \ref{defphir} &&\footnotesize coefficients of the expansion of $\omega(x)\phi(x)$ at $r_b$\\
$\gamma_{00}$ & \ref{def_gam00} &&\footnotesize $\sqrt{g_{00}}$\\
$\hat\delta$ & \ref{defdelta} &&\footnotesize $\delta^{-1}$\\
$\delta$ & \ref{defdeltahat} &&\footnotesize exponent in the asymptotic behaviour of $X$ at $r_b$\\
$\epsilon$ & \ref{defepsilon1} & &\footnotesize  sign of $e(r)$ in representation (a)\\
$\epsilon_N$ & \ref{novconst}, \ref{def_epsilonN_metric} &&\footnotesize $\epsilon$ for Novikov metric, sign of $\Lambda$ \\
$\epsilon(r)$ & \ref{defepsilon2} & &\footnotesize shorthand for $2e(r)$ in representation (b)\\
$\epsilon_\xi$ & \ref{defepsxi} &  &\footnotesize sign constant indicating the sheet of the universe\\
$\varepsilon$  & \ref{defu}  &  &\footnotesize  1 for a massive particle and 0 for a photon\\
$\vartheta$ & & \ref{sect_defgen} &\footnotesize polar coordinate\\
$\theta$ & \ref{def_lightangle} & &\footnotesize angle of velocity with respect to the radial direction\\
$\theta_S$, $\theta_N$ & \ref{def_thetaS_thetaN} & &\footnotesize $\theta$ in Schwarzschild-$\Lambda$ frame, in Novikov frame \\
$d\theta_{\rm o}$, $d\theta_{\rm e}$ &  & \ref{sect_angdistance_defs} &\footnotesize elementary angle at observer, emitter\\
$\kappa$ & \ref{defrho}, \ref{kapanum} &  &\footnotesize $4\pi G/c^2$ \\
$\lambda$ & \ref{defer} &&\footnotesize exponent in the asymptotic behaviour of $e(r)$ at $r_b$\\
$\lambda_S$, $\lambda_N$ & \ref{lambda_transfo} & &\footnotesize wavelength in Schwarzschild-$\Lambda$ frame, in Novikov frame \\
$\lambda_{\rm o}$, $\lambda_{\rm e}$&  &\ref{sect_surface brightness}&\footnotesize wavelength at observer, at emitter \\
$\lambda(R)$ &\ref{metric_static} &&\footnotesize log of coef. of $dR^2$ in a static spherical diagonal metric\\
$\mu$ & \ref{defmr} &&\footnotesize exponent in the asymptotic behaviour of $m(r)$ at $r_b$\\
$\nu$ & \ref{defphir} &&\footnotesize exponent in the asymptotic behaviour of $\omega(r)\phi(r)$ at $r_b$\\
$\nu(R)$ &\ref{metric_static} &&\footnotesize log coef. of ${d\tilde t}^2$ in a static spherical diagonal metric\\
$\xi$ & \ref{defxiOP}&  &\footnotesize regularized shell label\\
\hline\hline
\end{tabular}
\end{table}

\begin{table}[h]
\caption{Definition of Greek symbols (continued)}             
\centering                          
\begin{tabular}{l l l l}        
\hline\hline                 
Symbol & Equation & Section &  Name \\    
\hline                        
$\rho(r,t)$ & \ref{defrho} & &\footnotesize  local Lorentzian mass (dust) density of the universe\\
$\rho_c$ & \ref{rhocenter} & &\footnotesize  scaling for $\rho(r,t)$\\
$\rho_{c,{\rm sync}}$ & \ref{rho_0_iso} & &\footnotesize  $\rho_c$ for the synchronous universe\\
$\rho_{\rm crit}$ & \ref{def_rhoc} & &\footnotesize  critical mass density \\
$\rho_L(R)$  & \ref{Poisson} &  &\footnotesize  mass density of a Milky Way type galaxy, excluding DM\\
$\rho_{\rm isoth}$  & \ref{def_rho_isoth} &  &\footnotesize  mass density of an isothermal sphere\\
$\rho_\nu(R)$, $\tilde\rho_\nu(\tilde R)$ & & \ref{sect_field} & \footnotesize Fermi-Dirac particle density around a BH, normalized\\
$\rho_{\nu,\infty}$ & \ref{def_rhoprime} & &\footnotesize  neutrino mass density of the universe on galactic scales \\
$\sigma_i$ & & \ref{sect_collapse} &\footnotesize  surface density of thin shell $i$ \\
$\tau$ & \ref{def_h} &  &\footnotesize  proper time\\
$\varphi$ & & \ref{sect_defgen} &\footnotesize  longitudinal coordinate\\
$\phi(r)$ & \ref{soleqRtint}, \ref{defphir} & &\footnotesize  shell phase function \\
$\phi_c$  & \ref{defphir} & &\footnotesize  $\phi(0)$\\
$\chi$ & \ref{def_chi}, \ref{def_chiregular} & &\footnotesize  shell label that regularizes $X$\\
$\psi(r,t)$ &  \ref{def_psi} & &\footnotesize  state of shell $r$ @ $t$\\
$\psi_i(t)$ & \ref{shell_evol}  &  &\footnotesize   $\psi(r,t)$ for shell $i$\\ 
$\bar\psi(r,t)$, $\bar\psi$ & \ref{soleqRtint}, \ref{defcyc} & &\footnotesize  integral that defines cyc implicitly \\
$\bar\psi_{\rm init}(r)$ &  \ref{novRt}& &\footnotesize  zero radial velocity state of a Novikov shell\\
$\bar\psi_{\rm max}(r)$ &  \ref{defpmax}, \ref{defpsimax}& &\footnotesize  state of shell $r$ @ max. expansion\\
$\omega(r)$ & \ref{defomega} & &\footnotesize  shell frequency\\
$\tilde\omega(r)$ & \ref{def_tildeom} & &\footnotesize  normalized shell frequency\\
$\omega_c$ & \ref{omisoch} & &\footnotesize  central shell expansion freq. \\
$\omega_i$  & \ref{shell_evol}  &   &\footnotesize   $\omega(r)$ for shell $i$ \\
$\omega_{{\rm sync}}$ & \ref{omega_0_M} & &\footnotesize  $\omega$ in synchronous universe \\
\hline\hline
\end{tabular}
\end{table}

\end{document}